\documentclass{aa}  

\usepackage{amsmath}
\usepackage{graphicx}
\usepackage{graphics}
\usepackage{subfigure}
\usepackage{txfonts}
\usepackage{natbib}
\usepackage{xcolor}
\usepackage{soul}
\usepackage{tablefootnote}
\usepackage{pdflscape}
\usepackage{tabularx}
\usepackage{threeparttable}
\usepackage{rotating}
\usepackage{longtable}
\usepackage{wrapfig}

\bibpunct{(}{)}{;}{a}{}{,}

\newcommand{\hi}{H~{\sc i}}
\newcommand{\hii}{H~{\sc ii}}
\newcommand{\ha}{\ifmmode {\rm H}\alpha \else H$\alpha$\fi}
\newcommand{\hb}{\ifmmode {\rm H}\beta \else H$\beta$\fi}
\newcommand{\lya}{\ifmmode {\rm Ly}\alpha \else Ly$\alpha$\fi}

\newcommand{\heii}{He~{\sc ii}}
\newcommand{\Heiiuv}{He~{\sc ii} $\lambda$1640}

\newcommand{\ebv}{\ifmmode E_{\rm B-V} \else $E_{\rm B-V}$\fi}
\newcommand{\av}{\ifmmode A_{\rm V} \else $A_{\rm V}$\fi}

\def\msun{\ifmmode M_{\odot} \else M$_{\odot}$\fi}
\def\msunyr{\ifmmode M_{\odot} {\rm yr}^{-1} \else M$_{\odot}$ yr$^{-1}$\fi}
\def\zsun{\ifmmode Z_{\odot} \else Z$_{\odot}$\fi}
\def\lsun{\ifmmode L_{\odot} \else L$_{\odot}$\fi}
\def\mup{\ifmmode M_{\rm up} \else M$_{\rm up}$\fi}
\def\mlow{\ifmmode M_{\rm low} \else M$_{\rm low}$\fi}

\def\aap{A\&A}

\def\apjl{ApJL}
\def\apjs{ApJS}

\newcommand{\oh}{\ifmmode 12 + \log({\rm O/H}) \else$12 + \log({\rm O/H})$\fi}
\newcommand{\nii}{[N~{\sc ii}]}
\newcommand{\niii}{[N~{\sc iii}]}
\newcommand{\oi}{[O~{\sc i}]}

\newcommand{\oiii}{[O~{\sc iii}]}
\newcommand{\oiv}{[O~{\sc iv}]}
\newcommand{\sii}{[S~{\sc ii}]}
\newcommand{\siii}{[S~{\sc iii}]}
\newcommand{\siv}{[S~{\sc iv}]}
\newcommand{\cii}{[C~{\sc ii}]}
\newcommand{\civ}{[C~{\sc iv}]}
\newcommand{\neiii}{[Ne~{\sc iii}]}
\newcommand{\neii}{[Ne~{\sc ii}]}
\newcommand{\nev}{[Ne~{\sc v}]}
\newcommand{\nevi}{[Ne~{\sc vi}]}

\newcommand{\feii}{[Fe~{\sc ii}]}
\newcommand{\feiii}{[Fe~{\sc iii}]}
\newcommand{\feiv}{[Fe~{\sc iv}]}
\newcommand{\fev}{[Fe~{\sc v}]}
\newcommand{\fevi}{[Fe~{\sc vi}]}
\newcommand{\fevii}{[Fe~{\sc vii}]}
\newcommand{\silii}{[Si~{\sc ii}]}
\newcommand{\arii}{[Ar~{\sc ii}]}
\newcommand{\ariii}{[Ar~{\sc iii}]}

\newcommand{\mgiv}{[Mg~{\sc iv}]}

\def\Sii{[S~{\sc ii}]\,$\lambda\lambda$6716,6731$\AA$}
\def\Siii{[S~{\sc iii}]\,$\lambda\lambda$9068,9532$\AA$} 

\def\Oii{[O~{\sc ii}]\,$\lambda\lambda$3726, 3728$\AA$}
\def\Oiii{[O~{\sc iii}]\,$\lambda\lambda$4959,5007$\AA$}
\def\oiiil{[O~{\sc iii}]$\lambda$\,5007$\AA$}
\def\oiiill{[O~{\sc iii}]$\lambda$\,4959$\AA$}

\newcommand{\Civ}{C~{\sc iv} $\lambda$5808}

\defcitealias{LebouteillerRamambason2022}{LR22}

\def\fesc{\ifmmode f_{\rm esc} \else $f_{\rm esc}$\fi}
\def\feschii{\ifmmode f_{\rm esc,HII} \else $f_{\rm esc,HII}$\fi}

\begin{document}

    \title{Inferring the HII region escape fraction of ionizing photons from infrared emission lines in metal-poor star-forming dwarf galaxies}
  \subtitle{}
  \author{L. Ramambason\inst{1}, 
  V. Lebouteiller\inst{1}, 
  A. Bik\inst{2},
  C. T. Richardson\inst{3},
  F. Galliano\inst{1}, 
  D. Schaerer\inst{4},
  C. Morisset\inst{5},
  F. L. Polles\inst{6}, 
  S.\,C. Madden\inst{1},
  M. Chevance\inst{7,8},
  I. De Looze\inst{9,10}
}
  
  \institute{Université Paris Cité, Université Paris-Saclay, CEA, CNRS, AIM, F-91191, Gif-sur-Yvette, France
  \and Department of Astronomy, Oskar Klein Centre, Stockholm University, AlbaNova University Centre, 106 91 Stockholm, Sweden
  \and Physics Department, Elon University, 100 Campus Drive CB 2625, Elon, NC, 27244, USA
  \and Observatoire de Gen\`eve, Universit\'e de Gen\`eve, 51 Ch. des Maillettes, 1290 Versoix, Switzerland
  \and Instituto de Astronomía, Universidad Nacional Autónoma de México, AP 106, 22800 Ensenada, B. C., Mexico
  \and  SOFIA Science Center, USRA, NASA Ames Research Center, M.S. N232-12, Moffett Field, CA, 94035, USA
  \and Astronomisches Rechen-Institut, Zentrum für Astronomie der Universität Heidelberg, Mönchhofstraße 12-14, D-69120 Heidelberg, Germany
  \and Institut fur Theoretische Astrophysik, Zentrum für Astronomie, Universität Heidelberg, D-69120 Heidelberg, Germany
  \and Sterrenkundig Observatorium, Ghent University, Krijgslaan 281 – S9, B-9000 Ghent, Belgium
  \and Department of Physics \& Astronomy, University College London, Gower Street, London WC1E 6BT, UK
}

\date{Received 26 April 2022 / Accepted 11 July 2022}

\abstract{
Local metal-poor galaxies stand as ideal laboratories for probing the properties of the interstellar medium (ISM) in chemically unevolved conditions. Detailed studies of this primitive ISM can help gain insights into the physics of the first primordial galaxies that may be responsible for the reionization. Quantifying the ISM porosity to ionizing photons in nearby galaxies may improve our understanding of the mechanisms leading to Lyman continuum photon leakage from galaxies. The wealth of infrared (IR) tracers available in local galaxies and arising from different ISM phases allows us to constrain complex models in order to estimate physical quantities.}
{
Primitive galaxies with low metal and dust content have been shown to host a patchier and more porous ISM than their high-metallicity counterparts, with numerous density-bounded regions from where ionizing photons might leak out. To what extent this peculiar structure contributes to the leakage of ionizing photons remains to be quantitatively studied. Such effects can only be investigated by accounting for the complexity and inhomogeneity of the ISM. We aim to provide a new statistical framework to quantify various galactic observables by constraining a representative multiphase and multisector topology using a combination of 1D models. 
}
{
To address these questions, we built a refined grid of models that include density-bounded regions and a possible contribution of an X-ray source. Using MULTIGRIS, a new Bayesian code based on Monte Carlo sampling, we combined the models as sectors under various assumptions to extract the probability density functions of the parameters and infer the corresponding escape fractions from \hii\ regions  (\feschii). We applied this new code to a sample of 39 well-know local starbursting dwarf galaxies from the Dwarf Galaxy Survey \citep{Madden_2013}.
}
{
 We confirm previous results that hinted at an increased porosity to ionizing photons of the ISM in low-metallicity galaxies and provide, for the first time, quantitative predictions for \feschii. The predicted \feschii\ for low-metallicity objects span a large range of values, up to $\sim$60\%, while the values derived for more metal-rich galaxies are globally lower. We also examine the influence of other parameters on the escape fractions, and find that the specific star-formation rate correlates best with \feschii. Finally, we provide observational line ratios that could be used as tracers of the photons escaping from density-bounded regions. Among others, we discuss the possible caveats of diagnostics based on \cii 158$\mu$m in low-metallicity environments as we find a strong metallicity dependence of the fraction of \cii 158$\mu$m emitted in the different ISM phases.
}
{
The new framework presented in this paper allows us to use suites of unresolved IR emission lines to constrain various galactic parameters, including the escape fraction of ionizing photons from \hii\ regions. Although this multisector modeling remains too simple to fully capture the ISM complexity, it can be used to preselect galaxy samples with potential leakage of ionizing photons based on current and forthcoming spectral data in unresolved surveys of local and high-redshift galaxies. 
}

 \keywords{Galaxies: starburst -- Galaxies: dwarf -- ISM: structure -- radiative transfer -- infrared: ISM -- methods: numerical}

\authorrunning{L. Ramambason et al.}
\titlerunning{Escape fractions in the DGS}
\maketitle

\section{Introduction}
\label{section_intro}

Young stars that have just been formed irradiate the surrounding gas and ionize the interstellar medium (ISM). Those pockets of ionized gas (i.e., \hii\ regions) often dominate the emission at galactic scales in star-forming  galaxies. At equilibrium, \hii\ regions are surrounded by atomic neutral hydrogen, the photodissociation regions (PDRs), which may recombine further from the stars to form H$_2$ in dense molecular clouds. In some cases, part of the ionizing radiation (Lyman continuum below 912\AA; LyC) can leak out of \hii\ regions and irradiate a diffuse ionized gas (DIG) reservoir \citep{Zurita_2002, Weilbacher_2018, Herenz_2017, Bik_2018, Menacho_2019, Menacho_2021}. Depending on the exact morphology of the gas distribution, the ionizing photons can freely travel on large scales to escape in the surrounding circum-galactic medium (CGM) and potentially in the inter-galactic medium (IGM).

The contribution of LyC-leaking galaxies to the total ionizing budget in the epoch of reionization (EoR, z $\sim$ 6--9) is a key element in our understanding of the reionization process. Recent simulations indicate that such populations of numerous, low-mass, LyC-leaking galaxies with average escape fractions (\fesc(LyC)) of 10--20$\%$ would be sufficient to reionize the whole universe without invoking any other contribution from ionizing sources such as active galactic nuclei  \citep[AGN;][]{Robertson_2013, Robertson_2015}. Under favorable assumptions on ionizing photons production and accounting for a subdominant contribution from AGN, \cite{Finkelstein_2019} find that even an average escape fraction below 5\% throughout the bulk of the EoR would be enough to match observational constraints. \cite{Naidu_2020} propose an alternative model where reionization is not driven by the lower mass galaxies but by a few (<5\%) highly star-forming galaxies with stellar masses above 10$^8$\,M$_{\odot}$ and extreme escape fractions (the "oligarchs"). 

Regardless of which galaxy population drives the reionization process, the inclusion of binary stars, which provide ionizing photons at later stellar evolution stages than single stars, might also play a crucial role in providing energetic photons over large timescales. Simulations of \cite{Ma_2016}  and \cite{rosdahl_sphinx_2018} find that this leads to significantly higher time-averaged escape fractions of ionizing photons. In low-metallicity environments hosting very massive stars, stellar-mass black holes might also contribute significantly to the ionizing photon production \citep[e.g.,][]{Mirabel_2011}.
While escaping LyC photons have been directly observed in the UV domain at redshifts below 0.5 \citep[e.g.,][]{Bergvall_2006, Leitet_2013, Borthakur_2014, leitherer_direct_2016, Izotov_2016a,Izotov_2016b, Izotov_2018a,Izotov_2018b,wang_new_2019, Izotov_2021, Wang_2021, Flury_2022, Flury_2022b}, and at z$\sim$2--3 \citep[e.g.,][]{Vanzella_2015, Vanzella_2016,Vanzella_2018,Vanzella_2020,Shapley_2016, DeBarros_2016,Steidel_2018,Bian_2017,Fletcher_2019,Rivera-Thorsen_2019, Pahl_2021, Pahl_2022} with observed \fesc(LyC) ranging from 2 to 72$\%$, direct detections of leaking LyC radiation is not possible -- or extremely unlikely -- above $z\sim$ 4 due to the absorption by neutral hydrogen in the IGM, preventing any direct observation of potentially LyC-leaking galaxies directly within the EoR. This observational barrier makes it difficult to perform quantitative studies of primordial LyC-leaking galaxies, which are much needed to understand their role in the reionization process. Moreover, since LyC detections only probe a single line of sight, the measured values are sensitive to viewing angle dependences.

To overcome these constraints, several indirect methods to trace the escape fraction have been explored. They rely on the fact that photons escaping from \hii\ regions are quite sensitive to the structure and properties of the surrounding gas. This view is supported by hydrodynamical simulations of the ISM, which account for an inhomogeneous gas distribution produced by turbulence and/or stellar feedback \citep[e.g.,][]{Fujita_2003,Trebitsch_2017, Kimm_2017, Kimm_2019, Kim_2018, kakiichi_lyman_2019, yoo_origin_2020}. First, far-UV absorption lines can serve as a promising proxy to infer the covering fraction of \hi\ gas, which places an upper limit on the amount of escaping photons \citep[e.g.,][]{Reddy_2016, gazagnes_neutral_2018, gazagnes_origin_2020, chisholm_accurately_2018, Saldana-Lopez_2022}. Another approach consisting in looking at far-UV colors selection diagrams has been proposed in, for example \cite{Vanzella_2015} and \cite{Naidu_2017}. Finally, emission lines arising from different phases of the ISM can also be very useful probes of the global porosity to ionizing photons. In the UV domain, the Lyman alpha (Ly$\alpha$) line profile has proven to be an interesting proxy with the presence of double-peaked or triple-peaked profiles being associated with LyC and Ly$\alpha$ leakage \citep[e.g.,][]{verhamme_using_2015, Henry_2015, Verhamme_2017, izotov_diverse_2020,Maji_2022}. In particular, the velocity separation of the blue peak and red peak has been shown to strongly anti-correlate with the measured escape fraction of Ly$\alpha$ photons.
Recent studies have also highlighted the interest of using proxies such as weak helium lines \citep{Izotov_2017} or the Mg~{\sc ii}\,$\lambda\lambda$2796,2803\AA\ doublet \citep[e.g.,][]{Henry_2018_mgii, Chisholm_2020_mgii,Xu_Xinfeng_2022, Katz_2022} for probing the low-density lines of sight necessary to allow LyC-photons to escape.

In the optical range, several proxies can serve as indicators of LyC leakage. In particular, line ratios involving ions with different ionization potentials, produced at different depths, have been proposed as indicators to discriminate between radiation-bounded and density-bounded \hii\ regions. Radiation-bounded regions correspond to ionized spheres delimited by their Strömgren radii set by the equilibrium between production of photons by stars and ionization of the surrounding gas. Density-bounded regions are instead delimited by the lack of matter, which sets their outer radius before the Strömgren radius. Hence, they allow part of the LyC-photons produced by stars to escape from \hii\ regions. The oxygen line ratio \oiiil/\Oii\ (O32) proposed by \cite{Jaskot_2013} and \cite{Nakajima_2014} was successfully used to select LyC-leaking candidates but no strong correlation was found with the measured values of escape fraction \citep[see][and discussions therein]{Izotov_2018b,Naidu_2018,Bassett_2019,Nakajima_2020}. 
Based on a similar idea, the lack of emission from  ions with low ionization potentials like \Sii\ has also been proposed to target leaking candidates \citep{wang_new_2019, katz_new_2020, Wang_2021}. 
This lack of emission of some low ionization species was first interpreted as the signature of a density-bounded galaxy where the outer part of \hii\ regions were completely stripped out. However, using simple photoionization models, \cite{stasinska_excitation_2015} have shown that on average galaxies with high O32 cannot have massive escapes of ionizing photons, since low ionization lines like \oi6300\AA\ are often also detected in these galaxies, implying the presence of radiation-bounded regions.
Subsequently \cite{plat_constraints_2019} and \cite{ramambason_reconciling_2020} noted that several strong LyC-emitters show surprisingly strong \oi6300\AA\ emission, and proposed several explanations. While \cite{stasinska_excitation_2015} and \cite{plat_constraints_2019} suggested that such emission could be powered by the presence of AGN or radiative-shocks, we proposed in \cite{ramambason_reconciling_2020} a 2-component model combining both density- and ionization-bounded regions.  

A complementary picture has been provided by studies of galaxies in the local universe with resolved \hii\ regions. Recently, \cite{Della_Bruna_2021} estimated the average escape fraction of ionizing photons from \hii\ regions in NGC\,7793 to be 67$\%$ from MUSE observations with a $\sim 10$\,pc resolution. A large fraction of those escaping photons is, however, likely reabsorbed within galactic scales and contributes to create a DIG reservoir also seen with MUSE \citep{Della_Bruna_2020}.
This picture is in line with recent PHANGS-MUSE observations of resolved \hii\ regions in nearby spiral galaxies \citep{Belfiore_2022, Chevance_2022}.
Other indirect methods applied to local objects based on the mapping of the ionization parameter \citep[e.g.,][]{zastrow_ionization_2011, zastrow_new_2013}, on the estimation of the intrinsic ionizing photon production rate from resolved stars \citep{choi_mapping_2020}, and on the ionized gas kinematics \citep{Eggen_2021} are also suggestive of large escape fractions of ionizing photons from \hii\ regions.
Aditionally, results from \cite{polles_modeling_2019} on the local starbursting galaxy IC10 indicate that the derived porosity depends on the spatial scale, with most clouds being matter-bounded at small scales ($\sim$60 to 200pc), while larger regions become more and more radiation-bounded at galactic scales. This result highlights the complexity of the ISM in which the energy produced by feedback is deposited at various spatial scales and over different dynamical timescales, hence producing a highly inhomogeneous internal structure.

While the indirect methods relying on spatially resolved information can be applied in the local universe, they are not applicable in more distant unresolved galaxies. Disentangling the contribution from the ionized and neutral gas phase is a crucial step to better understand the interplay between ionizing photons and the surrounding ISM.
To do so, the infrared (IR) domain offers interesting tracers, not only arising from \hii\ regions but also from the PDR and molecular phases. Unfortunately, this part of the spectra is often inaccessible in samples of known LyC-leaking galaxies and only a few objects with LyC detections were also observed in the IR. 
As an alternative, recent studies \citep[e.g.,][]{2012_Cormier, Cormier_2015, cormier_herschel_2019} have tried to constrain the covering factor of neutral gas (PDR covering factor) in local galaxies to quantify the porosity to ionizing radiation by estimating the fraction of gas residing in neutral atomic and molecular phases.
This complex approach was first introduced in \cite{pequignot_heating_nodate} that provided an unprecedentedly detailed analysis of the proto-typical, low-metallicity galaxy I\,Zw\,18 by developing refined multisector topological models representing the contribution of each phase to the total emission. Similar models were adapted and successfully applied to local objects (e.g., Haro\,11: \citealt{2012_Cormier}, I\,Zw\,18: \citealt{lebouteiller_neutral_2017}, IC\,10: \citealt{polles_modeling_2019}), to a sample of local dwarf galaxies \citep{cormier_herschel_2019} drawn from the Dwarf Galaxy Survey \citep[DGS;][]{Madden_2013} and to resolved regions in the Small and Large Magellanic Clouds. \citep[SMC/LMC;][]{Lambert-Huygues_2021}.
The results from those studies indicate that nonunity PDR covering factors of neutral gas are necessary to reproduce the emission lines of most local objects. Such findings appear at odds with the few UV observations that detect little to no LyC-leakage in the local universe \citep[e.g.,][]{Bergvall_2006, Leitet_2013, Borthakur_2014, leitherer_direct_2016}.

In this context, it becomes crucial to understand what properties of the ISM are responsible for its porosity to ionizing radiation and determine if and how well integrated emission lines of unresolved galaxies can be used to constrain their escape fractions of ionizing photons.
The question is especially interesting in the context of high-redshift studies, as more and more galaxies are detected with facilities like ALMA and NOEMA at $z\sim$4--9 \citep[e.g.,][]{Inoue_2016, Carniani_2017, Walter_2018, De_Breuck_2019, Hashimoto_2019, harikane_large_2020, falkendal_alma_2020, Bakx_2020, Meyer_2022}. Such observatories and the advance of future ones such as the James Webb Space Telescope (JWST) are opening a new window to observe galaxies close to or within the EoR.

In this paper, we present a first application of MULTIGRIS \citep[][hereafter LR22]{LebouteillerRamambason2022}, a new Bayesian code designed to constrain multisector models using spectra of unresolved galaxies. Our work builds on previous studies \citep[e.g.,][]{2012_Cormier, lebouteiller_neutral_2017, polles_modeling_2019, cormier_herschel_2019, Lambert-Huygues_2021} in which multisector models were constrained using frequentist methods. In particular, this paper is a direct continuation of \cite{cormier_herschel_2019} that used a $\chi^2$ minimization method to select the best-fitting configurations between 1- and 2-sector models and derived PDR covering factors. We revisit those results using the same sample of galaxies but adopting a new method. Using a Bayesian framework allows us to overcome some major issues of the $\chi^2$ method: difficulty to derive errorbars, sensitivity to outliers, impossibility to include complex priors etc. Most importantly, it allows us to infer probability density functions (PDF) of various parameters, including for the first time the escape fraction of ionizing photons from \hii\ regions (\feschii). 

The paper is organized as follows. In Sect. \ref{section_observations} we present our sample and the tracers used in the analysis. The grid of models and the Bayesian code are presented in Sects. \ref{section_models} and \ref{section_mgris} and our results in Sects. \ref{section_prelimary_results} and \ref{section_fesc_results}. We discuss the limits and possible improvements of this new framework in Section \ref{discussion}. Our main conclusions are summarized in Sect. \ref{section_conclusion}.

\section{Sample}
\label{section_observations}

\subsection{Overview}
\label{section_overview}

\begin{table}[b]
    \centering
    \caption{IR tracers used as constraints and corresponding ionization potentials\protect\footnotemark[1] for ionic lines.}
    \begin{tabular}{p{0.15\textwidth}p{0.25\textwidth}}
         & Tracers \\
        \hline
        \hline
        Molecular hydrogen & H$_2$ S(0), H$_2$ S(1), H$_2$ S(2),
        H$_2$ S(3) \\
        Neutral and ionized & \oi$\lambda\lambda$63,145$\mu$m\\
        gas tracers & \feii$\lambda\lambda$17,25$\mu$m (7.9eV), \\
         & \silii$\lambda$34$\mu$m (8.2eV),\\
         & \cii$\lambda$158$\mu$m (11.3eV),\\  
         & Hu$\alpha \lambda$12$\mu$m (13.6eV),\\
         & \nii$\lambda\lambda$122,205$\mu$m (14.5eV),\\
         & \arii$\lambda$7$\mu$m (15.7eV),\\
         & \feiii$\lambda$23$\mu$m (16.2eV),\\
         & \neii$\lambda$12$\mu$m (21.6eV),\\
         & \siii$\lambda\lambda$18,33$\mu$m (23.3eV),\\
         & \ariii$\lambda\lambda$9,21$\mu$m (27.6eV),\\
         & \niii$\lambda$57$\mu$m (29.6eV), \\
         & \siv$\lambda$10$\mu$m (34.7eV),\\
         & \oiii$\lambda$88$\mu$m (35.1eV), \\
         & \neiii$\lambda$15$\mu$m (40.9eV)\\
         &\oiv$\lambda$26$\mu$m (54.9eV),\\ 
         & \nev$\lambda\lambda$14,24$\mu$m (97.1eV), \\
         Total IR luminosity & $L_{\rm TIR}$ (1$\mu$m-1000$\mu$m) \\
         \hline
    \end{tabular}
    \begin{tablenotes}
    \footnotesize
    \item $^1$We report the ionization potentials corresponding to the energy thresholds required to create the ion producing a given emission line, either by de-excitation or by recombination.
    \end{tablenotes}
    \label{tracers}
\end{table}

Our sample is drawn from the \textit{Herschel} Dwarf Galaxy Survey \citep[DGS;][]{Madden_2013}, which gathers photometric and spectroscopic observations of 50 nearby ($0.5-191$\,Mpc) galaxies in the far-infrared (FIR) and submillimeter domains performed with the \textit{Herschel} Space Telescope. All of these galaxies were also observed in the mid-infrared (MIR) domain with the \textit{Spitzer} observatory and spectrocopic measurements are available for all but 5 objects. We focus on a sub-sample used in \cite{cormier_herschel_2019} that selected 38 compact galaxies with at least three spectral lines detected in the MIR and FIR domain ($\sim$5 to 120$\mu$m) among the 50 observed galaxies. The DGS sample exhibits a wide range of physical properties, which makes it an ideal laboratory to study the variation of ISM properties over a range of physical and chemical conditions. In particular, this sample is ideally suited to study how the escape fraction evolves in local, low-metallicity galaxies whose ISM may resemble primordial galaxies from the early universe. More specifically, the intense sources of radiation (see Sect. \ref{subsect_radiation_sources}) as well as the low masses, compact sizes and metal-poor gas reservoirs (see Sect. \ref{subsect_gas_dust}) of these galaxies may, to some extent, resemble the physical and chemical conditions of the primordial dwarf galaxies. We note, however, that our selection criterion favors IR-bright galaxies hosting actively star-forming regions that have formed in a previously enriched ISM. The possible analogy with primordial galaxies should hence be taken with caution.

We use the line fluxes provided in \cite{cormier_herschel_2019} which combine \textit{Herschel}/PACS (60--210$\mu$m) data with \textit{Spitzer}/IRS (5--35$\mu$m) data. The latter are available for all galaxies in our sample except three (HS\,0017+1055, UGC\,4483 and UM\,133). 
The fluxes correspond to integrated measurements for galaxies that were fully covered by the instrumental apertures, except for one galaxy. The only exception is NGC\,4214, which was observed in two pointings (central and southern star-forming regions) that are studied separately in this study. The extraction procedures and corrections applied to extended sources are detailed in \cite{Cormier_2015}. We corrected one line flux from \cite{cormier_herschel_2019} (\nev14$\mu$m for NGC\,5253) that corresponded to a false detection due to an error in the fitting process. Finally, we instead use a detection upper limit at 2-$\sigma$ of 0.112$\times$10$^{16}$W\,m$^{-2}$. We also include the total IR luminosities ($L_{\rm TIR}$) derived by modeling the dust spectral energy distributions (SEDs) in \cite{remy-ruyer_linking_2015}. All the available upper limits were used in the analysis, contrary to \cite{cormier_herschel_2019} that manually selected a suite of classical emission lines arising from \hii\ regions and PDR. Our study aims to extend the multiphase picture of the ISM that was provided in \cite{cormier_herschel_2019} by including more lines arising from different phases of the ISM and tracing different physical processes. The list of observables used as constraints is summarized in Table \ref{tracers}.

Although it requires several lines to derive the various parameter values with reasonable uncertainties, the code can run properly as long as it is provided with at least one line and one upper limit (which are necessary to set the prior distributions). In this case, the given solution correspond to a realtively wide PDF (defined in Sect. \ref{rep_pdf}) and large errorbars. While little can be said about the parameter values of such individual objects, they do not, however, bias the analysis of the global trends in our sample. We thus decided to include all objects with at least one detection and one upper limit, which adds one more galaxy to the sample of \cite{cormier_herschel_2019} (HS\,2352+2733, detected in \oiii88$\mu$m with two upper limits on \cii158$\mu$m and $L_{\rm TIR}$), hence leading to a total of 39 galaxies.

In addition to including upper limits for the first time, we can also enlarge the selection of lines since our method is more robust to outliers than the previously used $\chi^2$ method. The number of constraints available for each galaxy varies from 1 single detection (in addition to two upper limits) up to 22 detected emission lines. The exact number of detections and instrumental upper limits available for each object is provided in Fig. \ref{nlines}. The choice of the suite of lines used in the analysis has an impact on the best solution that is selected by our code and, in turn, on the parameter values that are derived. Choosing the optimal suite of lines to be used and the minimal number needed to constrain a given parameter is a complex problem and we refer to \citetalias{LebouteillerRamambason2022} for a more detailed discussion. 

In this study we focus only on IR lines although most galaxies in our sample are also detected in the optical domain. We check a posteriori that our models predict values consistent with the available H$\alpha$ measurements (see Sect. \ref{results_sfr}). Combining optical and IR lines is possible with MULTIGRIS but would require an additional treatment of the dust attenuation and of systematic uncertainties due to instruments. Increasing the number of lines used as constraints can also raise specific issues regarding redundancies and the risk to over-constrain some parameters. We postpone this study to a future work and refer to LR22 for an example combining both IR and optical lines.

\subsection{Radiation sources and feedback}
\label{subsect_radiation_sources}

The DGS galaxies are starbursting galaxies with prominent MIR and FIR emission lines that hint at the presence of a population of young UV emitting stars, which strongly irradiate the ISM \citep{Madden_2013}. Some galaxies in our sample (15/39) have also been reported to host a population of Wolf-Rayet stars and signatures associated with massive stars have been reported in \cite{Schaerer_1999}.

Additionally, evidence suggests that several galaxies in our sample may host X-ray sources with the claimed detections reported in the following papers: HS\,1442+4250, VII\,Zw\,403 \citep{Papaderos_1994,Kaaret_2011,Brorby_2014}, NGC\,1569, NGC\,5253, NGC\,4214 \citep{Ott_2005,binder_2015, McQuinn_2018} and NGC\,625 \citep{McQuinn_2018}. Some of them even host Ultra-Luminous X-ray sources (ULX) with measured X-ray luminosities above $10^{39}$erg\,s$^{-1}$: Haro\,2 \citep{OtiFloranes_2012}, Haro\,11 \citep{Prestwich_2015, Gross_2021}, He\,2-10 \citep{Ott_2005, Reines_2011},  I\,Zw\,18 \citep{Thuan_2004,Ott_2005, Kaaret_2011, Kaaret_Feng_2013, Brorby_2014} and SBS\,0335-052 \citep{Thuan_2004, Prestwich_2013}. We note that in some objects, other physical mechanisms are considered to explain the X-ray emission and the claimed detections of compact objects have been actively debated (e.g., He\,2-10; \citealt{Cresci_2017}, HS\,1442+4250; \citealt{Senchyna_heii_2020}, and NGC\,5253; \citealt[][and references therein]{zastrow_ionization_2011}). Regardless of their exact nature (e.g., high-mass X-ray binaries, intermediate mass black hole or AGN), the contribution of such sources may have an important impact in low-metallicity, transparent environments in which high energy photons can travel freely over large scales. In I\,Zw\,18, the galaxy with the lowest metallicity in our sample, \cite{lebouteiller_physical_2019} have shown that the X-ray radiation emitted by a single point source ULX dominates the energy balance in the neutral gas over galactic scales.

We also expect X-ray heating of the ISM to produce specific spectroscopic signatures such as the presence of emission lines associated with very high ionization potentials (e.g., above 40eV) ions. Among the X-ray-sensitive tracers that we consider in this study (see Table \ref{tracers}), \neiii\ is detected in 33/39 sources and \oiv\ detected in 17/39 sources. While we only have upper limits on \nev$\lambda\lambda$24,14$\mu$m, other lines produced by ions with high ionization potentials have been detected in the optical domain \citep[e.g., \feiv, \fev, \nev\ and even \fevi, \fevii;][]{Izotov_2001, Izotov_2004_nev,Izotov_2004_tol}. We note, however, that the origin of such lines is debated with two main hypothesis being either the presence of X-ray sources \citep[e.g.,][]{lebouteiller_neutral_2017, schaerer_new_2019, Simmonds_2021} or the presence of fast radiative shocks \citep[e.g.,][]{Allen_2008, Izotov_2012}. In this study we do not take into account shocks since most of the lines that we consider arise from the \hii\ region and PDR where the dominant source of feedback is likely radiation pressure \citep{Lee_2016, Lee_2019}. In any case, although shocks may contribute to the emission of some of the lines we consider (e.g., lines associated with very high ionization potentials and H$_2$ lines), we do not have enough constraints to disentangle the contribution of shocks to the emission in the current study. For a similar reason, we do not study the impact of varying the cosmic rays (CR) rate in our models although CR might significantly contribute to the heating of neutral gas, especially at low-metallicity \citep{lebouteiller_neutral_2017}.

Even when nonstellar sources do not significantly contribute to the energy balance, their presence can be linked to mechanical feedback mechanisms that may facilitate the escape of LyC photons produced by stars. Any additional kind of feedback that modifies the gas reservoirs surrounding the stars (e.g., through fragmentation or creation of low-density channels) might affect the resulting global porosity to ionizing photons. The DGS sample exhibits a wide variety of feedback mechanisms, which impact the gas structure and kinematics. In particular, several indications of the ISM being disrupted have been discussed in previous studies that have reported signatures of outflows (in Haro\,2: \citealt{OtiFloranes_2012}, Haro\,11: \citealt{Menacho_2019}), ionization cones (NGC\,5253: \citealt{zastrow_ionization_2011}), superbubbles possibly associated with galactic winds (NGC\,1569: \citealt{Westmoquette_2008, Sanches_Cruces_2015}), and other signatures of strong feedback phenomena (NGC\,1705: \citealt{zastrow_new_2013}, NGC\,625: \citealt{Cannon_2004, McQuinn_2018}, UM\,461: \citealt{Carvalho_2018}, NGC\,4214: \citealt{Martin_1998, McQuinn_2018} and Pox\,186: \citealt{Eggen_2021}). Such dynamical effects strongly affect the chemical and physical conditions of the surrounding gas. They might even lower the metallicity of galaxies since metal-enriched outflows can remove newly formed metals from \hii\ regions \citep[e.g.,][]{Amorin_2010, hogarth_chemodynamics_2020}. Another direct effect might also be the creation of low-density channels formed by gas ejection that may favor escaping photons. 

\subsection{Gas and dust properties}
\label{subsect_gas_dust}

Our sample spans a large range of sub-solar\footnote{We use the solar value from \cite{Asplund_2009} of 12+log(O/H)=8.69} metallicities ranging from 12+log(O/H)=7.14 ($\sim$1/35\ Z$_\odot$) up to 8.43 ($\sim$1/2\ Z$_\odot$), derived from empirical strong line methods \citep{Madden_2013, Phd_RemyRuyer_2013}. Their dust-to-gas mass ratios also span a large range of values (0.07--0.33) and have been carefully investigated in \cite{remy-ruyer_gas--dust_2014,remy-ruyer_linking_2015} and \cite{galliano_nearby_2021} using continuum measurements from \textit{Spitzer} and \textit{Herschel}. 

\cite{madden_tracing_2020} have shown that in such environments, the UV photons can penetrate deeper in the clouds and photodissociate CO, hence creating a layer of CO-dark gas. While H$_2$ and CO are only detected in a few galaxies in our sample, evidence suggests that their molecular gas reservoir could be largely underestimated when using CO lines as a tracer. Using in particular the \cii158um line, \cite{madden_tracing_2020} has developed a method to estimate masses of CO-dark H$_2$ gas residing in the envelop where CO is photodissociated by strong radiation field and find that the DGS galaxies most likely host unseen molecular gas reservoir (>70\% of the total H$_2$ in all the galaxies in our sample) that can explain their high star-formation rates (SFR; $-2.2 < \rm log\,SFR < 1.4$) compared to the CO-based estimates of molecular gas content.

Additionally, some of the DGS galaxies are associated with large \hi\ reservoirs with masses ranging from $10^7$ to $\sim 10^{11}$\,M$_\odot$ \citep{remy-ruyer_gas--dust_2014} and specific gass masses (M$_{\rm HI}$/M$_*$) ranging from $0.03$ to $17.3$. 
Although this neutral component is important, little is known on the actual distribution of this gas in our sample. Some of the objects observed in absorption have been associated with large column densities reaching values greater than 10$^{21}$cm$^{-2}$ (see Table \ref{Nh_in_dgs} in the Appendix \ref{appendix} and associated references). Such high values completely rule out photons escaping along these lines of sight. However, previous studies have also put in evidence the inhomogeneous distribution of the neutral component and line-of-sight effects \citep[e.g.,][]{gazagnes_origin_2020}.

\section{Models}
\label{section_models}
\subsection{Modeling strategy}
\label{modelling_strat}

One of the main challenges driving our modeling strategy is the necessity to deal with spatially unresolved observations in which the structure of the ISM is not directly accessible. In such cases, emission arising from the different phases of the ISM (e.g., \hii\ regions, PDRs, molecular gas, DIG) is blended into one single beam. 

To overcome this problem, we recover the underlying gas distribution in the different ISM phases from unresolved spectra. Using a multisector topology as a representative view of the galaxy, such as proposed in \cite{pequignot_heating_nodate}, it becomes possible to disentangle the relative contribution of each phase. Each "sector" corresponds to a fraction of a sphere of gas where radiative transfers are performed in a 1D continuous way throughout the \hii\ region, PDR, and molecular zone. While this model simplifies the actual geometry of a galaxy, there is much to gain from combining sectors (even independent ones) as compared to a single model, especially when dealing with tracers coming from different phases or depths. The need to automate and generalize this multisector topological approach led to the development of a new code, MULTIGRIS, which allows a flexible combination of models within a grid. We briefly describe the code in Sect. \ref{section_mgris} and refer to \citetalias{LebouteillerRamambason2022} for a detailed description of the general strategy used in combining sectors.

The models used in the combination must account for the main physical and chemical mechanisms producing the line emission. Photoionization and photodissociation codes with complex chemical networks and refined prescriptions for radiative transfer are well suited to that purpose. They have been extensively used to study \hii\ regions (e.g., with Cloudy; \citealt{2017_Cloudy_v17}, MAPPINGS\,V; \citealt{2018_mappingV}) or PDR (e.g., with MeudonPDR; \citealt{2006_meudon_pdr}). We chose to use Cloudy models that allow a consistent treatment of the emission line physics throughout the ionized phase, PDR, and molecular phase. Combining such models into a representative topology somewhat compensates for their simplistic geometry, which, by itself, cannot capture the full complexity of the multiphase ISM. In particular, several studies pointed out the necessity to combine several regions having different physical depths to successfully reproduce the emission lines of local objects \cite[e.g.,][]{Binette_1996, pequignot_heating_nodate, lebouteiller_neutral_2017, 2012_Cormier, cormier_herschel_2019, polles_modeling_2019, ramambason_reconciling_2020, Lambert-Huygues_2021}. The present study is a direct continuation of the previous analysis from \cite{cormier_herschel_2019} of the DGS sample (presented in Sect. \ref{section_overview}) using topological models. Based on 1-sector and 2-sector models, with varying PDR covering factors, they selected the best fitting models using a $\chi^2$ minimization. In each model, at least one sector was computed until A$_V$=5 and the PDR scaling factor was defined a posteriori as a linear scaling, between 0 and 1, applied to the lines arising from the PDR. For 2-sector models, an additional sector stopping at the ionization front was added. Based on this definition, they find nonunity covering factors for most of the galaxies in our sample, and a strong correlation between metallicity and covering factor.

Similarly to \cite{cormier_herschel_2019}, we model each galaxy as a weighted combination of sectors where the mixing weights represent the contribution of each sector (see Sect. \ref{section_topo}). The models are computed with a fixed cluster luminosity and scaled to match the observed line fluxes used as constraints with a free scaling factor defined within the Bayesian model (see Sect. \ref{section_code}). The prescriptions used for this new grid are inspired by \cite{cormier_herschel_2019} with a few modifications. One major change is that the line luminosities are saved for each tracer in a cumulative mode, for each depth computed in Cloudy, allowing us to include density-bounded sectors (i.e., models stopping before the ionization front), which was not the case in \cite{cormier_herschel_2019}. This setting also allows a refined sampling of the parameter controlling the depth of each model (see Sect. \ref{geom_of_hii}). We discuss the changes in the model grid and their implications in the following section.

\subsection{Cloudy models}

\begin{table}[b!]
\begin{threeparttable}
\centering
\caption{Input parameters of Cloudy models.}
\label{table1}
\begin{tabular}{p{0.15\textwidth}p{0.25\textwidth}}
    \hline
     & Fixed parameters  \\
    \hline
    \hline
    Stellar population & BPASS v2.1\protect\footnotemark[1] \citep{eldridge_binary_2017} \\
     & Broken power-law IMF\\
     & $\alpha_{1}$ (0.1\,M$_{\odot}$ - 0.5\,M$_{\odot}$) = -1.30 \\
     & $\alpha_{2}$ (0.5\,M$_{\odot}$ - 100\,M$_{\odot}$) = -2.35 \\
     & Single burst, including binary stars \\
     X-ray component & Multicolor blackbody \citep{1984PASJ...36..741M}, T$_{\rm out}$=10$^3$K, variable T$_{\rm in}$ \\
     Other fields & Cosmic microwave background\\
      & Cosmic ray background 0.5log\\
      \hline
      Grains & SMC graphite/silicate model \citep{2001_smc_grains} \\
      Z$_{\rm dust}$, q$_{\rm PAH}$ & Values scaled with metallicity based on the analytical fits from \cite{galliano_nearby_2021}\\
      X/O profiles\protect\footnotemark[2] & N$^{(1)}$, C$^{(1)}$, Ne$^{(2)}$, S$^{(2)}$, Ar$^{(2)}$, Fe$^{(2)}$, Si$^{(3)}$, Cl$^{(4)}$\\
      Other elements & table ISM from Cloudy\\
      \hline
      Density law & n$_{\rm H}$ = n$_{\rm H, HII} \times$ (1+N(H)/[10$^{21}$ cm$^{-2}$]) \\
      Stopping criteria & A$_{\rm V}$=10 or T$_{\rm e}$=10K\\
    \hline
    \hline
     & 8 varied parameters \\
    \hline
    log(L$_{*}$/L$_{\odot}$) & [7, 9]\\
    L$_{\rm X}$/L$_{*}$ & [0, 0.001, 0.01, 0.1]\\
    log(T$_{\rm X, in}$) & [5, 6, 7] K\\
    log([O/H]) & [-2.190, -1.889, -1.190, -0.889, -0.667, -0.491, -0.190, 0.111]\\
    Age of burst & [1,2,3,4,5,6,8,10] Myrs \\
    log(n$_{\rm H}$) & [0,1,2,3,4] cm$^{-3}$\\
    log(U$_{\rm in}$)\protect\footnotemark[3] & [-4, -3, -2, -1, 0]\\
    cut\protect\footnotemark[4] & [0,4], step=0.25\\
    \hline
\end{tabular}

\begin{tablenotes}
   \footnotesize
   \item [1] Metallicity bins in BPASS are Z= 10$^{-5}$, 10$^{-4}$, 0.001, 0.002, 0.003, 0.004, 0.006, 0.008, 0.010, 0.014, 0.020, 0.030 and 0.040 (i.e. 0.05\% of solar to twice solar). We note that Z$_\odot$=0.02 while the default value we use here is Z$_\odot$=0.014. 
   \item [2] References for elemental abundances. (1): \citealt{nicholls_abundance_2017}, (2): \citealt{izotov_chemical_2006}, (3): \citealt{Izotov_Thuan_1999}, (4): fixed at -3.4.
   \item [3] The input ionization parameter, U$_{\rm in}$, is set by varying the inner radius for a given luminosity. It is described in Sect. \ref{geom_of_hii}.
   \item [4] The cut parameter controls the stopping depth of a given model. It is described in Sect. \ref{geom_of_hii}.
\end{tablenotes}

\end{threeparttable}
\end{table}

The grid used in this article was built using the photoionization and photodissociation code Cloudy v17.02 \citep{2017_Cloudy_v17}. Each model consists of a spherical shell of gas placed at a fixed inner radius of the incident radiation source. We use a closed, spherical geometry taking into account the transmitted and reflected radiation, assuming a unity covering factor of the gas. The radiative transfer is computed along each line of sight (1D) in a continuous way throughout the \hii\ region, PDR, and molecular zone. We summarize the main input parameters and the range of values spanned for each parameter in Table \ref{table1}. The grid contains 28\,800 models with an X-ray source and 3\,200 models without X-ray source, which results in a total of 32\,000 Cloudy models. Each model is then truncated a posteriori with 17 cuts to create 544 000 sub-models. The current grid includes predictions for 516 emission lines from UV to IR.

\subsubsection{Radiation field}
\begin{figure*}[htb]
\centering
\includegraphics[width=20cm]{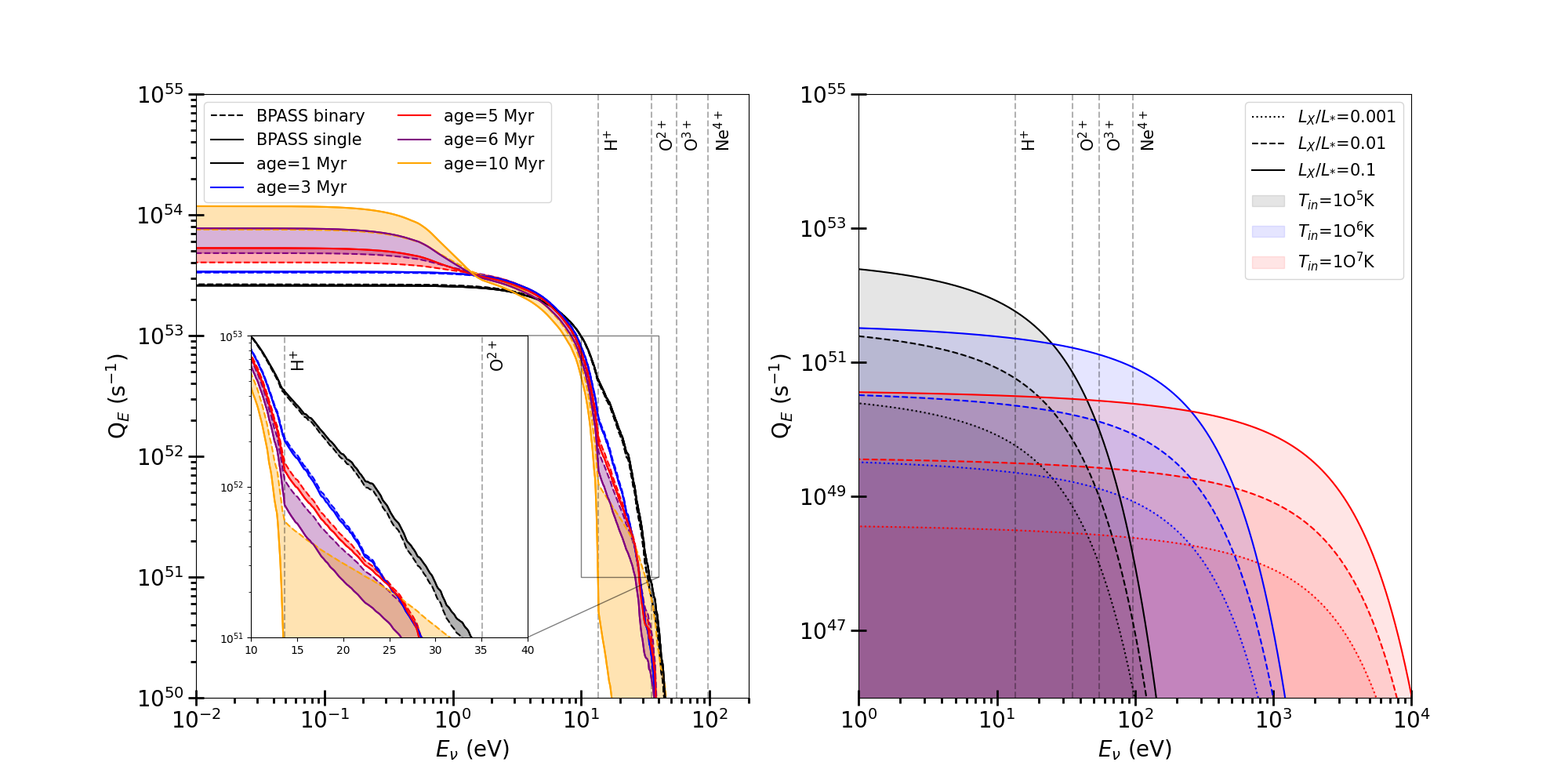}
\caption{Stellar (left) and X-ray (right) incident SEDs of Cloudy models for a solar metallicity and bolometric luminosity of 10$^9$\,L$_{\odot}$. $Q_E$ shows the number of photons produced above a given energy $E_{\nu}$. The vertical dashed lines represent the ionization potentials of H$^{+}$ (13.6eV), O$^{2+}$ (35.1eV), O$^{3+}$ (54.9eV) and Ne$^{4+}$ (97.1eV).
On the left-hand side, the dashed lines represent BPASS models with binary stars while the solid lines of the same color correspond to BPASS models without binary stars for a single stellar population of the same age. The change in the spectra due to the inclusion of binary stars is represented by the shaded area in between both lines. The insert shows a zoom around the Lyman edge at 912$\AA$ where the additional contribution to ionizing photons is visible for ages above $3$\,Myr and increases with the age of the burst.
On the right-hand side, the colors represent different inner temperature of the multicolor blackbody and the different linestyles to different percentage of stellar luminosity: 0.001\%, 0.01\% and 0.1\%.}
\label{SED}
\end{figure*}

In our grid we consider a stellar component and an X-ray component as represented in Fig. \ref{SED}. The stellar population consists of a single burst population from BPASSv2.1 \citep{eldridge_binary_2017} that includes the contribution from binary stars. We note that the stellar population in \cite{cormier_herschel_2019} was instead computed using a continuous star-formation history (SFH) for a 10\,Myr old cluster simulated with Starburst99 \citep{Starburst99_Leitherer_1999}. Switching to a continuous SFH instead of single-bursts would result in a change in our ionizing spectrum as young O/B stars would provide ionizing photons over longer timescales. Since we consider only one single burst, our grid spans ages below $10$\,Myr, after which the H$\alpha$ emission drops. We stress, however, that for the lowest metallicity models, the contribution from binary stars delays this drop in H$\alpha$ and that ages up to $\sim$30 Myrs \citep{2018_Xiao_bpass} should be explored. Considering that our solutions favor ages below 6 Myr and that the addition of a new bin in stellar age would significantly increase the size of our grid of models, we postpone this potential improvement to a future work. In practice, an older population of stars is present in many of the galaxies in our sample but single bursts of later ages alone would not match the spectral signatures that we study here. We note that considering a combination of bursts instead of single-burst model could have a substantial impact on the ionizing continuum, since mixed-age models, which allows contributions from extremely young stars, typically produce more ionizing photons and harder ionizing spectra than single-burst and constant star-formation rate models corresponding to similar ages \citep{Chisholm_2019}.

The stellar evolutionary tracks incorporate the effect of mass transfers between members of binary systems and the stellar initial mass function (IMF) includes a distribution of binaries tuned to reproduce the binary fractions observed in the local Universe. We use a broken power-law IMF with two indices of $-1.3$ and $-2.35$ with a change of slope at $0.5$\,M$_{\odot}$, which is the default IMF in \cite{eldridge_binary_2017}. Although very massive stars with masses above $100$\,M$_{\odot}$ may exist in local, low-metallicity galaxies \citep{Crowther_2010, wofford_2021}, such objects remain largely unconstrained in models. Since we do not need to invoke very massive stars to reproduce the IR lines considered in this study, we choose to use the BPASS default mass cut-off at $100$\,M$_{\odot}$.

In Fig. \ref{SED}, we illustrate the effects of including binary stars by comparing the single-star and binary-stars BPASS \citep{eldridge_binary_2017} SEDs. A clear difference is visible at ages above 3--4\,Myrs where models that include binary stars produce more ionizing photons. This feature has also been pointed out in \cite{2018_Xiao_bpass}: while single-star and binary-star populations produce fairly similar hydrogen- and helium-ionizing spectra at young ages, the inclusion of binaries produces a shallower drop in ionizing flux than for their single-star counterparts at later ages.
The inclusion of binaries with ages between 1--10 Myr has a most profound effect on lines with ionization potentials between 13.6 up to $\sim$54 eV (see Table \ref{table1}). Indeed, X-ray binaries provide high energy photons with an energy sufficient to power the emission of several species with very high ionization potentials. 

However, it is well-known that these high ionization lines cannot be reproduced by classical photoionization models \citep[e.g.,][]{stasinska_excitation_2015}, even when including effects of binary stars \citep[cf.][]{Stanway_Eldridge_2019}. Other sources producing a harder ionizing continuum need to be invoked to reproduce the high ionization potential lines observed in local and high-redshift galaxies (e.g., \heii: \citealt{2018_kehrig_SBS0335, schaerer_new_2019, Stanway_Eldridge_2019, Senchyna_heii_2020, senchyna_2021}; or \civ: \citealt{Stark_2015, senchyna_2021, Senchyna_2022}). For example, high-mass X-ray binaries have been proposed in \cite{schaerer_new_2019} to explain the presence of ubiquitous \heii\ emission. While \cite{Senchyna_heii_2020} find that the addition of an accretion disk associated with high-mass X-ray binaries is insufficient to fully explain the \heii\ emission, several recent studies have argued that the addition of an X-ray source is still needed to simultaneously reproduce the emission lines arising from different ISM phases, including the emission from ions with high ionization potential \citep[e.g.,][]{Simmonds_2021, Olivier_2021, Umeda_2022}. 

Similarly, we find that in order to produce both \oiv\ and \nev\ emission, we need to add an X-ray source that produces harder ionizing photons than BPASS models alone. As discussed in Sect. \ref{section_overview}, the exact nature of such compact objects is unknown (e.g., high-mass X-ray binaries, intermediate mass black hole or AGN). Moreover, the X-ray spectra themselves are widely unconstrained and rely on strong modeling assumptions \citep{Simmonds_2021}. Hence, we choose to use a general prescription to model a compact object surrounded by an accretion disk. To do so, we used a multicolor blackbody spectrum as defined in \cite{1984PASJ...36..741M}. This spectrum is defined by an outer temperature fixed at 10$^3$\,K and an inner temperature varying from 10$^5$ to 10$^7$\,K. The luminosity is set with respect to the stellar luminosity and varies from 0\% to 10\% of the stellar cluster luminosity. The right-hand side panel of Fig. \ref{SED} shows the X-ray component when varying the inner temperature of the disk and the relative luminosity. Although simplistic, our prescription for the X-ray source is more general than that used in previous studies that considered single blackbody spectra \citep[e.g.,][]{lebouteiller_neutral_2017, cormier_herschel_2019}.

Finally, we include the cosmic microwave background (CMB) and CR background. Following \cite{cormier_herschel_2019} we use a CR rate $\sim$3 times higher than the standard CR rate (2 $\times$ 10$^{-16}$ s$^{-1}$; \citealt{2007_indriolo_cosmics}) to account for the recent star-formation history in the DGS and match the observed \oi\ IR line ratio. We note that CRs and X-rays both impact the ionization and heating in the PDR but we do not have means to disentangle both effects. This somewhat arbitrary choice for the CR rate has a strong effect at low metallicities (below 1/10\,Z$_\odot$) where the emission of low ionization potential species and recombination lines is boosted even at large $A_V$, deep inside the molecular zone. However, in the range of $A_V$, density and temperature favored in the results presented here, the CR effects should not be significant. Because we have no means to discriminate between CR and X-ray effect, we do not further discuss their potential contribution in this paper.

\subsubsection{Metal and dust abundances}

Our grid is designed to match the abundance patterns of low-metallicity dwarf galaxies that belong to the group of blue compact dwarf (BCD) galaxies. The prescriptions used in our models are summarized in Table \ref{table1}.
The abundances for nitrogen and carbon are based on \cite{nicholls_abundance_2017} who derive analytical curves accounting for primary and secondary production. Their fit is derived from a large sample of stellar measurements in the Milky Way spanning a wide range of metallicities (6 $\leq$ 12+log(O/H) $\leq$ 9). Their analytical curves are compatible with the gas-phase measurements for carbon and nitrogen in the BCDs from \cite{Izotov_Thuan_1999}, which have little depletion due to their poor dust content. Neon, sulfur, argon, and iron abundance profiles follow the regressions of \cite{izotov_chemical_2006}, based on a sample of low-metallicity BCDs. To avoid extrapolating at high metallicities, we used flat profiles for metallicities above 8.2 for those four elements. Because of the low statistics of \cite{izotov_chemical_2006} for chlorine, we fixed the [Cl/O]\footnote{The brackets indicate values normalized by the solar abundance ratio.} value to the median of their sample (-3.4). The silicon abundance profile is based on \cite{Izotov_Thuan_1999}. For all the other metals we used the values from the Cloudy ISM table, assuming they scale linearly with the oxygen abundance. The profiles used in our grid are given in Fig. \ref{abund_profile}.

We apply the same scaling relative to the solar value to both the gas and stellar metallicity. In practice, since BPASS uses a slightly different solar value (Z$_\odot= 0.020$) than the Cloudy one (Z$_\odot= 0.014$), the metallicities are slightly offset. The helium abundance is fixed to the value provided in the ISM abundance set in Cloudy. The dust-to-gas mass ratio and abundance of polycyclic aromatic hydrocarbons (PAH) are computed following the metallicity-dependent prescription of \cite{galliano_nearby_2021} based on Bayesian dust SED fits of 798 galaxies (including our sample) using the code HerBIE \citep{Galliano_2018} with the THEMIS grain properties \citep{Jones_2017}. 
The dust-to-gas mass ratio hence follows a 4th degree polynomial relation at metallicity above 12+log(O/H) = 7.3 and scales linearly with metallicity below this threshold \citep[see Eqs. (8) and (9) from][]{galliano_nearby_2021}.

\cite{galliano_nearby_2021} also provide an analytical prediction for the mass fraction of aromatic features emitting grains (q$_{\rm AF}$), assuming those features are carried by small a-C(:H) grains. Here, we assume that such features are carried by PAHs instead and estimate the mass fraction of PAHs  (q$_{\rm PAH}$= M$_{\rm PAH}$/M$_{\rm dust}$) as q$_{\rm PAH} \sim$ q$_{\rm AF}$ / 2.2. The abundance of PAHs is known to strongly vary under different physical conditions: they are especially sensitive to metallicity effects and to the strength of the interstellar radiation field (ISRF). \cite{galliano_nearby_2021} find that q$_{\rm PAH}$ is primarly driven by metallicity effects while correlations with ISRF indicators are weaker. We adopt a single value of q$_{\rm PAH}$ per metallicity bin, corresponding to the analytical fit from \cite{galliano_nearby_2021}. We note that this prescription sets the total PAH abundance in our Cloudy models while the abundance profile follows the default one in Cloudy, that is, scaling as H$^0$/H. This results in a profile where PAHs are mostly present in the PDR but completely destroyed in the \hii\ regions and molecular zone. This assumption is consistent with observational studies that show that PAHs are absent in the ionized region \citep[e.g.,][]{Relano_2018, Chastenet_2019}; however, PAHs can also be present in the molecular phase \citep[e.g.,][]{Chastenet_2019}, which have not been considered in the Cloudy modeling. Considering that all the PAHs reside in the PDRs might in turn result in a slight overestimation of the PDR heating by photoelectric effect.

\subsubsection{Geometry of \hii\ regions}
\begin{figure}[htb]
    \centering
\includegraphics[width=9cm]{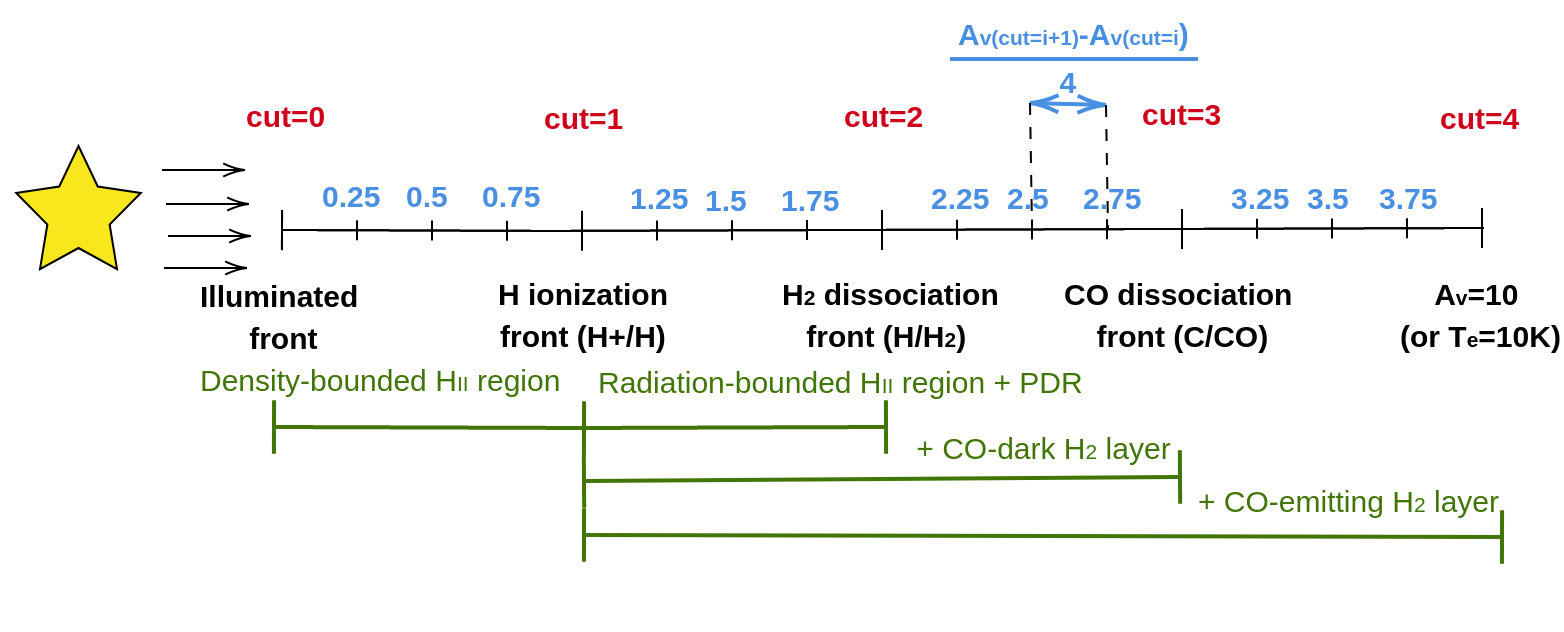}
\caption{Schematic view of the 17 cuts used to create sub-models.}
\label{cuts}
\end{figure}

\label{geom_of_hii}

Each of our models consists in a spherical shell of gas around a central ionizing source, whose inner radius is set to match the input ionization parameter defined as:
    \begin{equation}
        \centering
        U_{\rm in} = \frac{Q(\rm H^0)}{4\pi ncR_{\rm in}^2},
    \end{equation}
    where $n$ is the hydrogen density at the inner radius, $U_{\rm in}$ is the input ionization parameter, $c$ is the speed of light, and $Q$(H$^{0}$) is the total number ionizing photons produced by the central source defined as follows:
    \begin{equation}
        \centering
        Q(\rm H^0)= \int_{E_{\nu} = 13.6eV}^{\infty}\frac{\textit{L}_*(\nu) + \textit{L}_X(\nu)}{h\nu} d\nu,
    \end{equation}
where $L_*(\nu)$ is the stellar luminosity and $L_X(\nu)$ the luminosity of the X-ray source emitted with a given frequency $\nu$. We note that this definition of $U_{\rm in}$ only constrains the ionization parameter at the inner irradiated edge and that the volume-averaged ionization parameter could be significantly different depending on the model parameters. In particular, we note that this definition is not appropriate to use for thick shells of gas for which the inner ionization parameter ($U_{\rm in}$) at the illuminated front can be very different from the ionization parameter at the ionization front $U_{\rm out}$. As an illustration, the relation between log $U_{\rm in}$, log $U_{\rm out}$ and the volume-averaged log <$U$> is provided in Fig. \ref{logU_grid}. Since the inner radius is set automatically to match a given pair of input ionization parameter and gas density, the only possibility to change the geometry of the region is to change the incident luminosity. At fixed $U_{\rm in}$, a large cluster luminosity result in a more shell-like geometry while a lower cluster luminosity result in a more filled-sphere geometry in which the gas lies closer to stars. \cite{stasinska_excitation_2015} show that such geometrical effects can affect the low-ionization to high-ionization line ratio (e.g., \oi/\oiii) as the emission from outer regions is boosted in a compact configuration. Although this geometrical effect is only secondary for most lines arising from the \hii\ regions, it has a much stronger effect on lines emitted near the ionization front and in the PDR and molecular zone.

The bolometric luminosity of 10$^9\,$L$_{\odot}$ chosen by \cite{cormier_herschel_2019} corresponds to the case of a galaxy dominated by one single to a few giant \hii\ regions. For the range of ionization parameters and stellar ages covered in our grid this corresponds to a thick shell geometry with log H$\alpha$ between $39.8$ and $40.7$ (in erg\,s$^{-1}$). Such super-giant \hii\ regions with H$\alpha$ luminosities above 10$^{39}$ erg\,s$^{-1}$ are numerous amoung BCD and starburst galaxies, but not always present \citep{1999ApJ...519...55Y}. They populate the upper end of the observed H$\alpha$ luminosity function derived from resolved galaxies \citep{bradley_composite_2006} and are likely to be optically thin \citep{Pellegrini_2012}. We also consider a lower luminosity $L_{\rm bol}$=10$^7\,L_{\odot}$ ($37.8\,\leq\,\rm log\,H\alpha\,(\rm erg\,s^{-1})\,\leq 38.7$), which corresponds to a typical \hii\ region, similar to those that dominate the H$\alpha$ luminosity function just before the cut-off value (log\,H$\alpha$=38.6\,erg\,s$^{-1}$) from \cite{bradley_composite_2006}. This case mimics a bursty star-formation where a few hundred to a few thousand compact clusters are responsible for the total emission. 

We adopt the same density law as in \cite{cormier_herschel_2019} in which $n_{\rm H}$ is nearly constant in the \hii\ region and scales with column density above 10$^{21}$\,cm$^{-2}$. This law provides a simple first order prescription of a smoothly varying density, which can describe both the density profile expected in dynamically expanding \hii\ regions \citep{Hosokawa_Inutsuka_2005} and in the interior of turbulent molecular clouds \citep{Wolfire_2010}.

One major change as compared to \cite{cormier_herschel_2019} is that luminosities are compiled in a cumulative fashion, meaning that the we can access the intrinsic luminosity of each line at a given depth in the Cloudy model instead of considering only the total resulting luminosity, which obviously depends on the stopping criterion. This is a crucial step to study the escape fraction of ionizing photons as this parameter is sensitive to the stopping depth of the model, especially near the ionization front. As illustrated in Fig. \ref{cuts}, each initial Cloudy model is used to create 17 sub-models stopping at different $A_V$ controlled by the "cut" parameter. The original model is cut at the inner radius (cut=0), ionization front (cut=1), H$_2$ dissociation front (cut=2), CO dissociation front (cut=3) and outer radius (cut=4). To sample the different phases (\hii\ region, PDR, CO-dark H$_2$ region and CO-emitting H$_2$ region) defined by those cuts, three additional cuts are added between each integer $i$ (cut=$i$+0.25, $i$+0.5, $i$+0.75), equally spaced in $A_V$ between cut=$i$ and cut=$i$+1. We stress that stopping the model at a given cut and truncating it a posteriori are not strictly equivalent but our tests have shown that this is a secondary effect.

Our cut parameter is analogous to other parameters that were used to describe density-bounded models such as the H$\beta$ fraction used in \cite{stasinska_excitation_2015} and \cite{ramambason_reconciling_2020} or the zero-age optical depth to LyC photons ($\tau_\lambda$) used in \cite{lebouteiller_neutral_2017} and \cite{plat_constraints_2019}. Although the cut parameter does not have a physical meaning, it allows us to ensure a good sampling of the region near the ionization and provides a simple characterization of density-bounded regions (cut<1) vs. ionization-bounded models (cut>1). Additionally, it is defined consistently regardless of the metallicity (as opposed to A$_V$).  

Since each Cloudy model is cut a posteriori into sub-models, the stopping criterion  is not crucial in this study. However, to incorporate the emission from different phases, models should be deep enough to include the neutral and molecular zone for each set of parameters. This transition is metallicity-dependent and requires going deep enough in A$_V$ for the lowest metallicity models. Most of our Cloudy models are computed until they reach a maximum A$_V$=10. Only the densest models cannot reach A$_V$=10 because their electronic temperature drops below 10\,K before reaching this optical depth.

\subsubsection{Escape fraction from \hii\ regions}
\label{cloudy_observables}

We calculate the escape fraction of ionizing photons from \hii\ regions by using the ionizing continuum that is provided by Cloudy at each depth in the model. This continuum saves the number of photons per frequency bins, at a given depth, for all energies greater than 1 Rydberg. We compute the escape fraction in a given range of energy as follow:
    \begin{equation}
        f_{\rm esc,HII}(E_1 \leq h\nu \leq E_2)(R) = \frac{L(E_1 \leq h\nu \leq E_2)(R)}{L(E_1 \leq h\nu \leq E_2)(R_{\rm in})}, 
    \end{equation}
where the numerator is the luminosity produced by the central sources reaching the radius $R$ with energy between $E_1$ and $E_2$, and the denominator is the luminosity impinging the cloud at the inner radius $R_{\rm in}$, within the same energy range. The number of ionizing photons at a given radius is calculated as:

\begin{equation}
    L(E_1 \leq h\nu \leq E_2)(R) = 4 \pi R^2 \int_{\nu_1}^{\nu_2} F_{\nu}(r)d\nu,
\end{equation}

where $F_{\nu}$ in the flux of photons with frequency $\nu$ at a given radius. In practice we consider \feschii = \fesc(1Ryd\,$\leq$\,h$\nu$\,$\leq$\,$\infty$)($R_{\rm cut}$) that includes all ionizing photons even in the X-ray regime, with $R_{\rm cut}$ the radius at which the sub-model is cut. This definition is equivalent to the ratio between the total observed emission below 912$\AA$ and the intrinsic emission below 912$\AA$. We ensured that the sampling of the cut values results in a fine enough sampling for \feschii\ as well. This definition of \feschii\ consistently accounts for the absorption of photons in the gas and by dust. However, it assumes that the integrated emission of a galaxy is dominated by \hii\ regions and that photons reaching the edge of a density-bounded region freely escape in the IGM. In practice, this tends to overestimate the global galactic escape fraction as some photons are likely to get reabsorbed by clumps or diffuse gas after having escaped from \hii\ regions. This \feschii\ is to be considered as the escape fraction from \hii\ regions and cautiously compared to other measurements, except for the two regions in NGC\,4124 (see Sect. \ref{section_overview}). While this definition corresponds to the true escape fraction from our model, accounting for the total ionizing luminosity below the Lyman edge, it differs from the definition commonly used in observational studies to measure \fesc(LyC), which most often relies on the flux ratio at 912$\AA$ and 1500$\AA$. This difference should be taken into account when comparing with LyC measurements. This will be discussed in Sect. \ref{discussion}.

\section{MULTIGRIS runs}
\label{section_mgris}
In this section we recall some of the most important features of the code used in the current study. A more detailed description of the code can be found in \citetalias{LebouteillerRamambason2022}.

\subsection{Topological representation}
\label{section_topo}

The notion of topology introduced in Sect. \ref{modelling_strat} is especially important in the context of the escape fraction. Indeed, regardless of the nature of the physical mechanisms at the origin of escaping photons, it seems that the LyC escape fraction is tightly linked to the distribution of gas in the ISM. In particular, \cite{gazagnes_origin_2020} have shown that the escape fractions of LyC and Ly$\alpha$ photons are quite sensitive to the distribution of neutral gas and dust, both at galactic scale and on specific lines of sight. This effect is even more pronounced for LyC photons that are easily absorbed by the neutral gas while  Ly$\alpha$ photons scatter and are destroyed only by dust.

\begin{figure*}[htb]
    \centering
    \includegraphics[width=18cm]{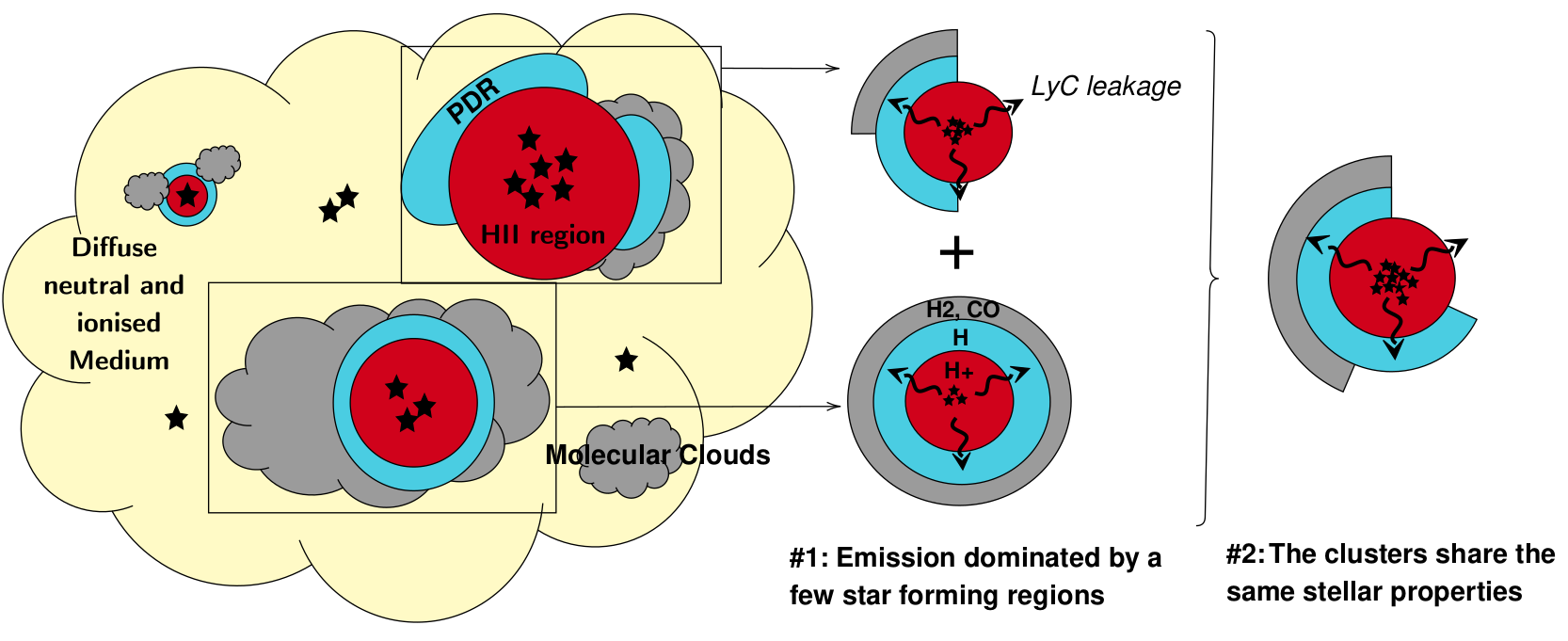}
    \caption{Schematic view of the ISM of a starburst galaxy and associated representative topology.}
    \label{topo}
\end{figure*}

One of the main advantages of our grid of models compared to previous topological studies (see Sect. \ref{section_intro}) is that it includes a free parameter that allows us to constrain the depth of each sector based on the observed line ratios (Sect. \ref{geom_of_hii}). This avoids having to fix an arbitrary stopping criteria that is not trivial to determine in studies that include PDRs and molecular regions \citep[e.g., as in][]{cormier_herschel_2019}. The topological models aim to provide representations of the distribution of matter (number and covering factors of sectors) and phases (stopping depth) in a galaxy. The different phases we consider are represented in Fig. \ref{topo}: an atomic ionized phase dominated by \hii\ regions near star clusters, a neutral atomic phase dominated by a warm component around \hii\ regions where hydrogen is photodissociated (PDR), and a neutral molecular phase concentrated in molecular clouds. Two phases are not accounted for in our Cloudy models: the DIG and the diffuse neutral gas (see discussion in Sect. \ref{discussion}). 

In essence, topological models are not designed to determine the spatial distribution of the gas but only the relative contribution of each phase and sector. As shown in Fig. \ref{topo}, we model a galaxy as a sum of representative star-forming regions that can be combined into one single representative model, assuming that the star clusters that dominate the emission share the same properties. In practice, we assume a single SED for each galaxy we model. This representation is especially well-suited for unresolved observations for which we precisely do not have access to spatial information. Additionally, most current photoionization codes cannot account for precise gas distribution that would require expensive 3D radiation transfer models. In our single cluster topological models, the emission lines are computed as a weighted linear combination of the emission from each sector as follows:
    \begin{equation}
        L= \sum_{\rm i=1}^{N_{\rm sectors}} w_{\rm i} L_{\rm i},
    \end{equation}
    where $w_{\rm i}$ is the mixing weight and $L_{\rm i}$ the predicted luminosity of a given line in the i$^{\rm th}$ sector.

This topological approach allows predictions for any other physical quantities available in Cloudy as long as it can be expressed as an analytical combination of the observables from individual sectors. We use the same formula for "extensive" quantities, meaning, that scale with luminosity (e.g., gas masses in the different phases, number of ionizing photons $Q$). In the current study, we consider configurations having a single cluster and different number of sectors (from 1 to 3). Under this hypothesis of a single cluster, the \feschii\ is also an extensive quantity and the averaged \feschii\ can be derive as follows:

    \begin{equation}
        f_{\rm esc,HII}= \sum_{\rm i=1}^{N_{\rm sectors}} w_{\rm i} f^{\rm i}_{\rm esc,HII}.
    \end{equation}
    
For a configuration with $N$ sectors, the free parameters correspond to $N$ times the 8 free parameters in the Cloudy models from Table \ref{table1} determined for each sector plus the mixing weights $w_i$. In addition, the global luminosity scaling factor of the cluster is free. This amounts to a total of 9$N$+1 parameters. The effective number of free parameters can be reduced by imposing priors that link some parameters together. To mimic the exposure of all sectors to a single cluster we impose that the stellar luminosity, X-ray luminosity, inner temperature of the of the X-ray-emitting accretion disk, metallicity, and age of the stellar burst are identical in all sectors. An additional constraint imposes that all $w_{\rm i}$ sum up to 1 (no hole in the model). The effective number of free parameters thus reduces to 4$N$+5 parameters.
    
The last constraint on the sum of $w_{\rm i}$ translates into one major modeling assumption: we assume that the cluster is fully surrounded by the different sectors and that we can constrain its total luminosity. If the cluster luminosity is unconstrained, the total luminosity scaling factor and the covering factors of each sector can be degenerate. In practice, this would result in quite different configurations that could all match a suite of emission lines but with different scaling factors. This caveat is particularly important for quantities that strongly vary with different configurations (in particular, the escape fraction) and will be discussed in Sect. \ref{discussion}. Hence, $L_{\rm TIR}$, which is used as a constraint (see Table \ref{tracers}), is essential in our analysis to constrain the scaling factor and match the observed total luminosity.

\subsection{Inference of the parameters}
\label{section_code}

MULTIGRIS can be used with different Markov chain Monte-Carlo (MCMC) samplers, which are compared in \citetalias{LebouteillerRamambason2022}. We used the Sequential Monte Carlo Sampler (SMC) that is adapted to multidimensional grids with several likelihood peaks. Interpolating on all parameters is computationally expensive and raises some issues for models located at the edges of our grid. Instead, we use a nearest neighbor interpolation on all the parameters of the grid (described in Table \ref{table1}) except for the metallicity, for which we perform a linear interpolation.

We define the likelihood of our data $p$($\overrightarrow{O}$|$\overrightarrow{\theta}, \mathcal{M}$), where $\overrightarrow{O}$ is the data and $\overrightarrow{\theta}$ the set of parameters of a model $\mathcal{M}$, by considering our suite of emission lines as independent identically distributed random variables (RV). Each RV is described as Gaussian distribution centered on the measured value and with a $\sigma$ corresponding to the uncertainty of the observation. Hence, the likelihood can be expressed as:
     \begin{equation}
        \mathcal{L} = p(\overrightarrow{O}|\overrightarrow{\theta}, \mathcal{M}) = \prod\limits_{i=0}^{N_{\rm obs}} \mathcal{N}(\mu=O_{\rm i}, \sigma^2=U_{\rm i}^2),    
    \end{equation}
    where $N$ is the number of emission lines with observed fluxes $O_{\rm i}$ and uncertainties $U_{\rm i}$. For undetected lines with instrumental upper limits, the Gaussian distribution is replaced by a half-Gaussian. We consider that upper limits correspond to a $2\sigma$ signal. To avoid possible biases due to lines detected with unrealistically small uncertainties, we force a minimal uncertainty of 10\% for all lines. 
    
In practice, single-sector models should be preferred if they are able to simultaneously reproduce all the emission lines of a given galaxy. However, galaxies having numerous lines are more likely to require a greater number of sectors. To which extent the addition of a supplementary sector improves the agreement with observations needs to be quantitatively studied. To compare models, one needs to estimate how well a model performs at predicting data. To do so, we estimate the marginal likelihood corresponding to each configuration by evaluating the model at the posterior distribution of the parameters. The marginal likelihood is defined as follows:
     \begin{equation}
        \mathcal{L_{\mathcal{M}}} = \int_\theta p(\overrightarrow{O}|\overrightarrow{\theta}, \mathcal{M}) d\overrightarrow{\theta}.
    \end{equation}
Because of the integration on all the free parameters $\overrightarrow{\theta}$, this metric penalizes models that require a large number of sectors without significantly improving the agreement with the data. Although the code can be used with any number of components, those models become less and less likely to be selected by the marginal likelihood criterion. Additionally, the computing time required for the MCMC sampling step scales with the number of free parameters. We hence limit this study to combinations involving 1, 2, or 3 sectors. An example of a 4-sector model for I\,Zw\,18, which includes additional constraints from the optical lines, is presented in \citetalias{LebouteillerRamambason2022}.

In Table \ref{table_lm}.1, we report the best (i.e., having the highest marginal likelihood) configuration for each galaxy, the corresponding accuracy at 3$\sigma$ and the marginal likelihood values in logarithm, for a varying number of sectors. In practice, we combine the results obtained for configurations having 1, 2, or 3 sectors by considering a weighted mean of the 3 configurations with the weights are defined as follows:
\begin{equation}
    \alpha_{\rm i\, sectors} = \frac{\mathcal{L}_{\mathcal{M},\rm i}}{\sum_{\rm j=1}^{\rm 3\, sectors} \mathcal{L}_{\mathcal{M},\rm j}},
\end{equation}
where $\mathcal{L}_{\mathcal{M},\rm i}$ is the marginal likelihood associated with the configuration having $i$ sectors (with 1 $\leq i \leq$ 3).
This weighted combination allows us to account for configurations having marginal likelihoods that are close to the best configuration. Since the weighted combination is performed directly on the MCMC draws, the uncertainties we derive for the parameters incorporate uncertainties on the combination of configurations, which can be considered as an uncertainty on the model itself. Other metrics could be used to define the weight, which are detailed in \citetalias{LebouteillerRamambason2022}. Using a different metric yields variations for a few individual objects in our sample but does not affect the global trends that we discuss in the following sections.

\begin{figure}[htb]
    \centering
    \includegraphics[width=4.4cm]{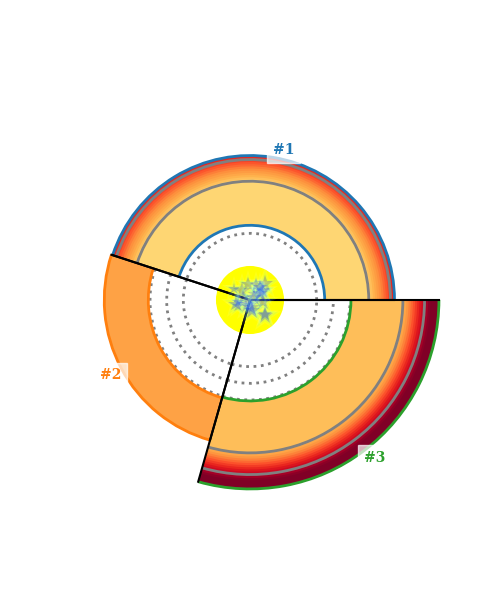}
    \includegraphics[width=4.4cm]{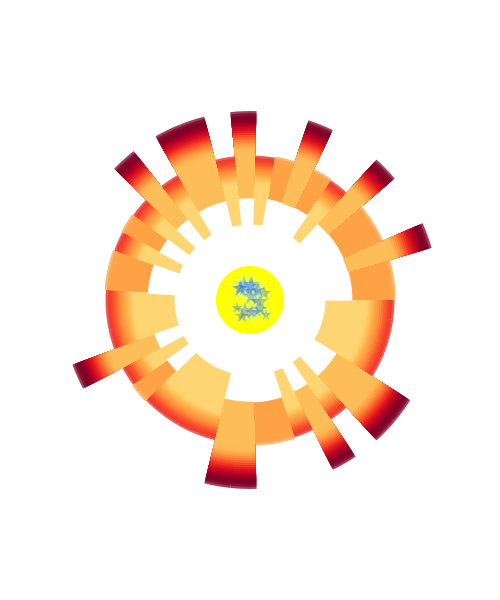}
    \caption{Schematic view of the single cluster 3-sector configuration for He2-10. The number of stars in the central cluster scales with the intensity of the radiation field, the proximity of the gas representing varying ionization parameters and the orange shading increases with density.}
    \label{sector_He210}
\end{figure}

As an illustration, Fig. \ref{sector_He210} shows the best configuration found by MULTIGRIS for He\,2-10. This is one of the galaxies for which we have the most constraints available (22 detections and 6 upper limits, see Fig. \ref{nlines}). For this galaxy, the 3-sector configuration is favored. The left-hand side plot represents the relative contribution of each of the 3 sectors and the right-hand side plot shows another version where the sectors have been randomly redistributed around the central source. We stress that the two topologies are completely equivalent in our study.

\subsection{Probability density functions}
\label{rep_pdf}

One of the advantages of using a Bayesian method is that the ensemble of solutions can be represented by a continuous distribution, which is well-suited to identify possible degeneracies or multimodal solution.
While frequentist methods like the $\chi^2$ only provide a group of the most likely values of a parameter, MULTIGRIS outputs a sample of draws from the posterior PDF $p(\overrightarrow{O}|\overrightarrow{\theta}, \mathcal{M})$ of any given parameter.  

The PDF can be described using various estimators. The mean, mode, and median values of the parameters can be significantly different in the case of multimodal or asymetric distribution. Ideally, one would like to represent the full posterior PDFs that are used in this study to derive trends and correlations. We also investigate the combined PDF concatenating all sources. To that purpose, we use kernel density estimate (KDE) plots, which provide a smoothed convolution (using a Gaussian kernel) of the 2D-PDF of our whole sample. Additionally, to represent 2D-PDFs of individual objects we adopt the skewed uncertainty ellipse (SUE) representation \citep[see e.g., Appendix F from][]{galliano_nearby_2021}. A SUE represents the 1$\sigma$ contour of a 2 dimensional split-normal distribution adjusted to have the same three first moments as the underlying PDF. The center of the SUE marks the location of the robust mean of the 2D-PDF. While the representation is convenient to locate the parameter space of highest probability (e.g., 1$\sigma$) for a given object, SUEs are not always representative of the underlying PDF, especially when several modes are present. In the following plots we show either the KDE of the full sample or the individual SUEs corresponding to each galaxy. While the KDE of the full sample should always be used to study the statistical trends and correlations, the individual SUEs can also be useful to visualize the region of maximal likelihood associated with each object.

\section{Consistency checks}
\label{section_prelimary_results}

The escape fraction of ionizing photons is a complex parameter with multiple dependences. Before diving into the interpretation of this complex observable (see Sect. \ref{section_fesc_results}), we perform consistency checks to see how the predictions from MULTIGRIS compare to classical diagnostics found in the literature. In particular, since the escape fraction of ionization photons is sensitive to both the gas and stellar content of a galaxy, we first examine some key parameters that may control the predicted values of \feschii: the metallicity and the star-formation rate.  

\subsection{Metallicity estimates}
\label{section_fesc_Z_SFR}
\begin{figure}[htb]
    \centering
\includegraphics[width=9cm]{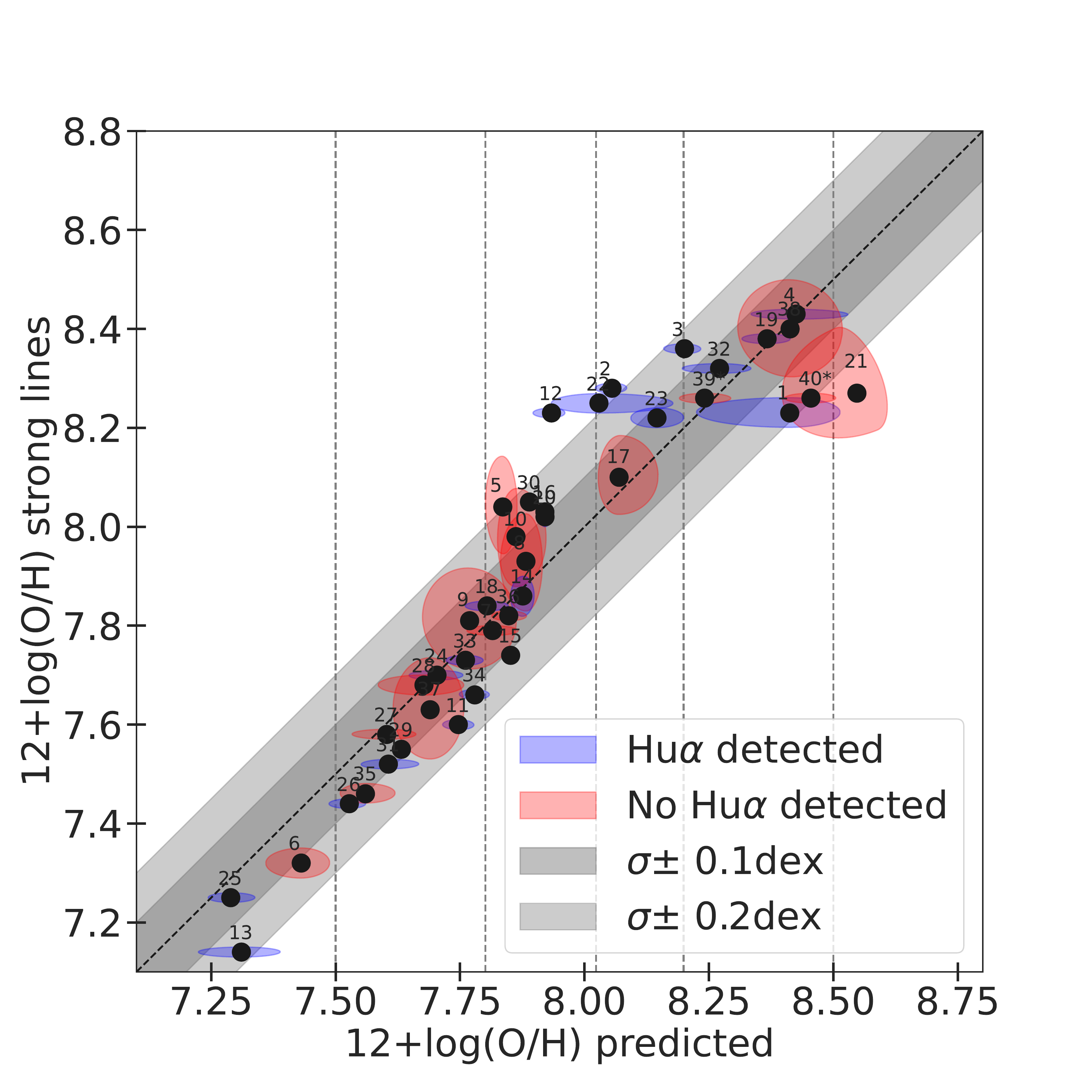}
    \caption{Predicted metallicity of MULTIGRIS vs. measured metallicity from the strong lines R23-P method from \cite{Pilyugin_Thuan_2005} based on optical oxygen lines ratio. The black dots represent the robust means of the posterior PDF. The 1$\sigma$ contours of the SUEs are calculated by simulating mock data for the y-axis with a dispersion corresponding to the measured uncertainty. The vertical dashed lines correspond to the metallicity bins of the grid. The galaxies are labeled with the numbers reported in table \ref{table_lm}. The numbers that are flagged with a star correspond to two pointings in NGC\,4214, which are excluded from the KDE. The color code indicates whether the Humphreys $\alpha$ line (7--6) is detected or not.}
    \label{compare_metallicities}
\end{figure}

To ensure that our metallicity estimates are not affected by the metallicity sampling of the model grid (see Sect. \ref{section_models}), we perform a linear interpolation on this parameter. For the other parameters, we take at each draw the nearest neighbor in the grid. Our approach differs from classical methods that usually rely on a single tracer or a combination of a few lines to derive the metallicity. Instead, our code uses the combined information from all emission lines to constrain model parameters, including the metallicity. Although this approach is more flexible (no need for a specific suite of emission lines), we need to rely on a particular abundance pattern because we do not have enough constraints on the abundance of all species. In this study, we use abundances tailored for compact, low-metallicity, dwarf galaxies, as described in Sect. \ref{cloudy_observables}.

The metallicity could be subject to multiple degeneracies. In particular, varying the cut parameter, that is, the A$_V$ at which each sector stops, impacts the emission of tracers used to estimate the metallicity. Previous works have reported that accounting for escaping ionizing photons tends to bias the metallicity estimates toward lower values \citep{2018_Xiao_bpass, Jiang_Tianxing_2019}. The metallicity is also degenerate with the stellar age parameter. As mentioned in \cite{2018_Xiao_bpass}, an old stellar population with no leakage can produce line ratios similar to a younger stellar population where part of the radiation is leaking through density-bounded regions.

To avoid such degeneracies between metallicity, stellar age and cut parameters, we allow the metallicity to vary under a weakly informative prior. This ensures the inference preferentially starts around the mean of the prior and prevents drawing too far from it. We use a Gaussian centered at the measured metallicity  with a large width parameter ($\sigma$ = 0.1+ $\sigma_{\rm mes}$ in logarithmic scale, where $\sigma_{\rm mes}$ is the uncertainty associated with the measured metallicity). We emphasize that although we use only constraints in the IR domain, the prior on metallicity incorporates information derived from optical lines since it is centered on the metallicity estimates derived from optical strong lines method. Galaxies that do not fall within the $\sigma$=0.1 dex envelop are galaxies for which the information given by all emission lines have pushed the posterior PDF away from the prior distribution. 

In Fig. \ref{compare_metallicities}, we compare this estimated metallicity to that measured from strong lines methods. The galaxies are labeled with the numbers reported in table \ref{table_lm}. The measured metallicities from the literature come from \cite{Madden_2013}, and references therein, based mainly on the optical oxygen lines ratio (R23-P) method from \cite{Pilyugin_Thuan_2005}. Their method is based on a two-parameter calibration involving the R23 ratio and an excitation parameter P, which corrects the estimates by accounting for the physical conditions in the \hii\ regions. The method used in \cite{Madden_2013} allows them to derive metallicity measurements for almost all the DGS galaxies except for three galaxies (HS 0017+1055, HS 0822+3542 and HS2351+2733) for which the R23-P method cannot be applied and the measured metallicity is derived using the T$_e$-direct method \citep{Ugryumov_2003, izotov_chemical_2006, Izotov_1994}. We note that the latter method is more reliable than the R23-P method \citep[e.g.,][]{maiolino_re_2019} that was chosen only because it could be applied in a consistent way to almost all objects in the DGS sample.

The uncertainties derived (1$\sigma$ uncertainties of SUEs defined in Sect. \ref{rep_pdf}) from our method are somewhat larger than those found in the literature, as they incorporate uncertainties on all emission lines and on the modeling (e.g., in particular our prescription for the abundance patterns vs. metallicity). When a hydrogen recombination line (e.g., the Humphreys $\alpha$ line (7--6); Hu$\alpha$) is available, the estimated metallicity tends to be more tightly constrained as the code can directly infer elemental abundances relative to hydrogen. This is illustrated in Fig. \ref{compare_metallicities} where galaxies that have Hu$\alpha$ detection tend to have smaller 1$\sigma$ contours than the ones without. Interestingly, our code is also able to estimate metallicities without hydrogen recombination line measurements, in some cases with a precision comparable to galaxies with Hu$\alpha$ measurements. As a consequence, the metallicity is inferred by constraining metal-to-oxygen abundances, which depend on metallicity in the grid (see Sect. \ref{section_models}). In this case the results we get are strongly dependent on the assumed metal-to-oxygen profiles, and in particular the N/O and C/O vs. O/H relations.

The metallicities are consistent with the corresponding strong line measurements within 0.1 dex for 31 out 40 galaxies and withing 0.2 dex for all galaxies except one (12: II\,Zw\,40), for which we predict a significantly smaller metallicity than that derive with the R23-P method. We note, however, that our measurement is consistent with the value measured for this galaxy (8.09 $\pm$ 0.02) using the $T_e$-direct method from \cite{izotov_chemical_2006}. The scatter we obtain is somewhat smaller than that derived for the DGS sample in \cite{Madden_2013} when comparing different calibration methods using the R23 ratio \citep{Pilyugin_Thuan_2005} and the T$_e$-direct method. \citep{izotov_chemical_2006}. We find that our predictions tend to systematically predict slightly higher metallicities than the R23-P method for metallicities below $7.8$. This might be due to the fact that we find galaxies with numerous density-bounded regions (i.e., with small cut parameters) among the lowest metallicities in our sample (see Sect. \ref{section_fesc_oh}). This effect might be linked to the fact that the R23-P diagnostic relies on an empirical calibration based on samples of galaxies, regardless of the presence or not of density-bounded \hii\ regions. If the calibration is dominated by radiation-bounded \hii\ regions, this diagnostic might slightly underestimate the metallicity of galaxies in which numerous density-bounded regions are present.
Nevertheless, based on the weak prior assumption and on the emission lines given as constraints, we find that our code infers metallicities in agreement with previous measurements for all galaxies  in our sample. This is a necessary condition to derive parameters such as \feschii\, which is expected to be metallicity-sensitive.

\subsection{Star-formation rate estimates}
\label{results_sfr}

\begin{figure}[h!]
    \centering
\includegraphics[width=9cm]{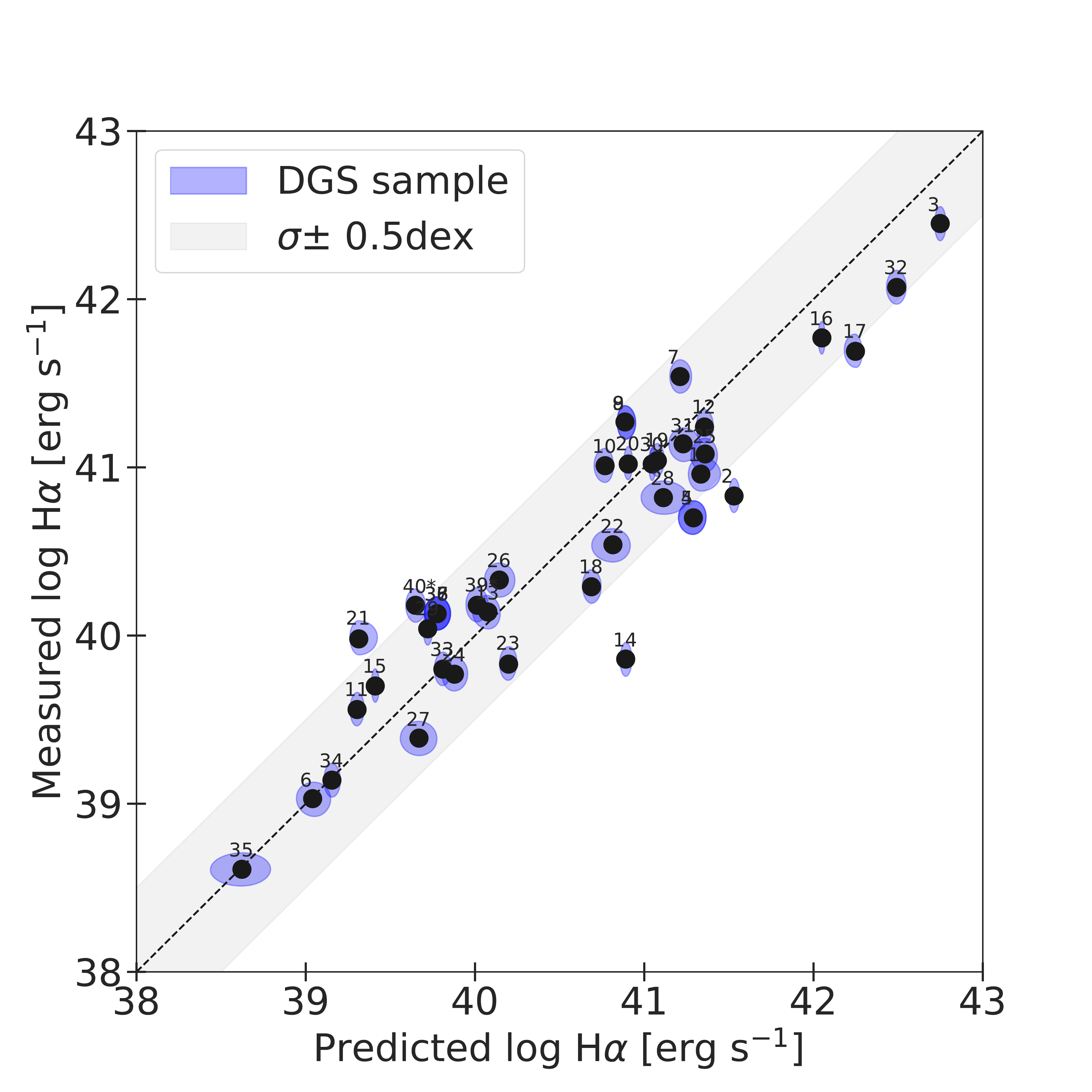}
\includegraphics[width=9cm]{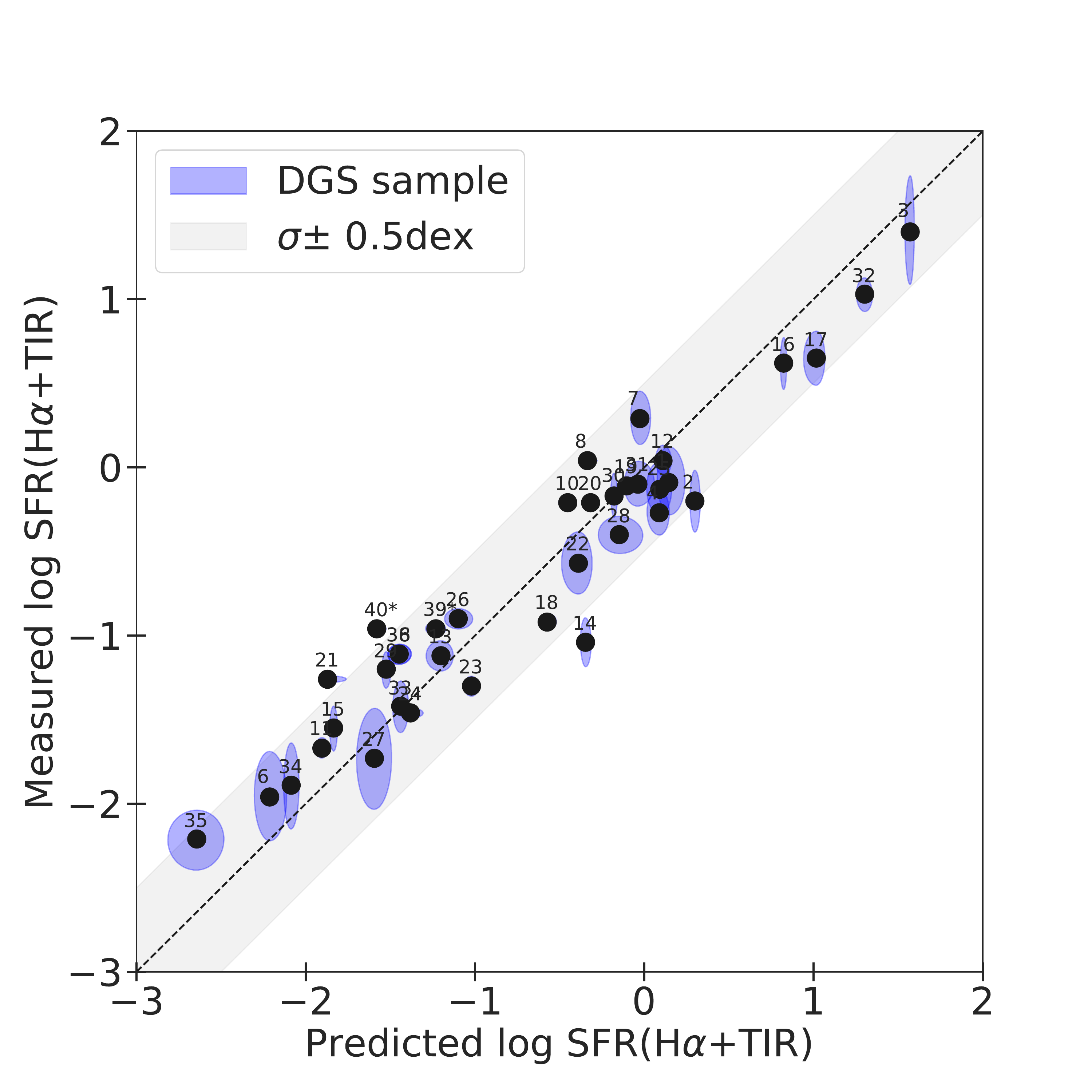}
    \caption{Agreement between \ha\ and SFR predictions with empirical measurements. \textbf{Top}: Predicted intrinsic H$\alpha$ emission vs. measured H$\alpha$ corrected for Galactic extinction from \citealt[][and references therein]{remy-ruyer_linking_2015}. \textbf{Bottom}: Predicted SFR vs. SFR(H$\alpha$+$L_{\rm TIR}$) from \cite{remy-ruyer_linking_2015}. The 1$\sigma$ contours of the SUEs are calculated by simulating mock data for the y-axis with a dispersion corresponding to the measured uncertainty. For 4 galaxies (HS\,0017+1055, HS\,0052+2536, HS\,1319+3224 and HS\,2352+2733) no H$\alpha$ measurements are available. For 2 of those galaxies (5: HS\,0052+2536 and 9: HS\,1319+3224), their measured SFR is derived in using diagnostics based on FUV.}
\label{compare_SFR}
\end{figure}

We now use the predictions provided by MULTIGRIS to infer SFR estimates in our sample and compare them to classical diagnostics. Although the SFR is not one of the primary (used for inference) or secondary (output from Cloudy) parameters, we use classical SFR proxies (H$\alpha$, $L_{\rm TIR}$ and  $Q$(H$^0$)), which we convert into SFR. We emphasize that H$\alpha$ observations are not used as constraints in our models but our code can provide PDFs of both observed and unobserved emission lines. In this work, we use the constraints from IR lines that, to first order, are not affected by extinction by dust, to estimate the number of ionizing photons produced by the central cluster, $Q$(H$^0$). The predicted H$\alpha$ luminosity depends of the incident flux $Q$(H$^0$) and on the geometry of the gas in our representative model. The main advantage of our method is that it takes into account information coming from all the emission lines available. However, like for other diagnostics, our estimates ultimately depend on the assumption made regarding the stellar population. 

We compare our predictions to a classical diagnostic based on composite diagnostics that combine tracers of dust-obscured (e.g., $L_{\rm TIR}$) and dust-unobscured (e.g., H$\alpha$) star-formation (see review from \citealt{Calzetti_2012}). First, we compare the predicted H$\alpha$ luminosities to the values from the literature gathered in \cite{remy-ruyer_linking_2015} in Fig. \ref{compare_SFR}. The latter correspond to H$\alpha$ luminosities corrected for underlying stellar absorption, \nii6548,6584A lines contamination, and foreground Galactic extinction. No correction has been applied to account for the intrinsic attenuation within galaxies. 
For consistency, we compare the corrected measurements to the predicted intrinsic H$\alpha$ emission i.e. what escapes from the galaxy without considering dust extinction. 

The predicted H$\alpha$ luminosity is compatible with observations within 0.5 dex for all galaxies in our sample except for two galaxies (14: Mrk 153 and 2: Haro\,3) for which our prediction is significantly larger. The fact that our models find H$\alpha$ values close to observed measurements is remarkable, especially since no optical lines were used as constraints. This means that the intrinsic values we predict are relatively robust even though we do not properly account for H$\alpha$ emitted by the DIG (see Sect. \ref{discussion}). It also indicates that taking into account the actual geometry of the gas by including density-bounded regions in our models only yields minor variations in the predicted intrinsic H$\alpha$ emission and remains compatible with observations using somewhat simple prescriptions to correct the observed H$\alpha$ emission for foreground Galactic extinction. We stress that the intrinsic fluxes predicted by our models do not account for internal attenuation within the galaxy. The fact that our predictions are in agreement with observations that were not corrected for internal attenuation suggest that this additional attenuation term has only a minor effect in low-metallicity galaxies. This could, however, be an important effect for more metal-rich sources and may also explain part of the scatter see in Fig. \ref{compare_SFR}.

We then use our line predictions to estimate the SFRs. To allow the comparison with the estimates provided in \cite{remy-ruyer_linking_2015}, we use the same empirical calibration from \cite{Kennicutt_2009} based on mixed tracers (H$\alpha$ and $L_{\rm TIR}$). This diagnostic corresponds to an SFR conversion assuming a Kroupa IMF\footnote{We apply a corrective factor of 0.945 (assuming that all the luminosity comes from stars with masses above 1M$_\odot$ and that all the mass comes from stars with masses below 8M$_\odot$) to our predicted SFR to compare them to measurements that assumed a Kroupa IMF.} and with coefficients calibrated on the \textit{Spitzer}-SINGS sample. For consistency, we use $L_{\rm TIR}$ luminosities predicted by our code. We note, however, that the observed $L_{\rm TIR}$ is used as a constraint and always well reproduced by our models within 0.1 dex. Similarly to the H$\alpha$ prediction, we find that our predictions for the SFR are consistent with measurements from \cite{remy-ruyer_linking_2015} within 0.5 dex for all galaxies but one (14: Mkr\,153), which is linked to our overestimation of H$\alpha$ emission. \footnote{There is also a good agreement with the measured SFR(UV+24$\mu$m). However, this comparison should be considered with caution due to the bursty nature of star formation in local metal-poor dwarfs \citep[e.g.,][]{de_looze_applicability_2014}.}

\begin{figure}[htb]
    \centering
    \includegraphics[width=9cm]{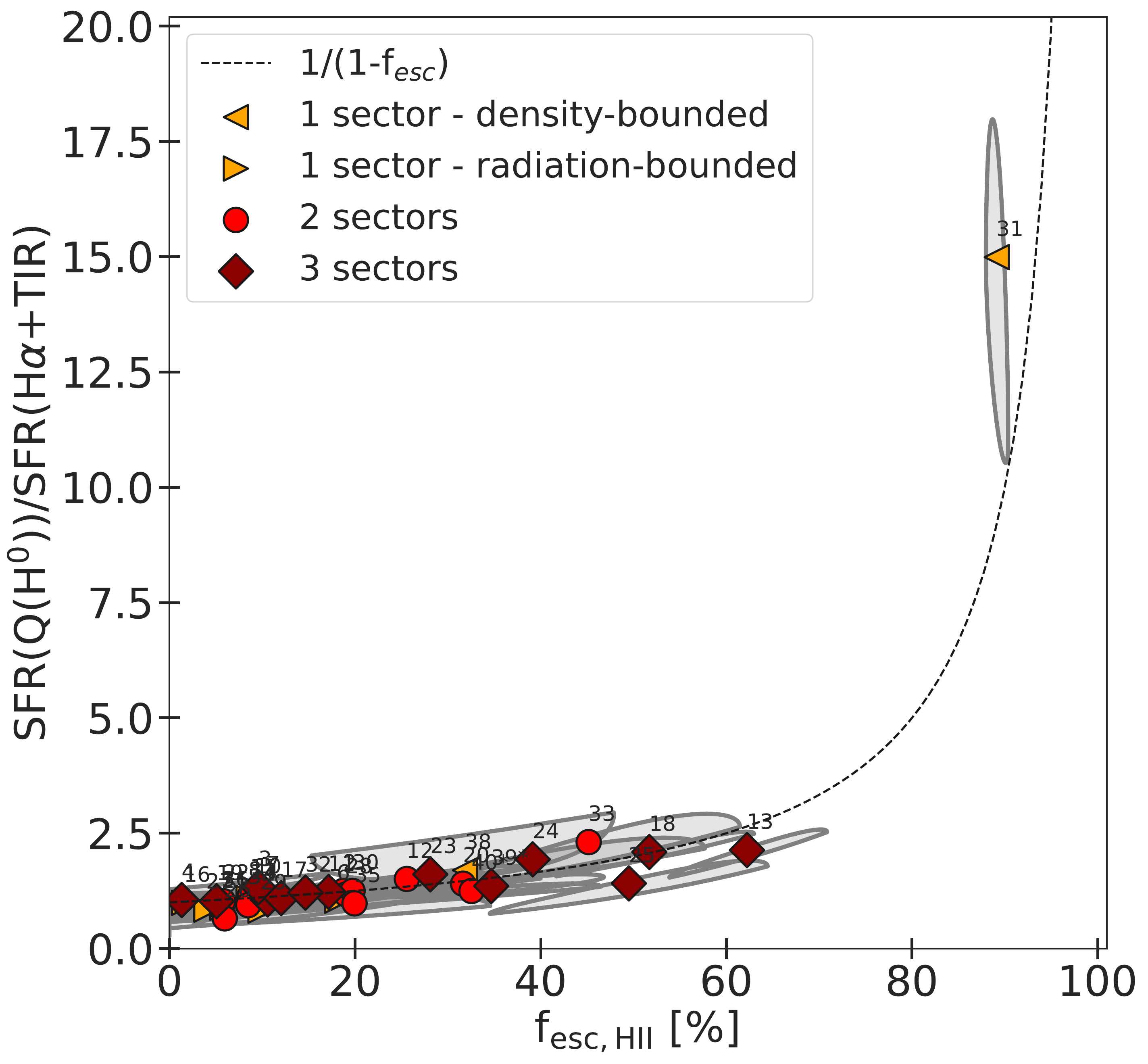}
\caption{Predicted SFR($Q$(H$^0$))/SFR(H$\alpha$+TIR) vs \feschii. The symbols represent the robust means of the posterior PDF and the gray contours show the 1$\sigma$ uncertainties of the SUEs. The dashed line corresponds to the analytical curve 1/(1-f$_{\rm esc})$.}
\label{compare_SFR_Q0_HaTIR}
\end{figure}

One advantage of our approach is that it allows us to account for the photon losses due to escape fractions in density-bounded sectors, which are ignored in empirical calibrations. In Fig. \ref{compare_SFR_Q0_HaTIR} we convert the incident flux of ionizing photon, $Q$(H$^0$), into SFR using the analytical formula provided in \cite{Calzetti_2012} assuming a Kroupa IMF. We then compare it to our prediction based on H$\alpha$+$L_{\rm TIR}$. Our SFR estimates based on $Q$(H$^0$) are in most cases shifted to higher values. If we consider SFR($Q$(H$^0$)) as the intrinsic SFR, the correction to apply to SFR(H$\alpha$+TIR) follows the analytical curve 1/(1-\feschii) with small deviations, which can be attributed to several effects. First, although H$\alpha$ emission is directly proportional to the number of LyC photons absorbed by the gas,(1-\feschii)Q(H$^0$), this conversion depends on the hydrogen recombination coefficient that varies with physical conditions (density and electronic temperature). Second, the $L_{\rm TIR}$ emission that is used to derive the SFR has more complex dependences; it is sensitive to the dust content of our models and can be partially powered by X-ray photons heating the PDR. Hence, the SFR estimates we derive from H$\alpha$+TIR should, in theory, be sensitive to variations of the average physical and chemical conditions in each galaxies of our sample. Nevertheless, we find a rather narrow dispersion around the 1/(1-\feschii) relation that may indicate that, to first order, such effects are small. Assuming that SFR(Q(H$^0$)) is the intrinsic SFR accounting for all the photons produced by the central source, the correction to apply to SFR(H$\alpha$+TIR) varies from a factor 1 to $\sim$2.5 for well constrained galaxies having at least two sectors, and reaches a factor $\sim$15 for one poorly-constrained galaxy (31: Tol\,1214-277) for which the best solution is a single-sector, completely density-bounded model (see Fig. \ref{compare_SFR_Q0_HaTIR}). 

\begin{figure}[htb]
    \centering
    \includegraphics[width=9cm]{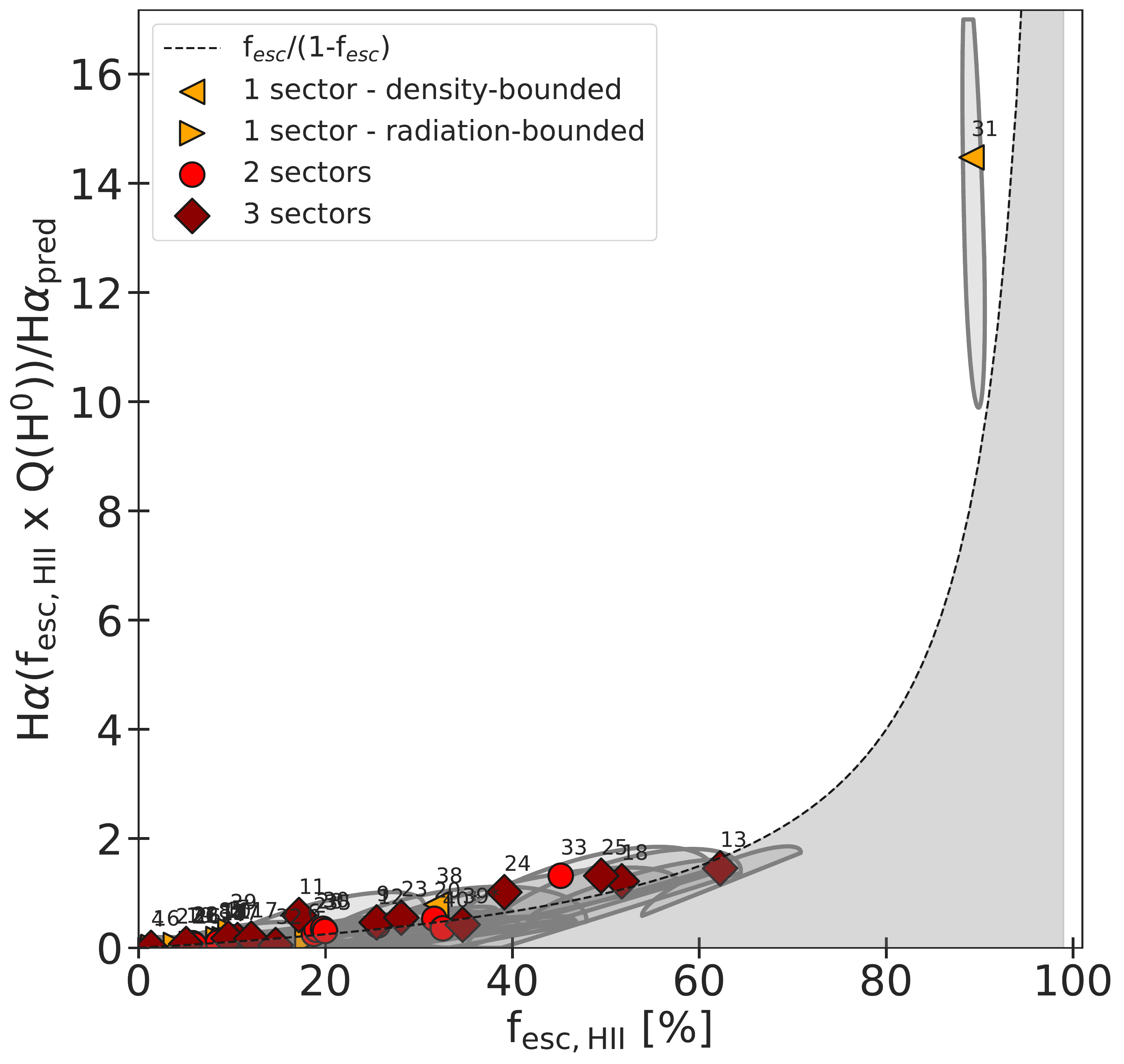}
\caption{Ratio of the maximum H$\alpha$ emission associated with photon leakage from \hii\ regions, assuming that all photons are reabsorbed and of the H$\alpha$ value predicted by our models vs. \feschii. We assumed Q(H$^0$)-to-H$\alpha$ conversion coefficient of 7.31 $\times$ 10$^{11}$ for a case B recombination with T$_e$=10\,000\,K and n=100cm$^{-3}$ from \cite{Kennicutt_1995} to estimate $\ha(\feschii \times$ Q(H$^0$)). The \ha$_{\rm pred}$ values are computed consistently in the Cloudy models. The dashed line correspond to the analytical curve f$_{\rm esc}$/(1-f$_{\rm esc})$.}
\label{missing_Ha}
\end{figure}

The difference between the predicted SFR(H$\alpha$+TIR) and SFR($Q$(H$^0$)) can therefore be explained by the presence of density-bounded regions leading to escaping photons. Indeed, photons that escape from density-bounded regions are accounted for in $Q$(H$^0$) but escape before ionizing the gas that produces H$\alpha$. Nevertheless, this effect is enhanced by our model assumptions (see Sect. \ref{discussion}) in which photons reaching the edge of density-bounded sectors can freely escape without further interaction. In principle, part of the leaking radiation should be reabsorbed and produce a DIG contributing to the global H$\alpha$ emission \citep{Mathis_1986, Sembach_2000, Wood_2010, Belfiore_2022}. This effect would likely reduce the correction factors that we predict here in order to obtain the intrinsic SFR($Q$(H$^0$)). Assuming that all the ionizing photons escaping from \hii\ regions are reabsorbed by the surrounding DIG and assuming a maximum value for the recombination coefficient of hydrogen (corresponding to densities typically found in \hii\ region conditions), we can provide an upper limit on the additional H$\alpha$ emission, which might be powered by leaky \hii\ regions. In Fig. \ref{missing_Ha}, we show the predictions for the maximum additional \ha\ emission powered by photon leakage, assuming a Q(H$^0$)-to-H$\alpha$ conversion coefficient of 7.31 $\times$ 10$^{11}$ for a case B recombination with T$_e$=10\,000\,K and n=100cm$^{-3}$ \citep{Kennicutt_1995} to derive $\ha(\feschii \times$ Q(H$^0$)). This quantity follows the analytical curve  \feschii/(1-\feschii). The small variations around this relation come from the fact that we assume a fixed recombination coefficient to estimate H$\alpha$(\feschii$\times$Q(H$^0$)) while H$\alpha_{\rm pred}$ is calculated by integrating H$\alpha$ emission at each depth, assuming a varying recombination coefficient that corresponds to the local physical conditions. We note that this upper limit was calculated with an unrealistically high recombination coefficient for DIG conditions and only yields an upper limit on the additional H$\alpha$ emission. The range of plausible values corresponding to the reabsorption of leaky LyC-photons by DIG is represented by the gray shaded area. Unsurprisingly, the large \feschii\ we derive translate into large upper limits of the H$\alpha$ emission powered by DIG; this maximum emission exceeds 50\% of the total H$\alpha$ value for the larger \feschii\ we infer (i.e., \feschii > 50\%). Although a significant fraction of the H$\alpha$ emission is expected to be powered by the DIG \citep[up to $\sim$50-60\%, see e.g.,][]{oey_survey_2007, Lacerda_2018}, this is not sufficient to explain the \feschii\ >50\% that are found for a few objects in our sample. Most importantly, this additional H$\alpha$ component can hardly be reconciled with the fact that our predicted H$\alpha$ values (which do not account for the DIG) are in good agreement with the observed values (see Fig. \ref{compare_SFR_Q0_HaTIR}). Possible explanations of this discrepancy are further discussed in Sect. \ref{discussion_missing_gas}.

\section{Escape fraction of ionizing photons}
\label{section_fesc_results}

We have seen in Sect. \ref{section_prelimary_results} that our new code is able to predict several galactic observables that are well constrained by the suite of lines given as inputs and compatible with previous observational studies. Those observables do not depend much on the inferred topology and can be estimated in a robust way, with little influence from, for example, the number of sectors considered (see \citetalias{LebouteillerRamambason2022}). The agreement of our predictions with empirical studies shows that simple prescriptions used in previous studies (e.g., single-component and no treatment of density-bounded regions) are sufficient, to first order, to estimate metallicities and SFR. Nevertheless, several other quantities can be predicted by our code, including more complex ones for which we have only a few, sometimes biased, observational proxies, and that strongly depend on the assumed topology. This possibility is especially interesting to study escape fractions for which direct observations are challenging and indirect proxies often yield different results (see Sect. \ref{section_intro}). We now present the results that we obtain for \feschii\ and discuss the observed trends and their possible physical interpretations.

\subsection{Escape fraction vs. metallicity}
\label{section_fesc_oh}

\begin{figure}[htb]
    \centering
    \includegraphics[width=9cm]{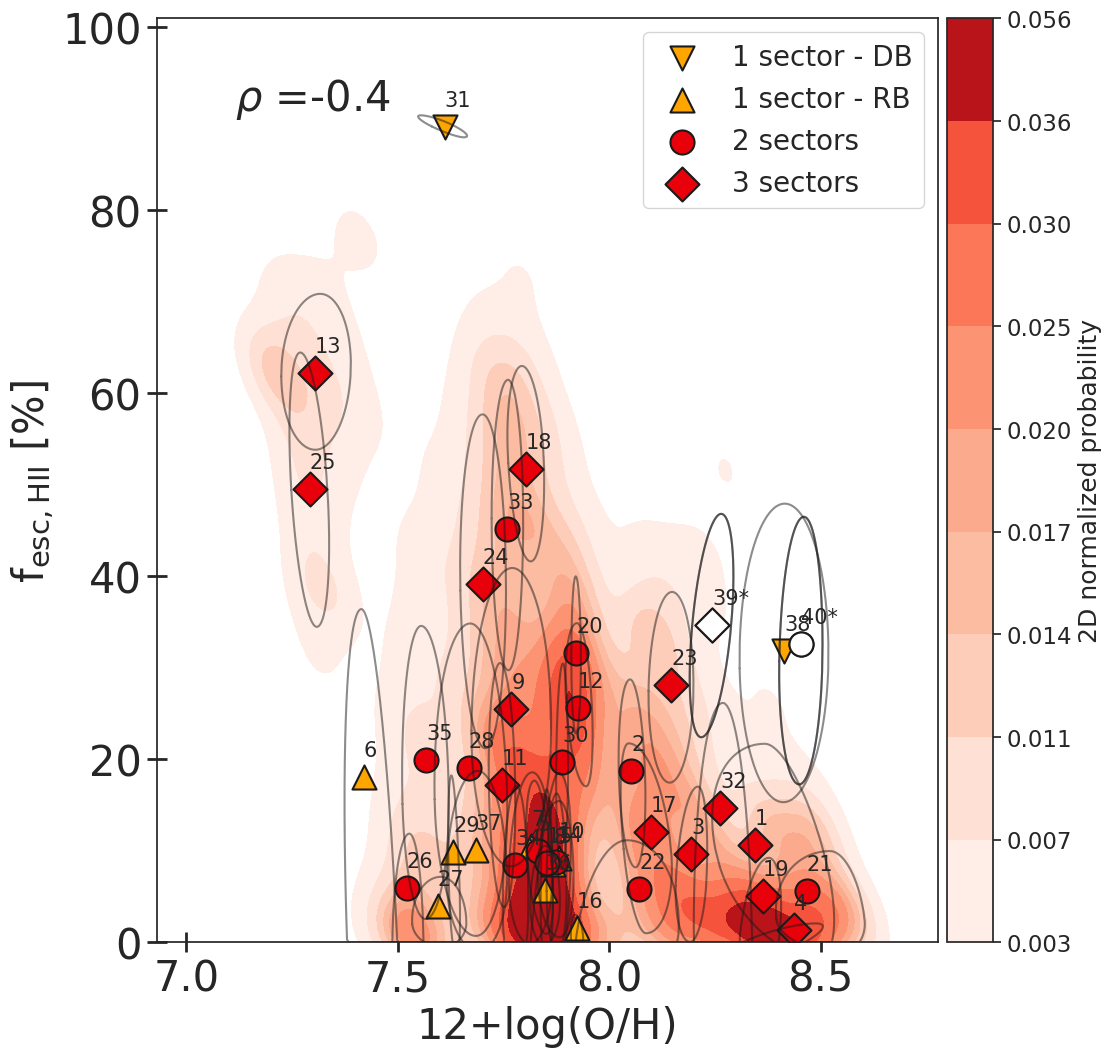}
    \includegraphics[width=9cm]{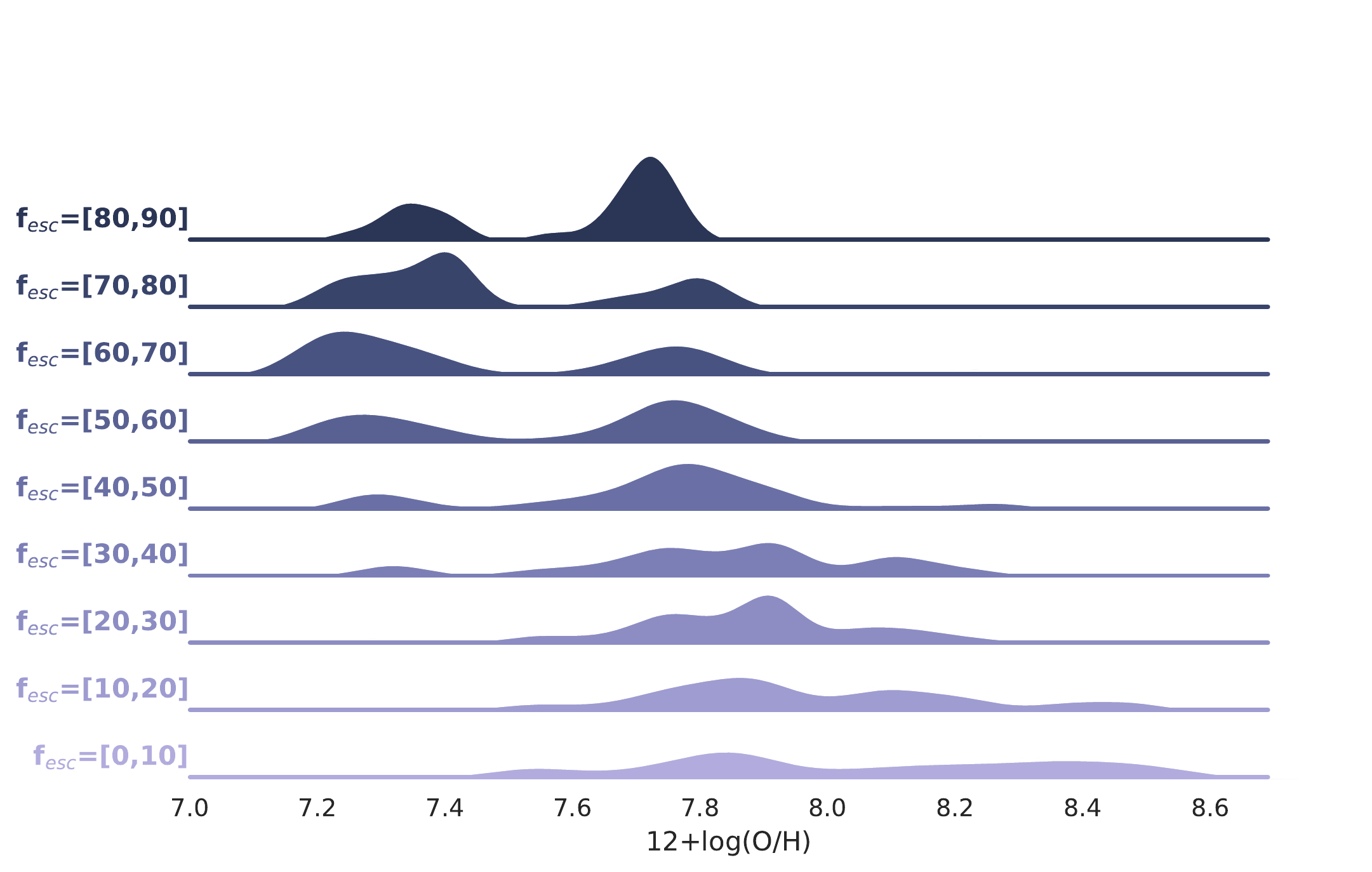}
    \caption{Escape fraction dependence on metallicity. \textbf{Top}: Escape fraction as a function of metallicity. The symbols and ellipses represent the robust means and associated uncertainties of the MCMC draws for individual galaxies, weighted by the marginal likelihood of each configuration (see Sect. \ref{rep_pdf}). The symbols and colors indicate the configuration having the highest marginal likelihood for each galaxy. The underlying KDE represents the distribution of all the MCMC draws for galaxies whose best model has N$_{\rm sectors} \geq 2$ only and $\rho$ the associated Spearman correlation coefficient. Orange triangles represent galaxies for which the best model is a single component either density-bounded (DB) or radiation-bounded (RB). Their values correspond respectively to upper and lower limits of \feschii. The white symbols that are flagged with a star correspond to two pointings in NGC\,4214, which are excluded from the KDE. \textbf{Bottom}: KDE per bins of \feschii\ for galaxies with at least two sectors.}
    \label{fesc_oh}
\end{figure}

In Fig. \ref{fesc_oh} (upper panel) we show the relation between \feschii\ (defined in Sect. \ref{cloudy_observables}) and metallicity. The inferred mean \feschii\ in our sample ranges from 0 up to around $\sim$60\%. Since we used the ionizing continuum above 13.6eV, it also includes high energy photons that may propagate unabsorbed throughout the neutral gas. Due to such photons, the inferred escape fraction is never exactly zero, although very low for some objects. In practice, even for completely radiation-bounded models, we predict escape fractions of the order of 10$^{-10}$--10$^{-6}$. We find that 11 galaxies are best modeled with one single-sector. In most of those cases (9 out of 11), the best solution (meaning, having the highest marginal likelihood; see Sect. \ref{section_code}) is radiation-bounded. When such single-sector, radiation-bounded models are favored, the resulting escape fraction is biased toward low \feschii\ values.

Density-bounded, single-sector models have been proved to not reproduce well the optical emission lines from \hii\ regions \citep[e.g.,][]{stasinska_excitation_2015, ramambason_reconciling_2020}. They are even less likely to match the observational constraints in the IR domain where some tracers arise from the PDR. For two  galaxies, however, our code favors single-sector solutions that are completely density-bounded (38: HS\,2352+2733 and 31: Tol\,1214-277). HS\,2352+2733 has only one detection (\oiii) and an upper limit on \cii, while for Tol\,1214-277 most of the detections are lines arising from the ionized region and only \cii\ and \oi\ constrains the PDR. However, the errorbars on its \cii\ and \oi\ detections are rather large, hence allowing the code to favor a completely density-bounded solution since there is not enough information to constrain the neutral gas properties in those galaxies. For such galaxies, the predicted LyC-leakage can reach high values since photons escape directly from one single density-bounded sector. Thus, for all the galaxies for which the best model has only a single-sector (whether completely radiation-bounded galaxies or completely density-bounded), the \feschii\ estimates that we derive are relatively uncertain, with a bias toward low values for radiation-bounded models and toward high values for density-bounded models. They can, however, be considered as lower and upper limits on the escape fraction. In the rest of the analysis, we focus on a subsample of 27 out of 40 galaxies for which the best models have at least two sectors and that correspond to galaxies covered in one single pointing to derive the main trends of \feschii\ with galactic parameters. In practice, we plot the KDEs and correlation coefficients for this subsample of galaxies while the individual SUEs are shown for the whole sample, including single-sector models and the two pointings in NGC\,4214. Although the KDE accounts for the full posterior probability distribution described by all the MCMC draws for each galaxy, we note that our sample remains stastically limited and that the trends derived should be taken with caution.

The predicted \feschii\ spans a wide range of values, which appear to be linked to the metallicity of the galaxy. By examining the KDE of the galaxies that are best constrained, we find a global trend of an increasing \feschii\ with decreasing metallicity. More specifically, in Fig. \ref{fesc_oh} (upper panel), the upper right part of the KDE is completely unpopulated meaning that high-metallicity galaxies have very low probability of reaching high values of \feschii. The dispersion increases with lower metallicity. While low-metallicity galaxies do not systematically exhibit leakage, if they do, they are more likely to reach high values of \feschii. In Fig. \ref{fesc_oh} (bottom panel), we show 1D-KDE plots of the metallicity by bins of \feschii\, for galaxies having at least 2 sectors. While this representation is quite sensitive to the intrinsic metallicity distribution of our sample, it provides a clear visual representation of the \feschii-metallicity relation. Although we find rather large distributions of metallicity for most of the \feschii\ bins, we see that the main peaks in metallicity are shifted toward lower metallicities when the escape fraction increases. We note that the current sample is poorly populated in the low-metallicity regime (only 3 galaxies below $\oh=7.5$) and does not allow us to derive a robust trend with metallicity. Nevertheless, we find enhanced values of \feschii\ in the low-metallicity regime, in which all the highest predicted \feschii\ are found.  Future observations of galaxies in the low-metallicity regime may reveal objects with small \feschii. The large scatter observed among galaxies that fall within a given metallicity bin may be linked to dependences in other parameters, which will be further examined in Sect. \ref{section_dependencies}. The metallicity-\feschii\ trend we present here remains clearly visible when we force the number of sectors to be the same for all galaxies (i.e., no selection of the best configuration) as shown in Fig. \ref{fesc_oh_nfixed}. 
By looking at the individual PDFs represented by the SUEs, we confirm that the highest mean \feschii\ are associated with the lowest metallicities in our sample. Specifically we infer $\feschii >40\%$ in 5 galaxies with metallicities below 7.8. and $\feschii >50\%$ in 2 galaxies with metallicities below 7.3. However, galaxies with \feschii\ below 10\% are found at all metallicities above 7.5 and are not preferentially associated with high metallicities. We also note that some objects exhibit somewhat high escape fractions on the high-metallicity part of our plot (e.g., 39: NGC\,4214-c and 40: NGC\,4214-s). However, those two objects correspond to separate pointings within one galaxy instead of an integrated measurement. They will be further examined in Sect.\,\ref{discussion}. 

The \feschii-metallicity relation presented in this section is reminiscent of the PDR covering factor vs. metallicity relation presented in \cite{cormier_herschel_2019}, that show that the PDR covering factor (calculated as a common scaling factor of PDR lines) correlates with metallicity. It is also globally consistent with observational studies in which LyC-leaking galaxies are predominantly found at low-metallicity, especially for the most extreme escape fractions. \citep[e.g.,][]{Leitet_2013, Izotov_2018a, Flury_2022b}. In the recent Low-redshift Lyman Continuum Survey, which gathers 89 LyC measurements and associated metallicity estimates, \cite{Flury_2022b} find that the strong LyC-emitters with measured $\fesc\ > 10\%$ are all found at metallicities below 8.1. However, they report a significant scatter in the \fesc-metallicity relation and do not observe any significant correlation for their whole sample. We further discuss possible explanations of this scatter in Sect. \ref{discussion}. 

Several, possibly combined, effects might lead to the \feschii-metallicity trend that we observe: first, the UV opacity of the diffuse gas in the ISM is strongly dependent on metallicity. This UV opacity is controlled, in particular, by the dust content of galaxies since dust grains have been shown to absorb a significant ($< 50\%$) part of the LyC radiation in \hii\ regions \citep{Inoue_2001}. Hence, in the low-metallicity regime where the dust-to-gas mass ratio drops steeply \citep{remy-ruyer_gas--dust_2014, galliano_nearby_2021}, the ISM becomes more transparent and photons can travel larger distances. Second, the escape of LyC photons from \hii\ regions also depends on the properties of the host molecular clouds in which stars form. Molecular clouds are optically thick in the UV at all metallicities. Therefore, the escape fraction strongly varies with their covering factor. Recent simulations \citep[e.g.,][]{Kimm_2019, yoo_origin_2020} have shown that the neutral gas covering factor is sensitive to both metallicity and stellar age effects with ionizing photons escaping more efficiently from their birth cloud when the metallicity is lower. \cite{yoo_origin_2020} emphasize that the metallicity of the cloud in which stars form seems to be an important driver of the cloud disruption timescales. They find that in the low-metallicity runs, stars break out of their birth cloud within a few megayears only while at higher metallicities they tend to stay enshrouded for a longer period.

\subsection{Dependences on other parameters}

We further examine the \feschii-metallicity relation by looking at possible secondary dependences that would explain the dispersion of this relation. In particular, we focus on examining the effect of various properties of the radiation sources: stellar mass, SFR, specific SFR (sSFR= SFR/M$_*$), age of the stellar burst, and X-ray-to-stellar luminosity ratio. We revisit the relation from Fig. \ref{fesc_oh} but split our sample by bins of values for secondary parameters (see Fig. \ref{dependencies_split_kde}). We note that the various parameters that we examine are all interdependent. Correlations between those fundamental parameters have been extensively studied in samples of local and high-redshift galaxies \citep[e.g.,][]{Nakajima_2014, Duarte_Puertas_2022} and, more specifically, among compact star-forming galaxies \citep{Izotov_2021,Flury_2022b}. In particular, the stellar mass, SFR, and sSFR are functions of the metallicity. Hence, we find that the highest \feschii\ that we infer are associated with low-metallicity, low stellar masses, and high sSFR. We find that those three parameters are the main drivers of large \feschii\ predicted in our sample. However, our analysis is limited by the observational bias of our sample in which only highly star-forming galaxies are populating the low-metallicity and low mass regime. With all those correlated parameters, it is not possible to infer which one has the strongest effect on \feschii. We also find that stellar populations with ages above 3Myr and an X-ray-to-stellar luminosity ratio above $0.5\%$ ($-2.3$ in log) seem to be associated with enhanced escape fractions (see panel 4 and 5 from Fig. \ref{dependencies_split_kde}). These results motivate the study of the \feschii\ dependences on galactic parameters other than metallicity, which we now examine in more details.

\label{section_dependencies}
\begin{figure*}[htb]
    \centering
    \includegraphics[width=6cm]{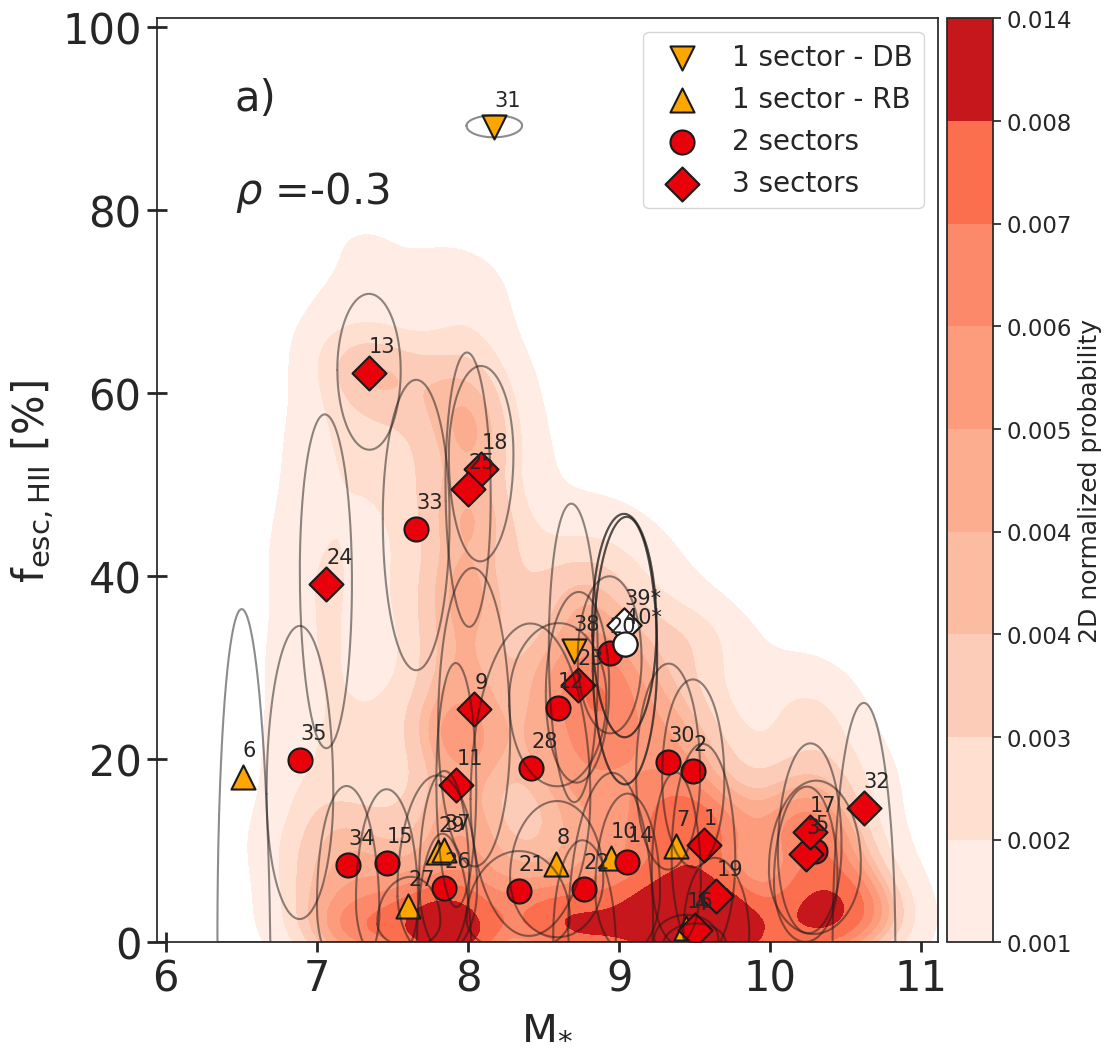}
    \includegraphics[width=6cm]{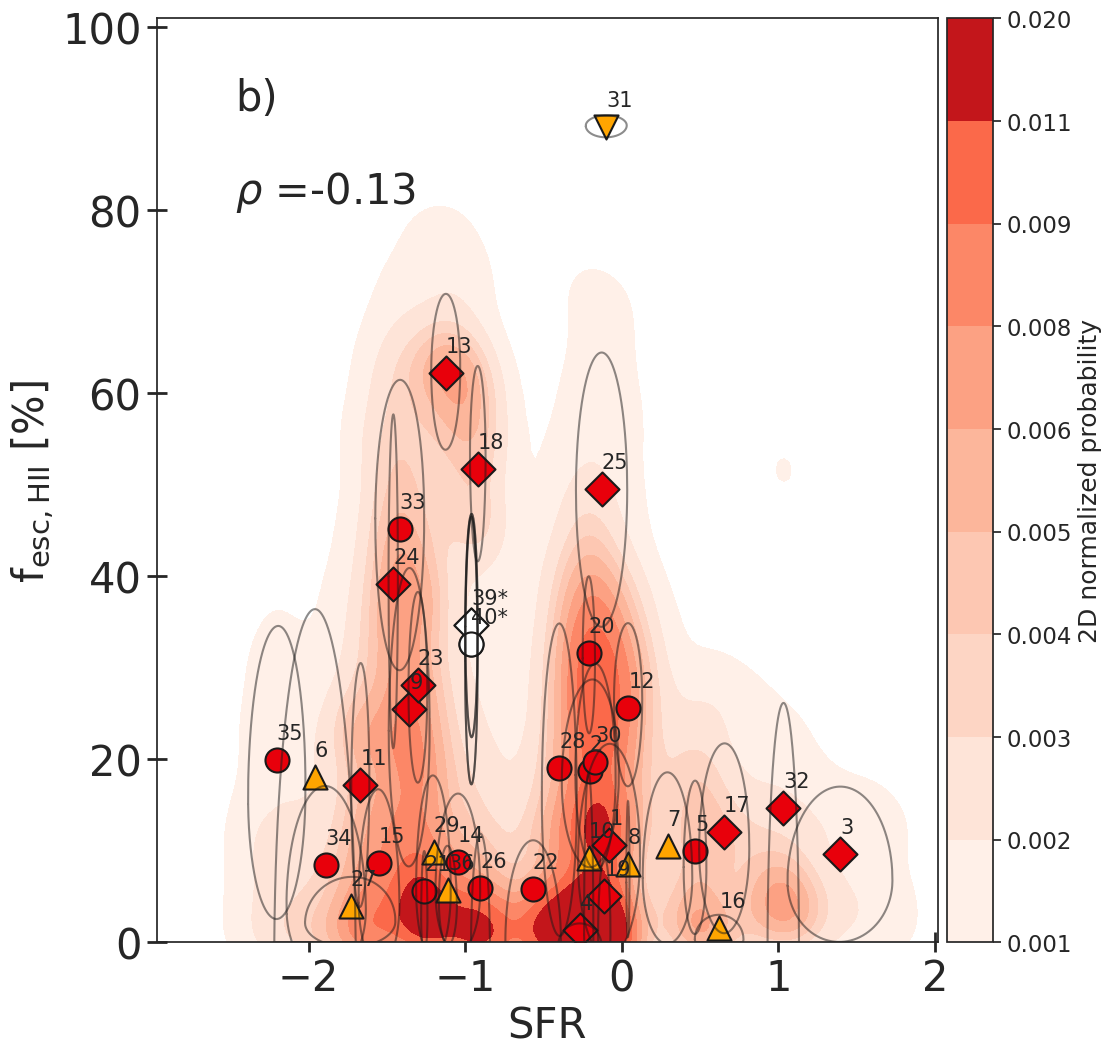}
    \includegraphics[width=6cm]{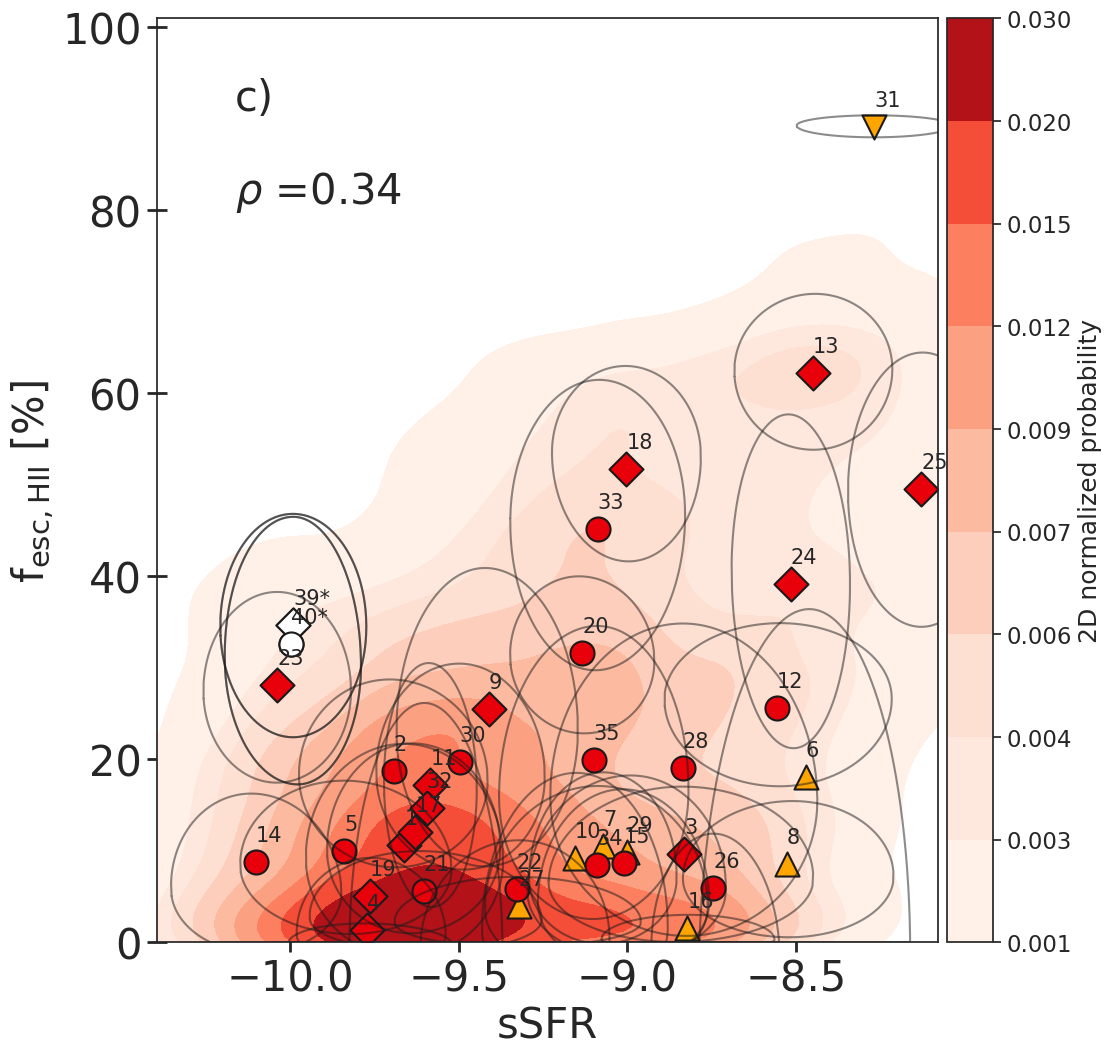}
    
    \includegraphics[width=6cm]{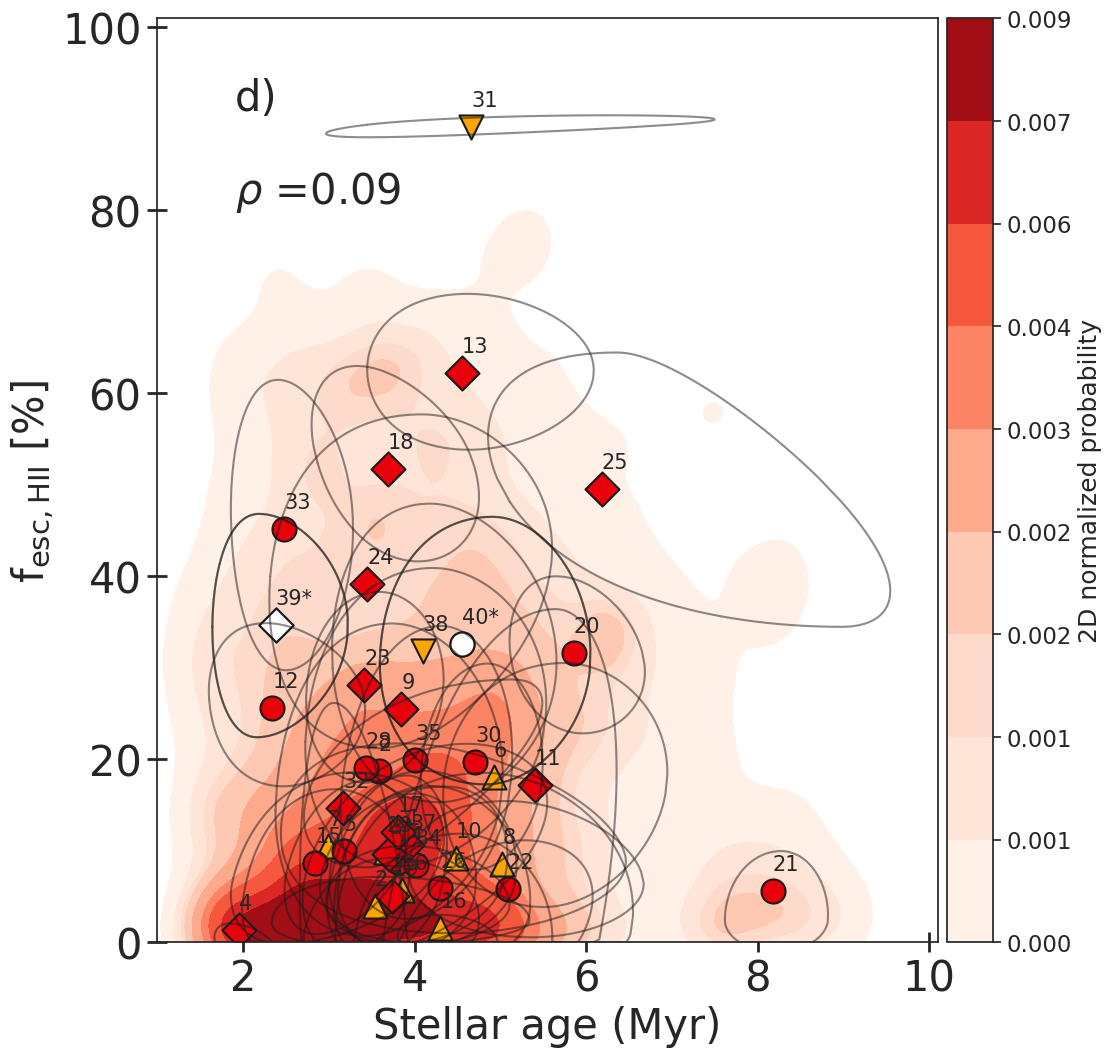}
    \includegraphics[width=6cm]{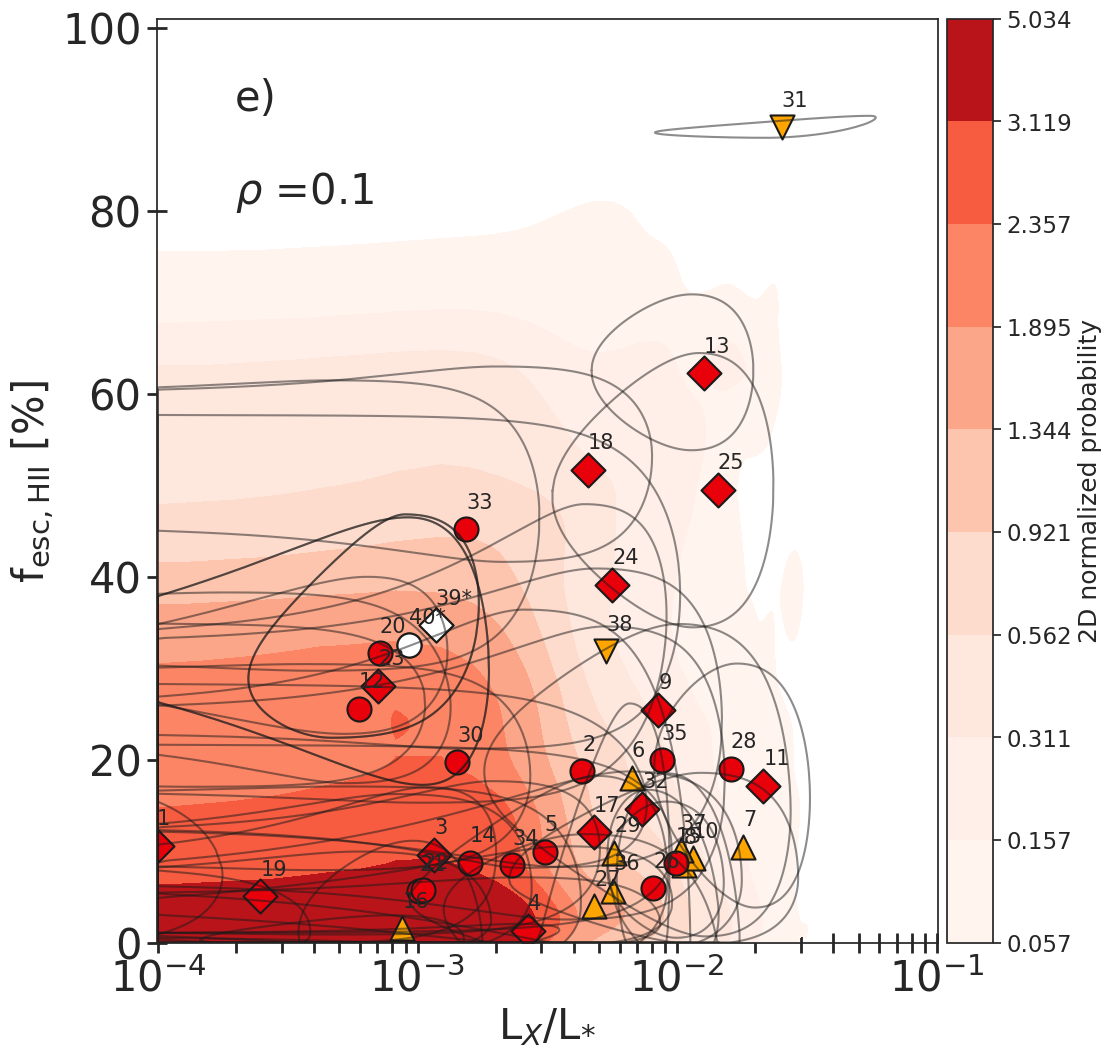}
    \includegraphics[width=6cm]{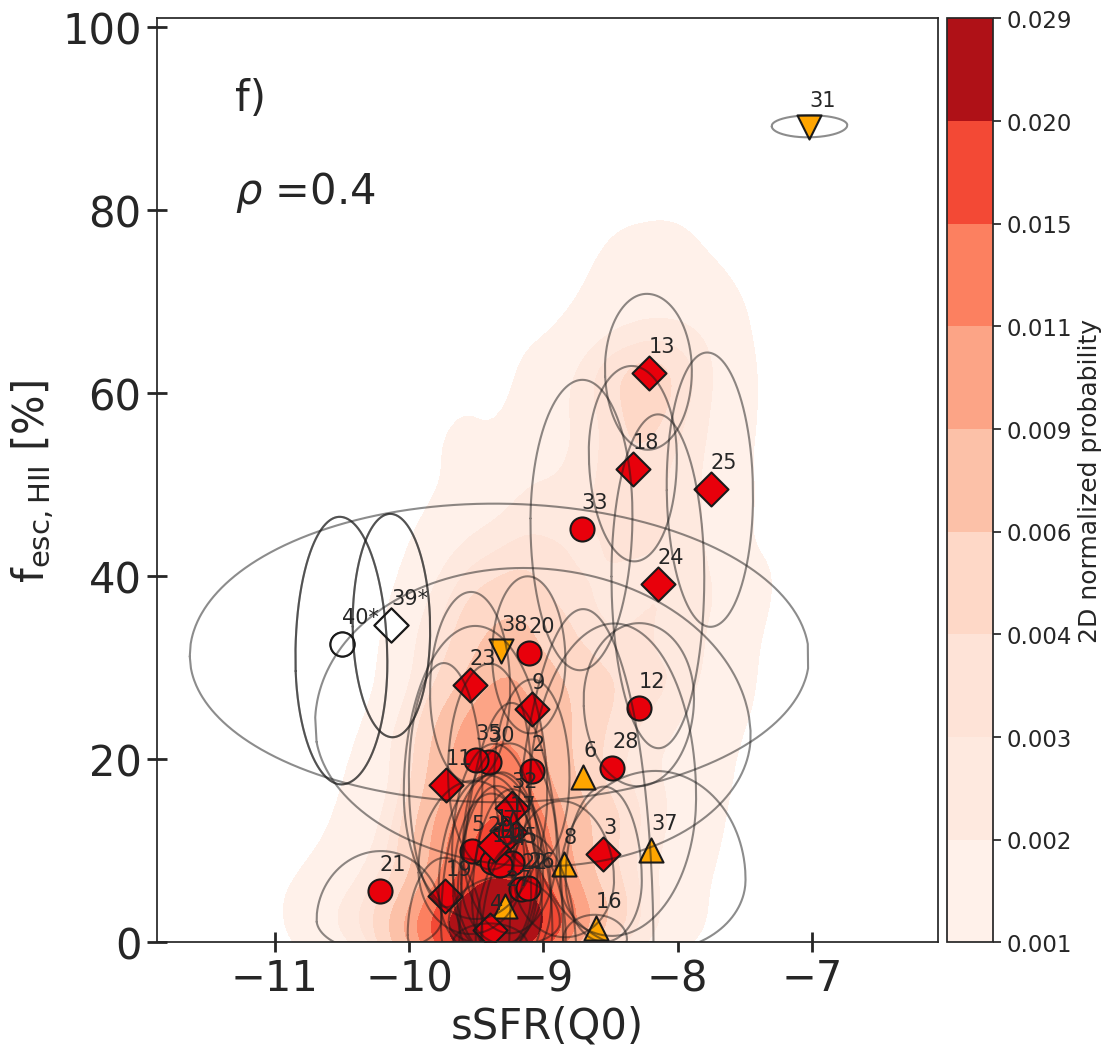}
    
    \caption{Escape fraction dependences on measured and inferred galactic parameters. \textbf{Top pannel}: Escape fraction vs. some measured galactic parameters: M$_{*}$, SFR and measured sSFR. \textbf{Bottom pannel}: Escape fraction vs. some inferred galactic parameters: stellar age, Lx, sSFR($Q$(H$^0$)). The underlying KDE represents the distribution of all the MCMC draws for galaxies with N$_{\rm sectors} \geq 2$ only and $\rho$ the associated Spearman correlation coefficient. Orange triangles represent galaxies for which the best model is a single component either density-bounded (DB) or radiation-bounded (RB). The white symbols that are flagged with a star correspond to two pointings in NGC\,4214, which are excluded from the KDE.}
    \label{dependencies_SUE}
\end{figure*}

\subsubsection{Stellar mass}
It has been suggested that galaxies with shallow gravitational wells, associated with low stellar masses, would be good candidates for potential LyC-leakage. In particular, this \fesc-M$_*$ trend has been observed in samples with masses down to 10$^8$\,M$_{\odot}$ \citep{Leitet_2013, Izotov_2018b}. However, a recent follow up study carried out in \cite{Izotov_2021} targeting galaxies with masses below 10$^8$\,M$_{\odot}$ finds no clear correlation between the global measured LyC escape fraction and the stellar masses.

In Fig. \ref{dependencies_SUE} (panel a), we find that large \feschii\ tend to be associated with low stellar masses, although with a large scatter. All the galaxies with masses above 10$^9$\,M$_{\odot}$ have mean escape fractions below 40\%. On the other hand, galaxies with masses below 10$^9$\,M$_{\odot}$ do not always have high escape fractions. We note that this effect might be dominated by the mass-metallicity distribution of our sample in which lower masses are associated with the lowest metallicity. The correlation between \feschii\ and metallicity is, nevertheless, much clearer. Globally we observe a weak anti-correlation between \feschii\ and M$_*$, although with a large scatter, especially in the low-mass regime. 

\subsubsection{Star-formation rates}

Recent simulations of photon leakage have shown that the process is a feedback driven mechanism \citep{Trebitsch_2017, Kimm_2017, Kimm_2019, Kim_2018,kim_2019, Kim_2021, kakiichi_lyman_2019, yoo_origin_2020}. Although the main driver, whether stellar or supernovae (SN) feedback, remains elusive, the escape fraction is predicted to increase drastically when the star-formation efficiency is enhanced. 

Surprisingly, we find no correlation between \feschii\ and SFR (see panel b from Fig. \ref{dependencies_SUE}). The highest values of \feschii\ are associated with moderately star-forming galaxies in our sample. This absence of trend can be understood by looking at the KDE in Fig. \ref{dependencies_split_kde} (1$^{\rm st}$ panel); the highest SFRs in our sample correspond to the highest metallicities, which do not allow high amounts of photon leakage. On the other hand, we find that \feschii\ increases with increasing sSFR. We show on Fig. \ref{dependencies_SUE}  the evolution of \feschii\ with respect to measured sSFR (panel c) and estimated sSFR from $Q$(H$^0$) (panel f) as defined in Sect. \ref{section_fesc_Z_SFR}. Although the trend is already visible when using the sSFR measurements from the literature, we find, not surprisingly, that applying a correction to account for the photon leakage through density-bounded sectors increases the dynamical range of the relation. Similarly to the M$_*$-\feschii\ relation and SFR-\feschii\ relation, the sSFR-\feschii\ relation is possibly driven by metallicity effects as the lower-metallicity sources are those with higher sSFR.

\subsubsection{Stellar ages}

Stellar evolution and the feedback mechanisms responsible for clouds dispersal may also impact the escape fraction. In our code, we only model very young bursts of ages below 10\,Myrs that represent the bulk of the \hii\ regions that dominate the emission. We expect both rotation and binary stars to affect our prediction for the \hii\ region escape fractions by producing more LyC photons for a longer period of time \citep[][and reference therein]{choi_mapping_2020}. In particular, \cite{choi_mapping_2020} finds that at low-metallicity (<1/3\,Z$_{\odot}$) models that include binary stars produce 60\% more ionizing flux compare to model without.

In Fig. \ref{dependencies_SUE} (panel d), we see that most of the stellar populations in our sample are dominated by young starbursts with ages of 2 to 6 Myr. These timescales suggest that leakage might occur at very early stage of stellar evolution. We find no clear trend of \feschii\ with stellar age, although large \feschii\ above $\sim$50\% seem to be associated with ages greater than 3--4\,Myrs. We emphasize, however, that large ages are not necessarily associated with leaking galaxies. As previously mentioned in Sect. \ref{section_fesc_Z_SFR}, we note that the age and metallicity are interdependent, with larger ages tending to be associated with the lowest metallicities.

While our findings are consistent with scenarios in which photons might escape at very young ages (e.g., \citealt{kakiichi_lyman_2019}), we emphasize that our modeling assumptions based on a single burst stellar population prevent us from examining the effect of an older underlying population. Other sources of feedback (e.g., turbulence, SN feedback) might also play a role at later stellar ages. While at early ages, the feedback is dominated by pressure and ionization mechanisms from young \hii\ regions, older stars can also contribute through stellar winds, especially in the Wolf-Rayet phase. We note that this stellar evolution also depends on metallicity effects with low-metallicity stars producing more ionizing photons over larger timescales \citep{eldridge_binary_2017, 2018_Xiao_bpass}. Finally, older generation of stars might be responsible for creating low-density channels in the ISM, either due to mechanical feedback of the starburst \citep[e.g.,][]{Ma_2020} or through SN explosions. The presence of such low-density channels carved by previous generations of stars might favor leakage associated with young bursts that formed in regions where gas has already been partially cleared out \citep[e.g.,][]{Ma_2020}.
Nevertheless, it is worth noting that the stellar ages we derive are consistent with the disruption timescales found in simulations from \cite{yoo_origin_2020}. They are also very close to the feedback timescales of massive O/B stars measured in resolved giant molecular clouds (GMCs) from nearby spiral galaxies \citep[e.g.,][]{Chevance_lifecycle_2020, Chevance_2022, Kim_Jaeyeon_2022}. \cite{Chevance_2022} also point out that the efficiency at which the injected energy couples with the GMCs is low (a few tens of percent) and conclude that most of the energy and momentum produced by young stars escape outside of the host GMC, which is consistent with the high \feschii\ we derive.

\subsubsection{X-ray sources}

As discussed in Sect. \ref{section_overview}, several bright X-ray sources have been observed in our sample. Due to our simple modeling assumption and the suite of lines used in this study, the X-ray luminosities we derive are not tightly constrained. One of the motivations to include additional feedback mechanisms is the presence in our sample of extreme emission lines emitted by species with very high ionization potentials (>54eV): \oiv\ or \nev. Such lines are, nevertheless, not numerous and not available for all our sample (see Fig. \ref{nlines}). As a consequence, for most of the galaxies, the 1$\sigma$ SUEs are large and the X-ray luminosity is not tightly constrained.

We note, however, that the X-ray-to-stellar luminosity ratio is also constrained by emission-lines in the neutral gas. This is due to the penetration of X-ray photons deep into the neutral and molecular gas because of their small photoionization cross-sections. Such photons are able to deposit their energy at large A$_V$, beyond the ionization front of \hii\ regions. Because of this effect, our code is able to place an upper limit on the X-ray-to-stellar luminosity ratio based only on the emission lines arising from the neutral gas (e.g., \feii\, \silii, \oi), even when no high ionization lines are detected. In particular, photoelectric heating on dust grains contributes to the heating of neutral gas in our models and is controlled by the assumed dust-to-gas mass ratio (fixed in our models) and total stellar luminosity (constrained by \hii\ region emission lines). Hence, the additional X-ray contribution to neutral gas heating is limited by the emission lines in the neutral gas. 

Although we derive uncertainties on the absolute log L$_X$ above 1 dex for most galaxies, we predict a relatively well-constrained X-ray luminosity for a few galaxies. Specifically, we obtain a total X-ray luminosity of log L$_X$= 40.33$_{-0.49}^{+0.53}$ erg\,s$^{-1}$ for I\,Zw\,18.\footnote{The interval corresponds to the  highest density interval at 94\%.} The value obtained for I\,Zw\,18 is slightly higher but compatible with the measurement of 1 $\times$ 10$^{40}$ erg\,s$^{-1}$ (intrinsic X-ray luminosity; 0.3--10keV) reported in \cite{Kaaret_Feng_2013}. In our case, the predictions are for the total X-ray luminosity integrated over the whole energy range covered by the multicolor blackbody assumed as an input.  While the X-ray-to-stellar luminosity ratio may allow us to pinpoint some of the most extreme escape fractions in our sample, we note, however, that no global trend seems to be visible, with some galaxies having little to no contribution coming from X-ray being associated with large escape fractions. 

\subsection{Tracers of  density-bounded regions and escaping photons}
\label{Section_tracers}

Our code allows us to disentangle the emission of tracers that emit throughout different phases of the ISM. This is in particular the case of the \cii 158$\mu$m line, which is emitted both in the ionized and neutral gas phase. This impacts the \oiii88$\mu$m/\cii158$\mu$m (O3C2) diagnostic as a tracer of \fesc\, which is quite popular in high-redshift studies \citep[e.g.,][]{harikane_large_2020, Katz_2021}, as it can be detected with ALMA up to redshifts above 6. This O3C2 ratio has been extensively studied with results indicating that high O3C2 values could be explained by specific physical conditions such as a high burstiness parameter \citep{Vallini_2021} or a low covering factor of PDR \citep{harikane_large_2020}. Such a trend has also been reported in \cite{Chevance_2016} and \cite{Chevance_2020c} that find that O3C2 is sensitive to the PDR filling factor defined for each pixel in their resolved analysis of 30Dor in the LMC. Nevertheless, at galactic scale, \cite{cormier_herschel_2019} find little correlation of O3C2 with their estimated PDR covering factors in the DGS.
We now discuss the complex origin of the \cii\ line and how it might affect the predictive power of the O3C2 line ratio. We then discuss the potential use of other line ratios to trace the presence of density-bounded regions and of associated LyC-photon leakage from \hii\ regions, both in the IR and optical domain.

\subsubsection{The complex origin of \cii 158$\mu$m emission}

\begin{figure}[htb]
    \centering
    \includegraphics[width=9cm]{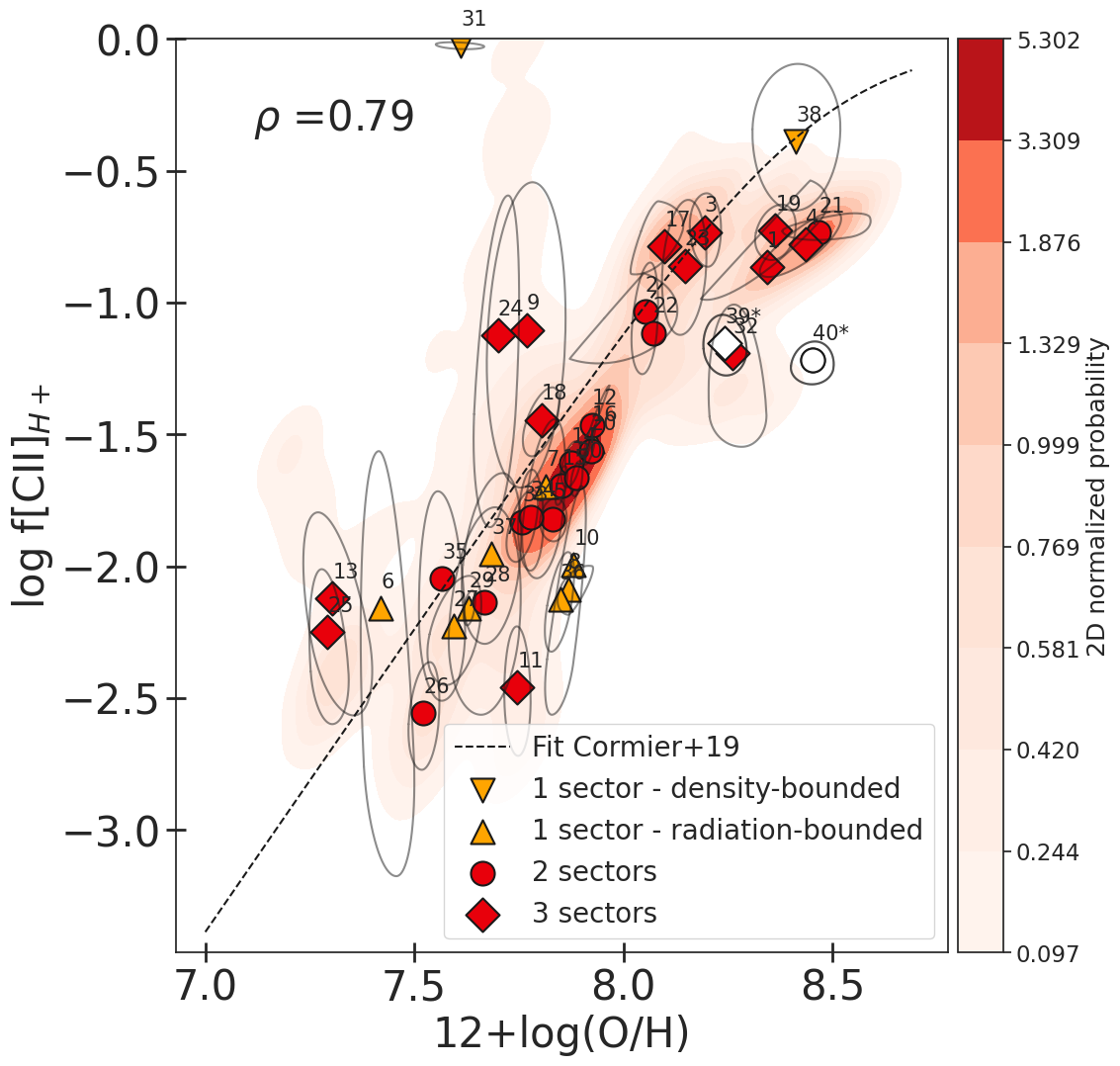}
    \caption{Fraction of \cii158$\mu$m in the ionized phase vs. metallicity. The single-sector, density-bounded models represented by orange stars pointing down are PDR-deprived; all their \cii\ emission arises from the ionized gas (log f\cii$_{\rm H+}\sim$ 0). The underlying KDE represents the distribution of all the MCMC draws for galaxies with N$_{\rm sectors} \geq 2$ only and $\rho$ the associated Spearman correlation coefficient. The dashed line represents the relation derived by \cite{cormier_herschel_2019} for the DGS, assuming a fixed electron density n$_e$=30cm$^{-3}$.}
    \label{fCII}
\end{figure}

The \cii158$\mu$m line is an important diagnostic, which constrains in principle the neutral gas reservoir and the potential presence of density-bounded regions. However, the interpretation of this tracer is challenging as \cii\ may be produced both throughout the neutral and ionized regions \citep[e.g.,][]{Oberst_2011, Ramos_Padilla_2021, Ramos_Padilla_2022}. In this section, we intend to disentangle the \cii\ contribution from different phases by examining predictions for the fraction of \cii\ luminosity arising from the ionized phase of the ISM (f\cii$_{\rm H+}$). 

To do so, we weigh the cumulative \cii\ line luminosity by the relative abundance profile of the hydrogen in each phase. For the ionized phase, we hence scale the cumulative line luminosity by the H$^{+}$ relative abundance profile and we integrate over the depth for each sector. The total \cii\ luminosity arising from the ionized phase is obtained by summing over all the sectors of the model. We then derive f\cii$_{\rm H+}$ by computing the ratio of the prediction for \cii\ in the ionized phase over the prediction for \cii\ total luminosity.

In Fig. \ref{fCII}, we note that for one of the 1-sector, density-bounded galaxies (31: HS\,2352+2733), log f\cii$_{\rm H+}=0$ since the gas in those models is completely ionized. For the other galaxies, log f\cii$_{\rm H+}$ varies between $-3$ to $-0.5$ and its value correlates with the metallicity of the galaxy. This confirms previous results from \cite{croxall_2017} and \cite{cormier_herschel_2019} that show that the fraction of \cii\ arising from the ionized phase decreases at low metallicities. We also find values in the same dynamical range as previously found in \cite{cormier_herschel_2019}  but with a somewhat smaller dispersion. According to \cite{croxall_2017}, this result might be linked to the presence of harder radiation fields in low-metallicity environments. As a result, the relative abundance of C$^{2+}$ in the ionized gas may increase with respect to the C$^{+}$ abundance, leading the \cii\ emission to be dominated by the neutral phases. We discuss the implications on the use of \cii\ as a potential tracer of the escape fraction in Sects. \ref{section_ir_tracers} and \ref{discussion}. 

\subsubsection{IR line ratios}
\label{section_ir_tracers}
\begin{figure*}[htb]
    \centering
    \includegraphics[width=6cm]{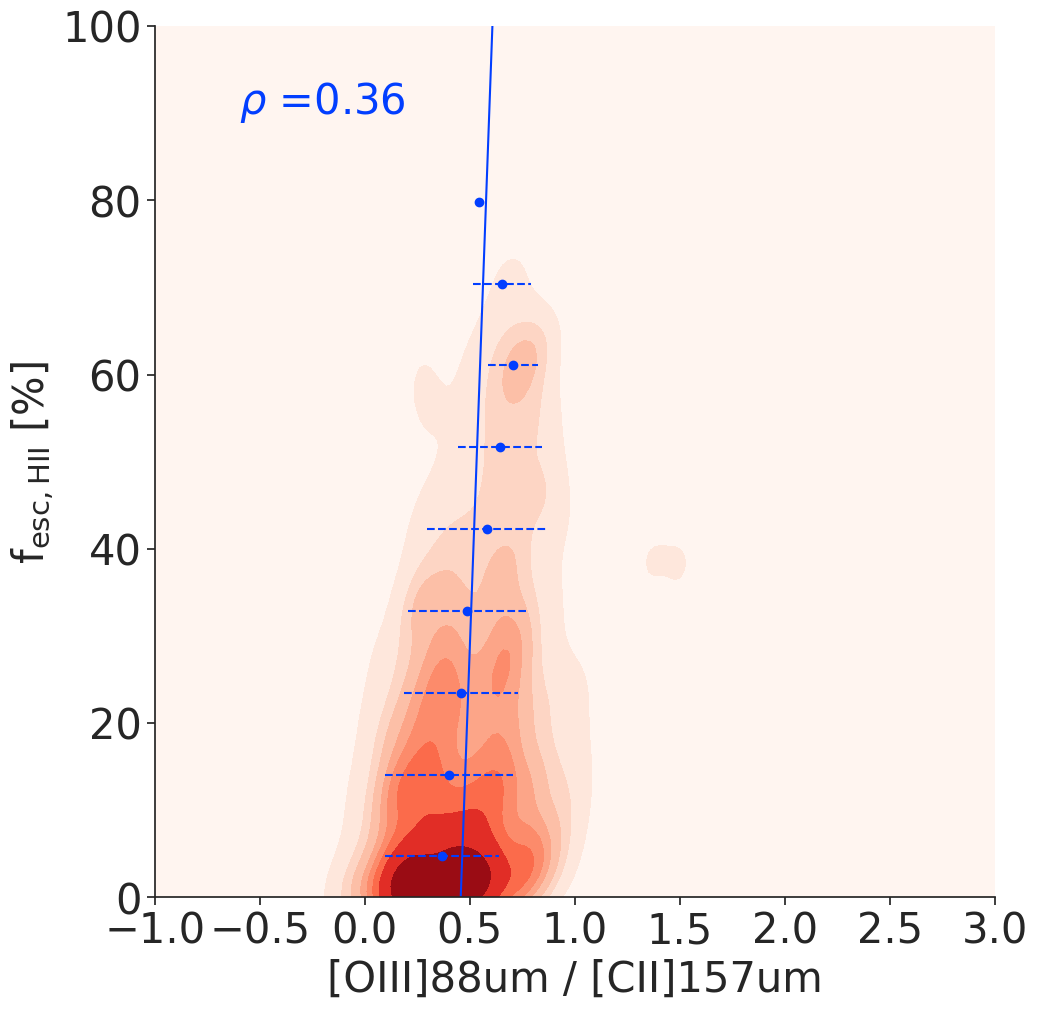}
    \includegraphics[width=6cm]{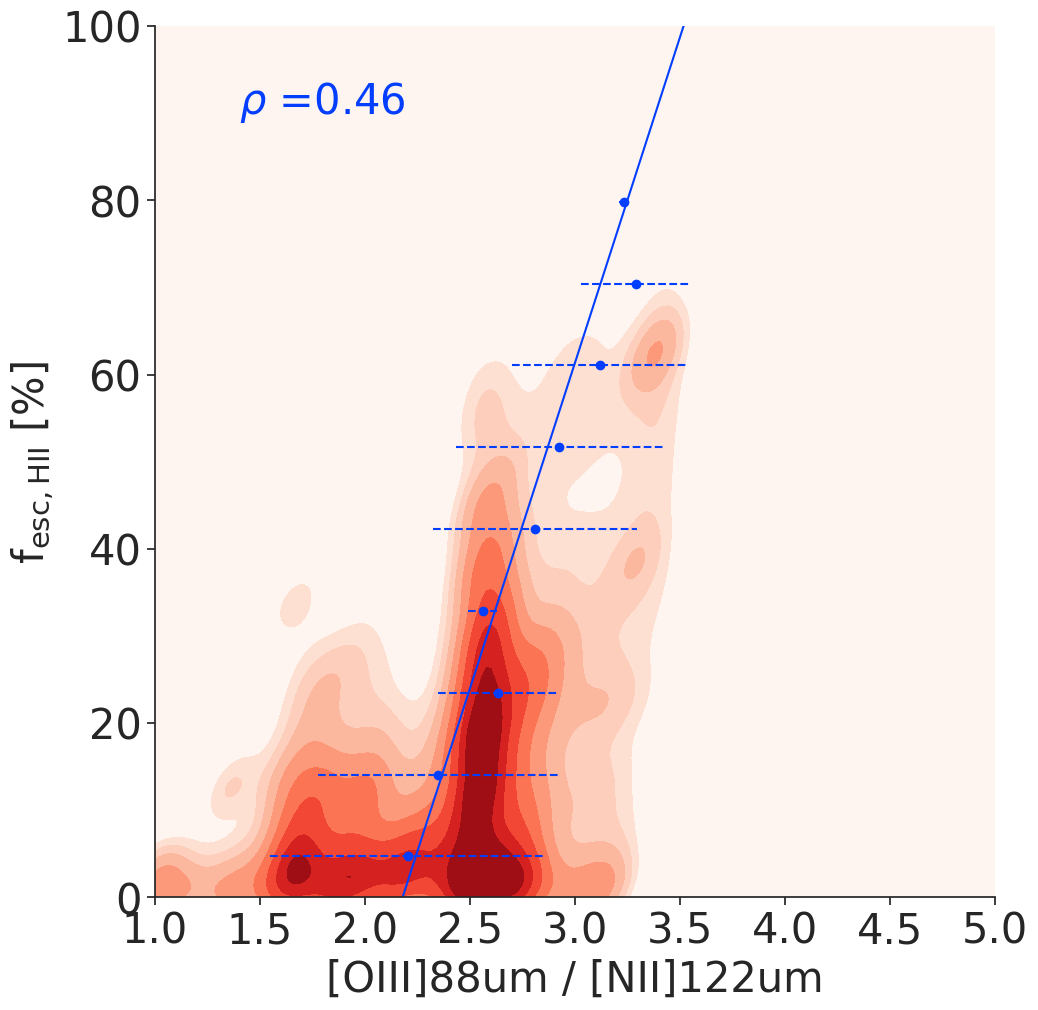}
    \includegraphics[width=6cm]{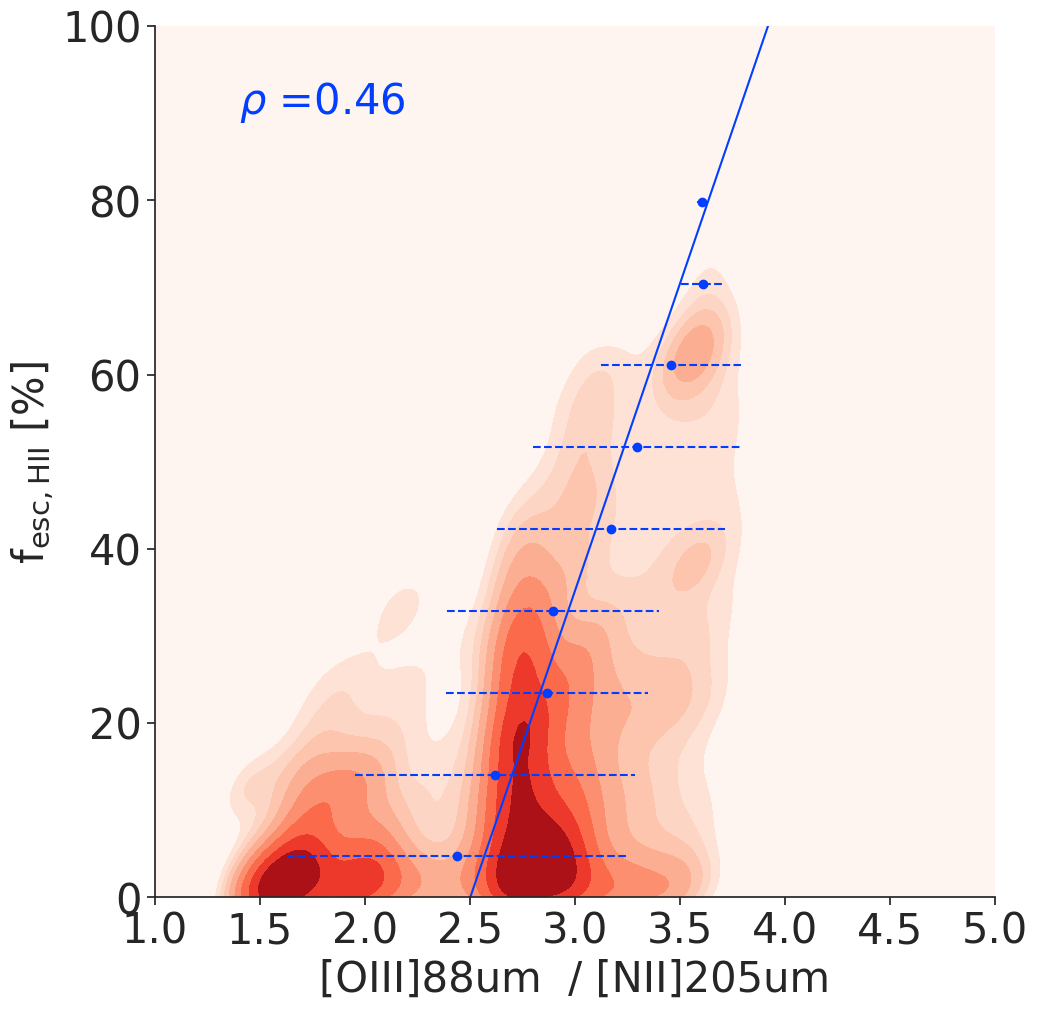}
    \caption{KDE showing \feschii\ as a function of IR ratios. The KDE represent the distribution of all the MCMC draws for galaxies with N $\geq$ 2 only and $\rho$ the associated Spearman correlation coefficient. For this, we group the MCMC draws for our selected sample by bins of \feschii\ and fit each bin with a gaussian $\mathcal{N}(x_{0}, \sigma^2)$. The blue dots and dashed lines respectively represent the gaussian mean $x_0$ and the gaussian width $\sigma$. The plain blue line shows a weighted linear regression applied to the gaussian centers, accounting for the widths of each bin. The corresponding fit values are provided in table \ref{fit_IR_lines}.}
    \label{fesc_ratio_alma_kde}
\end{figure*}
    
\begin{figure*}[htb]
    \centering
    \includegraphics[width=6cm]{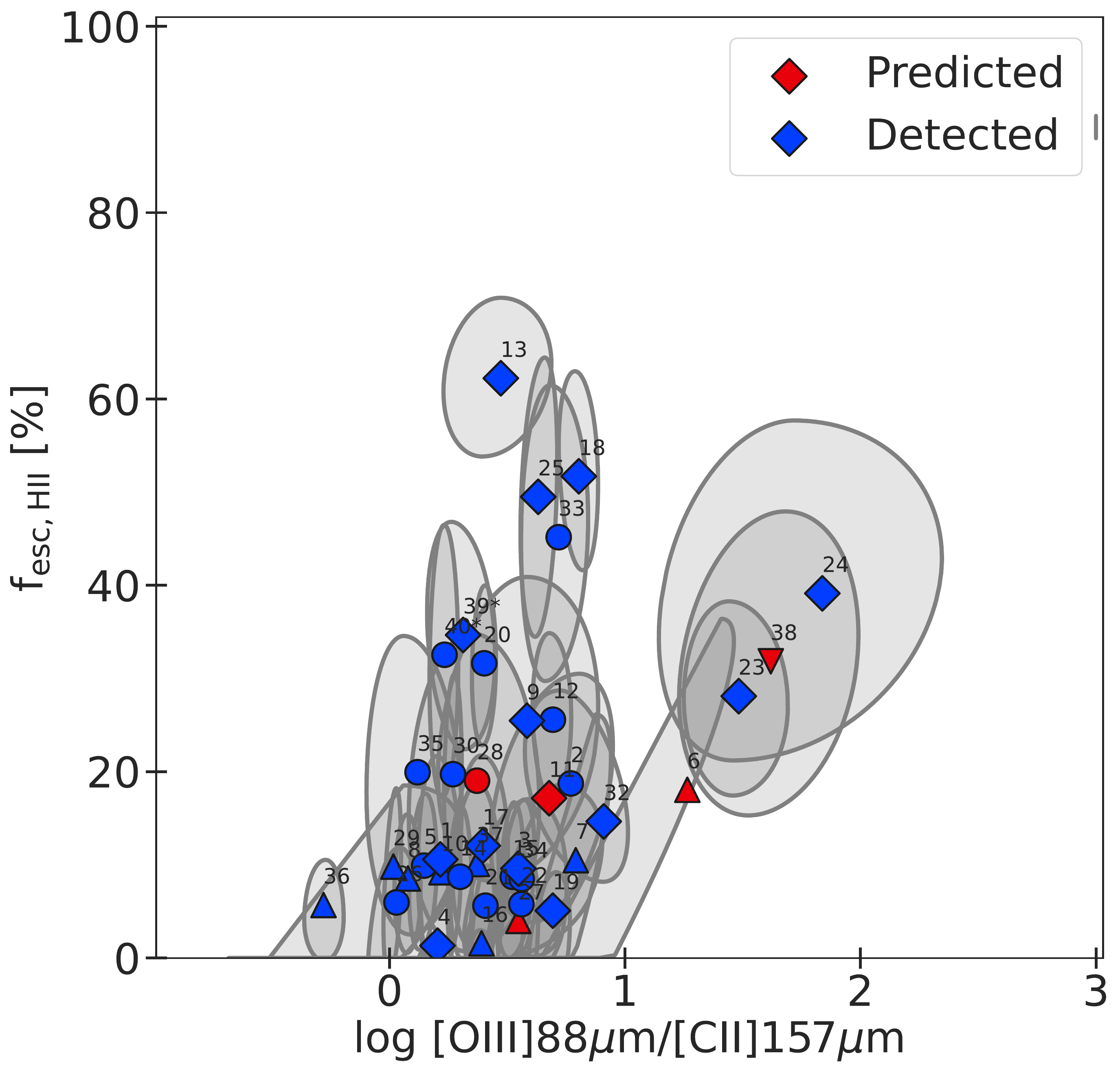}
    \includegraphics[width=6cm]{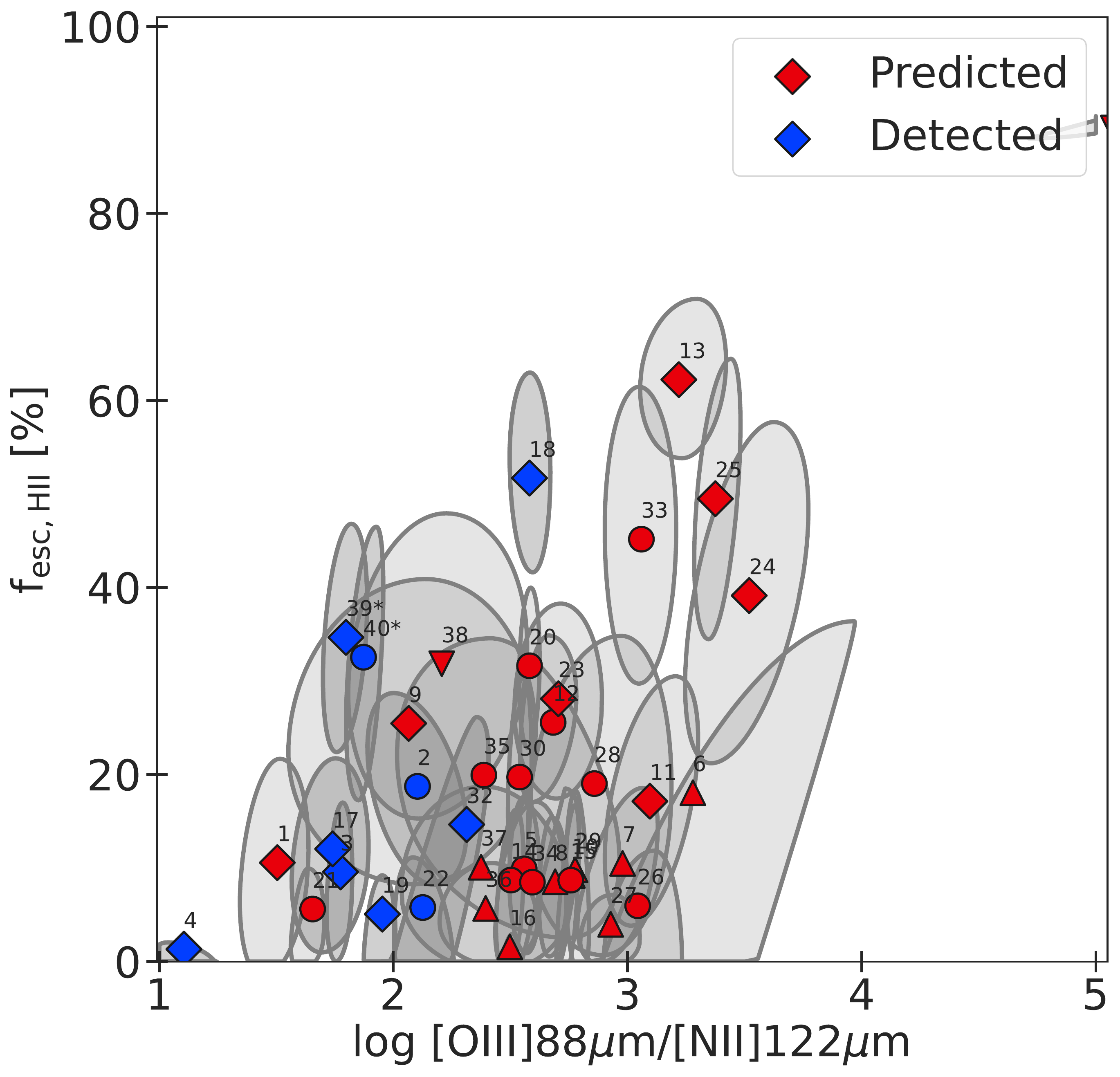}
    \includegraphics[width=6cm]{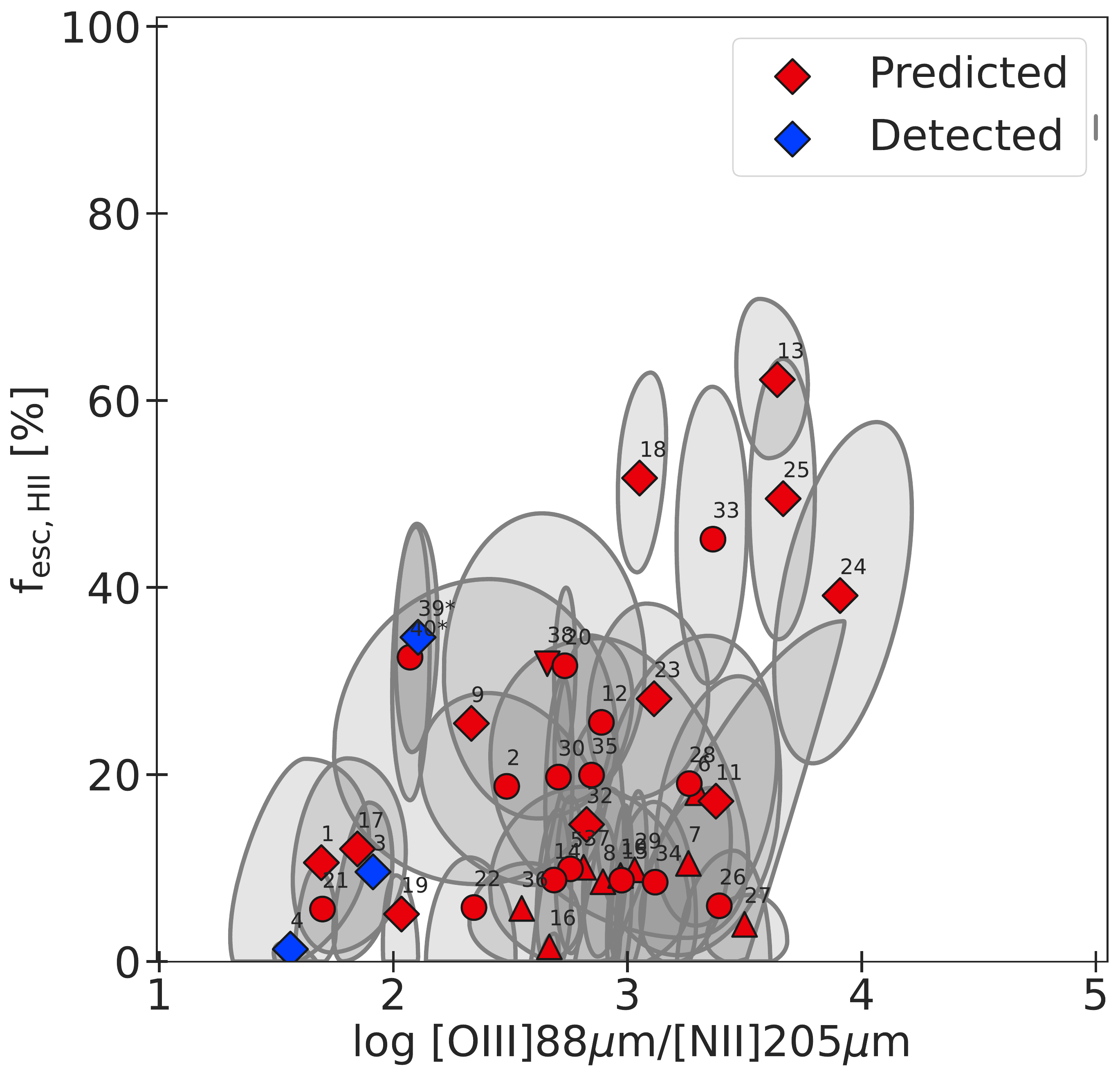}
    \caption{Individual SUEs showing \feschii\ as a function of IR ratios. We note that Tol\,1214-277 (31) most often lies outside of the range shown in x-axis. This is due to the lack of constraints on this galaxy for which the best model (one sector, completely density-bounded) predicts unrealistically low emission from ions with low ionisation potential.}
    \label{fesc_ratio_alma_sue}
\end{figure*}

\begin{figure*}[htb]
    \centering
    \includegraphics[width=6cm]{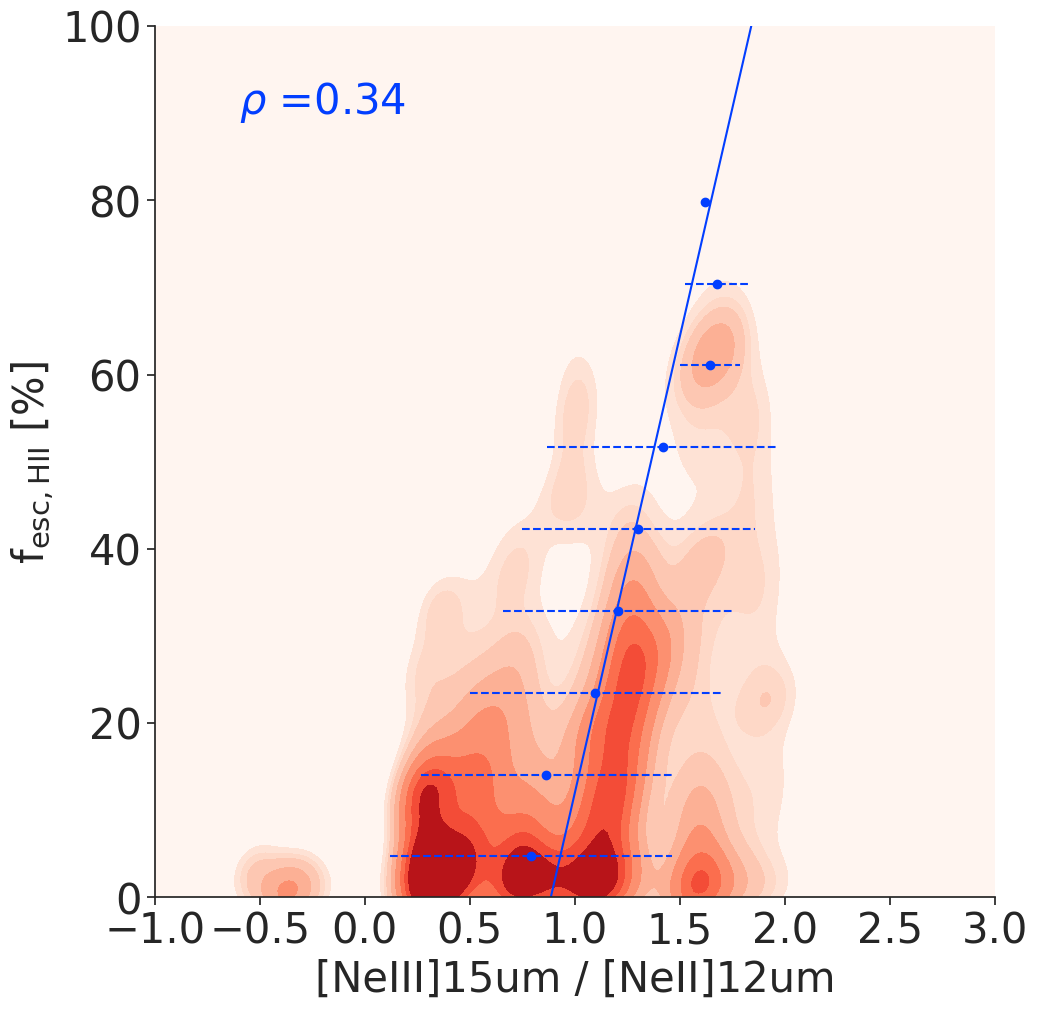}
    \includegraphics[width=6cm]{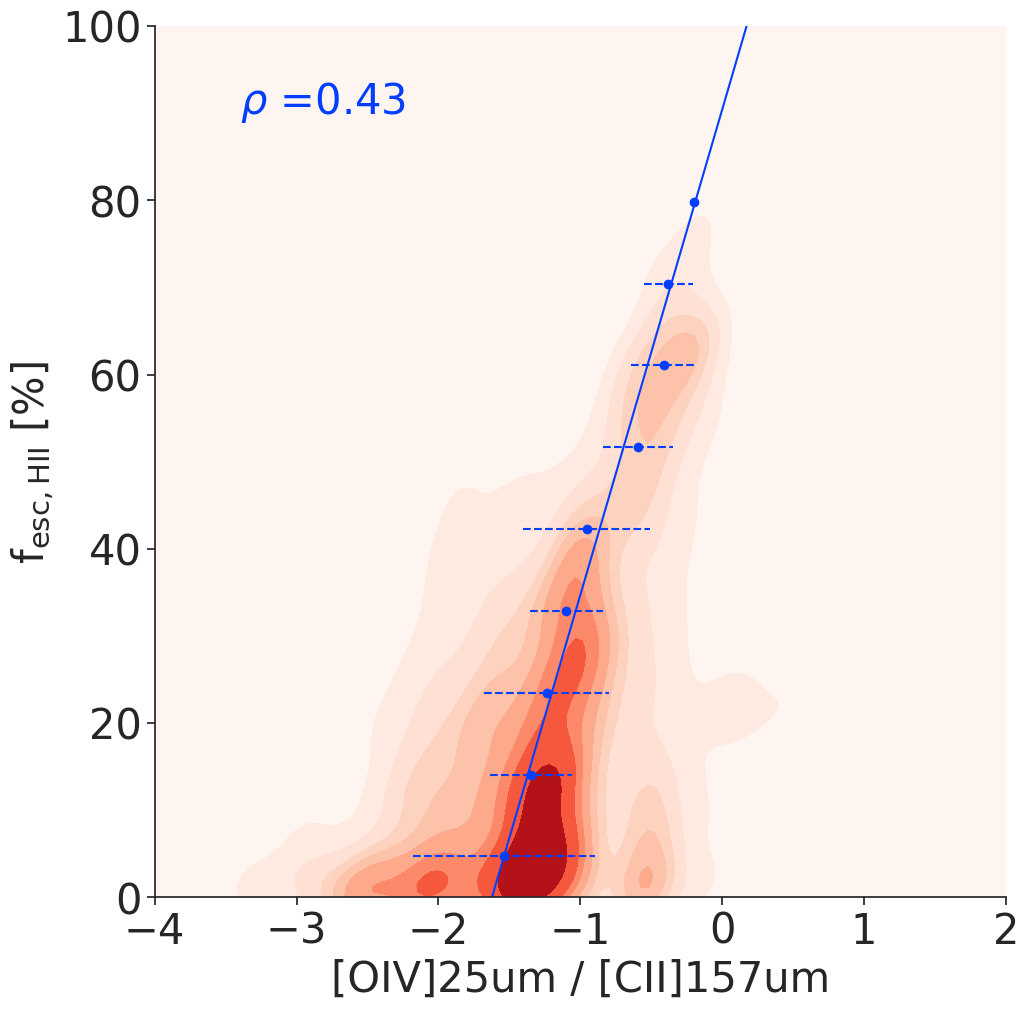}
    \includegraphics[width=6cm]{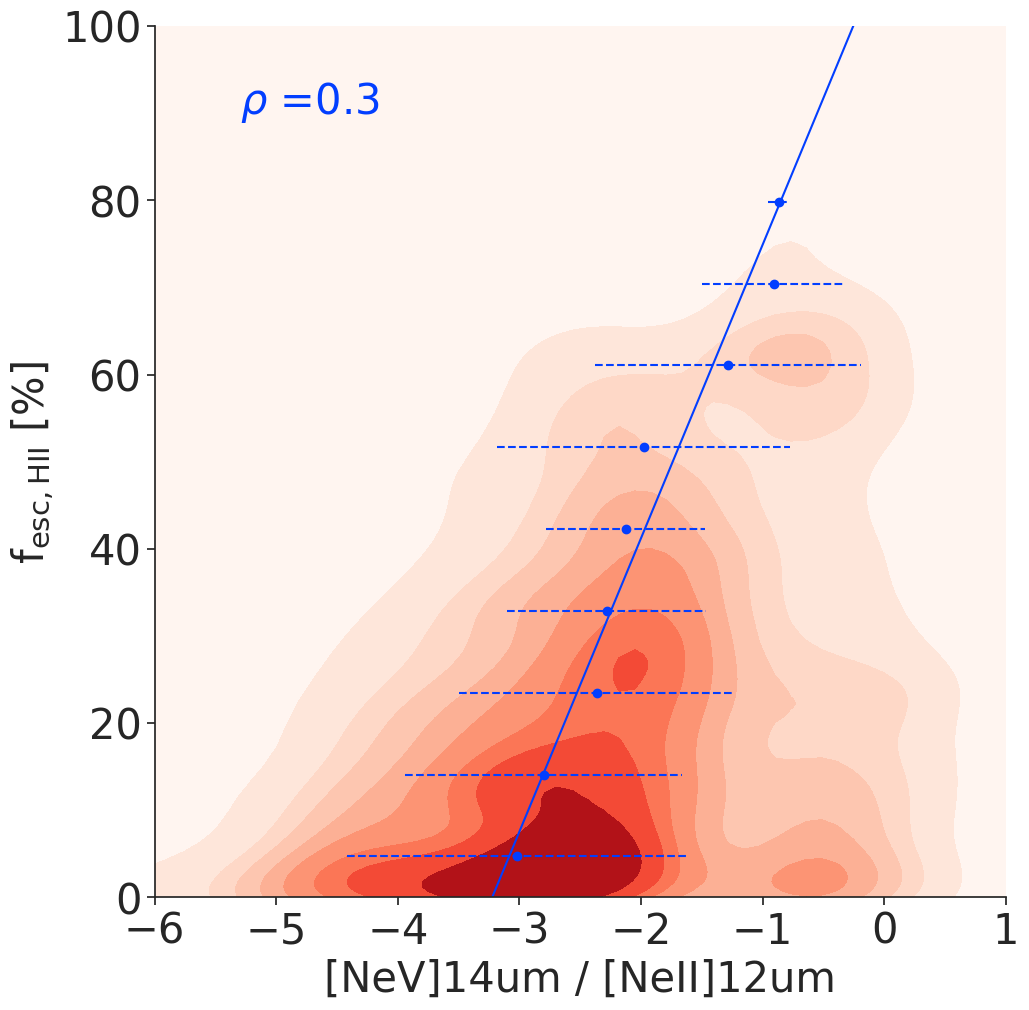}
    \caption{KDE showing \feschii\ as a function of line ratio involving ions with high ionization potentials.}
    \label{fesc_ratio_highIP_kde}
\end{figure*}

\begin{figure*}[htb]
    \centering
    \includegraphics[width=6cm]{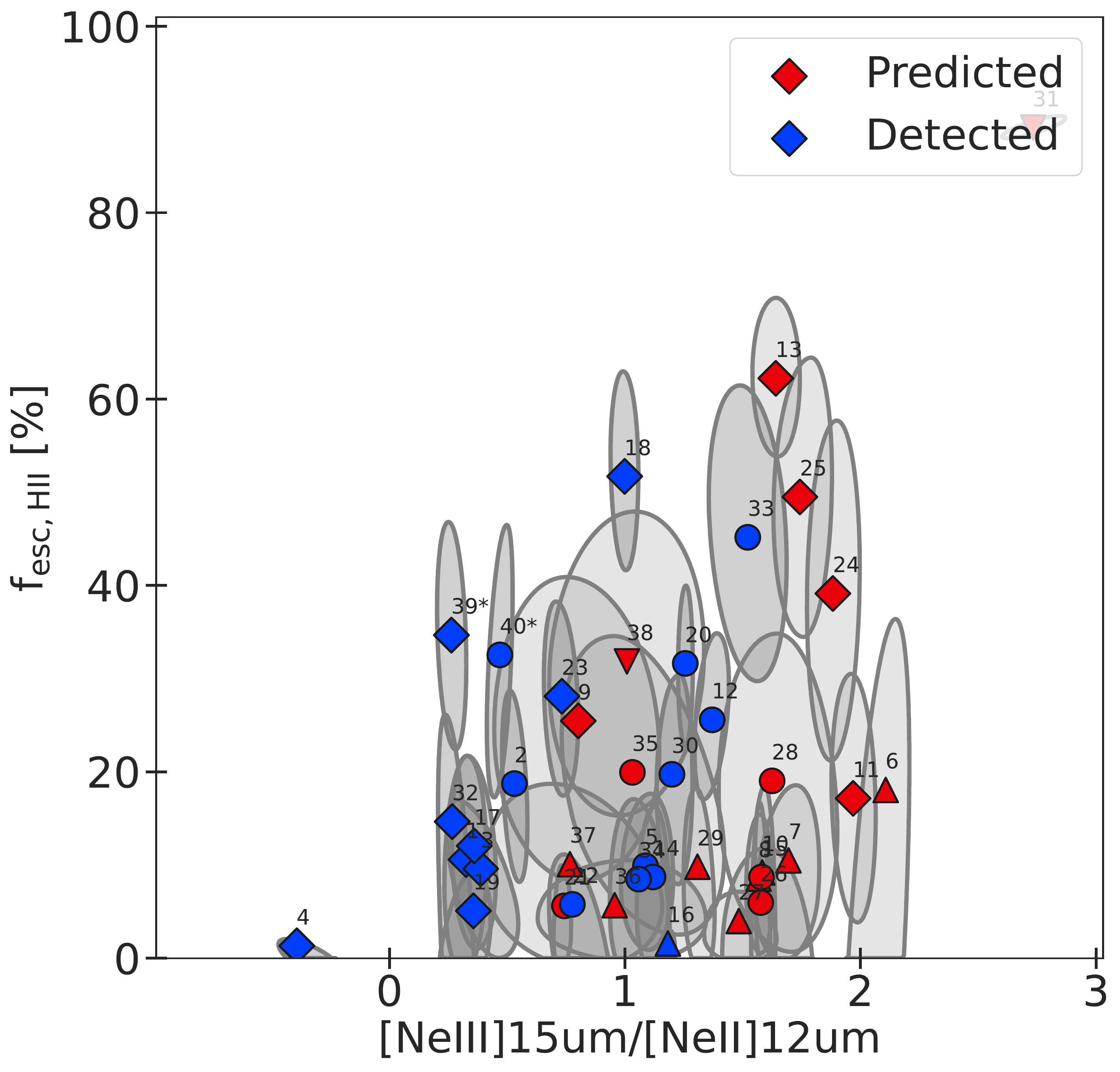}
    \includegraphics[width=6cm]{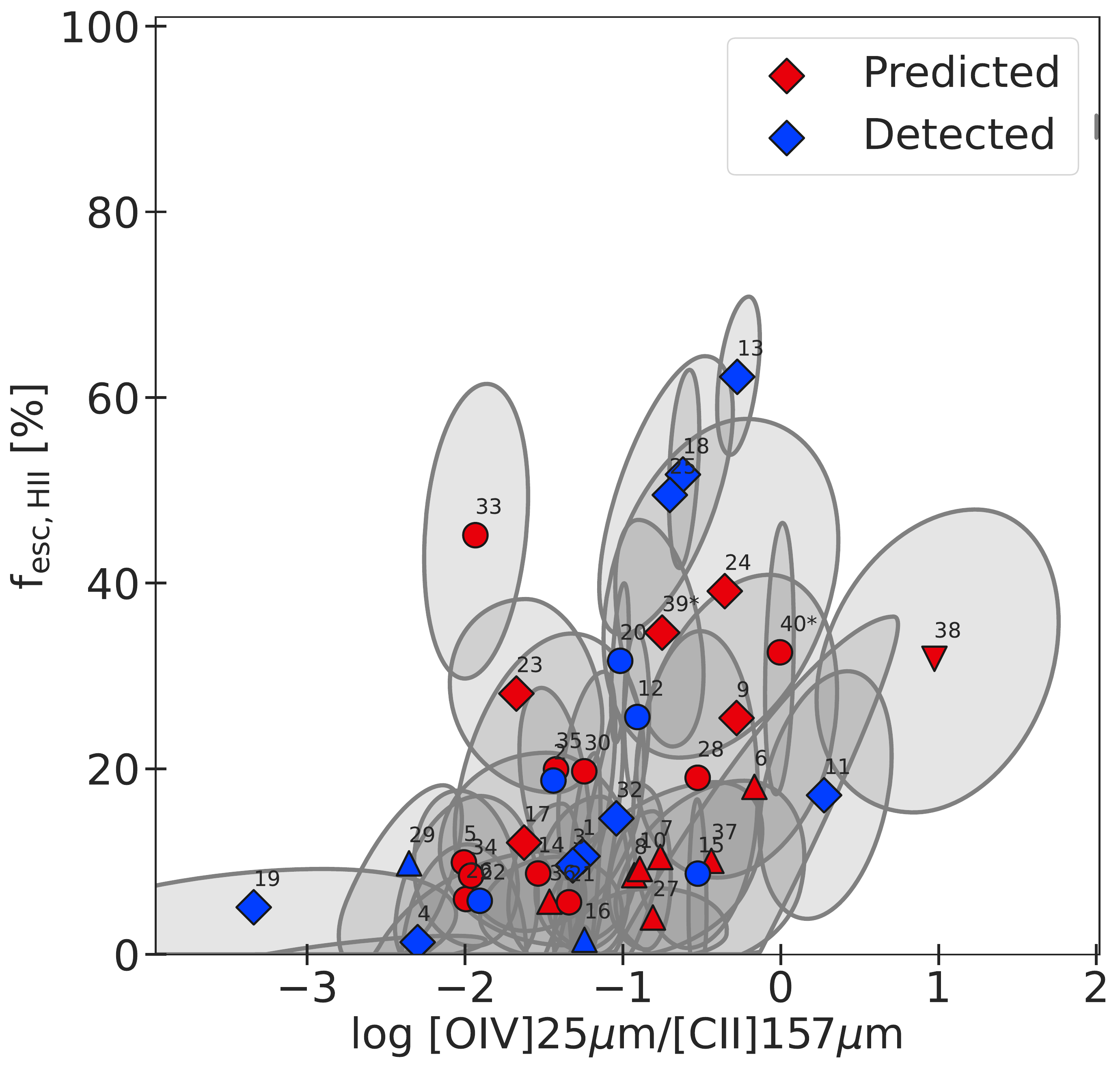}
    \includegraphics[width=6cm]{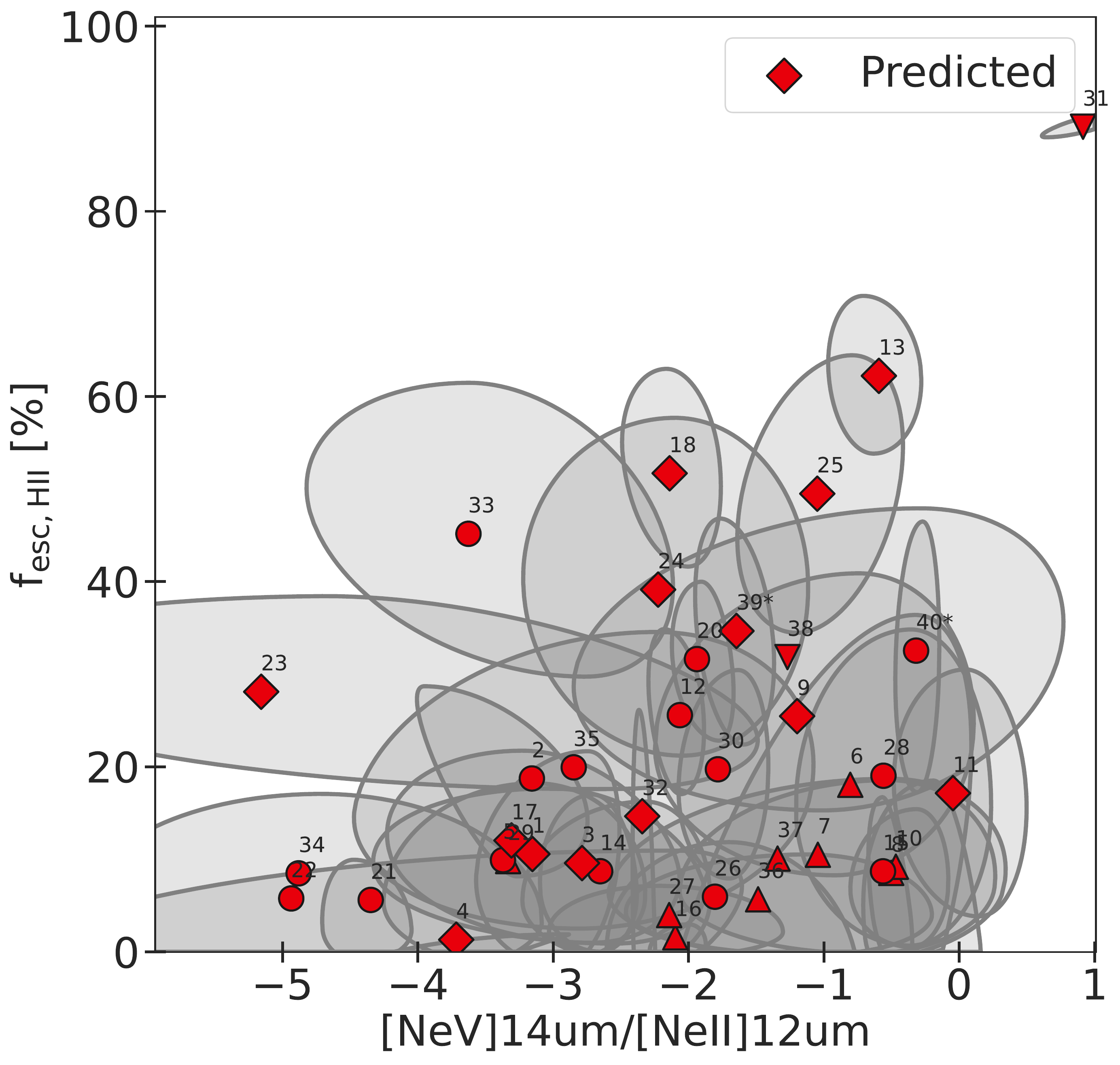}
    \caption{Individual SUEs showing \feschii\ as a function of line ratio involving ions with high ionization potentials.}
    \label{fesc_ratio_highIP_sue}
\end{figure*}

We now provide some observed or unobserved tracers of the escape fraction based on what is predicted by our multisector models. The MCMC algorithm used in our code (see Sect. \ref{section_code}) does not only sample the PDFs of primary parameters but can also infer the PDFs of secondary parameters and unobserved tracers. In particular, we can extract predictions for unobserved line fluxes and the associated uncertainties that we represent using the same formalism as for primary parameters (i.e., SUEs and KDE, see Sect. \ref{rep_pdf}).

Line emission produced by ions with different ionization potentials allow us to trace different physical depths. Hence, ratios involving lines emitted closer to the stars, in highly excited gas and lines emitted further away in regions near or past the ionization front provide an efficient way to probe the structure of the gas around a cluster of stars. This method can help identify galaxies in which numerous density-bounded regions are present. We expect such regions, which are bounded by the lack of gas, to have a reduced emission of low ionization lines emitted further away from the stars. We show in Figs. \ref{fesc_ratio_alma_kde}, \ref{fesc_ratio_alma_sue}, \ref{fesc_ratio_highIP_kde}, and \ref{fesc_ratio_highIP_sue} the line fluxes predicted by our models for all lines, regardless of whether or not they are detected. We note, however, than when a line is detected and used as a constraint, the prediction should be close to the observed value.

In Figs. \ref{fesc_ratio_alma_kde} and \ref{fesc_ratio_alma_sue} we show how classical ratios of IR lines relate with \feschii. 
In particular, we examine the O3C2 ratio, which involves the \cii 158$\mu$m whose complex origin was discussed in the previous subsection. In the framework of our multisector modeling, we do find a correlation between O3C2 and \feschii. Although this correlation is not clearly seen by looking at individual SUEs, it is visible on the KDE representing the whole sample (i.e., the concatenated MCMC draws for all the galaxies with at least 2 sectors). We emphasize, however, that for most galaxies the dynamical range of the O3C2 in our sample is rather narrow ($<2$dex), which do not allow us to easily identify leakers based on their O3C2 ratio.

Although the \nii\ lines are fainter and more challenging to detect, we find that \oiii88$\mu$m/\nii122$\mu$m and \oiii88$\mu$m/\nii205$\mu$m (O3N2) are better tracers of \feschii\ both because of the tightness of the relation and of their wider dynamical range (over 3 orders of magnitudes), which makes them potentially interesting diagnostics for observation-based classification. These correlations might be explained by the fact that both \oiii\ and \nii\ originate from \hii\ regions but are emitted by ions with different ionization potentials (35.1eV for O$^{2+}$ and 14.5eV for N$^{+}$). This means that \oiii\ is predominantly emitted in a highly irradiated region close to the stars while \nii\ arises from the outer part of the \hii\ region, and hence their ratio depends on the depth of each sector. On the other hand, \cii\ has a complex origin and arises both from \hii\ regions and PDRs.  As shown in Fig. \ref{fCII}, only a fraction of \cii\ arises from the ionized gas. As one goes deeper in the neutral gas phase beyond the ionization front, the \cii\ emission increases while \feschii\ remains unchanged. Hence, in low-metallicity environments in which \cii\ is predominantly emitted in the neutral and molecular phases, the O3N2 ratio could be a better tracer of density-bounded \hii\ regions as it is less contaminated by PDR emission than O3C2. We emphasize that, despite their interest for observational studies, the ratios presented here are strongly sensitive to the abundance profiles (C/O and N/O) assumed in this study. Other tracers, including ratios of the same element that are not sensitive to abundance profiles are provided in Table \ref{fit_IR_lines} where we report the best potential lines ratio having a Spearman coefficient above 0.3.

In Figs. \ref{fesc_ratio_highIP_kde} and \ref{fesc_ratio_highIP_sue}, we examine some ratios involving the lines emitted by ions with high ionization potentials: \neiii15$\mu$m, \oiv25$\mu$m and \nev14$\mu$m. The luminosity of these lines strongly depends on the properties of the potential X-ray source and can span a large range of values. Preliminary tests performed with single-star population have shown that our code always needed an additional contribution of an X-ray component to match the observed values of such highly ionized species (as well as neutral gas lines). In the current study, which includes a contribution from binary stars with BPASS, an additional contribution from X-ray is not systematically needed and the X-ray-to-stellar luminosity ratios span several orders of magnitudes (see panel e from Fig. \ref{dependencies_SUE}). We also note that typical ages found by MULTIGRIS are shifted to later ages when including effects from binary stars.

We find a correlation between \feschii\ and \neiii15$\mu$m/\neii12$\mu$m (Ne32), \oiv25$\mu$m/\cii158$\mu$m (O4C2) and \nev14$\mu$m/\neii12$\mu$m (Ne52). Nevertheless, we find that \oiv25$\mu$m/\cii158$\mu$m (O4C2) and \nev14$\mu$m/\neii12$\mu$m (Ne52) cover much larger dynamical ranges. In Fig. \ref{fesc_ratio_highIP_sue}, the large uncertainties of the ellipses reflect the uncertainty on the X-ray component, which is poorly constrained when no \oiv\ and \nev\ detections are provided. Although such ratios involving lines emitted by ions with high ionization potentials are subject to large uncertainties due to the unknown X-ray source properties, they could become of high interest if more constraints on the X-ray spectrum are available. In particular, a more complete statistical sample would provide interesting insights on whether or not leakage is linked to the presence of an X-ray source. Compared to our current predictions, it would also help pinpoint some of the missing physics in our models and might provide insights on the nature of X-ray sources. 

\subsubsection{Optical line ratios}

\begin{figure*}
    \centering
    \includegraphics[width=6cm]{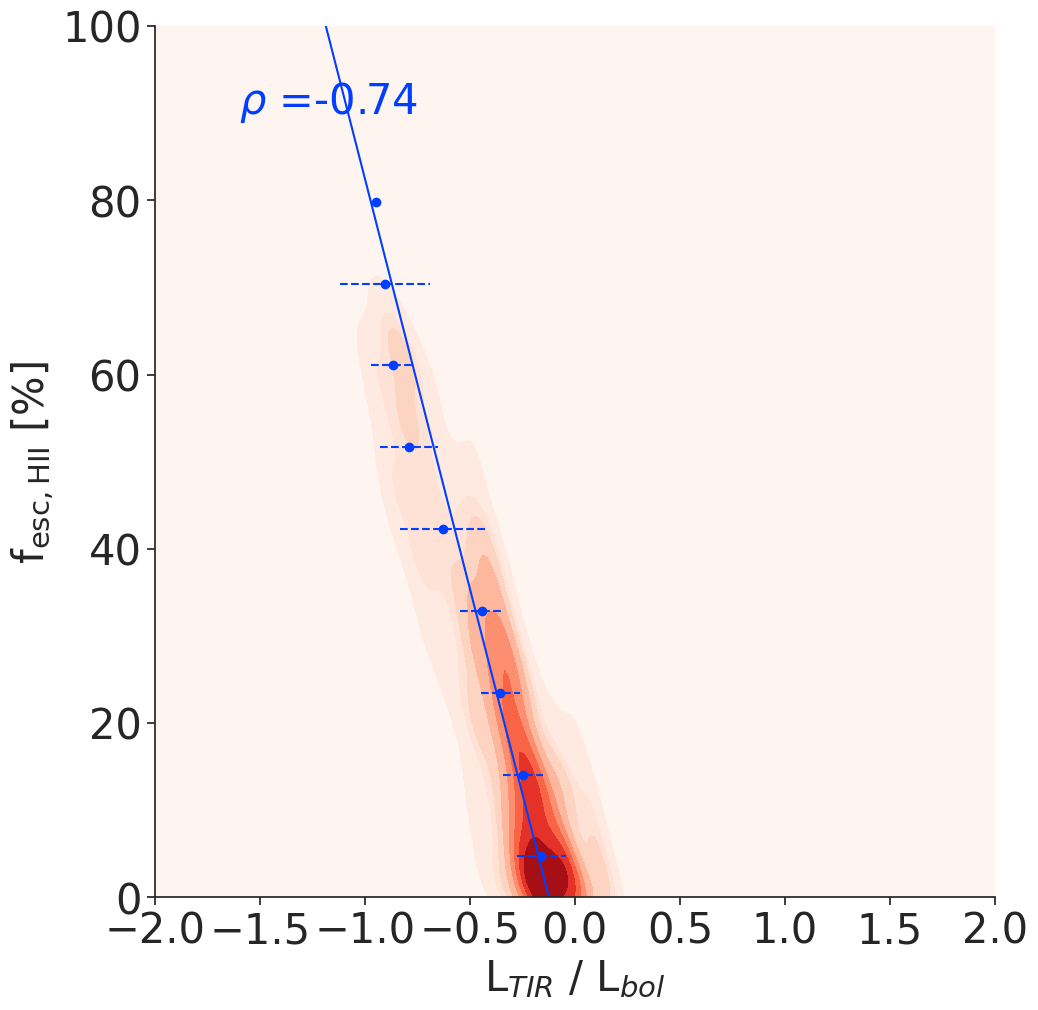}
    \includegraphics[width=6cm]{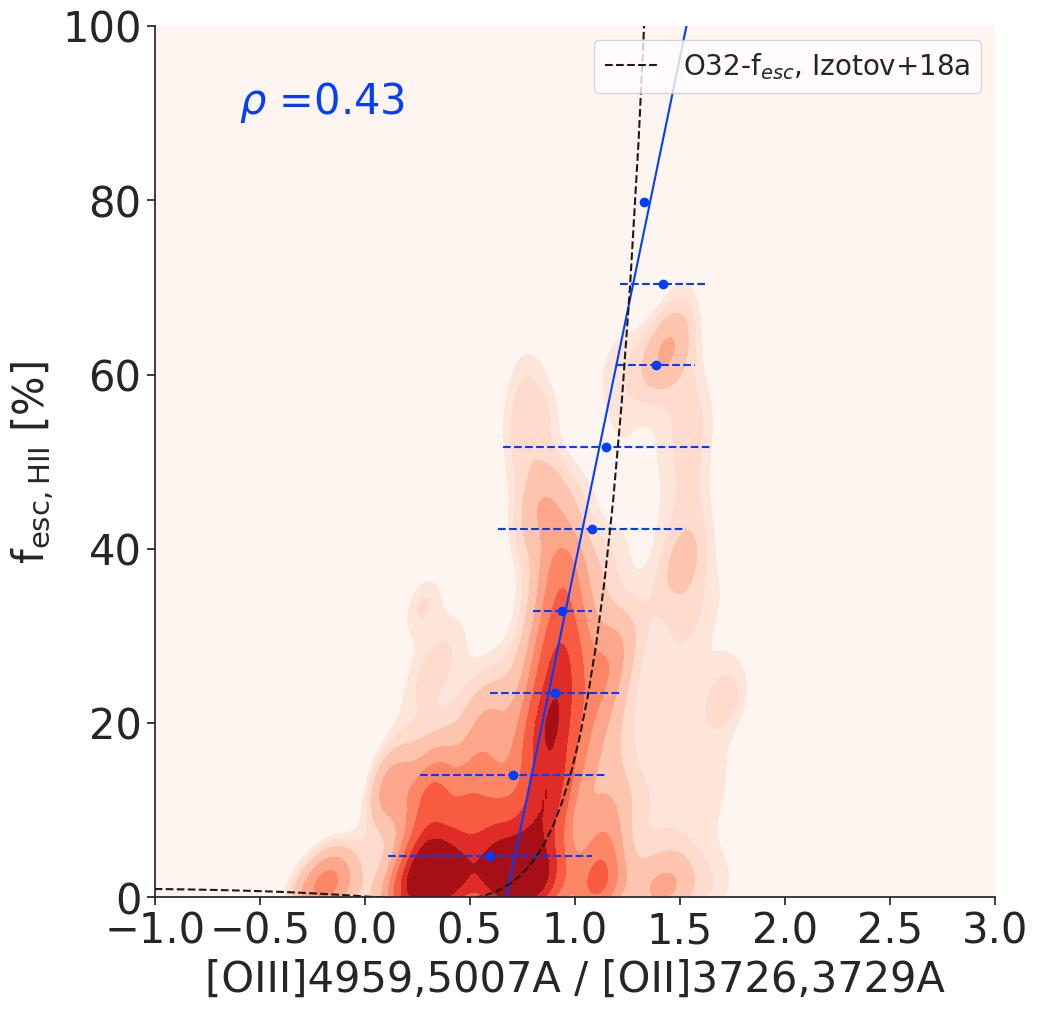}
    \includegraphics[width=6cm]{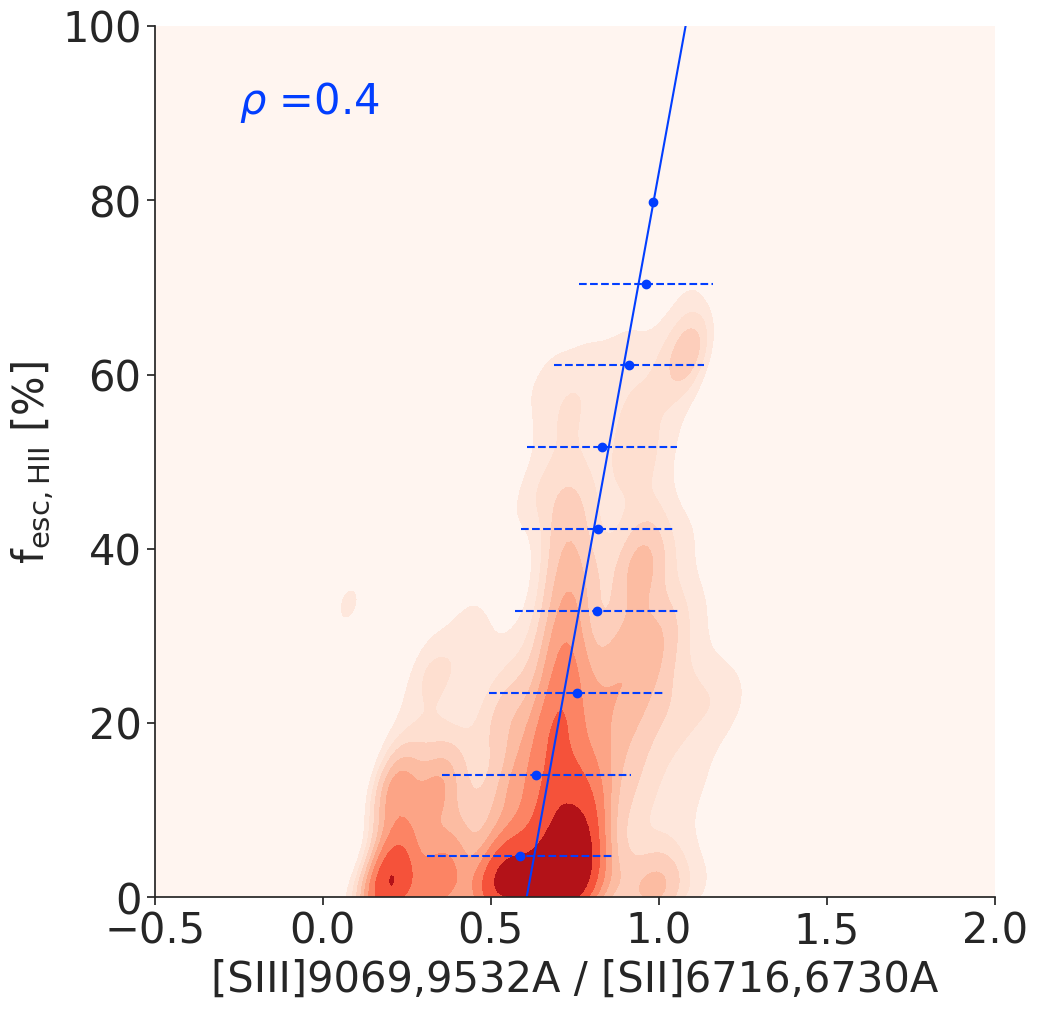}
    \caption{KDE showing \feschii\ as a function of IR and optical line ratios. The black dashed line on the second panel shows the relation O32-\fesc(LyC) from \cite{Izotov_2018a}, assuming that \oiiil/\oiiill $\approx$ 3.}
    \label{fesc_opt_kde}
\end{figure*}

\begin{figure*}
    \centering
    \includegraphics[width=6cm]{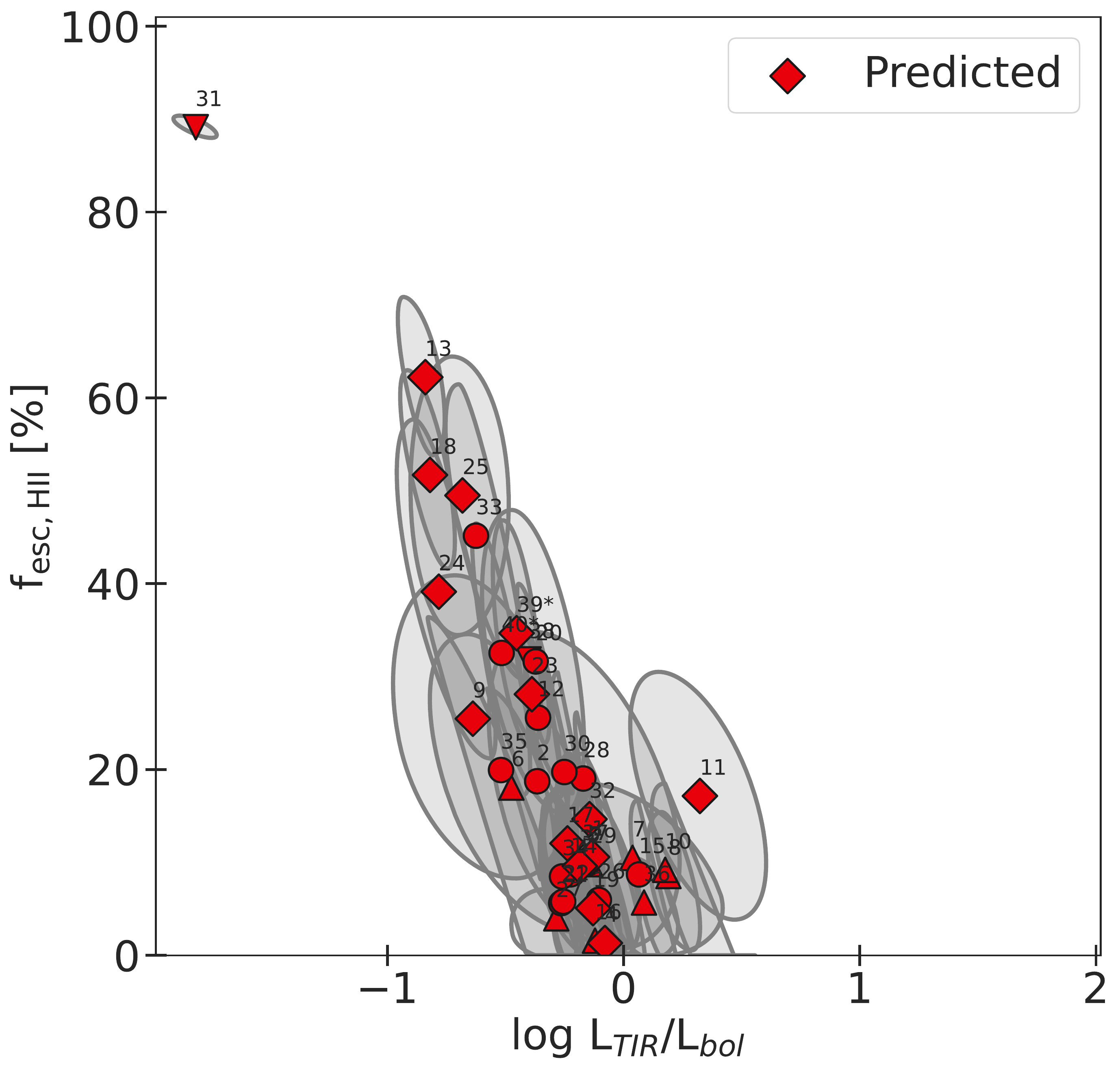}
    \includegraphics[width=6cm]{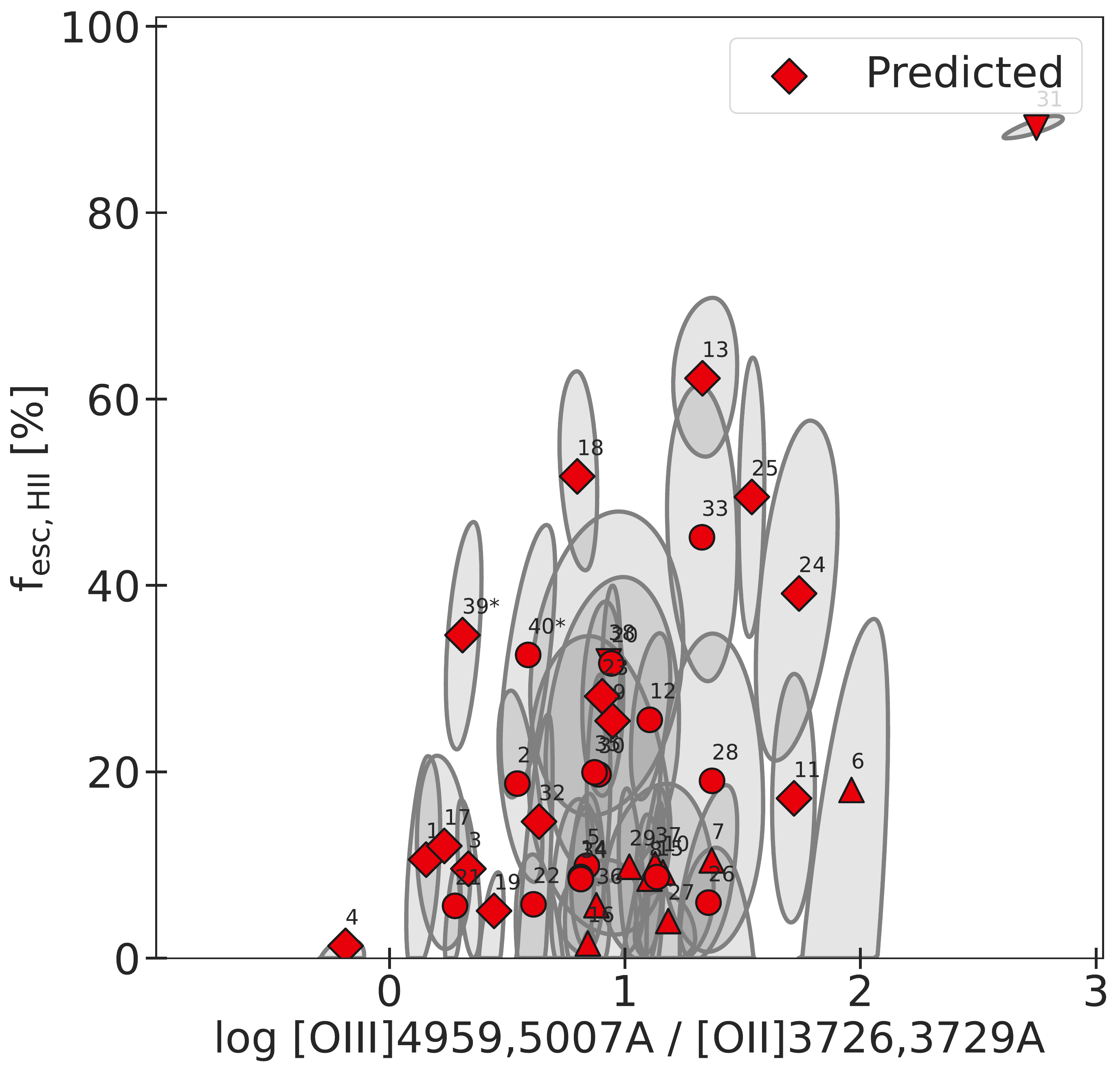}
    \includegraphics[width=6cm]{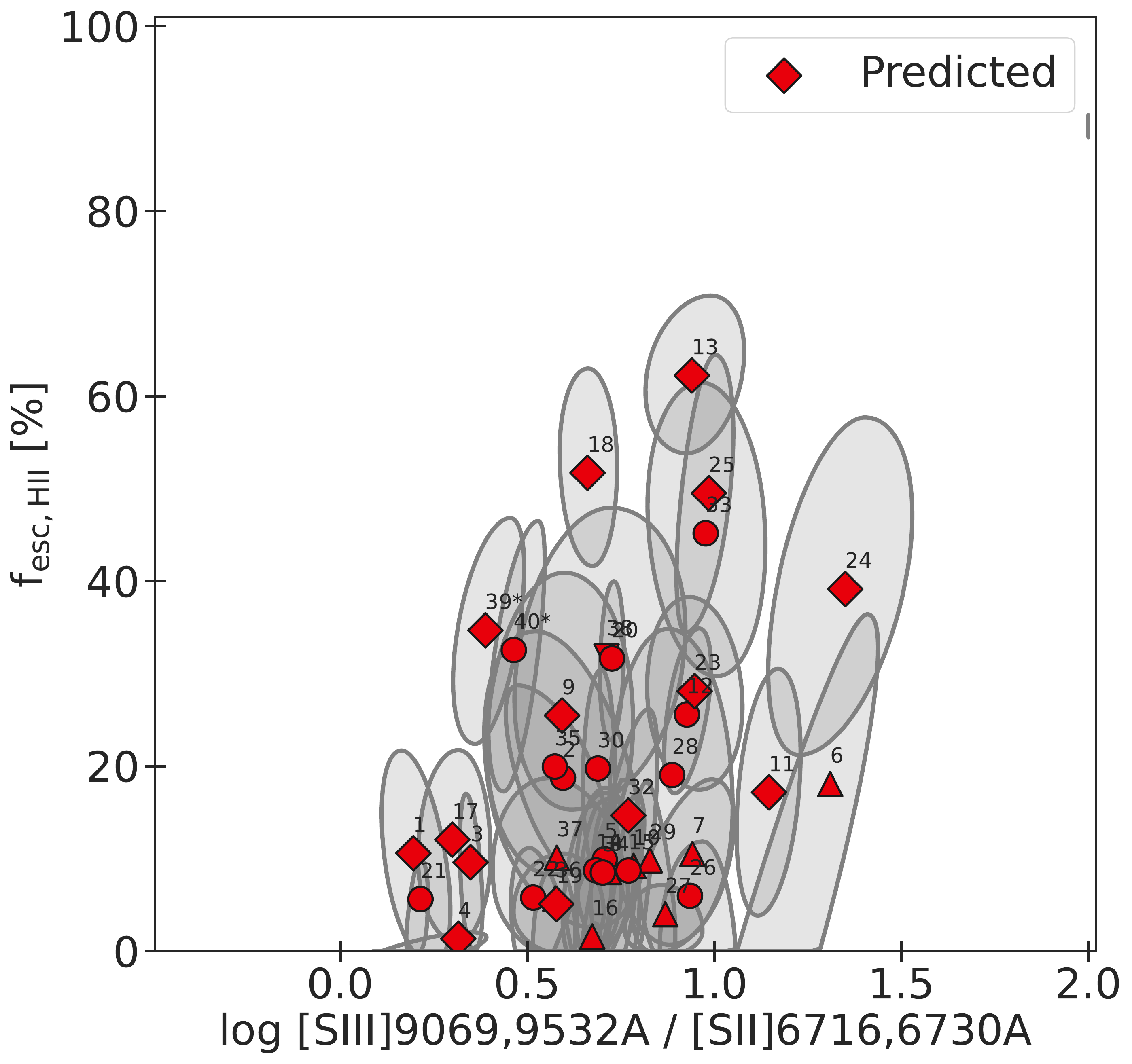}
    \caption{Individual SUEs showing \feschii\ as a function of IR and optical line ratios.}
    \label{fesc_opt_sue}
\end{figure*}
Although our study focuses on IR tracers, we also examined a few well-known optical and mixed tracers that have been discussed in the literature as possible tracers of LyC escape fractions (see Figs. \ref{fesc_opt_kde} and \ref{fesc_opt_sue}). We find that the $L_{\rm TIR}$/$L_{\rm bol}$ ratio anti-correlates tightly with \feschii. While $L_{\rm bol}$ accounts for the total luminosity produced by the central stellar cluster (and the additional X-ray contribution), $L_{\rm TIR}$ traces the photons that have been reprocessed by the surrounding gas and dust and re-emitted in the IR. Thus, this ratio traces well the escaping energetic photons that are produced by the stellar cluster but escape before being reprocessed by the gas. We note, however, that this method either requires modeling assumptions for the stellar population or fitting the continuum to derive $L_{\rm bol}$.
    
Alternatively, we also put to test the \Oiii/\Oii\ (O32) and \Siii/\Sii\ (S32) optical line ratios. The O32 ratio has been claimed to efficiently pick up some of the most extreme leakers \citep{Nakajima_2014, Izotov_2016a, Izotov_2016b, Izotov_2018a, Izotov_2018b, Flury_2022b}. In particular, \cite{Flury_2022b} find a significant correlation of the escape fraction detected in the UV with O32, although with large scatter. Nevertheless, the O32-\fesc\ relation is not straightforward; \cite{stasinska_excitation_2015} have examined galaxies with enhanced O32 ratio and find that a large fraction of them exhibit no sign of leakage. Additionally, recent simulations from \cite{barrow_lyman_2020} find that the O32 ratio fluctuates over timescales different than the dynamical time associated with escape fractions. They conclude that although high O32 tend to be associated with high \fesc, no systematic correlation is expected. Although less classically used than the O32 ratio, the S32 ratio has been shown to be a good tracer of the ionization parameter \citep{Kewley_2002, mingozzi_sdss_2020}. Because it involves the \Sii\ lines emission, it should also be sensitive to the presence of density-bounded regions with reduced \Sii\ emission \citep{wang_new_2019, Wang_2021, ramambason_reconciling_2020}. This ratio has been used, for example, in \cite{zastrow_ionization_2011, zastrow_new_2013} as a diagnostic of \hii\ region optical depths. Still, this trend has not yet been confirmed by any observational study since \Siii\ emission has only been detected in a few LyC-leaking galaxies.
Based on our models constrained by IR emission lines, we find that \feschii\ correlates well with the predicted O32 and S32. In particular, the fit we derive for the O32-\feschii\ relation is remarkably close to the relation from \cite{Izotov_2018b} derived from fitting 11 LyC-leaking galaxies, and compatible with the new sample of 89 galaxies (detection and nondetection) observed in the Low-$z$ Lyman Continuum Survey \citep{Flury_2022b}. We further discuss the reasons that may explain those trends, and why they might be associated with a large scatter in observational studies, in the following section.

\section{Discussion}
\label{discussion}

\subsection{From gas covering factors to \hii\ region escape fraction}

While it has been shown that the escape fraction of ionizing photons is tightly related to the structure of the neutral surrounding gas \citep[e.g.,][]{gazagnes_neutral_2018, gazagnes_origin_2020, chisholm_accurately_2018, Saldana-Lopez_2022}, quantitatively linking the gas distribution to the amount of escaping photons remains a complex task. Absorption lines are powerful proxies to probe the properties of the gas on a given line of sight. In particular, \hi\ absorption lines that saturate above a given column density have been used to derive \hi\ covering fractions, which describe the distribution of gas on global galactic scales. Due to the gas kinematics, the covering fractions that are derived are always a lower limit to the real geometric covering fraction \citep[e.g.,][]{gazagnes_origin_2020}.
Part of this neutral gas component can also be probed in emission where specific tracers emit in dense and irradiated PDRs. Such tracers are mostly accessible in the IR domain. (e.g., \oi, \cii, \silii).

Previous IR studies have focused on quantifying a PDR covering factor \citep{2012_Cormier, cormier_herschel_2019, harikane_large_2020}, which has been used to describe the apparent porosity to ionizing radiation at galactic scales. This PDR covering factor was calculated using models where all the lines arising from the PDR were simultaneously scaled by a common factor between 0 and 1. This grouping is somewhat arbitrary as some lines emit throughout both \hii\ regions and the PDRs. Additionally, these models did not account for possible density-bounded regions since the PDR zone (either total or partial) was always combined with a fully radiation-bounded \hii\ region. Hence, the interpretation of the PDR covering factors is difficult as the simplistic geometry does not account well for complex gas distribution and cannot be linked to potentially escaping photons. In particular, this quantity is very sensitive to modeling assumptions such as the stopping criteria for models (e.g., depth or A$_V$). Among other factors, accounting for a nonunity filling factor of the ISM or for the physical depth of the photodissociated layers changes the values derived for the PDR covering factor: a very thin PDR with a large covering factor can produce the same emission lines as a thick PDR with a lower one. 

In our topological models, we relax the assumption of a full PDR model scaled by a covering factor in favor of an agnostic combination of sectors (as described in Sect. \ref{topo}) where both the weights and the depths of each sector are free parameters. The Bayesian framework described in Sect. \ref{section_code} allows us to derive PDF and credibility intervals for secondary parameters and to calculate, for each sector, the amount of escaping photons. We then derive a global averaged \feschii\ (see Sect. \ref{cloudy_observables}) corresponding to the population of \hii\ regions present in a galaxy. This integrated quantity directly relates to the ISM gas distribution as opposed to the previously used PDR covering factor.

Nevertheless, as mentioned in Sect. \ref{section_topo}, our approach bears its own caveats. In particular, the interdependence between total cluster luminosity and escape fractions is not easy to solve. Nevertheless, although scaling the total cluster luminosity affects the absolute lines fluxes, it does not modify the line ratios that are tracing density-bounded regions (see Sect. \ref{Section_tracers}). Constraining the cluster luminosity through $L_{\rm TIR}$ is a crucial step that allows us to lift this degeneracy. Still, since we do not use any continuum emission, the UV luminosity we derive for our model depends on our modeling assumptions for the stellar population. The values we derive for the total bolometric luminosities and the escape fractions incorporate uncertainties on the stellar population. Another limit of our work comes from the assumed topology of galaxies; although the multisector combination introduces a layer of complexity that is usually unaccounted for when using single-component photoionization models, this representation remains simplistic when compared to the actual ISM morphology.

\subsection{From \hii\ regions to galaxies}

The \hii\ region escape fractions that we derive in this study are hardly comparable with the direct UV detection of LyC. Such detections are, first of all, very challenging at low-redshift because the sensitivity of current observatories (e.g., HST/COS) drops at short wavelengths. The few detections available are remarkably small in the local universe and only one galaxy in our sample has a LyC detection (Haro\,11, \fesc=3.2\%). We note that this value is compatible with the prediction we derive for Haro\,11 of \feschii=9.59$_{-9.51}^{+9.80}$\%.\footnote{The interval corresponds to the highest density interval at 94\%.} Similarly, other observations of a few galaxies in the Local Universe detect little to no LyC leakage \citep[e.g.,][]{Bergvall_2006, Leitet_2013, Borthakur_2014, leitherer_direct_2016}. Those observations remain very sparse and suffer uncertainties due to line-of-sight variations. They are hard to reconcile with predictions from models \citep{Fujita_2003} and resolved obervations of local starbursts, which suggest that they should be leaking a substantial amount of photons. Quantifying the amount of ionizing photons that are actually escaping at galactic scales remains an elusive question to which we cannot answer within the scope of this study. 

\subsubsection{Resolution effects}

Due to the simplified topology we consider (i.e., a galaxy being a weighted sum of star-forming regions), our predictions for \feschii\ represent an averaged value of the number of photons escaping from the \hii\ regions that populate one galaxy. Although this observable is not easily linked to direct galactic-scale observations, it is representative of the ISM structure within star-forming regions and provides a metric to quantify how many density-bounded regions contribute to the integrated emission. This approach is an attempt to generalize previous analysis performed at local scales with resolved studies in which escape fractions could directly be mapped at the scale of \hii\ regions. 

Focusing on resolved \hii\ regions in the local starburst galaxy IC10, \cite{polles_modeling_2019} introduced a method to take into account density-bounded regions by considering the cloud depth as a free parameter. While their approach is limited to single-sector models, they find that the observed spatial scale influences the predicted depth, with regions being almost all density-bounded at small scales ($\sim$ 200pc) but shifting to a radiation-bounded regime at large scales. At even smaller scales, hints of substantial escape fractions (above 80\%) have been reported in \cite{Chevance_2022} where they find very low feedback coupling efficiencies in GMCs from 9 nearby star-forming galaxies, meaning that most of the feedback energy injected by young stars is not dissipated and may escape.

Another technique based on ionization parameter mapping (IPM) using the \Oiii/\Sii\ line ratio was introduced by \cite{Pellegrini_2012} to study the optical depth of \hii\ regions in the LMC and SMC. This method has allowed them to classify the population of \hii\ regions of both galaxies and derive rough estimates of the average luminosity weighted \feschii\ ($\geq$ 40\% in both galaxies). These estimates were then corrected using DIG measurements to derive an global average escape fraction of 4--9\% in the SMC and 11--17\% in the LMC. A similar technique using IPM based on S32 was used on local starbursts (including three galaxies in our sample) by \cite{zastrow_ionization_2011, zastrow_new_2013} to identify ionizing cones and gas features that could be associated with enhanced escape fractions. Despite the very good resolution of their data ($\sim$ 60 to 200 pc\,arcsec$^{-1}$), the analysis could not provide any quantitative estimate of the escape fractions since individual \hii\ regions were not resolved.

Recent studies have been able to overcome this resolution limit and have provided quantitative estimates of the escape fraction for a few local galaxies. Using $\sim$83 000 resolved stars, \cite{choi_mapping_2020} find that the escape fractions of star-forming regions within NGC\,4214 exhibit a wide range of values going up to 40\%. Their results yield a global escape fraction of about $\sim$25\% for the whole galaxy. Although our study focuses on two different regions of this galaxy, the values we derive (34.67$_{-13.27}^{+12.29}$\% for NGC\,4214-c and 32.55$_{-15.40}^{+16.44}$\% for NGC\,4214-s) are in good agreement with their predictions. Another approach based on \hii\ region mapping was also made possible by the resolution of MUSE on the ELT. \cite{Weilbacher_2018} mapped 386 \hii\ regions from the central field of the Antennae nebula (having approximately a solar metallicity; \citealt{Bastian_2006, Lardo_2015}) and found 38 of them to exhibit large \feschii\ with a mean value of 72\%. They estimate the overall escape fraction of the galaxy to be around 7\%. Recently, \cite{Della_Bruna_2021} found an average escape fraction of 67\% for the population of \hii\ regions observed with MUSE in NGC\,7793 (which has an averaged metallicity close to the solar value; \citealt{Pilyugin_2014}). They find that the radiation leaking from \hii\ regions is sufficient to explain by itself the measured fraction of DIG in this galaxy.

Our results corroborate the picture drawn by these recent studies that find that a substantial amount of ionizing photons might be leaking from \hii\ regions. Connecting these escaping photons to the surrounding diffuse gas is a key element to understand whether or not photons might be leaking out at larger scales.

\subsubsection{Missing diffuse component}
\label{discussion_missing_gas}

In our models, a galaxy is represented as a combination of \hii\ regions connected to PDR and molecular zone (see Fig. \ref{topo}). 
Additionally, the potential presence of DIG is accounted for by allowing the code to add a sector with very low hydrogen density (as low as 1 cm$^{-3}$). This component is only added by our code if it is needed to reproduce the suite of constraints. In our solutions, however, such diffuse sectors are usually absent. Instead, the code favors configurations with hydrogen density typical of \hii\ regions (around 100 cm$^{-3}$). This is understandable as the IR tracers that we use to constrain our models predominantly trace \hii\ regions and PDRs. Additionally, our treatment of this DIG component is limited by the independent radiative transfers within each Cloudy model considered as a sector. Although the photon transfer is consistently computed throughout each phase, the different sectors are independent from one another and do not overlap, meaning that photons escaping from a given sector cannot be reabsorbed by a second component along the line of sight. 

This unaccounted for DIG could be problematic to reproduce the optical emission lines (e.g., \ha\ in Sect. \ref{section_prelimary_results} and optical line ratios in Sect. \ref{Section_tracers}) we derive in our analysis. Indeed, contamination by DIG has been extensively studied in the optical domain and is known to significantly contribute to some line ratios such as \sii/H$\alpha$, \nii/H$\alpha$ and \oiii/H$\beta$ \citep{oey_survey_2007, Zhang_2017, Kreckel_2016, Vale_Asari_2019, asari_importance_2020, Espinosa-Ponce_2020, Della_Bruna_2021, Belfiore_2022}. Nevertheless, we have seen in Sect. \ref{results_sfr} that our models predict \ha\ fluxes in agreement with observations. This agreement is expected since our code reproduces both the metallicity (see Fig.  \ref{compare_metallicities}) and the recombination line Hu$\alpha$ in the galaxies in which it is detected. Nevertheless, it appears at odds with the large \feschii\ we derive and could be explained by different scenarios. A first possibility is that the measured H$\alpha$ emission in the DGS galaxies is dominated by \hii\ regions with a negligible contribution from DIG. This would be in line with observational studies that find that, as opposed to nonstarbursting galaxies, the H$\alpha$ emission in starbursts is largely dominated by \hii\ regions in which the emission is boosted \citep{hanish_2010}. In that case, both our models and H$\alpha$ observations would be fairly independent of DIG. This would suggest that the set of observables used in this study is not sensitive to DIG and that additional tracers are needed to probe this diffuse component. 

Alternatively, if part of the detected H$\alpha$ emission is powered by DIG, the fact that our models reproduce it without including a diffuse component means that our code somehow compensates for the missing DIG contribution, although assuming a simplified DIG-free geometry. As opposed to the previous scenario, this would only be possible if some of the IR tracers we use are sensitive to the DIG contribution. Although, to our knowledge, no systematic studies are available regarding DIG contamination in the IR domain, we do have some observational evidence that it may marginally contribute to the emission of some of the tracers considered in our study \citep{2012_Cormier, cormier_herschel_2019}. All in all, we find that our models based on combination of \hii\ regions are sufficient to reproduce the observed H$\alpha$ emission, although they may be either missing or compensating for the DIG contribution. To disentangle those scenarios, further progress is needed to either correct the emission lines for DIG contamination before using them as constraints of MULTIGRIS or properly including in the models an additional DIG component. To that purpose, high spatial-resolution and sensitive telescopes in the far-IR would be useful to study the properties of the DIG.

Most importantly, we also do not account for the actual distribution of diffuse neutral gas. This is a major caveat considering that escaping photons has been shown to be very sensitive to the distribution of neutral hydrogen \citep[e.g.,][]{gazagnes_origin_2020}. Although our models make the most of the few IR constraints tracing the PDRs, we are only accounting for the dense, highly irradiated neutral gas, which is seen in emission. This might partly explain why our estimates of the total \hi\ masses (M(\hi)), shown in Fig. \ref{MHI_comp}, tend to be significantly smaller that the measurements from \cite{remy-ruyer_gas--dust_2014} even though we find that those values agree within 0.5dex for 16 out of 40 galaxies. However, comparing the \hi\ column densities of our sectors, we see that even the deepest sector of each model has a column density much lower than that measured in absorption (see Table \ref{Nh_in_dgs}). This is one of the main limitations of our models: even when our models successfully recover the total \hi\ gas mass, the distribution of sectors remains too simplistic. The actual distribution of the ISM likely resembles more the randomly distributed version presented in Fig. \ref{topo}. In particular, introducing holes or very low column-density sight-lines in our topological models would naturally push the code to predict deeper sectors with higher column densities. Because of the aforementioned limitations regarding the diffuse ionized and neutral gas, we cannot provide estimates for the global galactic escape fraction in our sample. Instead, we focus on analysing the trends with various parameters and their physical interpretation.

\subsection{Dependences of \feschii\ }
\label{discussion_fesc_dep}

\subsubsection{Correlation with galactic properties and observational limitations}

We find a clear anti-correlation of \feschii\ with metallicity and a positive correlation with sSFR. Such trends are reminiscent of those observed in the local sample of few direct measurements from \cite{Leitet_2013}. They are, however, not clearly seen in larger surveys using UV detection of the LyC \citep[e.g.,][]{Saxena_2021, Pahl_2022}, although the recently observed Low-redshift Lyman Continuum Survey (89 galaxies with LyC measurements) has revealed a similar trend of \fesc\ with sSFR \citep{Flury_2022b}.

The lack of observed correlation between galactic properties and measured LyC escape fractions might be due to a dilution effect of the localized \hii\ regions from where photons escape into the galactic ISM \citep{Saxena_2021}. This dilution may explain why we find clear trends of \feschii\ with several galactic parameters (see Sect. \ref{section_dependencies}) while such correlations are unseen at galactic scales. Another crucial parameter that might strongly limit the observation of empirical correlation is the unknown inclination of unresolved galaxies. For example, \cite{Bassett_2019} emphasize that different viewing angles can affect both the measured LyC values and the observed optical lines ratios (such as O32). Under the hypothesis of an anisotropic LyC leakage, \cite{Nakajima_2020} suggest that the viewing angle of porous \hii\ regions might be the main parameter controlling the detection or nondetection of LyC leakage in a sample of Ly$\alpha$ emitting galaxies from the LymAn Continuum Escape Survey (LACES). Although the dependence of emission line ratios on the viewing angle is somewhat mitigated by the use of optically thin tracers (e.g., IR lines), the exact impact of inclination on the observed emission from galaxies remains to be quantitatively examined. This effect was not examined in the current study since we cannot extract any information on the inclination in our sample of galaxies. 

Further progress on this issue requires the use of complex 3D radiative transfer codes as this effect is not accounted for when using 1D radiative transfer codes such as Cloudy. For example, \cite{Mauerhofer_2021} performed a detailed investigation of the viewing angle impact on simulated galaxies and find that the measured projected escape fractions are strongly impacted by orientation. (with variations from 0 to 47\%). This lack of correlation might also be due to propagation effects through the DIG, which are not taken into account in our \hii\ region modeling. Those effects remain largely unconstrained since they depend on the \hi\ gas 3D-distribution, which is unknown.

\subsubsection{Potential impact from nonstellar feedback}
\begin{figure*}[hbt]
    \centering
    \includegraphics[height=6cm,width=6cm]{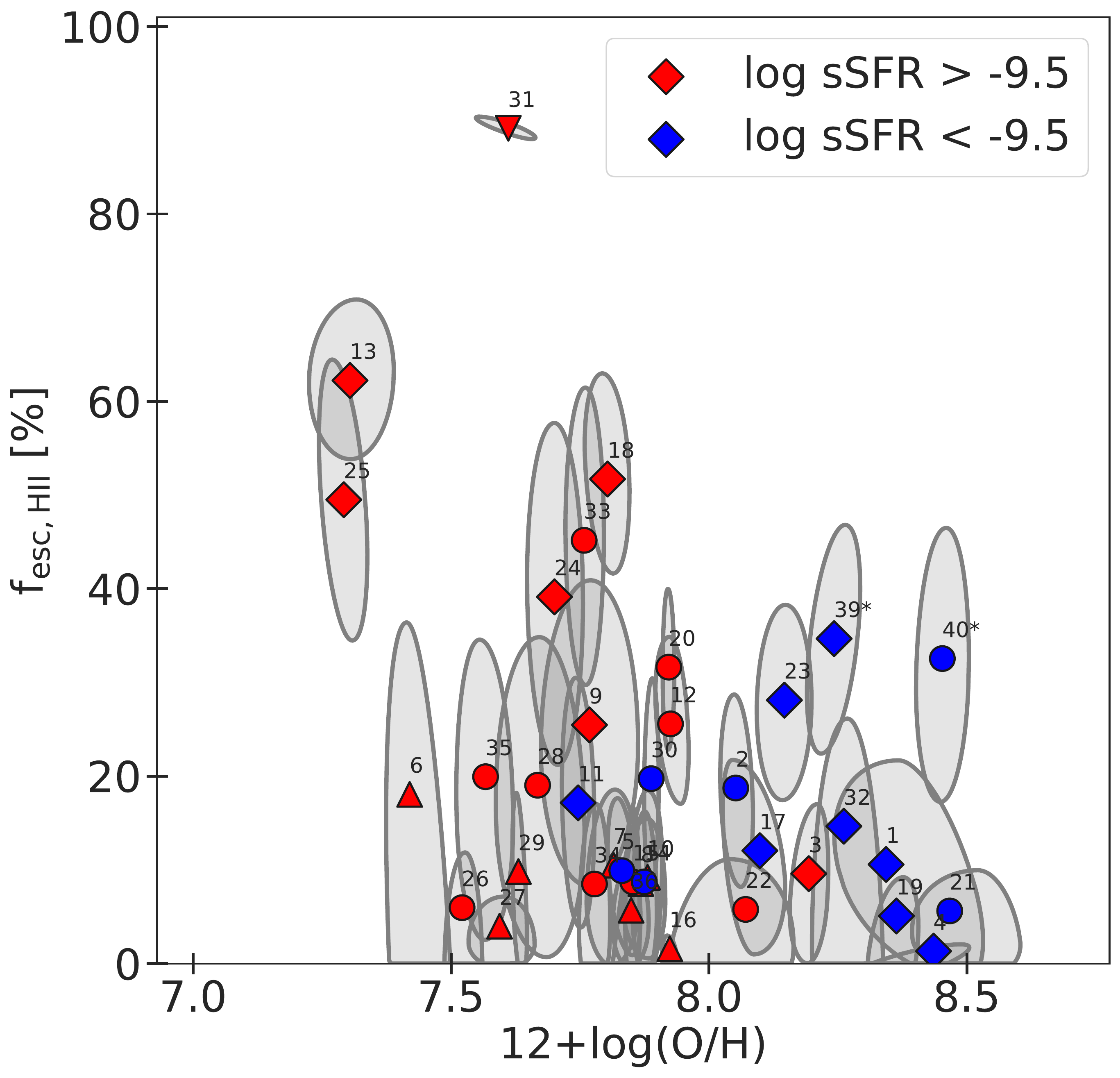}
    \includegraphics[height=6cm,width=6cm]{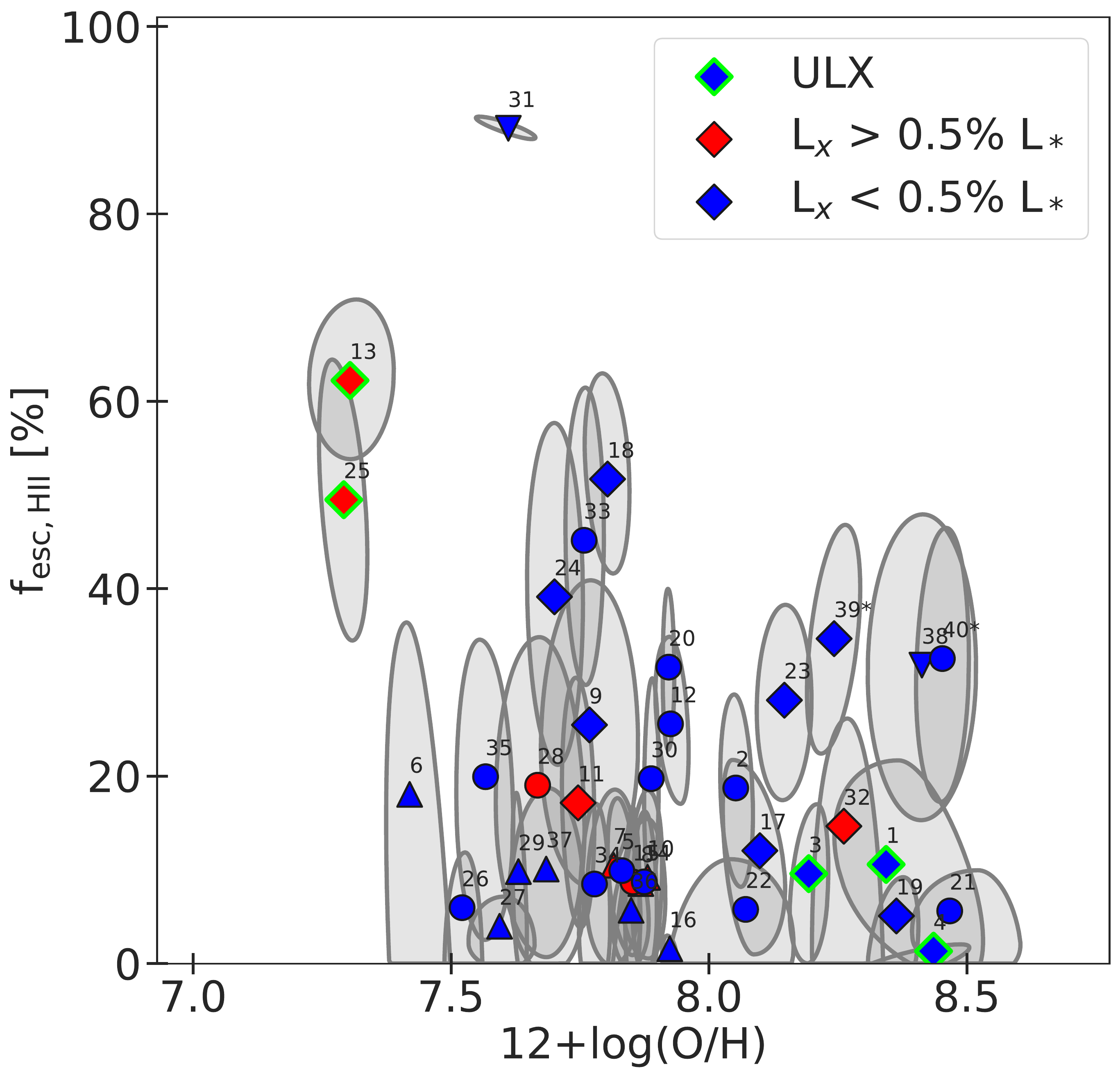}
    \includegraphics[height=6cm,width=6cm]{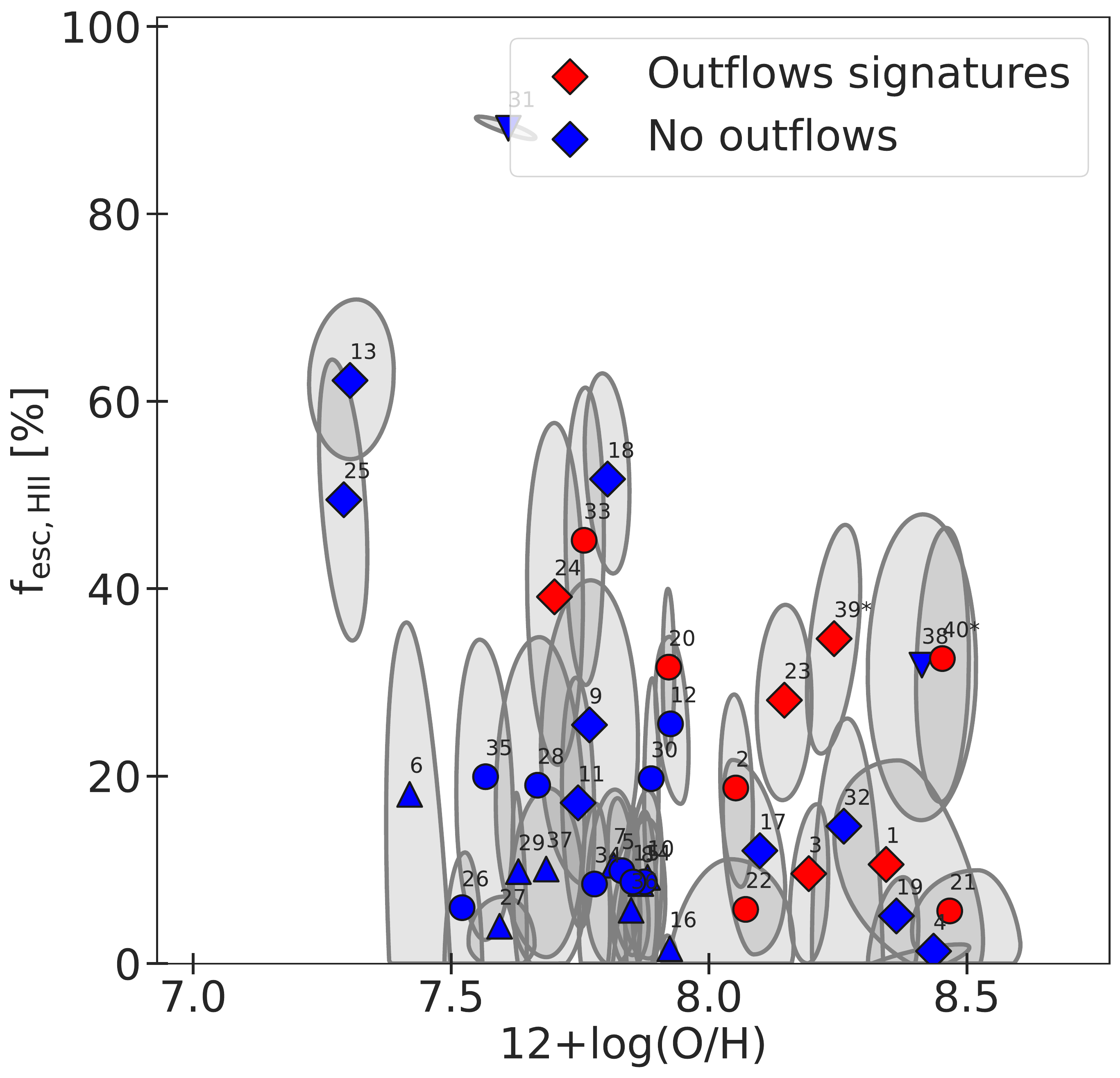}
    \caption{Impact of sSFR, X-ray sources and outflows on the escape fraction. \textbf{Left:} Impact of the measured sSFR from \cite{remy-ruyer_linking_2015} on the \feschii -metallicity relation. No sSFR measurement are available for galaxies 37: HS\,0017+1055 and 38: HS\,2352+2733. \textbf{Middle:} Impact of the X-ray-to-stellar luminosity ratio. The green annuli mark galaxies in which a potential ULX (L$_X$> 10$^{39}$ erg\,s$^{-1}$) have been reported.\protect\footnotemark[1] \textbf{Right:} Impact of outflowing gas signatures.\protect\footnotemark[2] }
    \label{lx_SFR_color}
    \begin{minipage}{\textwidth}
    \footnotesize
    \raggedright
    \vspace{0.1cm}
    \hrulefill\par
    \vspace{0.1cm}
    [1] 1: Haro\,2 \citep{OtiFloranes_2012}, 3: Haro\,11 \citep{Prestwich_2015, Gross_2021}, 4: He\,2-10 \citep{Ott_2005, Reines_2011},  13: I\,Zw\,18\citep{Thuan_2004,Ott_2005, Kaaret_2011, Kaaret_Feng_2013, Brorby_2014} and 25:SBS\,0335-052 \citep{Thuan_2004, Prestwich_2013}.
    [2] 1: Haro\,2 \cite{OtiFloranes_2012}, 3: Haro\,11 \citep{Menacho_2019},  20: NGC\,1569 \citep{Sanches_Cruces_2015}, 21: NGC\,1705 \citep{zastrow_new_2013}, 22: NGC\,5253 \cite{zastrow_ionization_2011}, 23: NGC\,625 \citep{Cannon_2004, McQuinn_2018}, 33: UM\,461 \citep{Carvalho_2018}, 39-40: NGC\,4214 \citep{McQuinn_2018}.
    \end{minipage}

\end{figure*}

Although the high \feschii\ in our sample are predominantly associated with low-metallicity, low mass, and high sSFR, a few objects in our sample exhibit surprisingly high \feschii\ with somewhat larger metallicities and moderate sSFR. We now discuss additional mechanisms that could explain the position of such objects on the \feschii-metallicity relation. In particular, we examine if other feedback mechanisms such as the potential presence of an ULX or outflows could be responsible for the position of outliers in our sample. In Fig. \ref{lx_SFR_color} (first panel), we mark the galaxies having a high measured sSFR (log sSFR > $-9.5$). This criterion is common to all the galaxies with inferred \feschii\,>\,40\% and shared by 10 out of 14 galaxies with \feschii\,>\,20\%. Nevertheless, 4 galaxies with \feschii\,>\,20\% exhibit only a moderate sSFR. We now examine other potential mechanisms that may be responsible for their enhanced \feschii.

On the middle panel from Fig. \ref{lx_SFR_color}, we examine the impact of the X-ray-to-stellar luminosity ratio. Although it has not been shown that the presence of an X-ray source have a direct impact on escape fractions, high L$_X$ tends to be associated with of intense star-formation rates \citep{Grimm_2003} and the emission of high-energy photons, which travel deep into the ISM may enhance the global escape fraction of ionizing photons.  
We find X-ray-to-stellar luminosity ratio above 0.5\% in 8 out of 40 sources. Two out of the five reported potential ULX in our sample are associated with L$_X$/L$_*$\,>\,0.5\%. 
We note that the two galaxies  (13: I\,Zw\,18 and 25: SBS\,0335-052) with the highest \feschii\ in our sample (>\,50\%) are associated with a known ULX. However, similar levels of X-ray contribution are also found in galaxies for which we infer little to no leakage. All in all, we find that large X-ray-to-stellar luminosities are not necessarily associated with enhanced leakage, although X-ray sources may contribute in some specific cases. Additionally, we note that a recent study from \cite{Marques-Chaves_2022} finds no correlation between the LyC escape fractions measured in the Low-$z$ Lyman Continuum Survey and the spectral hardness of the ionizing sources.

Finally, we expect dynamical effects disturbing the gas kinematics to also impact the \hii\ region escape fractions. Outflows may remove gas from the galactic disk and create low-density channel from where photons can freely travel \citep{Fujita_2003}. Recent studies have also examined how gas fragmentation induced by turbulence might play a role in producing the patchy structure of the ISM through which photons escape \citep[e.g.,][]{kakiichi_lyman_2019, hogarth_chemodynamics_2020}. Although outflow signatures are only detected in a few objects (11/40), they offer a plausible explanation to interpret the position of some outliers having large predicted \feschii\ but modest sSFR and little to no contribution from an additional X-ray source. Detailed studies of the gas kinematics are available for a few objects in our sample (see Sect. \ref{section_overview} for references) based on either resolved observations or analysis of emission line profiles. Since photoionization and photodissociation models do not account for dynamical effects, we only provide a qualitative discussion on the potential role of gas outflows. In Fig. \ref{lx_SFR_color} (third panel), we show that the presence of outflows may explain the position of several galaxies with relatively high metallicities (12+log(O/H) > 8) but \feschii\,>\,20\% (e.g., 2: Haro\,3, 23: NGC\,625, and 39--40: NGC\,4214). 

In summary, we find that although the presence of nonstellar feedback mechanisms such as a strong contribution of an X-ray source or outflowing gas may contribute to power photon leakage, such mechanisms seem to only have a secondary influence on the inferred \feschii. Instead, we find that variations of \feschii\ appear to be mainly controlled by global galactic properties such as the metallicity, stellar mass and sSFR. However, once photons have escaped from their \hii\ region, the additional feedback mechanisms that we have discussed in this section might play a role to convert \feschii\ into global galactic escape fractions by removing or ionizing the surrounding neutral gas.

\subsection{Prospects for future observations}
We provide in this study an observational framework to pinpoint galaxies with a substantial amount of \feschii. The presence of leaking \hii\ regions providing a substantial amount of ionizing photons is a necessary condition for a galaxy to be a potential LyC-leaker at larger scales. Hence, the tracers that we propose here could serve as a first basis to select galaxy samples with potential leakage.

We identify line ratios that could be used as tracers of the escape fraction. First, we find that among the classically used optical line ratio, O32 and S32 lines ratio correlate well with our predicted \feschii. This finding is at odds with observational studies finding little correlations with such diagnostics, and large scatters. In particular, the clear O32-\feschii\ correlation is unexpected in the view of several recent studies that reported numerous caveats and limits in the use the O32 ratio for studying \fesc\ \citep[e.g.,][]{stasinska_excitation_2015, plat_constraints_2019, Bassett_2019, Flury_2022, Flury_2022b}. One possible explanation is that the \feschii\ we derive is less sensitive to viewing angle dependences than the measured escape fractions. Alternatively, this result might also be due to our sample selection, which focuses on highly star-forming galaxies. Additionally, both O32 and S32 ratios have been shown to correlate with the ionization parameter \citep{Nakajima_2014, Kewley_2002}. Disentangling the effect of metallicity, ionization parameter, and escape fractions remains a difficult task, even within the Bayesian framework that we developed here.

Among the IR tracer, the well-know O3C2 ratio has been used in the analysis of high-$z$ galaxies with ALMA \citep{harikane_large_2020}. Nevertheless, the interpretation of high O3C2 ratio is still uncertain and debated. Recently, \cite{Katz_2021} have shown that this ratio is sensitive to a wide variety of other galactic parameters including the C/O ratio, IMF, ionization parameter, gas density and CMB attenuation. 
In our study, assuming that the C/O in the local universe is observationally well constrained, we do find a trend of increasing \hii\ region escape fraction associated with higher O3C2 ratios. However, we find that ratios involving the \nii\ lines (122 and 205 $\mu$m) provide better diagnostics to trace \feschii\ since \nii\ is less contaminated than \cii\ by PDR emission. Although fainter, these lines also fall in the detection range of ALMA and have already been detected in a few high-redshift galaxies above $z\sim$ 4. \citep[e.g.,][]{De_Breuck_2019}. Even for nondetections, upper limits on \nii\ could also be useful to set lower limits on the resulting escape fractions. 

For tracers such as \oiv\ and \nev, associated with ions of high ionization potentials, our interpretation is limited by the little number of constraints that we have on the X-ray component. The JWST will open a new window for the observation of galaxies with signatures of high energy sources and their associated spectral signatures (e.g., \mgiv $\lambda$4.49$\mu$m, \nevi $\lambda$7.64$\mu$m, and \nev$\lambda$14.32$\mu$m; \citealt{satyapal_diagnostic_2020, Richardson_2022}). The combination of such lines with the emission of neutral gas lines (e.g., \feii $\lambda$1.64$\mu$m) boosted by X-ray photons may help disentangle the feedback from potential compact objects from the one induced by turbulence and shocks \citep{Cresci_2010,Cresci_2017}. Those IR tracers could be used in combination with other tracers of high energy photons available in the optical range (e.g., \nev, \fev) and in the UV range (e.g., \Civ, \Heiiuv, \citealt{wofford_2021, berg_chemical_2019, Berg_2021}). A first example of the combination of optical and IR data is presented in \citetalias{LebouteillerRamambason2022}. Looking for counter-parts in the radio and X-ray domain with sensitive instruments (e.g., with the forthcoming \textit{Athena} telescope), may also allow progress on the identification of the exact nature of the different sources producing high energy photons. 

Finally, the numerous caveats mentioned in this section regarding the distribution of neutral gas might be lifted by incorporating into our models tracers of the neutral gas (e.g., the Ly$\alpha$ line, the hydrogen 21cm line) or by using priors calibrated on absorption lines studies. Such observations, both in the local and high-redshift universe, will provide a more complex picture of the interplay between low-metallicity ISM and the intense sources of radiation expected in the early universe. In a subsequent publication we will explore and model the optical emission lines of galaxies from the Low-redshift Lyman Continuum Survey with MULTIGRIS (Marques-Chaves et al., in prep).

\section{Conclusion}
\label{section_conclusion}

This paper presents the first application of a new Bayesian code, MULTIGRIS \citepalias{LebouteillerRamambason2022}, which opens new possibilities in terms of multiphase modeling of the ISM. We developed a framework to automatically combine photoionization and photodissociation models as different sectors representative of different physical and chemical ISM conditions, and including the presence of density-bounded regions from where ionizing photons can escape. We applied this new code to a sample of low-metallicity dwarf galaxies from the DGS \citep{Madden_2013} with extensive sets of IR emission line measurements and 
whose ISM properties may resemble those of galaxies in the early Universe.
We examined the impact of using multisector models on the predictions obtained for well-constrained physical quantities such as the metallicity and SFR. We then derived predictions for the \hii\ region escape fraction of ionizing photons in the DGS and studied how this observable varies with various galactic parameters. Our main results are summarized below:
\begin{itemize}
    \item Accounting for several, possibly density-bounded, sectors does not significantly affect the metallicity and SFR we derive. This confirms that empirical calibrations based on emission lines provide robust results that hold at galactic scales, even when accounting for a slightly more complex topology. We find, however, that not accounting for escaping photons may result in a moderate underestimation of the SFR. The correction factor we derive by estimating SFR based on $Q$(H$^0$) tightly follows the analytical curve 1/(1-\feschii).
    \item We confirm previous results from \cite{cormier_herschel_2019} that show that the ISM structure and porosity strongly varies with metallicity. Specifically, we find that \feschii\ tends to increase when metallicity decreases. In the global KDE derived for the whole DGS sample, we find that the high-metallicity and high-\feschii\ region of the parameter space is completely unpopulated, meaning that high-metallicity galaxies have a low probability of reaching high values of \feschii. Despite a somewhat large scatter around this relation, especially at low-metallicity, this trend is robust to changes in our modeling assumptions on the number of sectors and holds when forcing a fixed number of sectors. Our findings back up the picture of an increasingly porous ISM at low-metallicity, in which energetic photons can easily leak out of \hii\ regions and ionize their surrounding gas. 
    \item We also find that the fraction of \cii\ emitted in the ionized phase drops by $\sim$2 orders of magnitude when going from a solar metallicity down to a metallicity of $\sim$1/35Z$_\odot$. We explain this effect by the presence of enhanced radiation fields in low-metallicity environments, which favor the ionization of C$^{2+}$ ions with respect to C$^{+}$. Hence, we find that \cii\ emission arises predominantly from the neutral atomic and molecular gas in low-metallicity environments. A more detailed study of the neutral and molecular gas mass distribution in the DGS, including the CO-dark component, will be carried out in a future work (Ramambason et al., in prep).
    \item We examine secondary dependences of \feschii\ with galactic properties and find a weak anti-correlation with stellar mass and a clear correlation with sSFR. These findings fit in the picture of the escape fraction being a feedback-dominated process with complex dependences on the gas mass distribution and gas chemistry.
    \item Finally, we provide several line ratios that correlate with \feschii, both in the IR domain (e.g., O3N2, O3C2), including some tracers with high ionization potential (e.g., Ne32, O4C2, and Ne52) and in the optical (e.g., O32, S32), as well as mixed tracers such as $L_{\rm TIR}$/$L_{\rm bol}$.
\end{itemize}

The modeling framework presented in this paper is highly sensitive to the wealth of tracers that are used as inputs, especially when they trace gas in different physical and chemical conditions. With that in mind, potentially more complex and realistic models could be constrained by making use of multiwavelength data combining emission from X-ray to IR.

\begin{acknowledgements}
The authors thank the anonymous referee for constructive feedback and useful comments. VL, CR and LR thank the FACE Foundation Thomas Jefferson Fund (grant number TJF21\_053). CM thanks the grant UNAM / PAPIIT - IN101220. MC gratefully acknowledges funding from the Deutsche Forschungsgemeinschaft (DFG, German Research Foundation) through an Emmy Noether Research Group (grant numbers KR4801/1-1 and CH2137/1-1), as well as from the European Research Council (ERC) under the European Union’s Horizon 2020 research and innovation program via the ERC Starting Grant MUSTANG (grant agreement number 714907).
\end{acknowledgements}
\bibliographystyle{aa}
\bibliography{MyLibrary.bib}

\begin{thebibliography}{219}
\expandafter\ifx\csname natexlab\endcsname\relax\def\natexlab#1{#1}\fi

\bibitem[{{Allen} {et~al.}(2008){Allen}, {Groves}, {Dopita}, {Sutherland}, \&
  {Kewley}}]{Allen_2008}
{Allen}, M.~G., {Groves}, B.~A., {Dopita}, M.~A., {Sutherland}, R.~S., \&
  {Kewley}, L.~J. 2008, \apjs, 178, 20

\bibitem[{{Amor{\'\i}n} {et~al.}(2010){Amor{\'\i}n}, {P{\'e}rez-Montero}, \&
  {V{\'\i}lchez}}]{Amorin_2010}
{Amor{\'\i}n}, R.~O., {P{\'e}rez-Montero}, E., \& {V{\'\i}lchez}, J.~M. 2010,
  \apjl, 715, L128

\bibitem[{{Asplund} {et~al.}(2009){Asplund}, {Grevesse}, {Sauval}, \&
  {Scott}}]{Asplund_2009}
{Asplund}, M., {Grevesse}, N., {Sauval}, A.~J., \& {Scott}, P. 2009, \araa, 47,
  481

\bibitem[{{Bakx} {et~al.}(2020){Bakx}, {Tamura}, {Hashimoto}, {Inoue}, {Lee},
  {Mawatari}, {Ota}, {Umehata}, {Zackrisson}, {Hatsukade}, {Kohno}, {Matsuda},
  {Matsuo}, {Okamoto}, {Shibuya}, {Shimizu}, {Taniguchi}, \&
  {Yoshida}}]{Bakx_2020}
{Bakx}, T. J.~L.~C., {Tamura}, Y., {Hashimoto}, T., {et~al.} 2020, \mnras, 493,
  4294

\bibitem[{{Barrow} {et~al.}(2020){Barrow}, {Robertson}, {Ellis}, {Nakajima},
  {Saxena}, {Stark}, \& {Tang}}]{barrow_lyman_2020}
{Barrow}, K. S.~S., {Robertson}, B.~E., {Ellis}, R.~S., {et~al.} 2020, \apjl,
  902, L39

\bibitem[{{Bassett} {et~al.}(2019){Bassett}, {Ryan-Weber}, {Cooke}, {Diaz},
  {Nanayakkara}, {Yuan}, {Spitler}, {Me{\v{s}}tri{\'c}}, {Garel}, {Sawicki},
  {Gwyn}, \& {Golob}}]{Bassett_2019}
{Bassett}, R., {Ryan-Weber}, E.~V., {Cooke}, J., {et~al.} 2019, \mnras, 483,
  5223

\bibitem[{{Bastian} {et~al.}(2006){Bastian}, {Emsellem}, {Kissler-Patig}, \&
  {Maraston}}]{Bastian_2006}
{Bastian}, N., {Emsellem}, E., {Kissler-Patig}, M., \& {Maraston}, C. 2006,
  \aap, 445, 471

\bibitem[{{Belfiore} {et~al.}(2022){Belfiore}, {Santoro}, {Groves},
  {Schinnerer}, {Kreckel}, {Glover}, {Klessen}, {Emsellem}, {Blanc}, {Congiu},
  {Barnes}, {Boquien}, {Chevance}, {Dale}, {Diederik Kruijssen}, {Leroy},
  {Pan}, {Pessa}, {Schruba}, \& {Williams}}]{Belfiore_2022}
{Belfiore}, F., {Santoro}, F., {Groves}, B., {et~al.} 2022, \aap, 659, A26

\bibitem[{{Berg} {et~al.}(2021){Berg}, {Chisholm}, {Erb}, {Skillman}, {Pogge},
  \& {Olivier}}]{Berg_2021}
{Berg}, D.~A., {Chisholm}, J., {Erb}, D.~K., {et~al.} 2021, \apj, 922, 170

\bibitem[{Berg {et~al.}(2019)Berg, Erb, Henry, Skillman, \&
  McQuinn}]{berg_chemical_2019}
Berg, D.~A., Erb, D.~K., Henry, R. B.~C., Skillman, E.~D., \& McQuinn, K. B.~W.
  2019, ApJ, 874, 93

\bibitem[{{Bergvall} {et~al.}(2006){Bergvall}, {Zackrisson}, {Andersson},
  {Arnberg}, {Masegosa}, \& {{\"O}stlin}}]{Bergvall_2006}
{Bergvall}, N., {Zackrisson}, E., {Andersson}, B.~G., {et~al.} 2006, \aap, 448,
  513

\bibitem[{{Bian} {et~al.}(2017){Bian}, {Fan}, {McGreer}, {Cai}, \&
  {Jiang}}]{Bian_2017}
{Bian}, F., {Fan}, X., {McGreer}, I., {Cai}, Z., \& {Jiang}, L. 2017, \apjl,
  837, L12

\bibitem[{{Bik} {et~al.}(2018){Bik}, {{\"O}stlin}, {Menacho}, {Adamo}, {Hayes},
  {Herenz}, \& {Melinder}}]{Bik_2018}
{Bik}, A., {{\"O}stlin}, G., {Menacho}, V., {et~al.} 2018, \aap, 619, A131

\bibitem[{{Binder} {et~al.}(2015){Binder}, {Williams}, {Eracleous},
  {Plucinsky}, {Gaetz}, {Anderson}, {Skillman}, {Dalcanton}, {Kong}, \&
  {Weisz}}]{binder_2015}
{Binder}, B., {Williams}, B.~F., {Eracleous}, M., {et~al.} 2015, \aj, 150, 94

\bibitem[{{Binette} {et~al.}(1996){Binette}, {Wilson}, \&
  {Storchi-Bergmann}}]{Binette_1996}
{Binette}, L., {Wilson}, A.~S., \& {Storchi-Bergmann}, T. 1996, \aap, 312, 365

\bibitem[{{Borthakur} {et~al.}(2014){Borthakur}, {Heckman}, {Leitherer}, \&
  {Overzier}}]{Borthakur_2014}
{Borthakur}, S., {Heckman}, T.~M., {Leitherer}, C., \& {Overzier}, R.~A. 2014,
  Science, 346, 216

\bibitem[{{Bradley} {et~al.}(2006){Bradley}, {Knapen}, {Beckman}, \&
  {Folkes}}]{bradley_composite_2006}
{Bradley}, T.~R., {Knapen}, J.~H., {Beckman}, J.~E., \& {Folkes}, S.~L. 2006,
  \aap, 459, L13

\bibitem[{{Brorby} {et~al.}(2014){Brorby}, {Kaaret}, \&
  {Prestwich}}]{Brorby_2014}
{Brorby}, M., {Kaaret}, P., \& {Prestwich}, A. 2014, \mnras, 441, 2346

\bibitem[{{Calzetti} {et~al.}(2012){Calzetti}, {Liu}, \&
  {Koda}}]{Calzetti_2012}
{Calzetti}, D., {Liu}, G., \& {Koda}, J. 2012, \apj, 752, 98

\bibitem[{{Cannon} {et~al.}(2004){Cannon}, {McClure-Griffiths}, {Skillman}, \&
  {C{\^o}t{\'e}}}]{Cannon_2004}
{Cannon}, J.~M., {McClure-Griffiths}, N.~M., {Skillman}, E.~D., \&
  {C{\^o}t{\'e}}, S. 2004, \apj, 607, 274

\bibitem[{{Carniani} {et~al.}(2017){Carniani}, {Maiolino}, {Pallottini},
  {Vallini}, {Pentericci}, {Ferrara}, {Castellano}, {Vanzella}, {Grazian},
  {Gallerani}, {Santini}, {Wagg}, \& {Fontana}}]{Carniani_2017}
{Carniani}, S., {Maiolino}, R., {Pallottini}, A., {et~al.} 2017, \aap, 605, A42

\bibitem[{{Carvalho} \& {Plana}(2018)}]{Carvalho_2018}
{Carvalho}, M.~S. \& {Plana}, H. 2018, \mnras, 481, 122

\bibitem[{{Chastenet} {et~al.}(2019){Chastenet}, {Sandstrom}, {Chiang},
  {Leroy}, {Utomo}, {Bot}, {Gordon}, {Draine}, {Fukui}, {Onishi}, \&
  {Tsuge}}]{Chastenet_2019}
{Chastenet}, J., {Sandstrom}, K., {Chiang}, I.-D., {et~al.} 2019, \apj, 876, 62

\bibitem[{{Chevance} {et~al.}(2022){Chevance}, {Kruijssen}, {Krumholz},
  {Groves}, {Keller}, {Hughes}, {Glover}, {Henshaw}, {Herrera}, {Kim}, {Leroy},
  {Pety}, {Razza}, {Rosolowsky}, {Schinnerer}, {Schruba}, {Barnes}, {Bigiel},
  {Blanc}, {Dale}, {Emsellem}, {Faesi}, {Grasha}, {Klessen}, {Kreckel}, {Liu},
  {Longmore}, {Meidt}, {Querejeta}, {Saito}, {Sun}, \& {Usero}}]{Chevance_2022}
{Chevance}, M., {Kruijssen}, J.~M.~D., {Krumholz}, M.~R., {et~al.} 2022,
  \mnras, 509, 272

\bibitem[{{Chevance} {et~al.}(2020{\natexlab{a}}){Chevance}, {Kruijssen},
  {Vazquez-Semadeni}, {Nakamura}, {Klessen}, {Ballesteros-Paredes}, {Inutsuka},
  {Adamo}, \& {Hennebelle}}]{Chevance_lifecycle_2020}
{Chevance}, M., {Kruijssen}, J.~M.~D., {Vazquez-Semadeni}, E., {et~al.}
  2020{\natexlab{a}}, \ssr, 216, 50

\bibitem[{{Chevance} {et~al.}(2020{\natexlab{b}}){Chevance}, {Madden},
  {Fischer}, {Vacca}, {Lebouteiller}, {Fadda}, {Galliano}, {Indebetouw},
  {Kruijssen}, {Lee}, {Poglitsch}, {Polles}, {Cormier}, {Hony}, {Iserlohe},
  {Krabbe}, {Meixner}, {Sabbi}, \& {Zinnecker}}]{Chevance_2020c}
{Chevance}, M., {Madden}, S.~C., {Fischer}, C., {et~al.} 2020{\natexlab{b}},
  \mnras, 494, 5279

\bibitem[{{Chevance} {et~al.}(2016){Chevance}, {Madden}, {Lebouteiller},
  {Godard}, {Cormier}, {Galliano}, {Hony}, {Indebetouw}, {Le Bourlot}, {Lee},
  {Le Petit}, {Pellegrini}, {Roueff}, \& {Wu}}]{Chevance_2016}
{Chevance}, M., {Madden}, S.~C., {Lebouteiller}, V., {et~al.} 2016, \aap, 590,
  A36

\bibitem[{Chisholm {et~al.}(2018)Chisholm, Gazagnes, Schaerer, Verhamme, Rigby,
  Bayliss, Sharon, Gladders, \& Dahle}]{chisholm_accurately_2018}
Chisholm, J., Gazagnes, S., Schaerer, D., {et~al.} 2018, 12

\bibitem[{{Chisholm} {et~al.}(2020){Chisholm}, {Prochaska}, {Schaerer},
  {Gazagnes}, \& {Henry}}]{Chisholm_2020_mgii}
{Chisholm}, J., {Prochaska}, J.~X., {Schaerer}, D., {Gazagnes}, S., \& {Henry},
  A. 2020, \mnras, 498, 2554

\bibitem[{{Chisholm} {et~al.}(2019){Chisholm}, {Rigby}, {Bayliss}, {Berg},
  {Dahle}, {Gladders}, \& {Sharon}}]{Chisholm_2019}
{Chisholm}, J., {Rigby}, J.~R., {Bayliss}, M., {et~al.} 2019, \apj, 882, 182

\bibitem[{{Choi} {et~al.}(2020){Choi}, {Dalcanton}, {Williams}, {Skillman},
  {Fouesneau}, {Gordon}, {Sandstrom}, {Weisz}, \&
  {Gilbert}}]{choi_mapping_2020}
{Choi}, Y., {Dalcanton}, J.~J., {Williams}, B.~F., {et~al.} 2020, \apj, 902, 54

\bibitem[{Cormier {et~al.}(2019)Cormier, Abel, Hony, Lebouteiller, Madden,
  Polles, Galliano, De~Looze, Galametz, \&
  Lambert-Huyghe}]{cormier_herschel_2019}
Cormier, D., Abel, N.~P., Hony, S., {et~al.} 2019, A\&A, 626, A23

\bibitem[{{Cormier} {et~al.}(2012){Cormier}, {Lebouteiller}, {Madden}, {Abel},
  {Hony}, {Galliano}, {Baes}, {Barlow}, {Cooray}, {De Looze}, {Galametz},
  {Karczewski}, {Parkin}, {R{\'e}my}, {Sauvage}, {Spinoglio}, {Wilson}, \&
  {Wu}}]{2012_Cormier}
{Cormier}, D., {Lebouteiller}, V., {Madden}, S.~C., {et~al.} 2012, \aap, 548,
  A20

\bibitem[{{Cormier} {et~al.}(2015){Cormier}, {Madden}, {Lebouteiller}, {Abel},
  {Hony}, {Galliano}, {R{\'e}my-Ruyer}, {Bigiel}, {Baes}, {Boselli},
  {Chevance}, {Cooray}, {De Looze}, {Doublier}, {Galametz}, {Hughes},
  {Karczewski}, {Lee}, {Lu}, \& {Spinoglio}}]{Cormier_2015}
{Cormier}, D., {Madden}, S.~C., {Lebouteiller}, V., {et~al.} 2015, \aap, 578,
  A53

\bibitem[{{Cresci} {et~al.}(2010){Cresci}, {Vanzi}, {Sauvage}, {Santangelo}, \&
  {van der Werf}}]{Cresci_2010}
{Cresci}, G., {Vanzi}, L., {Sauvage}, M., {Santangelo}, G., \& {van der Werf},
  P. 2010, \aap, 520, A82

\bibitem[{{Cresci} {et~al.}(2017){Cresci}, {Vanzi}, {Telles}, {Lanzuisi},
  {Brusa}, {Mingozzi}, {Sauvage}, \& {Johnson}}]{Cresci_2017}
{Cresci}, G., {Vanzi}, L., {Telles}, E., {et~al.} 2017, \aap, 604, A101

\bibitem[{{Crowther} {et~al.}(2010){Crowther}, {Schnurr}, {Hirschi}, {Yusof},
  {Parker}, {Goodwin}, \& {Kassim}}]{Crowther_2010}
{Crowther}, P.~A., {Schnurr}, O., {Hirschi}, R., {et~al.} 2010, \mnras, 408,
  731

\bibitem[{{Croxall} {et~al.}(2017){Croxall}, {Smith}, {Pellegrini}, {Groves},
  {Bolatto}, {Herrera-Camus}, {Sandstrom}, {Draine}, {Wolfire}, {Armus},
  {Boquien}, {Brandl}, {Dale}, {Galametz}, {Hunt}, {Kennicutt}, {Kreckel},
  {Rigopoulou}, {van der Werf}, \& {Wilson}}]{croxall_2017}
{Croxall}, K.~V., {Smith}, J.~D., {Pellegrini}, E., {et~al.} 2017, \apj, 845,
  96

\bibitem[{{de Barros} {et~al.}(2016){de Barros}, {Vanzella}, {Amor{\'\i}n},
  {Castellano}, {Siana}, {Grazian}, {Suh}, {Balestra}, {Vignali}, {Verhamme},
  {Zamorani}, {Mignoli}, {Hasinger}, {Comastri}, {Pentericci},
  {P{\'e}rez-Montero}, {Fontana}, {Giavalisco}, \& {Gilli}}]{DeBarros_2016}
{de Barros}, S., {Vanzella}, E., {Amor{\'\i}n}, R., {et~al.} 2016, \aap, 585,
  A51

\bibitem[{{De Breuck} {et~al.}(2019){De Breuck}, {Wei{\ss}}, {B{\'e}thermin},
  {Cunningham}, {Apostolovski}, {Aravena}, {Archipley}, {Chapman}, {Chen},
  {Fu}, {Jarugula}, {Malkan}, {Mangian}, {Phadke}, {Reuter}, {Stacey},
  {Strandet}, {Vieira}, \& {Vishwas}}]{De_Breuck_2019}
{De Breuck}, C., {Wei{\ss}}, A., {B{\'e}thermin}, M., {et~al.} 2019, \aap, 631,
  A167

\bibitem[{De~Looze {et~al.}(2014)De~Looze, Cormier, Lebouteiller, Madden, Baes,
  Bendo, Boquien, Boselli, Clements, Cortese, Cooray, Galametz, Galliano,
  Graciá-Carpio, Isaak, Karczewski, Parkin, Pellegrini, Rémy-Ruyer,
  Spinoglio, Smith, \& Sturm}]{de_looze_applicability_2014}
De~Looze, I., Cormier, D., Lebouteiller, V., {et~al.} 2014, A\&A, 568, A62

\bibitem[{{Della Bruna} {et~al.}(2020){Della Bruna}, {Adamo}, {Bik},
  {Fumagalli}, {Walterbos}, {{\"O}stlin}, {Bruzual}, {Calzetti}, {Charlot},
  {Grasha}, {Smith}, {Thilker}, \& {Wofford}}]{Della_Bruna_2020}
{Della Bruna}, L., {Adamo}, A., {Bik}, A., {et~al.} 2020, \aap, 635, A134

\bibitem[{{Della Bruna} {et~al.}(2021){Della Bruna}, {Adamo}, {Lee}, {Smith},
  {Krumholz}, {Bik}, {Calzetti}, {Fox}, {Fumagalli}, {Grasha}, {Messa},
  {{\"O}stlin}, {Walterbos}, \& {Wofford}}]{Della_Bruna_2021}
{Della Bruna}, L., {Adamo}, A., {Lee}, J.~C., {et~al.} 2021, \aap, 650, A103

\bibitem[{{Duarte Puertas} {et~al.}(2022){Duarte Puertas}, {Vilchez},
  {Iglesias-P{\'a}ramo}, {Moll{\'a}}, {P{\'e}rez-Montero}, {Kehrig},
  {Pilyugin}, \& {Zinchenko}}]{Duarte_Puertas_2022}
{Duarte Puertas}, S., {Vilchez}, J.~M., {Iglesias-P{\'a}ramo}, J., {et~al.}
  2022, arXiv e-prints, arXiv:2205.01203

\bibitem[{{Eggen} {et~al.}(2021){Eggen}, {Scarlata}, {Skillman}, \&
  {Jaskot}}]{Eggen_2021}
{Eggen}, N.~R., {Scarlata}, C., {Skillman}, E., \& {Jaskot}, A. 2021, \apj,
  912, 12

\bibitem[{{Eldridge} {et~al.}(2017){Eldridge}, {Stanway}, {Xiao}, {McClelland},
  {Taylor}, {Ng}, {Greis}, \& {Bray}}]{eldridge_binary_2017}
{Eldridge}, J.~J., {Stanway}, E.~R., {Xiao}, L., {et~al.} 2017, \pasa, 34, e058

\bibitem[{{Espinosa-Ponce} {et~al.}(2020){Espinosa-Ponce}, {S{\'a}nchez},
  {Morisset}, {Barrera-Ballesteros}, {Galbany}, {Garc{\'\i}a-Benito},
  {Lacerda}, \& {Mast}}]{Espinosa-Ponce_2020}
{Espinosa-Ponce}, C., {S{\'a}nchez}, S.~F., {Morisset}, C., {et~al.} 2020,
  \mnras, 494, 1622

\bibitem[{{Falkendal} {et~al.}(2021){Falkendal}, {Lehnert}, {Vernet}, {De
  Breuck}, \& {Wang}}]{falkendal_alma_2020}
{Falkendal}, T., {Lehnert}, M.~D., {Vernet}, J., {De Breuck}, C., \& {Wang}, W.
  2021, \aap, 645, A120

\bibitem[{{Ferland} {et~al.}(2017){Ferland}, {Chatzikos}, {Guzm{\'a}n},
  {Lykins}, {van Hoof}, {Williams}, {Abel}, {Badnell}, {Keenan}, {Porter}, \&
  {Stancil}}]{2017_Cloudy_v17}
{Ferland}, G.~J., {Chatzikos}, M., {Guzm{\'a}n}, F., {et~al.} 2017, \rmxaa, 53,
  385

\bibitem[{{Finkelstein} {et~al.}(2019){Finkelstein}, {D'Aloisio},
  {Paardekooper}, {Ryan}, {Behroozi}, {Finlator}, {Livermore}, {Upton
  Sanderbeck}, {Dalla Vecchia}, \& {Khochfar}}]{Finkelstein_2019}
{Finkelstein}, S.~L., {D'Aloisio}, A., {Paardekooper}, J.-P., {et~al.} 2019,
  \apj, 879, 36

\bibitem[{{Fletcher} {et~al.}(2019){Fletcher}, {Tang}, {Robertson}, {Nakajima},
  {Ellis}, {Stark}, \& {Inoue}}]{Fletcher_2019}
{Fletcher}, T.~J., {Tang}, M., {Robertson}, B.~E., {et~al.} 2019, \apj, 878, 87

\bibitem[{{Flury} {et~al.}(2022{\natexlab{a}}){Flury}, {Jaskot}, {Ferguson},
  {Worseck}, {Makan}, {Chisholm}, {Saldana-Lopez}, {Schaerer}, {McCandliss},
  {Wang}, {Ford}, {Heckman}, {Ji}, {Giavalisco}, {Amorin}, {Atek}, {Blaizot},
  {Borthakur}, {Carr}, {Castellano}, {Cristiani}, {De Barros}, {Dickinson},
  {Finkelstein}, {Fleming}, {Fontanot}, {Garel}, {Grazian}, {Hayes}, {Henry},
  {Mauerhofer}, {Micheva}, {Oey}, {Ostlin}, {Papovich}, {Pentericci},
  {Ravindranath}, {Rosdahl}, {Rutkowski}, {Santini}, {Scarlata}, {Teplitz},
  {Thuan}, {Trebitsch}, {Vanzella}, {Verhamme}, \& {Xu}}]{Flury_2022}
{Flury}, S.~R., {Jaskot}, A.~E., {Ferguson}, H.~C., {et~al.}
  2022{\natexlab{a}}, \apjs, 260, 1

\bibitem[{{Flury} {et~al.}(2022{\natexlab{b}}){Flury}, {Jaskot}, {Ferguson},
  {Worseck}, {Makan}, {Chisholm}, {Saldana-Lopez}, {Schaerer}, {McCandliss},
  {Xu}, {Wang}, {Oey}, {Ford}, {Heckman}, {Ji}, {Giavalisco}, {Amor{\'\i}n},
  {Atek}, {Blaizot}, {Borthakur}, {Carr}, {Castellano}, {Barros}, {Dickinson},
  {Finkelstein}, {Fleming}, {Fontanot}, {Garel}, {Grazian}, {Hayes}, {Henry},
  {Mauerhofer}, {Micheva}, {Ostlin}, {Papovich}, {Pentericci}, {Ravindranath},
  {Rosdahl}, {Rutkowski}, {Santini}, {Scarlata}, {Teplitz}, {Thuan},
  {Trebitsch}, {Vanzella}, \& {Verhamme}}]{Flury_2022b}
{Flury}, S.~R., {Jaskot}, A.~E., {Ferguson}, H.~C., {et~al.}
  2022{\natexlab{b}}, \apj, 930, 126

\bibitem[{{Fujita} {et~al.}(2003){Fujita}, {Martin}, {Mac Low}, \&
  {Abel}}]{Fujita_2003}
{Fujita}, A., {Martin}, C.~L., {Mac Low}, M.-M., \& {Abel}, T. 2003, \apj, 599,
  50

\bibitem[{{Galliano}(2018)}]{Galliano_2018}
{Galliano}, F. 2018, \mnras, 476, 1445

\bibitem[{{Galliano} {et~al.}(2021){Galliano}, {Nersesian}, {Bianchi}, {De
  Looze}, {Roychowdhury}, {Baes}, {Casasola}, {Cassar{\'a}}, {Dobbels},
  {Fritz}, {Galametz}, {Jones}, {Madden}, {Mosenkov}, {Xilouris}, \&
  {Ysard}}]{galliano_nearby_2021}
{Galliano}, F., {Nersesian}, A., {Bianchi}, S., {et~al.} 2021, \aap, 649, A18

\bibitem[{{Gazagnes} {et~al.}(2020){Gazagnes}, {Chisholm}, {Schaerer},
  {Verhamme}, \& {Izotov}}]{gazagnes_origin_2020}
{Gazagnes}, S., {Chisholm}, J., {Schaerer}, D., {Verhamme}, A., \& {Izotov}, Y.
  2020, \aap, 639, A85

\bibitem[{{Gazagnes} {et~al.}(2018){Gazagnes}, {Chisholm}, {Schaerer},
  {Verhamme}, {Rigby}, \& {Bayliss}}]{gazagnes_neutral_2018}
{Gazagnes}, S., {Chisholm}, J., {Schaerer}, D., {et~al.} 2018, \aap, 616, A29

\bibitem[{{Grimm} {et~al.}(2003){Grimm}, {Gilfanov}, \& {Sunyaev}}]{Grimm_2003}
{Grimm}, H.~J., {Gilfanov}, M., \& {Sunyaev}, R. 2003, \mnras, 339, 793

\bibitem[{{Gross} {et~al.}(2021){Gross}, {Prestwich}, \& {Kaaret}}]{Gross_2021}
{Gross}, A.~C., {Prestwich}, A., \& {Kaaret}, P. 2021, \mnras, 505, 610

\bibitem[{{Hanish} {et~al.}(2010){Hanish}, {Oey}, {Rigby}, {de Mello}, \&
  {Lee}}]{hanish_2010}
{Hanish}, D.~J., {Oey}, M.~S., {Rigby}, J.~R., {de Mello}, D.~F., \& {Lee},
  J.~C. 2010, \apj, 725, 2029

\bibitem[{{Harikane} {et~al.}(2020){Harikane}, {Ouchi}, {Inoue}, {Matsuoka},
  {Tamura}, {Bakx}, {Fujimoto}, {Moriwaki}, {Ono}, {Nagao}, {Tadaki}, {Kojima},
  {Shibuya}, {Egami}, {Ferrara}, {Gallerani}, {Hashimoto}, {Kohno}, {Matsuda},
  {Matsuo}, {Pallottini}, {Sugahara}, \& {Vallini}}]{harikane_large_2020}
{Harikane}, Y., {Ouchi}, M., {Inoue}, A.~K., {et~al.} 2020, \apj, 896, 93

\bibitem[{{Hashimoto} {et~al.}(2019){Hashimoto}, {Inoue}, {Mawatari}, {Tamura},
  {Matsuo}, {Furusawa}, {Harikane}, {Shibuya}, {Knudsen}, {Kohno}, {Ono},
  {Zackrisson}, {Okamoto}, {Kashikawa}, {Oesch}, {Ouchi}, {Ota}, {Shimizu},
  {Taniguchi}, {Umehata}, \& {Watson}}]{Hashimoto_2019}
{Hashimoto}, T., {Inoue}, A.~K., {Mawatari}, K., {et~al.} 2019, \pasj, 71, 71

\bibitem[{{Henry} {et~al.}(2018){Henry}, {Berg}, {Scarlata}, {Verhamme}, \&
  {Erb}}]{Henry_2018_mgii}
{Henry}, A., {Berg}, D.~A., {Scarlata}, C., {Verhamme}, A., \& {Erb}, D. 2018,
  \apj, 855, 96

\bibitem[{{Henry} {et~al.}(2015){Henry}, {Scarlata}, {Martin}, \&
  {Erb}}]{Henry_2015}
{Henry}, A., {Scarlata}, C., {Martin}, C.~L., \& {Erb}, D. 2015, \apj, 809, 19

\bibitem[{{Herenz} {et~al.}(2017){Herenz}, {Hayes}, {Papaderos}, {Cannon},
  {Bik}, {Melinder}, \& {{\"O}stlin}}]{Herenz_2017}
{Herenz}, E.~C., {Hayes}, M., {Papaderos}, P., {et~al.} 2017, \aap, 606, L11

\bibitem[{{Hogarth} {et~al.}(2020){Hogarth}, {Amor{\'\i}n}, {V{\'\i}lchez},
  {H{\"a}gele}, {Cardaci}, {P{\'e}rez-Montero}, {Firpo}, {Jaskot}, \&
  {Ch{\'a}vez}}]{hogarth_chemodynamics_2020}
{Hogarth}, L., {Amor{\'\i}n}, R., {V{\'\i}lchez}, J.~M., {et~al.} 2020, \mnras,
  494, 3541

\bibitem[{{Hosokawa} \& {Inutsuka}(2005)}]{Hosokawa_Inutsuka_2005}
{Hosokawa}, T. \& {Inutsuka}, S.-i. 2005, \apj, 623, 917

\bibitem[{{Indriolo} {et~al.}(2007){Indriolo}, {Geballe}, {Oka}, \&
  {McCall}}]{2007_indriolo_cosmics}
{Indriolo}, N., {Geballe}, T.~R., {Oka}, T., \& {McCall}, B.~J. 2007, \apj,
  671, 1736

\bibitem[{{Inoue} {et~al.}(2001){Inoue}, {Hirashita}, \& {Kamaya}}]{Inoue_2001}
{Inoue}, A.~K., {Hirashita}, H., \& {Kamaya}, H. 2001, \apj, 555, 613

\bibitem[{{Inoue} {et~al.}(2016){Inoue}, {Tamura}, {Matsuo}, {Mawatari},
  {Shimizu}, {Shibuya}, {Ota}, {Yoshida}, {Zackrisson}, {Kashikawa}, {Kohno},
  {Umehata}, {Hatsukade}, {Iye}, {Matsuda}, {Okamoto}, \&
  {Yamaguchi}}]{Inoue_2016}
{Inoue}, A.~K., {Tamura}, Y., {Matsuo}, H., {et~al.} 2016, Science, 352, 1559

\bibitem[{{Izotov} {et~al.}(2001){Izotov}, {Chaffee}, \& {Green}}]{Izotov_2001}
{Izotov}, Y.~I., {Chaffee}, F.~H., \& {Green}, R.~F. 2001, \apj, 562, 727

\bibitem[{{Izotov} {et~al.}(2004{\natexlab{a}}){Izotov}, {Noeske}, {Guseva},
  {Papaderos}, {Thuan}, \& {Fricke}}]{Izotov_2004_nev}
{Izotov}, Y.~I., {Noeske}, K.~G., {Guseva}, N.~G., {et~al.} 2004{\natexlab{a}},
  \aap, 415, L27

\bibitem[{{Izotov} {et~al.}(2016{\natexlab{a}}){Izotov}, {Orlitov{\'a}},
  {Schaerer}, {Thuan}, {Verhamme}, {Guseva}, \& {Worseck}}]{Izotov_2016a}
{Izotov}, Y.~I., {Orlitov{\'a}}, I., {Schaerer}, D., {et~al.}
  2016{\natexlab{a}}, Nature, 529, 178

\bibitem[{{Izotov} {et~al.}(2004{\natexlab{b}}){Izotov}, {Papaderos}, {Guseva},
  {Fricke}, \& {Thuan}}]{Izotov_2004_tol}
{Izotov}, Y.~I., {Papaderos}, P., {Guseva}, N.~G., {Fricke}, K.~J., \& {Thuan},
  T.~X. 2004{\natexlab{b}}, \aap, 421, 539

\bibitem[{{Izotov} {et~al.}(2016{\natexlab{b}}){Izotov}, {Schaerer}, {Thuan},
  {Worseck}, {Guseva}, {Orlitov{\'a}}, \& {Verhamme}}]{Izotov_2016b}
{Izotov}, Y.~I., {Schaerer}, D., {Thuan}, T.~X., {et~al.} 2016{\natexlab{b}},
  MNRAS, 461, 3683

\bibitem[{{Izotov} {et~al.}(2018{\natexlab{a}}){Izotov}, {Schaerer}, {Worseck},
  {Guseva}, {Thuan}, {Verhamme}, {Orlitov{\'a}}, \& {Fricke}}]{Izotov_2018a}
{Izotov}, Y.~I., {Schaerer}, D., {Worseck}, G., {et~al.} 2018{\natexlab{a}},
  MNRAS, 474, 4514

\bibitem[{{Izotov} {et~al.}(2020){Izotov}, {Schaerer}, {Worseck}, {Verhamme},
  {Guseva}, {Thuan}, {Orlitov{\'a}}, \& {Fricke}}]{izotov_diverse_2020}
{Izotov}, Y.~I., {Schaerer}, D., {Worseck}, G., {et~al.} 2020, \mnras, 491, 468

\bibitem[{Izotov {et~al.}(2006)Izotov, Stasińska, Meynet, Guseva, \&
  Thuan}]{izotov_chemical_2006}
Izotov, Y.~I., Stasińska, G., Meynet, G., Guseva, N.~G., \& Thuan, T.~X. 2006,
  A\&A, 448, 955

\bibitem[{{Izotov} \& {Thuan}(1999)}]{Izotov_Thuan_1999}
{Izotov}, Y.~I. \& {Thuan}, T.~X. 1999, \apj, 511, 639

\bibitem[{{Izotov} {et~al.}(2017){Izotov}, {Thuan}, \& {Guseva}}]{Izotov_2017}
{Izotov}, Y.~I., {Thuan}, T.~X., \& {Guseva}, N.~G. 2017, \mnras, 471, 548

\bibitem[{{Izotov} {et~al.}(1994){Izotov}, {Thuan}, \&
  {Lipovetsky}}]{Izotov_1994}
{Izotov}, Y.~I., {Thuan}, T.~X., \& {Lipovetsky}, V.~A. 1994, \apj, 435, 647

\bibitem[{{Izotov} {et~al.}(2012){Izotov}, {Thuan}, \& {Privon}}]{Izotov_2012}
{Izotov}, Y.~I., {Thuan}, T.~X., \& {Privon}, G. 2012, \mnras, 427, 1229

\bibitem[{{Izotov} {et~al.}(2021){Izotov}, {Worseck}, {Schaerer}, {Guseva},
  {Chisholm}, {Thuan}, {Fricke}, \& {Verhamme}}]{Izotov_2021}
{Izotov}, Y.~I., {Worseck}, G., {Schaerer}, D., {et~al.} 2021, \mnras, 503,
  1734

\bibitem[{{Izotov} {et~al.}(2018{\natexlab{b}}){Izotov}, {Worseck}, {Schaerer},
  {Guseva}, {Thuan}, {Fricke}, \& {Orlitov{\'a}}}]{Izotov_2018b}
{Izotov}, Y.~I., {Worseck}, G., {Schaerer}, D., {et~al.} 2018{\natexlab{b}},
  MNRAS, 478, 4851

\bibitem[{{James} {et~al.}(2014){James}, {Aloisi}, {Heckman}, {Sohn}, \&
  {Wolfe}}]{James_2014}
{James}, B.~L., {Aloisi}, A., {Heckman}, T., {Sohn}, S.~T., \& {Wolfe}, M.~A.
  2014, \apj, 795, 109

\bibitem[{{Jaskot} \& {Oey}(2013)}]{Jaskot_2013}
{Jaskot}, A.~E. \& {Oey}, M.~S. 2013, \apj, 766, 91

\bibitem[{{Jiang} {et~al.}(2019){Jiang}, {Malhotra}, {Rhoads}, \&
  {Yang}}]{Jiang_Tianxing_2019}
{Jiang}, T., {Malhotra}, S., {Rhoads}, J.~E., \& {Yang}, H. 2019, \apj, 872,
  145

\bibitem[{{Jones} {et~al.}(2017){Jones}, {K{\"o}hler}, {Ysard}, {Bocchio}, \&
  {Verstraete}}]{Jones_2017}
{Jones}, A.~P., {K{\"o}hler}, M., {Ysard}, N., {Bocchio}, M., \& {Verstraete},
  L. 2017, \aap, 602, A46

\bibitem[{{Kaaret} \& {Feng}(2013)}]{Kaaret_Feng_2013}
{Kaaret}, P. \& {Feng}, H. 2013, in AAS/High Energy Astrophysics Division,
  Vol.~13, AAS/High Energy Astrophysics Division \#13, 402.06

\bibitem[{{Kaaret} {et~al.}(2011){Kaaret}, {Schmitt}, \&
  {Gorski}}]{Kaaret_2011}
{Kaaret}, P., {Schmitt}, J., \& {Gorski}, M. 2011, \apj, 741, 10

\bibitem[{{Kakiichi} \& {Gronke}(2021)}]{kakiichi_lyman_2019}
{Kakiichi}, K. \& {Gronke}, M. 2021, \apj, 908, 30

\bibitem[{Katz {et~al.}(2022)Katz, Garel, Rosdahl, Mauerhofer, Kimm, Blaizot,
  Michel-Dansac, Devriendt, Slyz, \& Haehnelt}]{Katz_2022}
Katz, H., Garel, T., Rosdahl, J., {et~al.} 2022, \mnras, stac1437

\bibitem[{{Katz} {et~al.}(2022){Katz}, {Rosdahl}, {Kimm}, {Garel}, {Blaizot},
  {Haehnelt}, {Michel-Dansac}, {Martin-Alvarez}, {Devriendt}, {Slyz},
  {Teyssier}, {Ocvirk}, {Laporte}, \& {Ellis}}]{Katz_2021}
{Katz}, H., {Rosdahl}, J., {Kimm}, T., {et~al.} 2022, \mnras, 510, 5603

\bibitem[{{Katz} {et~al.}(2020){Katz}, {{\v{D}}urov{\v{c}}{\'\i}kov{\'a}},
  {Kimm}, {Rosdahl}, {Blaizot}, {Haehnelt}, {Devriendt}, {Slyz}, {Ellis}, \&
  {Laporte}}]{katz_new_2020}
{Katz}, H., {{\v{D}}urov{\v{c}}{\'\i}kov{\'a}}, D., {Kimm}, T., {et~al.} 2020,
  \mnras, 498, 164

\bibitem[{{Kehrig} {et~al.}(2018){Kehrig}, {V{\'\i}lchez}, {Guerrero},
  {Iglesias-P{\'a}ramo}, {Hunt}, {Duarte-Puertas}, \&
  {Ramos-Larios}}]{2018_kehrig_SBS0335}
{Kehrig}, C., {V{\'\i}lchez}, J.~M., {Guerrero}, M.~A., {et~al.} 2018, \mnras,
  480, 1081

\bibitem[{{Kennicutt} {et~al.}(1995){Kennicutt}, {Bresolin}, {Bomans},
  {Bothun}, \& {Thompson}}]{Kennicutt_1995}
{Kennicutt}, Robert~C., J., {Bresolin}, F., {Bomans}, D.~J., {Bothun}, G.~D.,
  \& {Thompson}, I.~B. 1995, \aj, 109, 594

\bibitem[{{Kennicutt} {et~al.}(2009){Kennicutt}, {Hao}, {Calzetti},
  {Moustakas}, {Dale}, {Bendo}, {Engelbracht}, {Johnson}, \&
  {Lee}}]{Kennicutt_2009}
{Kennicutt}, Robert~C., J., {Hao}, C.-N., {Calzetti}, D., {et~al.} 2009, \apj,
  703, 1672

\bibitem[{{Kewley} \& {Dopita}(2002)}]{Kewley_2002}
{Kewley}, L.~J. \& {Dopita}, M.~A. 2002, \apjs, 142, 35

\bibitem[{{Kim} {et~al.}(2022){Kim}, {Chevance}, {Kruijssen}, {Leroy},
  {Schruba}, {Barnes}, {Bigiel}, {Blanc}, {Cao}, {Congiu}, {Dale}, {Faesi},
  {Glover}, {Grasha}, {Groves}, {Hughes}, {Klessen}, {Kreckel}, {McElroy},
  {Pan}, {Pety}, {Querejeta}, {Razza}, {Rosolowsky}, {Saito}, {Schinnerer},
  {Sun}, {Tomi{\v{c}}i{\'c}}, {Usero}, \& {Williams}}]{Kim_Jaeyeon_2022}
{Kim}, J., {Chevance}, M., {Kruijssen}, J.~M.~D., {et~al.} 2022, arXiv
  e-prints, arXiv:2206.09857

\bibitem[{{Kim} {et~al.}(2018){Kim}, {Kim}, \& {Ostriker}}]{Kim_2018}
{Kim}, J.-G., {Kim}, W.-T., \& {Ostriker}, E.~C. 2018, \apj, 859, 68

\bibitem[{{Kim} {et~al.}(2019){Kim}, {Kim}, \& {Ostriker}}]{kim_2019}
{Kim}, J.-G., {Kim}, W.-T., \& {Ostriker}, E.~C. 2019, \apj, 883, 102

\bibitem[{{Kim} {et~al.}(2021){Kim}, {Ostriker}, \& {Filippova}}]{Kim_2021}
{Kim}, J.-G., {Ostriker}, E.~C., \& {Filippova}, N. 2021, \apj, 911, 128

\bibitem[{{Kimm} {et~al.}(2019){Kimm}, {Blaizot}, {Garel}, {Michel-Dansac},
  {Katz}, {Rosdahl}, {Verhamme}, \& {Haehnelt}}]{Kimm_2019}
{Kimm}, T., {Blaizot}, J., {Garel}, T., {et~al.} 2019, \mnras, 486, 2215

\bibitem[{{Kimm} {et~al.}(2017){Kimm}, {Katz}, {Haehnelt}, {Rosdahl},
  {Devriendt}, \& {Slyz}}]{Kimm_2017}
{Kimm}, T., {Katz}, H., {Haehnelt}, M., {et~al.} 2017, \mnras, 466, 4826

\bibitem[{{Kreckel} {et~al.}(2016){Kreckel}, {Blanc}, {Schinnerer}, {Groves},
  {Adamo}, {Hughes}, \& {Meidt}}]{Kreckel_2016}
{Kreckel}, K., {Blanc}, G.~A., {Schinnerer}, E., {et~al.} 2016, \apj, 827, 103

\bibitem[{{Kunth} {et~al.}(1998){Kunth}, {Mas-Hesse}, {Terlevich}, {Terlevich},
  {Lequeux}, \& {Fall}}]{Kunth_1998}
{Kunth}, D., {Mas-Hesse}, J.~M., {Terlevich}, E., {et~al.} 1998, \aap, 334, 11

\bibitem[{{Lacerda} {et~al.}(2018){Lacerda}, {Cid Fernandes}, {Couto},
  {Stasi{\'n}ska}, {Garc{\'\i}a-Benito}, {Vale Asari}, {P{\'e}rez},
  {Gonz{\'a}lez Delgado}, {S{\'a}nchez}, \& {de Amorim}}]{Lacerda_2018}
{Lacerda}, E.~A.~D., {Cid Fernandes}, R., {Couto}, G.~S., {et~al.} 2018,
  \mnras, 474, 3727

\bibitem[{{Lambert-Huyghe} {et~al.}(2022){Lambert-Huyghe}, {Madden},
  {Lebouteiller}, {Galliano}, {Abel}, {Hu}, {Ramambason}, \&
  {Polles}}]{Lambert-Huygues_2021}
{Lambert-Huyghe}, A., {Madden}, S.~C., {Lebouteiller}, V., {et~al.} 2022, arXiv
  e-prints, arXiv:2206.15417

\bibitem[{{Lardo} {et~al.}(2015){Lardo}, {Davies}, {Kudritzki}, {Gazak},
  {Evans}, {Patrick}, {Bergemann}, \& {Plez}}]{Lardo_2015}
{Lardo}, C., {Davies}, B., {Kudritzki}, R.~P., {et~al.} 2015, \apj, 812, 160

\bibitem[{{Le Petit} {et~al.}(2006){Le Petit}, {Nehm{\'e}}, {Le Bourlot}, \&
  {Roueff}}]{2006_meudon_pdr}
{Le Petit}, F., {Nehm{\'e}}, C., {Le Bourlot}, J., \& {Roueff}, E. 2006, \apjs,
  164, 506

\bibitem[{{Lebouteiller} {et~al.}(2019){Lebouteiller}, {Cormier}, {Madden},
  {Galametz}, {Hony}, {Galliano}, {Chevance}, {Lee}, {Braine}, {Polles},
  {Reque{\~n}a-Torres}, {Indebetouw}, {Hughes}, \&
  {Abel}}]{lebouteiller_physical_2019}
{Lebouteiller}, V., {Cormier}, D., {Madden}, S.~C., {et~al.} 2019, \aap, 632,
  A106

\bibitem[{Lebouteiller {et~al.}(2017)Lebouteiller, Péquignot, Cormier, Madden,
  Pakull, Kunth, Galliano, Chevance, Heap, Lee, \&
  Polles}]{lebouteiller_neutral_2017}
Lebouteiller, V., Péquignot, D., Cormier, D., {et~al.} 2017, A\&A, 602, A45

\bibitem[{{Lebouteiller} \& {Ramambason}(2022)}]{LebouteillerRamambason2022}
{Lebouteiller}, V. \& {Ramambason}, L. 2022, arXiv e-prints, arXiv:2207.05657

\bibitem[{{Lee} {et~al.}(2019){Lee}, {Madden}, {Le Petit}, {Gusdorf},
  {Lesaffre}, {Wu}, {Lebouteiller}, {Galliano}, \& {Chevance}}]{Lee_2019}
{Lee}, M.~Y., {Madden}, S.~C., {Le Petit}, F., {et~al.} 2019, \aap, 628, A113

\bibitem[{{Lee} {et~al.}(2016){Lee}, {Madden}, {Lebouteiller}, {Gusdorf},
  {Godard}, {Wu}, {Galametz}, {Cormier}, {Le Petit}, {Roueff}, {Bron},
  {Carlson}, {Chevance}, {Fukui}, {Galliano}, {Hony}, {Hughes}, {Indebetouw},
  {Israel}, {Kawamura}, {Le Bourlot}, {Lesaffre}, {Meixner}, {Muller}, {Nayak},
  {Onishi}, {Roman-Duval}, \& {Sewi{\l}o}}]{Lee_2016}
{Lee}, M.~Y., {Madden}, S.~C., {Lebouteiller}, V., {et~al.} 2016, \aap, 596,
  A85

\bibitem[{{Leitet} {et~al.}(2013){Leitet}, {Bergvall}, {Hayes}, {Linn{\'e}}, \&
  {Zackrisson}}]{Leitet_2013}
{Leitet}, E., {Bergvall}, N., {Hayes}, M., {Linn{\'e}}, S., \& {Zackrisson}, E.
  2013, \aap, 553, A106

\bibitem[{Leitherer {et~al.}(2016)Leitherer, Hernandez, Lee, \&
  Oey}]{leitherer_direct_2016}
Leitherer, C., Hernandez, S., Lee, J.~C., \& Oey, M.~S. 2016, ApJ, 823, 64

\bibitem[{{Leitherer} {et~al.}(1999){Leitherer}, {Schaerer}, {Goldader},
  {Delgado}, {Robert}, {Kune}, {de Mello}, {Devost}, \&
  {Heckman}}]{Starburst99_Leitherer_1999}
{Leitherer}, C., {Schaerer}, D., {Goldader}, J.~D., {et~al.} 1999, \apjs, 123,
  3

\bibitem[{{Ma} {et~al.}(2016){Ma}, {Hopkins}, {Kasen}, {Quataert},
  {Faucher-Gigu{\`e}re}, {Kere{\v{s}}}, {Murray}, \& {Strom}}]{Ma_2016}
{Ma}, X., {Hopkins}, P.~F., {Kasen}, D., {et~al.} 2016, \mnras, 459, 3614

\bibitem[{{Ma} {et~al.}(2020){Ma}, {Quataert}, {Wetzel}, {Hopkins},
  {Faucher-Gigu{\`e}re}, \& {Kere{\v{s}}}}]{Ma_2020}
{Ma}, X., {Quataert}, E., {Wetzel}, A., {et~al.} 2020, \mnras, 498, 2001

\bibitem[{{Madden} {et~al.}(2020){Madden}, {Cormier}, {Hony}, {Lebouteiller},
  {Abel}, {Galametz}, {De Looze}, {Chevance}, {Polles}, {Lee}, {Galliano},
  {Lambert-Huyghe}, {Hu}, \& {Ramambason}}]{madden_tracing_2020}
{Madden}, S.~C., {Cormier}, D., {Hony}, S., {et~al.} 2020, \aap, 643, A141

\bibitem[{{Madden} {et~al.}(2013){Madden}, {R{\'e}my-Ruyer}, {Galametz},
  {Cormier}, {Lebouteiller}, {Galliano}, {Hony}, {Bendo}, {Smith}, {Pohlen},
  {Roussel}, {Sauvage}, {Wu}, {Sturm}, {Poglitsch}, {Contursi}, {Doublier},
  {Baes}, {Barlow}, {Boselli}, {Boquien}, {Carlson}, {Ciesla}, {Cooray},
  {Cortese}, {de Looze}, {Irwin}, {Isaak}, {Kamenetzky}, {Karczewski}, {Lu},
  {MacHattie}, {O'Halloran}, {Parkin}, {Rangwala}, {Schirm}, {Schulz},
  {Spinoglio}, {Vaccari}, {Wilson}, \& {Wozniak}}]{Madden_2013}
{Madden}, S.~C., {R{\'e}my-Ruyer}, A., {Galametz}, M., {et~al.} 2013, \pasp,
  125, 600

\bibitem[{{Maiolino} \& {Mannucci}(2019)}]{maiolino_re_2019}
{Maiolino}, R. \& {Mannucci}, F. 2019, \aapr, 27, 3

\bibitem[{{Maji} {et~al.}(2022){Maji}, {Verhamme}, {Rosdahl}, {Garel},
  {Blaizot}, {Mauerhofer}, {Pittavino}, {Victoria Feser}, {Chuniaud}, {Kimm},
  {Katz}, \& {Haehnelt}}]{Maji_2022}
{Maji}, M., {Verhamme}, A., {Rosdahl}, J., {et~al.} 2022, \aap, 663, A66

\bibitem[{{Marques-Chaves} {et~al.}(2022){Marques-Chaves}, {Schaerer},
  {Amor{\'\i}n}, {Atek}, {Borthakur}, {Chisholm}, {Fern{\'a}ndez}, {Flury},
  {Giavalisco}, {Grazian}, {Hayes}, {Heckman}, {Henry}, {Izotov}, {Jaskot},
  {Ji}, {McCandliss}, {Oey}, {{\"O}stlin}, {Ravindranath}, {Rutkowski},
  {Saldana-Lopez}, {Teplitz}, {Thuan}, {Verhamme}, {Wang}, {Worseck}, \&
  {Xu}}]{Marques-Chaves_2022}
{Marques-Chaves}, R., {Schaerer}, D., {Amor{\'\i}n}, R.~O., {et~al.} 2022,
  \aap, 663, L1

\bibitem[{{Martin}(1998)}]{Martin_1998}
{Martin}, C.~L. 1998, \apj, 506, 222

\bibitem[{{Mathis}(1986)}]{Mathis_1986}
{Mathis}, J.~S. 1986, \apj, 301, 423

\bibitem[{{Mauerhofer} {et~al.}(2021){Mauerhofer}, {Verhamme}, {Blaizot},
  {Garel}, {Kimm}, {Michel-Dansac}, \& {Rosdahl}}]{Mauerhofer_2021}
{Mauerhofer}, V., {Verhamme}, A., {Blaizot}, J., {et~al.} 2021, \aap, 646, A80

\bibitem[{{McQuinn} {et~al.}(2018){McQuinn}, {Skillman}, {Heilman}, {Mitchell},
  \& {Kelley}}]{McQuinn_2018}
{McQuinn}, K. B.~W., {Skillman}, E.~D., {Heilman}, T.~N., {Mitchell}, N.~P., \&
  {Kelley}, T. 2018, \mnras, 477, 3164

\bibitem[{{Menacho} {et~al.}(2021){Menacho}, {{\"O}stlin}, {Bik}, {Adamo},
  {Bergvall}, {Della Bruna}, {Hayes}, {Melinder}, \&
  {Rivera-Thorsen}}]{Menacho_2021}
{Menacho}, V., {{\"O}stlin}, G., {Bik}, A., {et~al.} 2021, \mnras, 506, 1777

\bibitem[{{Menacho} {et~al.}(2019){Menacho}, {{\"O}stlin}, {Bik}, {Della
  Bruna}, {Melinder}, {Adamo}, {Hayes}, {Herenz}, \& {Bergvall}}]{Menacho_2019}
{Menacho}, V., {{\"O}stlin}, G., {Bik}, A., {et~al.} 2019, \mnras, 487, 3183

\bibitem[{{Meyer} {et~al.}(2022){Meyer}, {Walter}, {Cicone}, {Cox}, {Decarli},
  {Neri}, {Novak}, {Pensabene}, {Riechers}, \& {Weiss}}]{Meyer_2022}
{Meyer}, R.~A., {Walter}, F., {Cicone}, C., {et~al.} 2022, \apj, 927, 152

\bibitem[{{Mingozzi} {et~al.}(2020){Mingozzi}, {Belfiore}, {Cresci}, {Bundy},
  {Bershady}, {Bizyaev}, {Blanc}, {Boquien}, {Drory}, {Fu}, {Maiolino},
  {Riffel}, {Schaefer}, {Storchi-Bergmann}, {Telles}, {Tremonti}, {Zakamska},
  \& {Zhang}}]{mingozzi_sdss_2020}
{Mingozzi}, M., {Belfiore}, F., {Cresci}, G., {et~al.} 2020, \aap, 636, A42

\bibitem[{{Mirabel} {et~al.}(2011){Mirabel}, {Dijkstra}, {Laurent}, {Loeb}, \&
  {Pritchard}}]{Mirabel_2011}
{Mirabel}, I.~F., {Dijkstra}, M., {Laurent}, P., {Loeb}, A., \& {Pritchard},
  J.~R. 2011, \aap, 528, A149

\bibitem[{{Mitsuda} {et~al.}(1984){Mitsuda}, {Inoue}, {Koyama}, {Makishima},
  {Matsuoka}, {Ogawara}, {Shibazaki}, {Suzuki}, {Tanaka}, \&
  {Hirano}}]{1984PASJ...36..741M}
{Mitsuda}, K., {Inoue}, H., {Koyama}, K., {et~al.} 1984, \pasj, 36, 741

\bibitem[{{Naidu} {et~al.}(2018){Naidu}, {Forrest}, {Oesch}, {Tran}, \&
  {Holden}}]{Naidu_2018}
{Naidu}, R.~P., {Forrest}, B., {Oesch}, P.~A., {Tran}, K.-V.~H., \& {Holden},
  B.~P. 2018, \mnras, 478, 791

\bibitem[{{Naidu} {et~al.}(2017){Naidu}, {Oesch}, {Reddy}, {Holden}, {Steidel},
  {Montes}, {Atek}, {Bouwens}, {Carollo}, {Cibinel}, {Illingworth},
  {Labb{\'e}}, {Magee}, {Morselli}, {Nelson}, {van Dokkum}, \&
  {Wilkins}}]{Naidu_2017}
{Naidu}, R.~P., {Oesch}, P.~A., {Reddy}, N., {et~al.} 2017, \apj, 847, 12

\bibitem[{{Naidu} {et~al.}(2020){Naidu}, {Tacchella}, {Mason}, {Bose}, {Oesch},
  \& {Conroy}}]{Naidu_2020}
{Naidu}, R.~P., {Tacchella}, S., {Mason}, C.~A., {et~al.} 2020, \apj, 892, 109

\bibitem[{{Nakajima} {et~al.}(2020){Nakajima}, {Ellis}, {Robertson}, {Tang}, \&
  {Stark}}]{Nakajima_2020}
{Nakajima}, K., {Ellis}, R.~S., {Robertson}, B.~E., {Tang}, M., \& {Stark},
  D.~P. 2020, \apj, 889, 161

\bibitem[{{Nakajima} \& {Ouchi}(2014)}]{Nakajima_2014}
{Nakajima}, K. \& {Ouchi}, M. 2014, \mnras, 442, 900

\bibitem[{Nicholls {et~al.}(2017)Nicholls, Sutherland, Dopita, Kewley, \&
  Groves}]{nicholls_abundance_2017}
Nicholls, D.~C., Sutherland, R.~S., Dopita, M.~A., Kewley, L.~J., \& Groves,
  B.~A. 2017, 20

\bibitem[{{Oberst} {et~al.}(2011){Oberst}, {Parshley}, {Nikola}, {Stacey},
  {L{\"o}hr}, {Lane}, {Stark}, \& {Kamenetzky}}]{Oberst_2011}
{Oberst}, T.~E., {Parshley}, S.~C., {Nikola}, T., {et~al.} 2011, \apj, 739, 100

\bibitem[{Oey {et~al.}(2007)Oey, Meurer, Yelda, Furst, Caballero‐Nieves,
  Hanish, Levesque, Thilker, Walth, Bland‐Hawthorn, Dopita, Ferguson,
  Heckman, Doyle, Drinkwater, Freeman, Kennicutt, Kilborn, Knezek, Koribalski,
  Meyer, Putman, Ryan‐Weber, Smith, Staveley‐Smith, Webster, Werk, \&
  Zwaan}]{oey_survey_2007}
Oey, M.~S., Meurer, G.~R., Yelda, S., {et~al.} 2007, ApJ, 661, 801

\bibitem[{{Olivier} {et~al.}(2021){Olivier}, {Berg}, {Chisholm}, {Erb},
  {Pogge}, \& {Skillman}}]{Olivier_2021}
{Olivier}, G.~M., {Berg}, D.~A., {Chisholm}, J., {et~al.} 2021, arXiv e-prints,
  arXiv:2109.06725

\bibitem[{{Ot{\'\i}-Floranes} {et~al.}(2012){Ot{\'\i}-Floranes}, {Mas-Hesse},
  {Jim{\'e}nez-Bail{\'o}n}, {Schaerer}, {Hayes}, {{\"O}stlin}, {Atek}, \&
  {Kunth}}]{OtiFloranes_2012}
{Ot{\'\i}-Floranes}, H., {Mas-Hesse}, J.~M., {Jim{\'e}nez-Bail{\'o}n}, E.,
  {et~al.} 2012, \aap, 546, A65

\bibitem[{{Ott} {et~al.}(2005){Ott}, {Walter}, \& {Brinks}}]{Ott_2005}
{Ott}, J., {Walter}, F., \& {Brinks}, E. 2005, \mnras, 358, 1453

\bibitem[{{Pahl} {et~al.}(2021){Pahl}, {Shapley}, {Steidel}, {Chen}, \&
  {Reddy}}]{Pahl_2021}
{Pahl}, A.~J., {Shapley}, A., {Steidel}, C.~C., {Chen}, Y., \& {Reddy}, N.~A.
  2021, \mnras, 505, 2447

\bibitem[{{Pahl} {et~al.}(2022){Pahl}, {Shapley}, {Steidel}, {Reddy}, \&
  {Chen}}]{Pahl_2022}
{Pahl}, A.~J., {Shapley}, A., {Steidel}, C.~C., {Reddy}, N.~A., \& {Chen}, Y.
  2022, \mnras, stac1767

\bibitem[{{Papaderos} {et~al.}(1994){Papaderos}, {Fricke}, {Thuan}, \&
  {Loose}}]{Papaderos_1994}
{Papaderos}, P., {Fricke}, K.~J., {Thuan}, T.~X., \& {Loose}, H.~H. 1994, \aap,
  291, L13

\bibitem[{{Pellegrini} {et~al.}(2012){Pellegrini}, {Oey}, {Winkler}, {Points},
  {Smith}, {Jaskot}, \& {Zastrow}}]{Pellegrini_2012}
{Pellegrini}, E.~W., {Oey}, M.~S., {Winkler}, P.~F., {et~al.} 2012, \apj, 755,
  40

\bibitem[{{P{\'e}quignot}(2008)}]{pequignot_heating_nodate}
{P{\'e}quignot}, D. 2008, \aap, 478, 371

\bibitem[{{Pilyugin} {et~al.}(2014){Pilyugin}, {Grebel}, {Zinchenko}, \&
  {Kniazev}}]{Pilyugin_2014}
{Pilyugin}, L.~S., {Grebel}, E.~K., {Zinchenko}, I.~A., \& {Kniazev}, A.~Y.
  2014, \aj, 148, 134

\bibitem[{{Pilyugin} \& {Thuan}(2005)}]{Pilyugin_Thuan_2005}
{Pilyugin}, L.~S. \& {Thuan}, T.~X. 2005, \apj, 631, 231

\bibitem[{{Plat} {et~al.}(2019){Plat}, {Charlot}, {Bruzual}, {Feltre},
  {Vidal-Garc{\'\i}a}, {Morisset}, {Chevallard}, \&
  {Todt}}]{plat_constraints_2019}
{Plat}, A., {Charlot}, S., {Bruzual}, G., {et~al.} 2019, \mnras, 490, 978

\bibitem[{Polles {et~al.}(2019)Polles, Madden, Lebouteiller, Cormier, Abel,
  Galliano, Hony, Karczewski, Lee, Chevance, Galametz, \&
  Lianou}]{polles_modeling_2019}
Polles, F.~L., Madden, S.~C., Lebouteiller, V., {et~al.} 2019, A\&A, 622, A119

\bibitem[{{Prestwich} {et~al.}(2015){Prestwich}, {Jackson}, {Kaaret}, {Brorby},
  {Roberts}, {Saar}, \& {Yukita}}]{Prestwich_2015}
{Prestwich}, A.~H., {Jackson}, F., {Kaaret}, P., {et~al.} 2015, \apj, 812, 166

\bibitem[{{Prestwich} {et~al.}(2013){Prestwich}, {Tsantaki}, {Zezas},
  {Jackson}, {Roberts}, {Foltz}, {Linden}, \& {Kalogera}}]{Prestwich_2013}
{Prestwich}, A.~H., {Tsantaki}, M., {Zezas}, A., {et~al.} 2013, \apj, 769, 92

\bibitem[{{Ramambason} {et~al.}(2020){Ramambason}, {Schaerer}, {Stasi{\'n}ska},
  {Izotov}, {Guseva}, {V{\'\i}lchez}, {Amor{\'\i}n}, \&
  {Morisset}}]{ramambason_reconciling_2020}
{Ramambason}, L., {Schaerer}, D., {Stasi{\'n}ska}, G., {et~al.} 2020, \aap,
  644, A21

\bibitem[{{Ramos Padilla} {et~al.}(2021){Ramos Padilla}, {Wang}, {Ploeckinger},
  {van der Tak}, \& {Trager}}]{Ramos_Padilla_2021}
{Ramos Padilla}, A.~F., {Wang}, L., {Ploeckinger}, S., {van der Tak}, F.~F.~S.,
  \& {Trager}, S.~C. 2021, \aap, 645, A133

\bibitem[{{Ramos Padilla} {et~al.}(2022){Ramos Padilla}, {Wang}, {van der Tak},
  \& {Trager}}]{Ramos_Padilla_2022}
{Ramos Padilla}, A.~F., {Wang}, L., {van der Tak}, F.~F.~S., \& {Trager}, S.
  2022, arXiv e-prints, arXiv:2205.11955

\bibitem[{{Reddy} {et~al.}(2016){Reddy}, {Steidel}, {Pettini},
  {Bogosavljevi{\'c}}, \& {Shapley}}]{Reddy_2016}
{Reddy}, N.~A., {Steidel}, C.~C., {Pettini}, M., {Bogosavljevi{\'c}}, M., \&
  {Shapley}, A.~E. 2016, \apj, 828, 108

\bibitem[{{Reines} {et~al.}(2011){Reines}, {Sivakoff}, {Johnson}, \&
  {Brogan}}]{Reines_2011}
{Reines}, A.~E., {Sivakoff}, G.~R., {Johnson}, K.~E., \& {Brogan}, C.~L. 2011,
  \nat, 470, 66

\bibitem[{{Rela{\~n}o} {et~al.}(2018){Rela{\~n}o}, {De Looze}, {Kennicutt},
  {Lisenfeld}, {Dariush}, {Verley}, {Braine}, {Tabatabaei}, {Kramer},
  {Boquien}, {Xilouris}, \& {Gratier}}]{Relano_2018}
{Rela{\~n}o}, M., {De Looze}, I., {Kennicutt}, R.~C., {et~al.} 2018, \aap, 613,
  A43

\bibitem[{{R{\'e}my-Ruyer}(2013)}]{Phd_RemyRuyer_2013}
{R{\'e}my-Ruyer}, A. 2013, PhD thesis

\bibitem[{{Richardson} {et~al.}(2022){Richardson}, {Simpson}, {Polimera},
  {Kannappan}, {Bellovary}, {Greene}, \& {Jenkins}}]{Richardson_2022}
{Richardson}, C.~T., {Simpson}, C., {Polimera}, M.~S., {et~al.} 2022, \apj,
  927, 165

\bibitem[{{Rivera-Thorsen} {et~al.}(2019){Rivera-Thorsen}, {Dahle}, {Chisholm},
  {Florian}, {Gronke}, {Rigby}, {Gladders}, {Mahler}, {Sharon}, \&
  {Bayliss}}]{Rivera-Thorsen_2019}
{Rivera-Thorsen}, T.~E., {Dahle}, H., {Chisholm}, J., {et~al.} 2019, Science,
  366, 738

\bibitem[{{Robertson} {et~al.}(2015){Robertson}, {Ellis}, {Furlanetto}, \&
  {Dunlop}}]{Robertson_2015}
{Robertson}, B.~E., {Ellis}, R.~S., {Furlanetto}, S.~R., \& {Dunlop}, J.~S.
  2015, \apjl, 802, L19

\bibitem[{{Robertson} {et~al.}(2013){Robertson}, {Furlanetto}, {Schneider},
  {Charlot}, {Ellis}, {Stark}, {McLure}, {Dunlop}, {Koekemoer}, {Schenker},
  {Ouchi}, {Ono}, {Curtis-Lake}, {Rogers}, {Bowler}, \&
  {Cirasuolo}}]{Robertson_2013}
{Robertson}, B.~E., {Furlanetto}, S.~R., {Schneider}, E., {et~al.} 2013, \apj,
  768, 71

\bibitem[{{Rosdahl} {et~al.}(2018){Rosdahl}, {Katz}, {Blaizot}, {Kimm},
  {Michel-Dansac}, {Garel}, {Haehnelt}, {Ocvirk}, \&
  {Teyssier}}]{rosdahl_sphinx_2018}
{Rosdahl}, J., {Katz}, H., {Blaizot}, J., {et~al.} 2018, \mnras, 479, 994

\bibitem[{Rémy-Ruyer {et~al.}(2014)Rémy-Ruyer, Madden, Galliano, Galametz,
  Takeuchi, Asano, Zhukovska, Lebouteiller, Cormier, Jones, Bocchio, Baes,
  Bendo, Boquien, Boselli, DeLooze, Doublier-Pritchard, Hughes, Karczewski, \&
  Spinoglio}]{remy-ruyer_gas--dust_2014}
Rémy-Ruyer, A., Madden, S.~C., Galliano, F., {et~al.} 2014, A\&A, 563, A31

\bibitem[{Rémy-Ruyer {et~al.}(2015)Rémy-Ruyer, Madden, Galliano,
  Lebouteiller, Baes, Bendo, Boselli, Ciesla, Cormier, Cooray, Cortese,
  De~Looze, Doublier-Pritchard, Galametz, Jones, Karczewski, Lu, \&
  Spinoglio}]{remy-ruyer_linking_2015}
Rémy-Ruyer, A., Madden, S.~C., Galliano, F., {et~al.} 2015, A\&A, 582, A121

\bibitem[{{Saldana-Lopez} {et~al.}(2022){Saldana-Lopez}, {Schaerer},
  {Chisholm}, {Flury}, {Jaskot}, {Worseck}, {Makan}, {Gazagnes}, {Mauerhofer},
  {Verhamme}, {Amor{\'\i}n}, {Ferguson}, {Giavalisco}, {Grazian}, {Hayes},
  {Heckman}, {Henry}, {Ji}, {Marques-Chaves}, {McCandliss}, {Oey},
  {{\"O}stlin}, {Pentericci}, {Thuan}, {Trebitsch}, {Vanzella}, \&
  {Xu}}]{Saldana-Lopez_2022}
{Saldana-Lopez}, A., {Schaerer}, D., {Chisholm}, J., {et~al.} 2022, \aap, 663,
  A59

\bibitem[{{S{\'a}nchez-Cruces} {et~al.}(2015){S{\'a}nchez-Cruces}, {Rosado},
  {Rodr{\'\i}guez-Gonz{\'a}lez}, \& {Reyes-Iturbide}}]{Sanches_Cruces_2015}
{S{\'a}nchez-Cruces}, M., {Rosado}, M., {Rodr{\'\i}guez-Gonz{\'a}lez}, A., \&
  {Reyes-Iturbide}, J. 2015, \apj, 799, 231

\bibitem[{{Satyapal} {et~al.}(2021){Satyapal}, {Kamal}, {Cann}, {Secrest}, \&
  {Abel}}]{satyapal_diagnostic_2020}
{Satyapal}, S., {Kamal}, L., {Cann}, J.~M., {Secrest}, N.~J., \& {Abel}, N.~P.
  2021, \apj, 906, 35

\bibitem[{{Saxena} {et~al.}(2022){Saxena}, {Pentericci}, {Ellis}, {Guaita},
  {Calabr{\`o}}, {Schaerer}, {Vanzella}, {Amor{\'\i}n}, {Bolzonella},
  {Castellano}, {Fontanot}, {Hathi}, {Hibon}, {Llerena}, {Mannucci},
  {Saldana-Lopez}, {Talia}, \& {Zamorani}}]{Saxena_2021}
{Saxena}, A., {Pentericci}, L., {Ellis}, R.~S., {et~al.} 2022, \mnras, 511, 120

\bibitem[{{Schaerer} {et~al.}(1999){Schaerer}, {Contini}, \&
  {Pindao}}]{Schaerer_1999}
{Schaerer}, D., {Contini}, T., \& {Pindao}, M. 1999, \aaps, 136, 35

\bibitem[{{Schaerer} {et~al.}(2019){Schaerer}, {Fragos}, \&
  {Izotov}}]{schaerer_new_2019}
{Schaerer}, D., {Fragos}, T., \& {Izotov}, Y.~I. 2019, \aap, 622, L10

\bibitem[{{Sembach} {et~al.}(2000){Sembach}, {Howk}, {Ryans}, \&
  {Keenan}}]{Sembach_2000}
{Sembach}, K.~R., {Howk}, J.~C., {Ryans}, R. S.~I., \& {Keenan}, F.~P. 2000,
  \apj, 528, 310

\bibitem[{{Senchyna} {et~al.}(2021){Senchyna}, {Stark}, {Charlot},
  {Chevallard}, {Bruzual}, \& {Vidal-Garc{\'\i}a}}]{senchyna_2021}
{Senchyna}, P., {Stark}, D.~P., {Charlot}, S., {et~al.} 2021, \mnras, 503, 6112

\bibitem[{{Senchyna} {et~al.}(2022){Senchyna}, {Stark}, {Charlot}, {Plat},
  {Chevallard}, {Chen}, {Jones}, {Sanders}, {Rudie}, {Cooper}, \&
  {Bruzual}}]{Senchyna_2022}
{Senchyna}, P., {Stark}, D.~P., {Charlot}, S., {et~al.} 2022, \apj, 930, 105

\bibitem[{{Senchyna} {et~al.}(2020){Senchyna}, {Stark}, {Mirocha}, {Reines},
  {Charlot}, {Jones}, \& {Mulchaey}}]{Senchyna_heii_2020}
{Senchyna}, P., {Stark}, D.~P., {Mirocha}, J., {et~al.} 2020, \mnras, 494, 941

\bibitem[{{Shapley} {et~al.}(2016){Shapley}, {Steidel}, {Strom},
  {Bogosavljevi{\'c}}, {Reddy}, {Siana}, {Mostardi}, \& {Rudie}}]{Shapley_2016}
{Shapley}, A.~E., {Steidel}, C.~C., {Strom}, A.~L., {et~al.} 2016, \apjl, 826,
  L24

\bibitem[{{Simmonds} {et~al.}(2021){Simmonds}, {Schaerer}, \&
  {Verhamme}}]{Simmonds_2021}
{Simmonds}, C., {Schaerer}, D., \& {Verhamme}, A. 2021, \aap, 656, A127

\bibitem[{{Stanway} \& {Eldridge}(2019)}]{Stanway_Eldridge_2019}
{Stanway}, E.~R. \& {Eldridge}, J.~J. 2019, \aap, 621, A105

\bibitem[{{Stark} {et~al.}(2015){Stark}, {Walth}, {Charlot}, {Cl{\'e}ment},
  {Feltre}, {Gutkin}, {Richard}, {Mainali}, {Robertson}, {Siana}, {Tang}, \&
  {Schenker}}]{Stark_2015}
{Stark}, D.~P., {Walth}, G., {Charlot}, S., {et~al.} 2015, \mnras, 454, 1393

\bibitem[{Stasińska {et~al.}(2015)Stasińska, Izotov, Morisset, \&
  Guseva}]{stasinska_excitation_2015}
Stasińska, G., Izotov, Y., Morisset, C., \& Guseva, N. 2015, A\&A, 576, A83

\bibitem[{{Steidel} {et~al.}(2018){Steidel}, {Bogosavljevi{\'c}}, {Shapley},
  {Reddy}, {Rudie}, {Pettini}, {Trainor}, \& {Strom}}]{Steidel_2018}
{Steidel}, C.~C., {Bogosavljevi{\'c}}, M., {Shapley}, A.~E., {et~al.} 2018,
  \apj, 869, 123

\bibitem[{{Sutherland} {et~al.}(2018){Sutherland}, {Dopita}, {Binette}, \&
  {Groves}}]{2018_mappingV}
{Sutherland}, R., {Dopita}, M., {Binette}, L., \& {Groves}, B. 2018, {MAPPINGS
  V: Astrophysical plasma modeling code}, Astrophysics Source Code Library,
  record ascl:1807.005

\bibitem[{{Thuan} {et~al.}(2004){Thuan}, {Bauer}, {Papaderos}, \&
  {Izotov}}]{Thuan_2004}
{Thuan}, T.~X., {Bauer}, F.~E., {Papaderos}, P., \& {Izotov}, Y.~I. 2004, \apj,
  606, 213

\bibitem[{{Trebitsch} {et~al.}(2017){Trebitsch}, {Blaizot}, {Rosdahl},
  {Devriendt}, \& {Slyz}}]{Trebitsch_2017}
{Trebitsch}, M., {Blaizot}, J., {Rosdahl}, J., {Devriendt}, J., \& {Slyz}, A.
  2017, \mnras, 470, 224

\bibitem[{{Ugryumov} {et~al.}(2003){Ugryumov}, {Engels}, {Pustilnik},
  {Kniazev}, {Pramskij}, \& {Hagen}}]{Ugryumov_2003}
{Ugryumov}, A.~V., {Engels}, D., {Pustilnik}, S.~A., {et~al.} 2003, \aap, 397,
  463

\bibitem[{{Umeda} {et~al.}(2022){Umeda}, {Ouchi}, {Nakajima}, {Isobe},
  {Aoyama}, {Harikane}, {Ono}, \& {Matsumoto}}]{Umeda_2022}
{Umeda}, H., {Ouchi}, M., {Nakajima}, K., {et~al.} 2022, \apj, 930, 37

\bibitem[{{Vale Asari} {et~al.}(2019){Vale Asari}, {Couto}, {Cid Fernandes},
  {Stasi{\'n}ska}, {de Amorim}, {Ruschel-Dutra}, {Werle}, \&
  {Florido}}]{Vale_Asari_2019}
{Vale Asari}, N., {Couto}, G.~S., {Cid Fernandes}, R., {et~al.} 2019, \mnras,
  489, 4721

\bibitem[{{Vale Asari} {et~al.}(2020){Vale Asari}, {Wild}, {de Amorim},
  {Werle}, {Zheng}, {Kennicutt}, {Johnson}, {Galametz}, {Pellegrini},
  {Klessen}, {Reissl}, {Glover}, \& {Rahner}}]{asari_importance_2020}
{Vale Asari}, N., {Wild}, V., {de Amorim}, A.~L., {et~al.} 2020, \mnras, 498,
  4205

\bibitem[{{Vallini} {et~al.}(2021){Vallini}, {Ferrara}, {Pallottini},
  {Carniani}, \& {Gallerani}}]{Vallini_2021}
{Vallini}, L., {Ferrara}, A., {Pallottini}, A., {Carniani}, S., \& {Gallerani},
  S. 2021, \mnras, 505, 5543

\bibitem[{{Vanzella} {et~al.}(2020){Vanzella}, {Caminha}, {Calura}, {Cupani},
  {Meneghetti}, {Castellano}, {Rosati}, {Mercurio}, {Sani}, {Grillo}, {Gilli},
  {Mignoli}, {Comastri}, {Nonino}, {Cristiani}, {Giavalisco}, \&
  {Caputi}}]{Vanzella_2020}
{Vanzella}, E., {Caminha}, G.~B., {Calura}, F., {et~al.} 2020, \mnras, 491,
  1093

\bibitem[{{Vanzella} {et~al.}(2015){Vanzella}, {de Barros}, {Castellano},
  {Grazian}, {Inoue}, {Schaerer}, {Guaita}, {Zamorani}, {Giavalisco}, {Siana},
  {Pentericci}, {Giallongo}, {Fontana}, \& {Vignali}}]{Vanzella_2015}
{Vanzella}, E., {de Barros}, S., {Castellano}, M., {et~al.} 2015, \aap, 576,
  A116

\bibitem[{{Vanzella} {et~al.}(2016){Vanzella}, {de Barros}, {Vasei}, {Alavi},
  {Giavalisco}, {Siana}, {Grazian}, {Hasinger}, {Suh}, {Cappelluti}, {Vito},
  {Amorin}, {Balestra}, {Brusa}, {Calura}, {Castellano}, {Comastri}, {Fontana},
  {Gilli}, {Mignoli}, {Pentericci}, {Vignali}, \& {Zamorani}}]{Vanzella_2016}
{Vanzella}, E., {de Barros}, S., {Vasei}, K., {et~al.} 2016, \apj, 825, 41

\bibitem[{{Vanzella} {et~al.}(2018){Vanzella}, {Nonino}, {Cupani},
  {Castellano}, {Sani}, {Mignoli}, {Calura}, {Meneghetti}, {Gilli}, {Comastri},
  {Mercurio}, {Caminha}, {Caputi}, {Rosati}, {Grillo}, {Cristiani}, {Balestra},
  {Fontana}, \& {Giavalisco}}]{Vanzella_2018}
{Vanzella}, E., {Nonino}, M., {Cupani}, G., {et~al.} 2018, \mnras, 476, L15

\bibitem[{{Verhamme} {et~al.}(2017){Verhamme}, {Orlitov{\'a}}, {Schaerer},
  {Izotov}, {Worseck}, {Thuan}, \& {Guseva}}]{Verhamme_2017}
{Verhamme}, A., {Orlitov{\'a}}, I., {Schaerer}, D., {et~al.} 2017, \aap, 597,
  A13

\bibitem[{Verhamme {et~al.}(2015)Verhamme, Orlitová, Schaerer, \&
  Hayes}]{verhamme_using_2015}
Verhamme, A., Orlitová, I., Schaerer, D., \& Hayes, M. 2015, A\&A, 578, A7

\bibitem[{{Walter} {et~al.}(2018){Walter}, {Riechers}, {Novak}, {Decarli},
  {Ferkinhoff}, {Venemans}, {Ba{\~n}ados}, {Bertoldi}, {Carilli}, {Fan},
  {Farina}, {Mazzucchelli}, {Neeleman}, {Rix}, {Strauss}, {Uzgil}, \&
  {Wang}}]{Walter_2018}
{Walter}, F., {Riechers}, D., {Novak}, M., {et~al.} 2018, \apjl, 869, L22

\bibitem[{{Wang} {et~al.}(2021){Wang}, {Heckman}, {Amor{\'\i}n}, {Borthakur},
  {Chisholm}, {Ferguson}, {Flury}, {Giavalisco}, {Grazian}, {Hayes}, {Henry},
  {Jaskot}, {Ji}, {Makan}, {McCandliss}, {Oey}, {{\"O}stlin}, {Saldana-Lopez},
  {Schaerer}, {Thuan}, {Worseck}, \& {Xu}}]{Wang_2021}
{Wang}, B., {Heckman}, T.~M., {Amor{\'\i}n}, R., {et~al.} 2021, \apj, 916, 3

\bibitem[{{Wang} {et~al.}(2019){Wang}, {Heckman}, {Leitherer}, {Alexandroff},
  {Borthakur}, \& {Overzier}}]{wang_new_2019}
{Wang}, B., {Heckman}, T.~M., {Leitherer}, C., {et~al.} 2019, \apj, 885, 57

\bibitem[{{Weilbacher} {et~al.}(2018){Weilbacher}, {Monreal-Ibero}, {Verhamme},
  {Sandin}, {Steinmetz}, {Kollatschny}, {Krajnovi{\'c}}, {Kamann}, {Roth},
  {Erroz-Ferrer}, {Marino}, {Maseda}, {Wendt}, {Bacon}, {Dreizler}, {Richard},
  \& {Wisotzki}}]{Weilbacher_2018}
{Weilbacher}, P.~M., {Monreal-Ibero}, A., {Verhamme}, A., {et~al.} 2018, \aap,
  611, A95

\bibitem[{{Weingartner} \& {Draine}(2001)}]{2001_smc_grains}
{Weingartner}, J.~C. \& {Draine}, B.~T. 2001, ApJ, 548, 296

\bibitem[{{Westmoquette} {et~al.}(2008){Westmoquette}, {Smith}, \&
  {Gallagher}}]{Westmoquette_2008}
{Westmoquette}, M.~S., {Smith}, L.~J., \& {Gallagher}, J.~S. 2008, \mnras, 383,
  864

\bibitem[{{Wofford} {et~al.}(2021){Wofford}, {Vidal-Garc{\'\i}a}, {Feltre},
  {Chevallard}, {Charlot}, {Stark}, {Herenz}, \& {Hayes}}]{wofford_2021}
{Wofford}, A., {Vidal-Garc{\'\i}a}, A., {Feltre}, A., {et~al.} 2021, \mnras,
  500, 2908

\bibitem[{{Wolfire} {et~al.}(2010){Wolfire}, {Hollenbach}, \&
  {McKee}}]{Wolfire_2010}
{Wolfire}, M.~G., {Hollenbach}, D., \& {McKee}, C.~F. 2010, \apj, 716, 1191

\bibitem[{{Wood} {et~al.}(2010){Wood}, {Hill}, {Joung}, {Mac Low}, {Benjamin},
  {Haffner}, {Reynolds}, \& {Madsen}}]{Wood_2010}
{Wood}, K., {Hill}, A.~S., {Joung}, M.~R., {et~al.} 2010, \apj, 721, 1397

\bibitem[{{Xiao} {et~al.}(2018){Xiao}, {Stanway}, \&
  {Eldridge}}]{2018_Xiao_bpass}
{Xiao}, L., {Stanway}, E.~R., \& {Eldridge}, J.~J. 2018, \mnras, 477, 904

\bibitem[{{Xu} {et~al.}(2022){Xu}, {Henry}, {Heckman}, {Chisholm}, {Worseck},
  {Gronke}, {Jaskot}, {McCandliss}, {Flury}, {Giavalisco}, {Ji}, {Amor{\'\i}n},
  {Berg}, {Borthakur}, {Bouche}, {Carr}, {Erb}, {Ferguson}, {Garel}, {Hayes},
  {Makan}, {Marques-Chaves}, {Rutkowski}, {{\"O}stlin}, {Rafelski},
  {Saldana-Lopez}, {Scarlata}, {Schaerer}, {Trebitsch}, {Tremonti}, {Verhamme},
  \& {Wang}}]{Xu_Xinfeng_2022}
{Xu}, X., {Henry}, A., {Heckman}, T., {et~al.} 2022, \apj, 933, 202

\bibitem[{{Yoo} {et~al.}(2020){Yoo}, {Kimm}, \& {Rosdahl}}]{yoo_origin_2020}
{Yoo}, T., {Kimm}, T., \& {Rosdahl}, J. 2020, \mnras, 499, 5175

\bibitem[{{Youngblood} \& {Hunter}(1999)}]{1999ApJ...519...55Y}
{Youngblood}, A.~J. \& {Hunter}, D.~A. 1999, \apj, 519, 55

\bibitem[{Zastrow {et~al.}(2013)Zastrow, Oey, Veilleux, \&
  McDonald}]{zastrow_new_2013}
Zastrow, J., Oey, M.~S., Veilleux, S., \& McDonald, M. 2013, ApJ, 779, 76

\bibitem[{Zastrow {et~al.}(2011)Zastrow, Oey, Veilleux, McDonald, \&
  Martin}]{zastrow_ionization_2011}
Zastrow, J., Oey, M.~S., Veilleux, S., McDonald, M., \& Martin, C.~L. 2011,
  ApJ, 741, L17

\bibitem[{{Zhang} {et~al.}(2017){Zhang}, {Yan}, {Bundy}, {Bershady}, {Haffner},
  {Walterbos}, {Maiolino}, {Tremonti}, {Thomas}, {Drory}, {Jones}, {Belfiore},
  {S{\'a}nchez}, {Diamond-Stanic}, {Bizyaev}, {Nitschelm}, {Andrews},
  {Brinkmann}, {Brownstein}, {Cheung}, {Li}, {Law}, {Roman Lopes}, {Oravetz},
  {Pan}, {Storchi Bergmann}, \& {Simmons}}]{Zhang_2017}
{Zhang}, K., {Yan}, R., {Bundy}, K., {et~al.} 2017, \mnras, 466, 3217

\bibitem[{{Zurita} {et~al.}(2002){Zurita}, {Beckman}, {Rozas}, \&
  {Ryder}}]{Zurita_2002}
{Zurita}, A., {Beckman}, J.~E., {Rozas}, M., \& {Ryder}, S. 2002, \aap, 386,
  801

\end{thebibliography}

\begin{appendix}

\section{Additional notes and figures}
\label{appendix}
In this section we gather additional figures that complement the main plots presented in this article. Those figures are mentioned throughout the article to illustrate some of the discussions but do not display essential results. We recall in the captions the corresponding sections to which they are related.

\begin{figure}[h!]
    \centering
    \includegraphics[width=9cm]{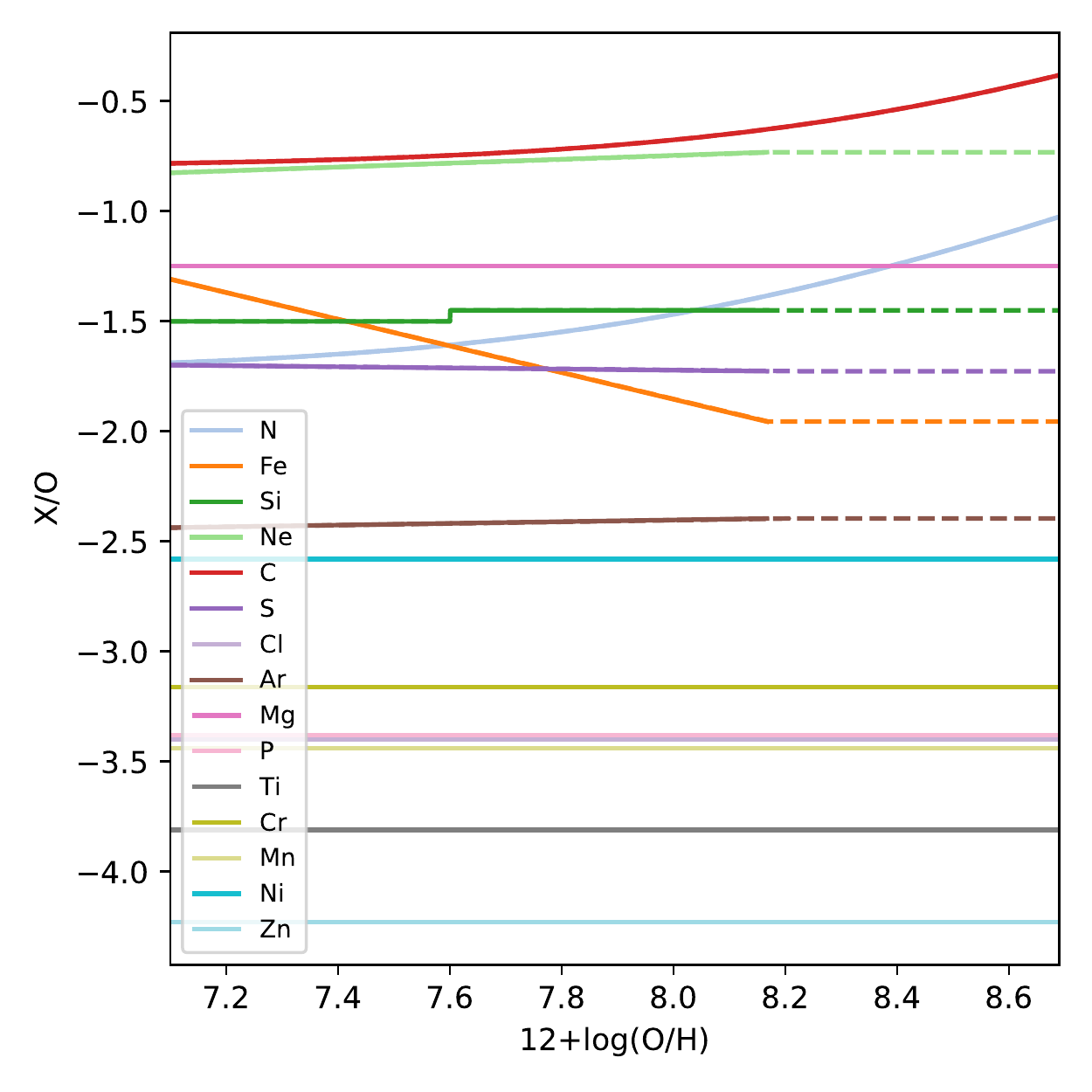}
    \caption{Abundance profile used to compute the grid of Cloudy models. This plot complements the description of the grid in Sect. \ref{section_models} and illustrate the prescription described in Table \ref{cloudy_observables}.}
    \label{abund_profile}
\end{figure}

\begin{figure}[h!]
    \centering
    \includegraphics[width=8cm]{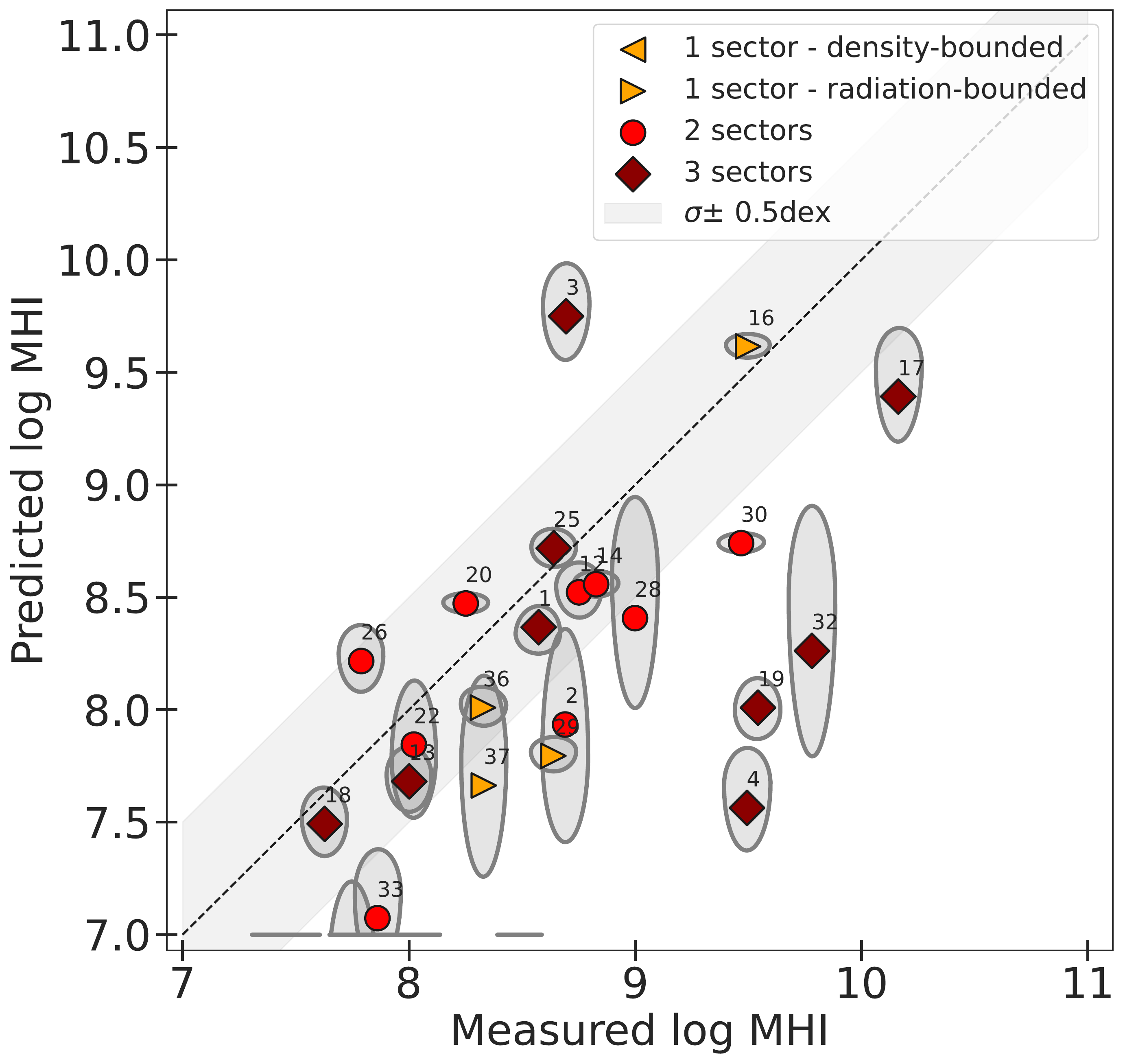}
    \caption{Predicted vs. measured \hi\ masses. As discussed in Sect. \ref{discussion_missing_gas}, the predicted \hi\ masses for the galaxies in our sample are consistent with the observed \hi\ masses from \cite{remy-ruyer_gas--dust_2014}. This agreement is at odds with the underprediction of \hi\ column densities (see Table \ref{Nh_in_dgs}) and is discussed in Sect. \ref{discussion}.}
    \label{MHI_comp}
\end{figure}

\begin{figure}[h!]
    \centering
    \includegraphics[width=7cm]{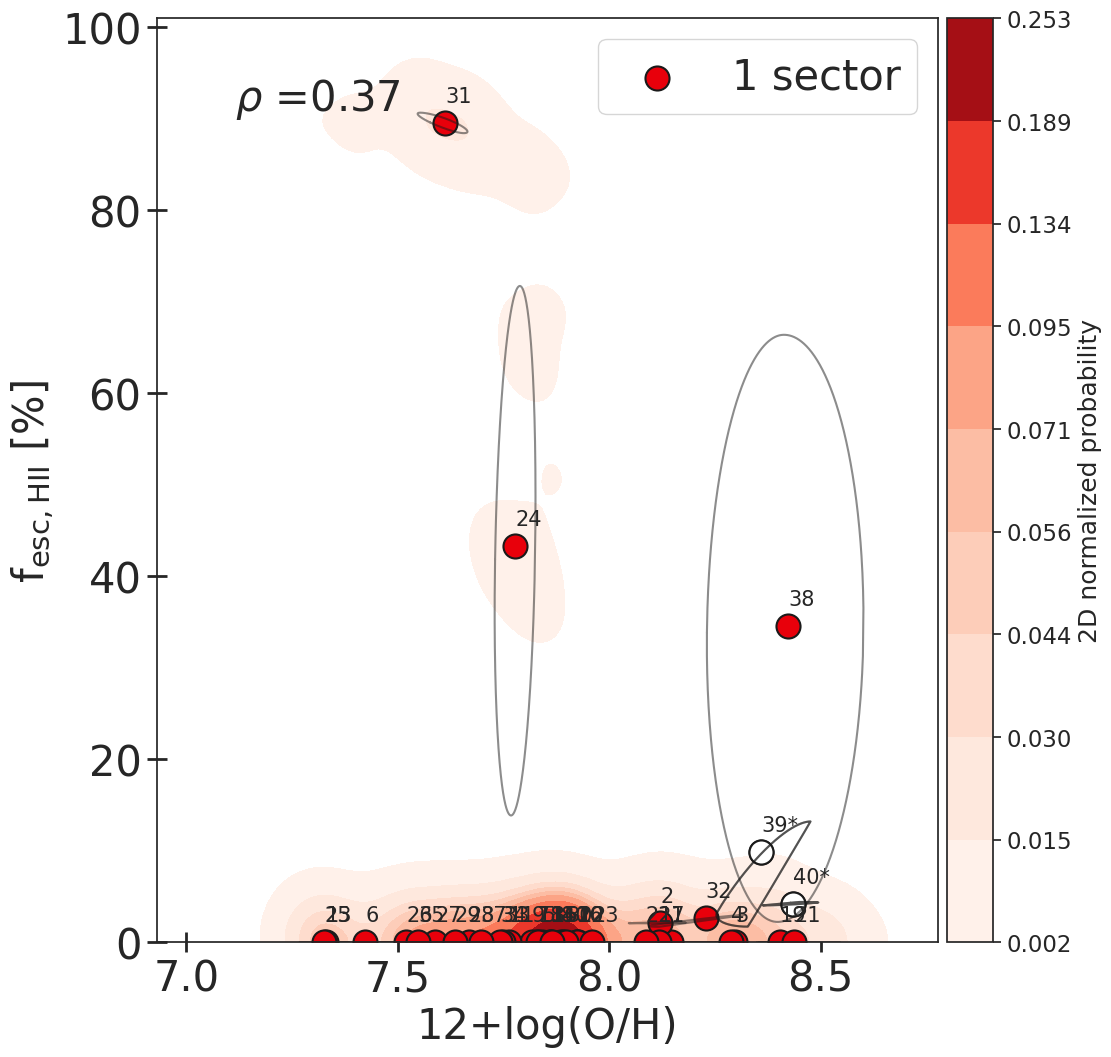}
    \includegraphics[width=7cm]{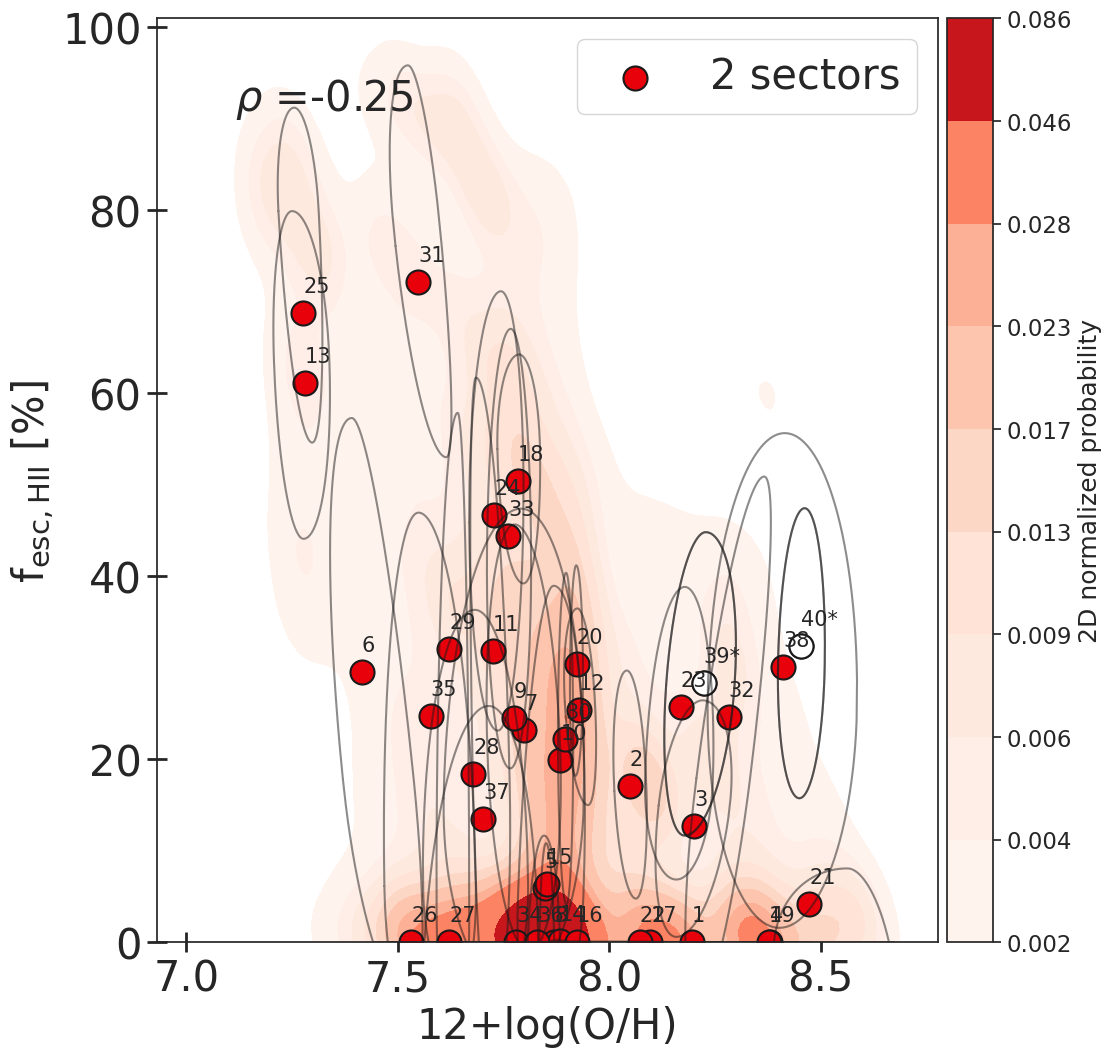}
    \includegraphics[width=7cm]{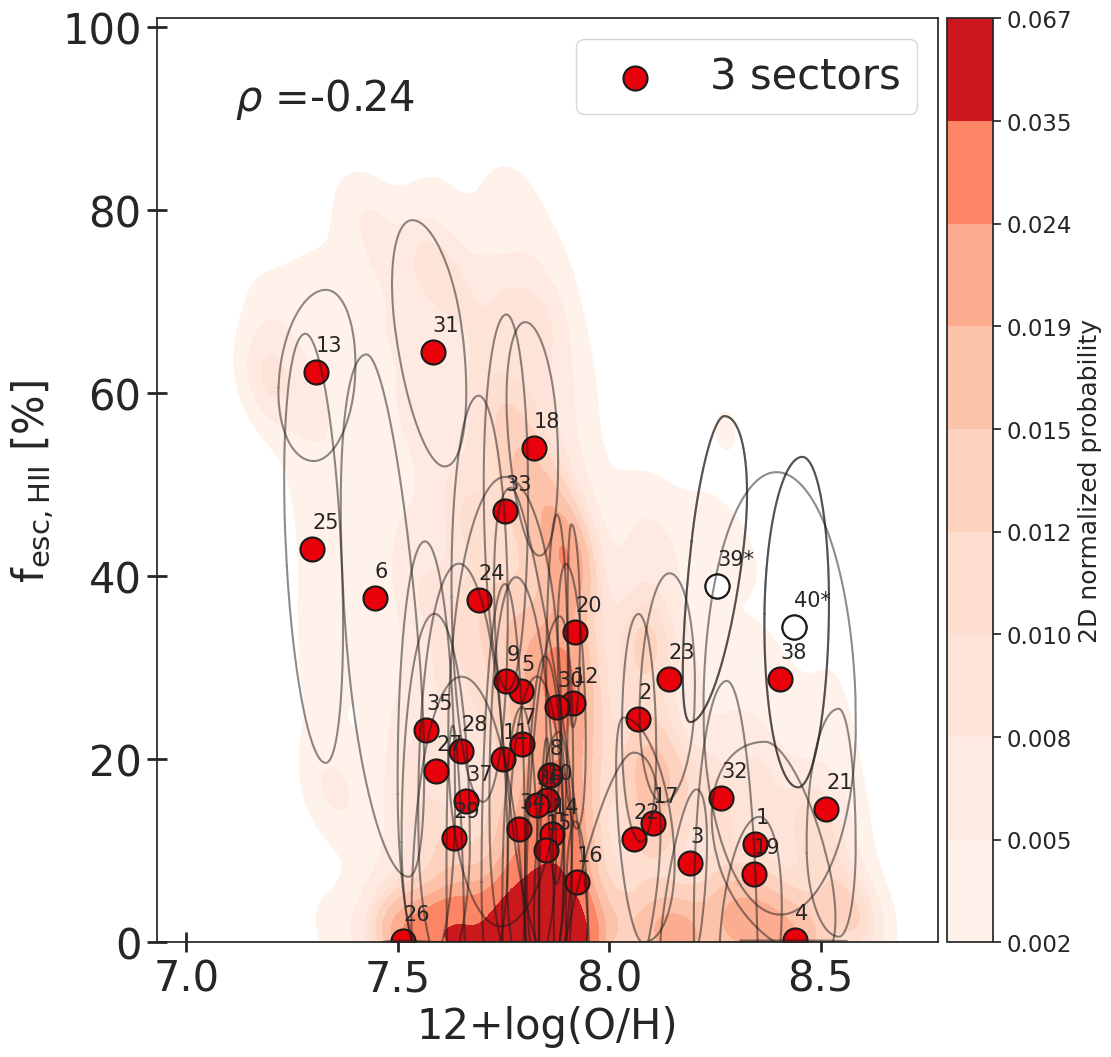}
    \caption{Metallicity-\feschii\  relation for a fixed number of sectors for all galaxies. These plots complement the Fig. \ref{fesc_oh} presented in Sect. \ref{section_fesc_oh} where the optimal number of sectors for a galaxy is chosen based on the score criteria presented in Sect. \ref{section_mgris}. We show here that the \feschii\-metallicity trend presented in Sect. \ref{section_fesc_oh} is robust and holds when we force an arbitrary number of sectors. We note that for single sector models, most solutions are completely radiation-bounded leading to escape fractions of 0\%.}
    \label{fesc_oh_nfixed}
\end{figure}

\begin{center}
\onecolumn
\begin{longtable}{|l|l|l|l|l|l|l|l|}

\caption{Best configuration selected for each galaxy and corresponding values of some indicators used to check model accuracy.} \label{table_lm} \\

\hline \multicolumn{1}{|c|}{\#} & \multicolumn{1}{c|}{Galaxy} & \multicolumn{1}{c|}{N$_{\rm sectors, best}$}& \multicolumn{1}{c|}{p-3$\sigma$ [\%]$^{1}$} & {$\mathcal{L}_{\rm M,best}$$^{2}$} & \multicolumn{1}{c|}{$\Delta \mathcal{L}_{M,1}^{3}$} & \multicolumn{1}{c|}{$\Delta \mathcal{L}_{M,2}^{3}$}& \multicolumn{1}{c|}{$\Delta \mathcal{L}_{M,3}^{3}$}  \\ \hline 
\endfirsthead

\multicolumn{8}{c}%
{{\bfseries \tablename\ \thetable{} -- continued}} \\
\hline \multicolumn{1}{|c|}{\#} & \multicolumn{1}{c|}{Galaxy} & \multicolumn{1}{c|}{N$_{\rm sectors, best}$}& \multicolumn{1}{c|}{p-3$\sigma$ [\%]$^{1}$} & {$\mathcal{L}_{\rm M,best}$$^{2}$} & \multicolumn{1}{c|}{$\Delta \mathcal{L}_{M,1}^{3}$} & \multicolumn{1}{c|}{$\Delta \mathcal{L}_{M,2}^{3}$}& \multicolumn{1}{c|}{$\Delta \mathcal{L}_{M,3}^{3}$}  \\ \hline 
\endhead

\hline
\endfoot

\hline \hline
\endlastfoot
1&Haro2&3.0&87.01&-45.49&-12.98&-4.41&0.0\\
2&Haro3&2.0&74.67&-60.51&-14.58&0.0&-1.28\\
3&Haro11&3.0&86.35&-51.95&-16.09&-1.32&0.0\\
4&He2-10&3.0&75.03&-93.15&-17.21&-2.59&0.0\\
5&HS0052+2536&2.0&86.21&-31.15&-0.17&0.0&-0.48\\
6&HS0822+3542&1.0&84.76&-13.19&0.0&-1.4&-3.62\\
7&HS1222+3741&1.0&88.61&-10.64&0.0&-0.96&-1.14\\
8&HS1304+3529&1.0&78.53&-18.49&0.0&-0.36&-0.75\\
9&HS1319+3224&3.0&94.67&-3.02&-1.25&-0.88&0.0\\
10&HS1330+3651&1.0&89.63&-24.35&0.0&-2.07&-0.64\\
11&HS1442+4250&3.0&74.18&-21.04&-0.23&-0.38&0.0\\
12&IIZw40&2.0&96.07&-39.4&-13.23&0.0&-1.6\\
13&IZw18&3.0&87.91&-31.75&-10.68&-2.24&0.0\\
14&Mrk153&2.0&97.54&-17.36&-1.42&0.0&-0.72\\
15&Mrk209&2.0&88.04&-24.83&-2.81&0.0&-0.64\\
16&Mrk930&1.0&98.09&-26.55&0.0&-0.67&-2.18\\
17&Mrk1089&3.0&80.25&-43.25&-10.14&-1.52&0.0\\
18&Mrk1450&3.0&87.57&-45.83&-9.31&-0.07&0.0\\
19&NGC1140&3.0&76.21&-73.47&-0.72&-3.14&0.0\\
20&NGC1569&2.0&74.77&-62.13&-5.98&0.0&-0.5\\
21&NGC1705&2.0&74.48&-51.53&-1.65&0.0&-1.45\\
22&NGC5253&2.0&76.91&-51.4&-4.36&0.0&-0.84\\
23&NGC625&3.0&74.99&-57.43&-4.6&-1.31&0.0\\
24&Pox186&3.0&92.09&-12.49&-3.46&-1.45&0.0\\
25&SBS0335-052&3.0&80.58&-34.49&-2.05&-0.34&0.0\\
26&SBS1159+545&2.0&78.39&-22.8&-0.14&0.0&-1.57\\
27&SBS1211+540&1.0&89.23&-16.74&0.0&-1.5&-1.94\\
28&SBS1249+493&2.0&89.72&-8.14&-2.76&0.0&-0.66\\
29&SBS1415+437&1.0&77.33&-36.14&0.0&-1.13&-0.01\\
30&SBS1533+574&2.0&77.95&-32.51&-0.83&0.0&-0.05\\
31&Tol1214-277&1.0&61.86&-21.46&0.0&-4.99&-4.56\\
32&UM448&3.0&75.5&-67.58&-2.34&-4.63&0.0\\
33&UM461&2.0&84.32&-40.5&-10.49&0.0&-0.47\\
34&VIIZw403&2.0&83.39&-34.71&-1.59&0.0&-0.05\\
35&UGC4483&2.0&75.98&-11.14&-0.81&0.0&-0.14\\
36&UM133&1.0&99.15&-2.47&0.0&-1.09&-0.99\\
37&HS0017+1055&1.0&96.12&-2.77&0.0&-0.05&-0.09\\
38&HS2352+2733&1.0&96.6&-1.08&0.0&-0.17&-0.27\\
39&NGC4214-c&3.0&90.66&-17.95&-9.85&-0.39&0.0\\
40&NGC4214-s&2.0&95.49&-11.49&-19.3&0.0&-2.45\\
\end{longtable}
\begin{tablenotes}
   \footnotesize
   \raggedright
   \item (1): Probability that the best model predictions are within 3$\sigma$ of the observations, for all the lines used as constraints.
   \item (2): Absolute value of the logarithm of the marginal likelihood for the best configuration corresponding to N$_{\rm sectors, best}$.
  \item (3): $\Delta \mathcal{L}_{M,\rm i}$ = ln($\mathcal{L}_{M, i}$/$\mathcal{L}_{M,\rm best}$)\\
\end{tablenotes}
\end{center}

\begin{sidewaysfigure*}[bt]
    \centering
    \includegraphics[height=0.55\textwidth]{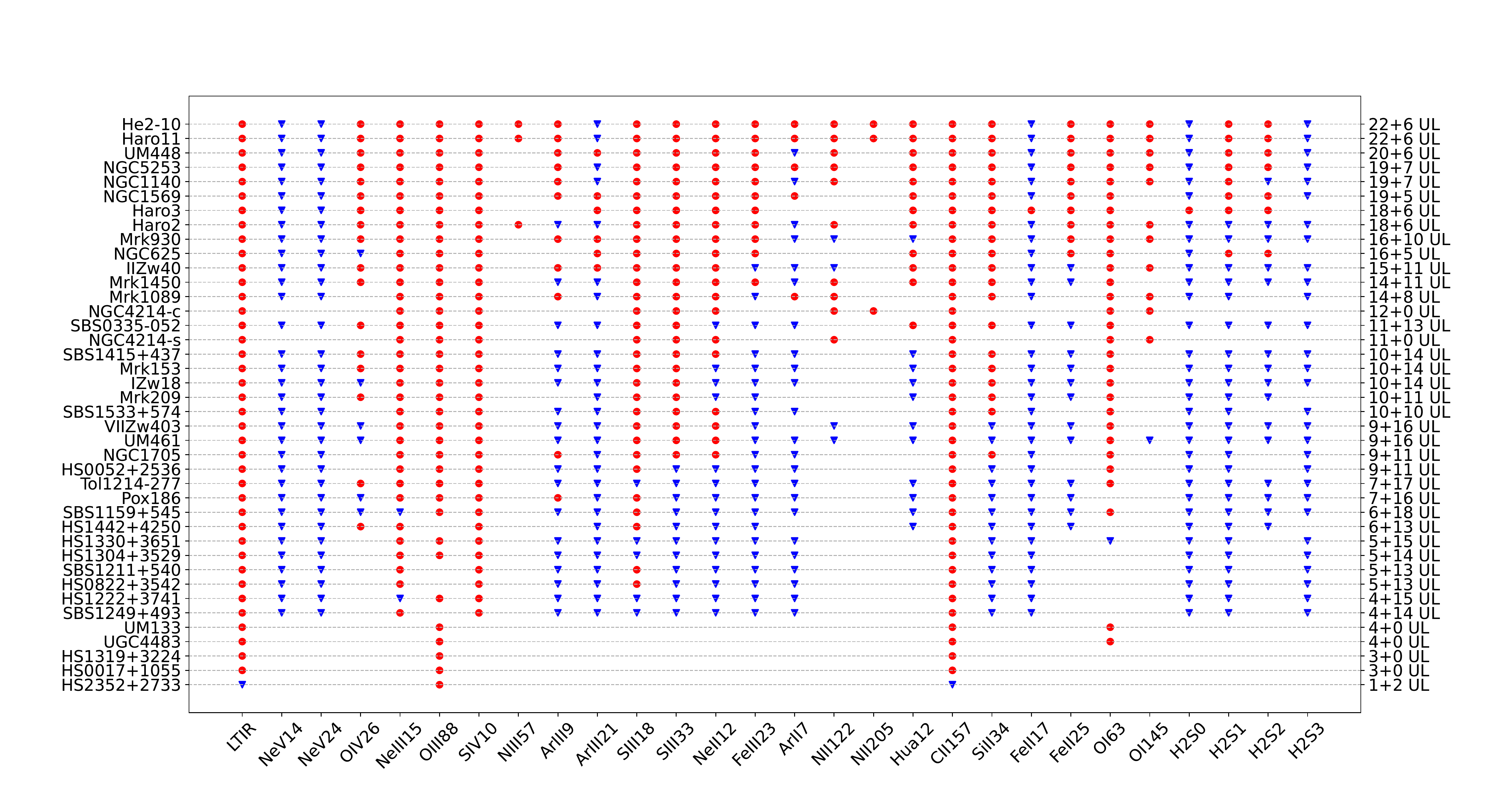}
    \caption{Detected lines (red dots) and upper limits (blue triangle) used in this paper. This figure illustrates the description of the available constraints for each galaxy in our sample that is presented in Sect. \ref{section_observations}. The total number of detections and upper limits used for each object are reported on the right-hand side y-axis.}
    \label{nlines}
\end{sidewaysfigure*}

\begin{figure*}[h!]
    \centering
    \includegraphics[width=16cm]{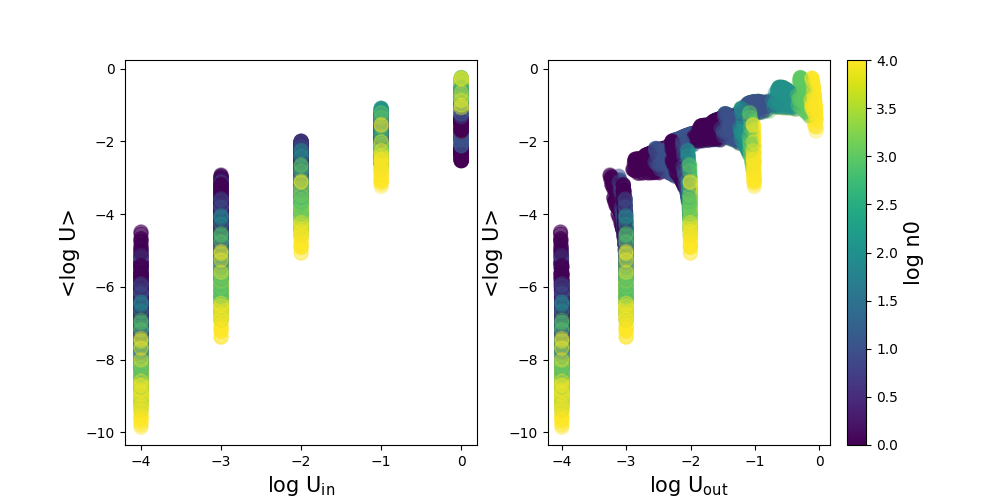}
    \includegraphics[width=16cm]{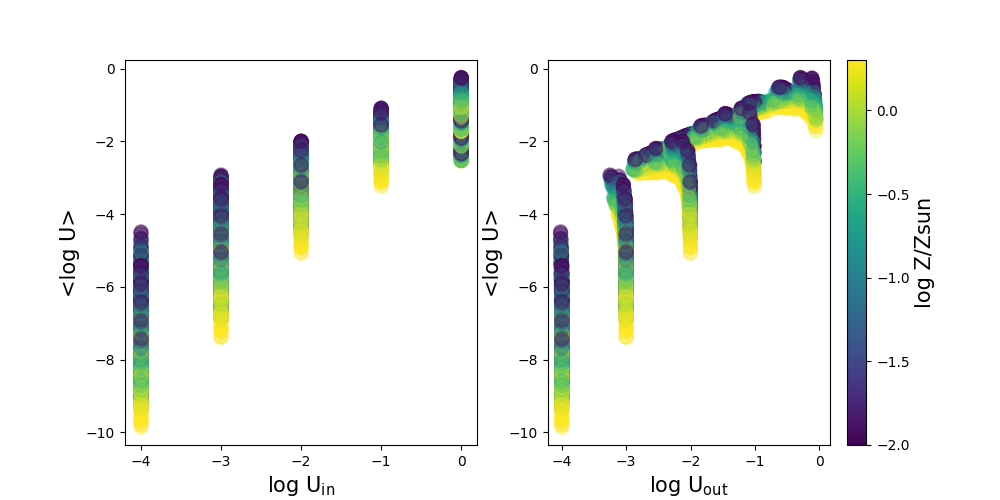}
    \caption{Evolution of the volume-averaged ionization parameter log <U> as a function log U$_{\rm in}$ and log U$_{\rm out}$ in the model grid. We show only a subsample corresponding to models with no X-ray source and stopping at the ionization front (cut=1). This figure complements the discussion on the possible caveats induced by the use of U$_{\rm in}$ as a primary parameter in our grid (see Sect. \ref{geom_of_hii}). We note that U$_{\rm in}$ only sets an upper limit on the actual volume averaged log <U>. Variations in terms of metallicity and initial density n$_0$ yield different geometries, which cover a large range of volume averaged log <U> for the same U$_{\rm in}$.}
    \label{logU_grid}
\end{figure*}

\begin{figure*}[h!]
    \centering
    \includegraphics[width=15cm]{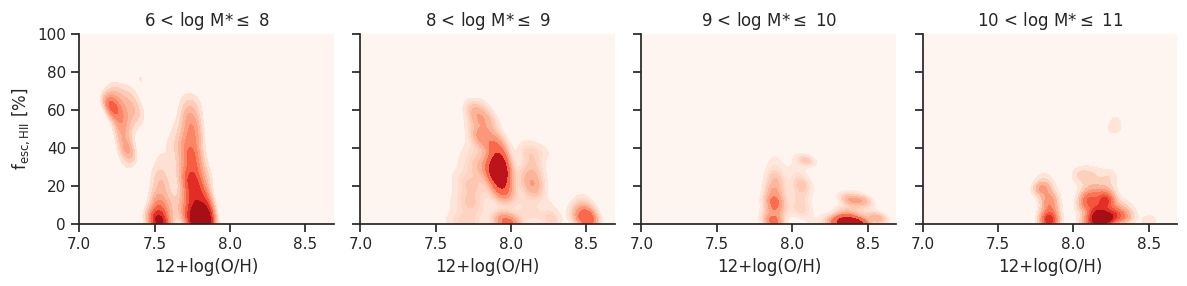}
    \includegraphics[width=15cm]{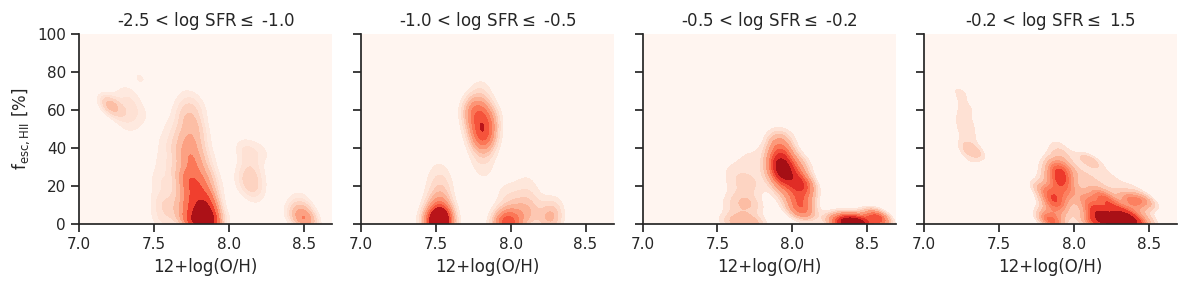}
    \includegraphics[width=15cm]{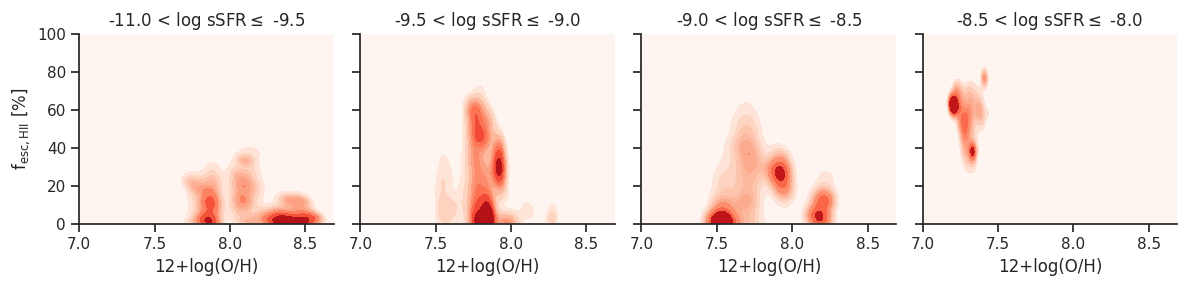}
    \includegraphics[width=15cm]{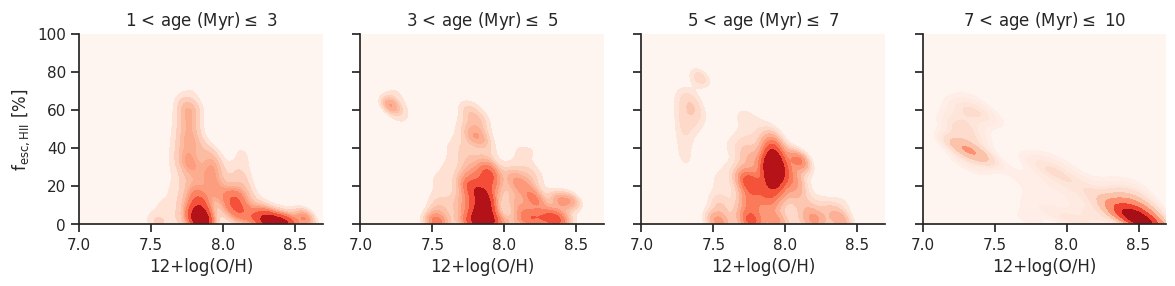}
    \includegraphics[width=15cm]{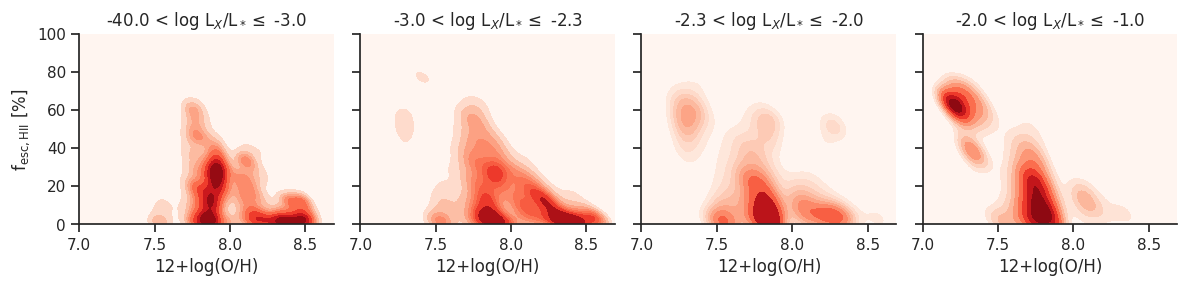}
    \includegraphics[width=15cm]{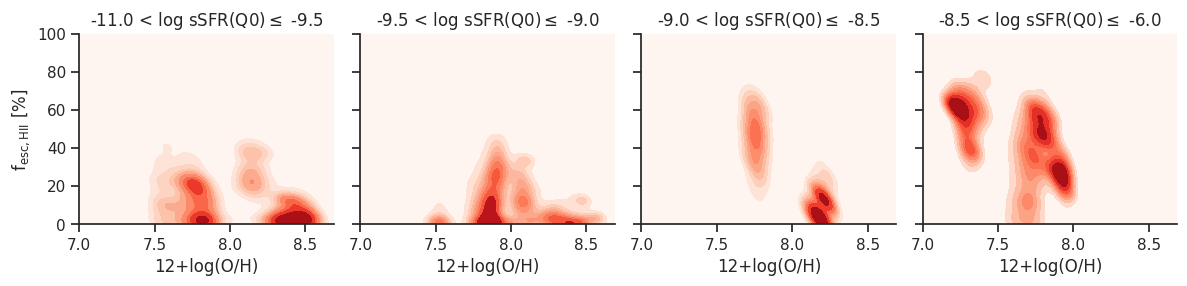}

    \caption{Impact of measured and inferred galactic parameters on the escape fraction vs metallicity relation. \textbf{Panels 1 to 3}:Impact of measured galactic parameters on the escape fraction vs metallicity relation:M$_{*}$, SFR, sSFR. \textbf{Panels 3 to 6}:Impact of inferred galactic parameters on the escape fraction vs metallicity relation: stellar age, Lx, sSFR(Q0). This figure complements the results presented in Sect. \ref{section_fesc_oh}. It provides a reinterpretation of Fig. \ref{fesc_oh} but accounting for the additional dependences on galactic parameters by splitting the sample by bins of values for secondary parameters. Those secondary parameters are further examined in Sect. \ref{section_dependencies}.}
    \label{dependencies_split_kde}
\end{figure*}

\clearpage
\begin{center}
\begin{longtable}{|l|l|l|l|}
\caption{Spearman correlation and values of the fit for the best IR line ratio tracing \feschii\ ($\rho \geq$0.3).} \label{fit_IR_lines} \\

\hline \multicolumn{1}{|c|}{\textbf{Line ratio}} & \multicolumn{1}{c|}{\textbf{$\rho$}} & \multicolumn{1}{c|}{\textbf{A}}& \multicolumn{1}{c|}{\textbf{B}} \\ \hline 
\endfirsthead

\multicolumn{4}{c}%
{{\bfseries \tablename\ \thetable{} -- continued}} \\
\hline \multicolumn{1}{|c|}{\textbf{Line ratio}} & \multicolumn{1}{c|}{\textbf{$\rho$}} & \multicolumn{1}{c|}{\textbf{A}}& \multicolumn{1}{c|}{\textbf{B}} \\ \hline 
\endhead

\hline
\endfoot

\hline \hline
\endlastfoot
H1\_12.3684m / Si2\_34.8046m&0.527&130.42&185.563\\
S4\_10.5076m / LTIR500.500m&0.519&149.899&456.264\\
H1\_12.3684m / Fe2\_25.9811m&0.486&98.124&30.459\\
H1\_12.3684m / LTIR500.500m&0.485&98.586&456.934\\
S4\_10.5076m / C2\_157.636m&0.482&354.963&-6.656\\
H1\_12.3684m / H2\_17.0300m&0.471&35.689&22.384\\
H1\_12.3684m / Fe2\_17.9314m&0.47&114.504&-61.946\\
\textbf{O3\_88.3323m / N2\_121.767m$^{(1)}$}&0.464&74.78&-163.043\\
H1\_12.3684m / O1\_63.1679m&0.464&672.149&1004.348\\
S4\_10.5076m / Si2\_34.8046m&0.464&183.689&-4.094\\
\textbf{O3\_88.3323m / N2\_205.244m$^{(1)}$}&0.461&70.567&-176.577\\
S4\_10.5076m / H2\_17.0300m&0.461&40.522&-39.822\\
H1\_12.3684m / N2\_121.767m&0.46&52.458&-3.56\\
H1\_12.3684m / C2\_157.636m&0.459&162.834&258.149\\
H1\_12.3684m / N2\_205.244m&0.449&50.224&-24.644\\
S3\_33.4704m / N2\_121.767m&0.448&88.792&-155.97\\
Ar3\_21.8253m / N2\_121.767m&0.447&100.214&9.157\\
H1\_12.3684m / Fe3\_22.9190m&0.446&115.686&29.741\\
S4\_10.5076m / N2\_121.767m&0.445&65.545&-110.542\\
S4\_10.5076m / N2\_205.244m&0.444&57.369&-108.528\\
S4\_10.5076m / H2\_28.2130m&0.442&26.491&-28.109\\
N3\_57.3238m / N2\_121.767m&0.442&122.575&-157.575\\
O4\_25.8832m / H2\_17.0300m&0.439&58.881&-4.892\\
O4\_25.8832m / N2\_205.244m&0.439&28.946&-3.069\\
Ne3\_15.5509m / N2\_121.767m&0.433&86.891&-156.961\\
S4\_10.5076m / Ne3\_15.5509m&0.433&390.741&12.354\\
Ar3\_8.98898m / N2\_121.767m&0.432&99.306&-104.741\\
\textbf{O4\_25.8832m / C2\_157.636m$^{(2)}$}&0.432&55.734&90.425\\
O4\_25.8832m / H2\_28.2130m&0.431&45.988&-12.765\\
O4\_25.8832m / N2\_121.767m&0.43&31.011&1.593\\
H1\_12.3684m / Ar3\_8.98898m&0.43&114.229&109.369\\
S3\_18.7078m / N2\_121.767m&0.428&85.685&-135.148\\
N3\_57.3238m / N2\_205.244m&0.428&99.587&-149.68\\
H1\_12.3684m / S3\_18.7078m&0.425&126.548&194.561\\
S3\_33.4704m / N2\_205.244m&0.424&76.793&-153.13\\
H1\_12.3684m / Ar3\_21.8253m&0.423&110.31&-18.305\\
Ne3\_15.5509m / N2\_205.244m&0.423&74.743&-152.017\\
O3\_88.3323m / N3\_57.3238m&0.422&297.947&-316.997\\
Ar3\_21.8253m / N2\_205.244m&0.419&83.176&-10.571\\
Fe3\_22.9190m / N2\_121.767m&0.418&90.861&-28.354\\
O3\_88.3323m / S3\_33.4704m&0.416&593.081&-270.111\\
S4\_10.5076m / Ar3\_8.98898m&0.414&206.794&-133.089\\
O4\_25.8832m / Si2\_34.8046m&0.41&45.209&78.238\\
S4\_10.5076m / S3\_18.7078m&0.41&320.263&-37.828\\
H1\_12.3684m / S3\_33.4704m&0.409&118.66&205.23\\
H1\_12.3684m / O1\_145.495m&0.408&-1864.77&-718.005\\
H2\_9.66228m / H2\_17.0300m&0.406&82.504&30.48\\
Ar3\_8.98898m / N2\_205.244m&0.406&84.64&-108.995\\
S4\_10.5076m / S3\_33.4704m&0.406&285.169&17.998\\
S3\_18.7078m / N2\_205.244m&0.406&73.539&-133.295\\
O3\_88.3323m / Ar2\_6.98337m&0.405&110.162&-273.012\\
H2\_12.2752m / H2\_17.0300m&0.405&101.622&69.297\\
H1\_12.3684m / Ne2\_12.8101m&0.405&60.832&47.995\\
S4\_10.5076m / O1\_63.1679m&0.404&-1237.19&-50.8\\
S4\_10.5076m / Fe2\_25.9811m&0.403&135.003&-158.851\\
O4\_25.8832m / Fe2\_25.9811m&0.403&41.974&22.719\\
S4\_10.5076m / Ar2\_6.98337m&0.403&91.98&-184.136\\
O3\_88.3323m / Si2\_34.8046m&0.403&227.883&-107.856\\
O4\_25.8832m / Fe2\_17.9314m&0.4&42.076&-9.965\\
S4\_10.5076m / Fe2\_17.9314m&0.4&187.512&-392.413\\
O3\_88.3323m / H2\_17.0300m&0.399&766.153&-1578.728\\
S4\_10.5076m / Ar3\_21.8253m&0.399&203.557&-364.099\\
S4\_10.5076m / N3\_57.3238m&0.399&164.224&-77.437\\
S4\_10.5076m / Fe3\_22.9190m&0.392&185.443&-234.334\\
O4\_25.8832m / Ar2\_6.98337m&0.392&36.116&-9.663\\
O3\_88.3323m / Ar3\_8.98898m&0.387&284.905&-318.54\\
S4\_10.5076m / Ne2\_12.8101m&0.385&78.214&-62.4\\
S3\_33.4704m / Ar2\_6.98337m&0.384&126.295&-249.556\\
Ne3\_15.5509m / H2\_17.0300m&0.38&-1741.288&2919.958\\
S3\_18.7078m / Ar2\_6.98337m&0.378&124.451&-226.135\\
O4\_25.8832m / H2\_12.2752m&0.377&244.608&-264.916\\
O3\_88.3323m / Ne2\_12.8101m&0.376&104.286&-149.435\\
O3\_88.3323m / Ar3\_21.8253m&0.376&293.241&-666.594\\
H1\_12.3684m / N3\_57.3238m&0.375&86.295&108.161\\
Fe3\_22.9190m / N2\_205.244m&0.374&83.949&-49.791\\
Ne3\_15.5509m / Ar2\_6.98337m&0.372&124.621&-254.015\\
O4\_25.8832m / Ne2\_12.8101m&0.37&34.232&31.129\\
S4\_10.5076m / O3\_88.3323m&0.368&532.903&278.691\\
Ne3\_15.5509m / Si2\_34.8046m&0.368&400.122&-32.15\\
O3\_88.3323m / Fe3\_22.9190m&0.368&310.557&-561.035\\
Ne5\_24.2065m / H2\_28.2130m&0.365&81.471&104.607\\
O4\_25.8832m / Fe3\_22.9190m&0.365&45.404&20.622\\
Ne3\_15.5509m / Ar3\_8.98898m&0.364&572.285&-415.788\\
Ne5\_24.2065m / H2\_17.0300m&0.363&86.667&152.264\\
Ne5\_14.3228m / H2\_28.2130m&0.362&77.986&107.966\\
O4\_25.8832m / O1\_63.1679m&0.362&82.694&120.192\\
Ne5\_14.3228m / H2\_17.0300m&0.36&80.987&151.057\\
O3\_88.3323m / LTIR500.500m&0.36&224.753&564.26\\
O3\_88.3323m / H2\_28.2130m&0.359&31.542&-53.397\\
\textbf{O3\_88.3323m / C2\_157.636m$^{(1)}$}&0.356&663.323&-302.962\\
S4\_10.5076m / H2\_12.2752m&0.355&-120.152&293.134\\
S3\_33.4704m / Ne2\_12.8101m&0.352&100.569&-78.356\\
O3\_88.3323m / Fe2\_17.9314m&0.351&212.062&-530.383\\
LTIR500.500m / N2\_121.767m&0.351&110.802&-517.864\\
Ne3\_15.5509m / H2\_28.2130m&0.35&32.395&-40.951\\
Ne3\_15.5509m / Ar3\_21.8253m&0.35&613.279&-1152.203\\
O4\_25.8832m / O1\_145.495m&0.349&96.075&35.9\\
S3\_18.7078m / Ne2\_12.8101m&0.347&100.916&-64.499\\
H2\_9.66228m / H2\_28.2130m&0.346&64.117&9.046\\
Ar3\_21.8253m / Ne2\_12.8101m&0.346&137.09&130.134\\
\textbf{Ne\_315.5509m / Ne2\_12.8101m$^{(2)}$}&0.344&104.82&-92.819\\
Ar3\_21.8253m / Ar3\_8.98898m&0.344&4298.962&5038.792\\
Ne3\_15.5509m / Fe2\_17.9314m&0.342&436.726&-948.463\\
O4\_25.8832m / Ar3\_8.98898m&0.34&45.886&51.343\\
O3\_88.3323m / S3\_18.7078m&0.337&591.496&-356.255\\
Ne3\_15.5509m / S3\_33.4704m&0.337&5021.612&-209.836\\
O4\_25.8832m / N3\_57.3238m&0.334&42.652&57.234\\
Ne3\_15.5509m / LTIR500.500m&0.334&385.81&1115.446\\
O4\_25.8832m / LTIR500.500m&0.334&47.755&219.676\\
Ar3\_8.98898m / Ne2\_12.8101m&0.334&124.763&-16.237\\
S4\_10.5076m / O1\_145.495m&0.333&-291.079&323.948\\
O4\_25.8832m / Ar3\_21.8253m&0.331&46.423&-1.869\\
Ne3\_15.5509m / Fe3\_22.9190m&0.33&496.239&-680.146\\
Ne3\_15.5509m / N3\_57.3238m&0.33&334.664&-182.823\\
O4\_25.8832m / H2\_9.66228m&0.328&490.167&-363.316\\
O4\_25.8832m / S3\_33.4704m&0.328&50.857&89.695\\
Ne5\_24.2065m / N2\_205.244m&0.326&33.235&58.843\\
Ne5\_24.2065m / N2\_205.244m&0.326&33.235&58.843\\
O4\_25.8832m / S3\_18.7078m&0.326&52.269&80.948\\
N3\_57.3238m / Ar2\_6.98337m&0.324&211.817&-319.099\\
Ne5\_14.3228m / N2\_205.244m&0.324&32.284&60.815\\
Ne5\_24.2065m / N2\_121.767m&0.322&35.963&69.431\\
Ne5\_14.3228m / N2\_121.767m&0.32&35.07&71.342\\
N3\_57.3238m / Ne2\_12.8101m&0.32&145.588&-44.665\\
O3\_88.3323m / Fe2\_25.9811m&0.318&171.32&-287.367\\
S3\_18.7078m / Ar3\_8.98898m&0.318&572.867&-293.951\\
Si2\_34.8046m / N2\_205.244m&0.316&75.291&-127.086\\
Si2\_34.8046m / N2\_121.767m&0.313&89.742&-135.462\\
H1\_12.3684m / Ne3\_15.5509m&0.312&131.48&227.804\\
H2\_12.2752m / H2\_28.2130m&0.312&77.74&32.316\\
Ne5\_24.2065m / Ar2\_6.98337m&0.311&38.156&71.118\\
Ne5\_24.2065m / H2\_12.2752m&0.311&266.128&297.313\\
S3\_18.7078m / Fe3\_22.9190m&0.311&541.475&-634.385\\
S3\_18.7078m / H2\_17.0300m&0.31&-15169.963&20996.747\\
Ne5\_14.3228m / Ar2\_6.98337m&0.308&37.532&73.659\\
S3\_18.7078m / Fe2\_17.9314m&0.307&357.808&-684.135\\
Ne5\_14.3228m / H2\_12.2752m&0.307&226.448&275.349\\
S3\_33.4704m / Fe3\_22.9190m&0.299&531.155&-699.26\\
Ne5\_24.2065m / Ne2\_12.8101m&0.299&34.38&107.357\\
Ne3\_15.5509m / S3\_18.7078m&0.298&-20086.043&4185.186\\
S3\_33.4704m / H2\_17.0300m&0.298&1494.699&-2272.294\\
\textbf{Ne5\_14.3228m / Ne2\_12.8101m$^{(2)}$}&0.297&33.683&108.606\\
Ne3\_15.5509m / C2\_157.636m&0.296&-1935.617&158.812\\
O4\_25.8832m / Ne3\_15.5509m&0.295&54.356&92.73\\
\hline
\textbf{Lbol / LTIR500.500m$^{(3)}$}&0.744&94.186&-7.617\\
Lbol / O2\_3726+9A&0.657&83.845&-199.221\\
Lbol / S2\_6716+30A&0.654&89.291&-289.073\\
Lbol / S3\_9069+532A&0.643&166.175&-442.796\\
Lbol / N2\_6548+84A&0.633&56.068&-182.48\\
Lbol / O1\_6300+63A&0.626&90.88&-388.582\\
Lbol / O2\_7320+30A&0.561&99.197&-416.007\\
He1\_4471.49A / LTIR500.500m&0.485&101.138&389.835\\
O3\_5007+4959A / LTIR500.500m&0.482&129.647&218.446\\
He2\_4685.64A / O1\_6300+63A&0.474&30.351&37.215\\
O3\_4363.21A / LTIR500.500m&0.474&89.804&330.476\\
H1\_6562.81A / LTIR500.500m&0.467&90.408&185.749\\
He1\_4471.49A / S2\_6716+30A&0.46&98.497&63.539\\
He1\_4471.49A / O1\_6300+63A&0.459&100.364&-38.5\\
He1\_4471.49A / O2\_3726+9A&0.456&85.291&133.629\\
He1\_7065.22A / LTIR500.500m&0.455&91.203&365.114\\
He2\_4685.64A / S2\_6716+30A&0.452&31.702&66.998\\
He1\_4471.49A / N2\_6548+84A&0.448&61.434&35.734\\
H1\_6562.81A / O1\_6300+63A&0.446&89.915&-196.585\\
He2\_4685.64A / N2\_6548+84A&0.445&26.832&54.478\\
H1\_6562.81A / S2\_6716+30A&0.441&76.735&-77.414\\
O3\_4363.21A / O1\_6300+63A&0.441&78.211&-35.368\\
H1\_6562.81A / O2\_3726+9A&0.439&73.687&-12.733\\
O3\_4363.21A / N2\_6548+84A&0.435&56.741&23.599\\
H1\_6562.81A / N2\_6548+84A&0.433&55.108&-64.184\\
O3\_5007+4959A / N2\_6548+84A&0.432&74.729&-121.723\\
He2\_4685.64A / O2\_3726+9A&0.431&31.665&92.547\\
He1\_7065.22A / O1\_6300+63A&0.431&84.558&-15.205\\
S3\_9069+532A / N2\_6548+84A&0.43&90.149&-58.788\\
O3\_4363.21A / S2\_6716+30A&0.43&104.103&26.316\\
He1\_7065.22A / N2\_6548+84A&0.43&53.519&44.727\\
He1\_7065.22A / S2\_6716+30A&0.429&85.351&71.304\\
O3\_5007+4959A / S2\_6716+30A&0.428&130.992&-201.584\\
O3\_4363.21A / O2\_3726+9A&0.428&83.402&108.7\\
O3\_5007+4959A / O1\_6300+63A&0.428&122.712&-307.378\\
He1\_7065.22A / O2\_3726+9A&0.428&74.339&130.894\\
O3\_4363.21A / O2\_7320+30A&0.426&94.94&-46.28\\
\textbf{O3\_5007+4959A / O2\_3726+9A$^{(3)}$}&0.425&116.705&-78.682\\
He2\_4685.64A / O2\_7320+30A&0.417&34.159&36.552\\
He1\_4471.49A / S3\_9069+532A&0.417&165.225&216.826\\
O2\_3726+9A / N2\_6548+84A&0.415&201.998&-192.16\\
O2\_7320+30A / N2\_6548+84A&0.414&162.477&132.045\\
O3\_5007+4959A / S3\_9069+532A&0.41&371.586&-352.117\\
He2\_4685.64A / LTIR500.500m&0.409&33.221&177.064\\
O3\_4363.21A / S3\_9069+532A&0.406&144.841&156.083\\
He1\_4471.49A / O2\_7320+30A&0.404&112.029&-6.214\\
\textbf{S3\_9069+532A / S2\_6716+30A$^{(3)}$}&0.402&211.532&-128.321\\
S3\_9069+532A / O2\_3726+9A&0.396&185.958&44.773\\
He1\_7065.22A / O2\_7320+30A&0.392&89.482&-7.637\\
He1\_7065.22A / S3\_9069+532A&0.39&121.28&187.334\\
H1\_6562.81A / S3\_9069+532A&0.39&136.392&-66.812\\
O3\_5007+4959A / O2\_7320+30A&0.385&154.476&-389.291\\
O3\_4363.21A / O3\_5006.84A&0.382&254.49&482.617\\
O3\_4363.21A / O3\_5007+4959A&0.382&249.591&505.516\\
He2\_4685.64A / S3\_9069+532A&0.378&36.926&103.908\\
H1\_6562.81A / O2\_7320+30A&0.374&98.921&-209.783\\
S2\_6716+30A / N2\_6548+84A&0.373&157.312&-5.541\\
O2\_7320+30A / S2\_6716+30A&0.347&840.02&810.936\\
O2\_7320+30A / O2\_3726+9A&0.343&437.593&811.678\\
LTIR500.500m / N2\_6548+84A&0.326&151.553&-489.71\\
S3\_9069+532A / O1\_6300+63A&0.319&199.35&-316.807\\
O3\_4363.21A / He1\_4471.49A&0.306&628.672&-88.671\\
He2\_4685.64A / He1\_4471.49A&0.305&51.199&68.665\\
He2\_4685.64A / H1\_6562.81A&0.304&54.035&170.776\\
\hline
\end{longtable}

\end{center}

\begin{tablenotes}
   \footnotesize
   \raggedright
   \item The fits are of the form Y = AX + B where Y = \feschii\ [\%] and X = log$_{10}$(R) for a given line ratio R. 
   \item (1): See Figs. \ref{fesc_ratio_alma_kde} and \ref{fesc_ratio_alma_sue}.
   \item (2): See Figs. \ref{fesc_ratio_highIP_kde} and \ref{fesc_ratio_highIP_sue}.
   \item (3): See Figs. \ref{fesc_opt_kde} and \ref{fesc_opt_sue}.\\
\end{tablenotes}

\clearpage
\begin{center}

\begin{longtable}{|l|l|l|l|l|l|l|}

\caption{Table comparing the predicted \hi\ column densities and \hi\ gas masses to values from the literature.}
\label{Nh_in_dgs} \\

\hline \multicolumn{1}{|c|}{\#} & \multicolumn{1}{c|}{Galaxy} & \multicolumn{1}{c|}{log MH~{\sc i}$_{\rm pred}^{(0)}$}& \multicolumn{1}{c|}{log MH~{\sc i}$_{\rm obs}^{(1)}$} & {log N(H~{\sc i})$_{\rm max, pred}$} & \multicolumn{1}{c|}{log N(H~{\sc i})$_{\rm min, pred}$} & \multicolumn{1}{c|}{log N(H~{\sc i})$_{\rm obs}$} \\ \hline 
\endfirsthead

\multicolumn{7}{c}%
{{\bfseries \tablename\ \thetable{} -- continued}} \\
\hline \multicolumn{1}{|c|}{\#} & \multicolumn{1}{c|}{Galaxy} & \multicolumn{1}{c|}{log MH~{\sc i}$_{\rm pred}^{(0)}$}& \multicolumn{1}{c|}{log MH~{\sc i}$_{\rm obs}^{(1)}$} & {log N(H~{\sc i})$_{\rm max, pred}$} & \multicolumn{1}{c|}{log N(H~{\sc i})$_{\rm min, pred}$} & \multicolumn{1}{c|}{log N(H~{\sc i})$_{\rm obs}$} \\ \hline 
\endhead

\hline
\endfoot

\hline \hline
\endlastfoot

1&Haro2&8.37$_{-0.07}^{+0.09}$&8.58$_{-0.08}^{+0.06}$&19.47&18.79&19.6--20.5$^{2}$\\
2&Haro3&7.93$_{-0.46}^{+0.70}$&9.05$_{-0.01}^{+0.01}$&18.92&18.29& \\
3&Haro11&9.75$_{-0.15}^{+0.21}$&8.70$_{-0.30}^{+0.18}$&19.08&18.73& \\
4&He2-10&7.56$_{-0.33}^{+0.18}$&8.49$_{-0.03}^{+0.03}$&19.33&18.74& \\
5&HS0052+2536&9.79$_{-0.06}^{+0.06}$&<10.68&20.0&18.54& \\
6&HS0822+3542&4.94$_{-1.71}^{+0.84}$&7.75$_{-0.11}^{+0.09}$&21.67&21.67& \\
7&HS1222+3741&8.39$_{-0.35}^{+0.17}$&&20.16&20.16& \\
8&HS1304+3529&8.83$_{-0.07}^{+0.08}$&&20.42&20.42& \\
9&HS1319+3224&6.4$_{-1.25}^{+1.07}$&&18.46&17.86& \\
10&HS1330+3651&8.43$_{-1.05}^{+0.46}$&&20.37&20.37& \\
11&HS1442+4250&5.48$_{-0.56}^{+0.59}$&8.49$_{-0.01}^{+0.01}$&20.84&19.29& \\
12&IIZw40&8.52$_{-0.11}^{+0.11}$&8.75$_{-0.08}^{+0.06}$&19.18&18.96& \\
13&IZw18&7.68$_{-0.11}^{+0.13}$&8.0$_{-0.08}^{+0.06}$&20.27&19.12&21.0–-21.5$^{2}$; 21.28$\pm$0.03$^{3}$\\
14&Mrk153&8.56$_{-0.06}^{+0.06}$&<8.84&19.61&18.4& \\
15&Mrk209&6.87$_{-0.10}^{+0.08}$&7.44$_{-0.03}^{+0.03}$&19.87&18.59& \\
16&Mrk930&9.62$_{-0.06}^{+0.04}$&9.5$_{-0.05}^{+0.05}$&19.54&19.54& \\
17&Mrk1089&9.39$_{-0.26}^{+0.21}$&10.17$_{-0.04}^{+0.03}$&18.96&18.33& \\
18&Mrk1450&7.49$_{-0.17}^{+0.15}$&7.63$_{-0.08}^{+0.06}$&19.27&18.81& \\
19&NGC1140&8.01$_{-0.14}^{+0.15}$&9.54$_{-0.13}^{+0.10}$&18.43&18.28& \\
20&NGC1569&8.47$_{-0.04}^{+0.04}$&8.25$_{-0.08}^{+0.06}$&19.26&18.98& \\
21&NGC1705&6.63$_{-0.09}^{+0.12}$&7.88$_{-0.06}^{+0.05}$&19.1&17.97& \\
22&NGC5253&7.85$_{-0.30}^{+0.35}$&8.03$_{-0.03}^{+0.02}$&18.83&18.79& \\
23&NGC625&5.46$_{-0.63}^{+0.66}$&8.04$_{-0.09}^{+0.07}$&19.16&18.72& \\
24&Pox186&5.3$_{-1.71}^{+1.14}$&<6.30&19.75&19.3& \\
25&SBS0335-052&8.72$_{-0.09}^{+0.09}$&8.64$_{-0.05}^{+0.04}$&21.29&18.85&21.4–-21.7$^{2}$; 21.7$\pm$0.05$^{3}$\\
26&SBS1159+545&8.22$_{-0.15}^{+0.11}$&<7.80&21.04&20.13& \\
27&SBS1211+540&6.69$_{-0.51}^{+0.40}$&7.75$_{-0.08}^{+0.06}$&20.56&20.56& \\
28&SBS1249+493&8.41$_{-0.45}^{+0.37}$&9.0$_{-0.29}^{+0.17}$&20.11&19.65& \\
29&SBS1415+437&7.79$_{-0.10}^{+0.07}$&8.64$_{-0.05}^{+0.04}$&20.8&20.8& \\
30&SBS1533+574&8.74$_{-0.04}^{+0.04}$&9.48$_{-0.05}^{+0.05}$&19.57&19.08& \\
31&Tol1214-277&2.41$_{-0.23}^{+0.73}$&<8.51&19.93&19.93& \\
32&UM448&8.26$_{-1.05}^{+0.31}$&9.78$_{-0.14}^{+0.11}$&19.26&18.68& \\
33&UM461&7.07$_{-0.30}^{+0.24}$&7.86$_{-0.02}^{+0.02}$&20.01&19.48& \\
34&VIIZw403&6.06$_{-0.35}^{+0.33}$&7.51$_{-0.04}^{+0.03}$&19.88&19.61& \\
35&UGC4483&6.0$_{-0.51}^{+0.38}$&7.40$_{-0.02}^{+0.02}$&20.44&19.42& \\
36&UM133&8.01$_{-0.09}^{+0.08}$&8.33$_{-0.02}^{+0.02}$&20.38&20.38& \\
37&HS0017+1055&7.66$_{-0.41}^{+0.39}$&<8.24&20.43&20.43& \\
38&HS2352+2733&4.53$_{-0.78}^{+0.70}$&<10.36&18.77&18.77& \\
39&NGC4214-c&7.17$_{-0.42}^{+0.36}$&<8.58&18.95&17.43&21.12$\pm$0.03$^{3}$ \\
40&NGC4214-s&6.61$_{-0.13}^{+0.11}$&<8.58&19.02&18.34&21.12$\pm$0.03$^{3}$ \\
\end{longtable}
\end{center}

\begin{tablenotes}
   \footnotesize
   \item  This table complements the discussion from Sect. \ref{discussion_missing_gas} and the \hi\ gas mass comparison presented in Fig. \ref{MHI_comp}. 
   \item (0): The interval is the highest density interval at 94\%.
   \item (1): \cite{remy-ruyer_gas--dust_2014}
   \item (2): \cite{Kunth_1998}
   \item (3): \cite{James_2014}
\end{tablenotes}

\end{appendix}
\end{document}